\documentclass[twoside,a4paper,12pt]{mybook}
\usepackage{fancyheadings}
\usepackage{times}
\usepackage{amsmath}
\usepackage{amsfonts}
\usepackage{amssymb,amscd}
\usepackage[dvips]{epsfig}
\usepackage{delarray}
\usepackage[german,english]{babel}

\renewcommand{\chaptermark}[1]{\markboth{#1}{}}

\newcommand{\T}{\textstyle}
\newcommand{\Dc}{\displaystyle}
\newcommand{\Sc}{\scriptstyle}
\newcommand{\be}{\begin{equation}}
\newcommand{\ee}{\end{equation}}
\newcommand{\bea}{\begin{eqnarray}}
\newcommand{\eea}{\end{eqnarray}}
\newcommand{\bp}{\begin{pmatrix}}
\newcommand{\ep}{\end{pmatrix}}
\newcommand{\half}{{\textstyle\frac{1}{2}}}
\newcommand{\third}{{\textstyle\frac{1}{3}}}
\newcommand{\quarter}{{\textstyle\frac{1}{4}}}
\newcommand{\Slash}[1]{#1 \hspace{-.5em}/\hspace{.11em}} 
\newcommand{\fourint}[1]{\int\!\frac{d^4 #1}{(2\pi)^4}}
\newcommand{\fourinta}[1]{\int\! d^4 #1}

\newcommand{\vect}[1]{{\mbox{\boldmath $#1$}}}
\newcommand{\dpartial}[1]{\frac{\partial}{\partial #1}}
\newcommand{\dpart}[2]{\frac{\partial #1}{\partial #2}}
\newcommand{\trisp}[1]{#1 _{\alpha\beta\gamma}}
\newcommand{\twosp}[1]{#1 _{\beta\gamma}}
\newcommand{\cdt}{\!\cdot\!}

\begin{document}
\selectlanguage{english}
\newlength{\figurewidth}
\setlength{\figurewidth}{6.8cm}

\thispagestyle{empty}
\vspace*{4cm}
\begin{center}{\Huge \bf Baryons as 
Relativistic Bound States \\[3mm] of Quark and Diquark}
\end{center}

\vspace*{4cm}

\centerline{\large \scshape Dissertation}
\begin{center}
T\"ubingen University
\end{center}

\vspace*{4cm}

\begin{center} 
{\large \scshape Martin Oettel} 
\end{center}

\newpage

\thispagestyle{empty}





\tableofcontents
\contentsline{chapter}{\numberline{}Bibliography}{149}
\markboth{Contents}{Contents}
 \chapter{Introduction}
 \label{model-chap}
  
\section{General remarks}

Up to now, no convincing solution for the problem
of describing baryons and mesons in terms of a few
underlying fields has been found, yet it is widely believed
that quarks and gluons are the basic entities which build
the hadrons. The associated theory is called 
Quantum Chromodynamics (QCD). The complexity of this theory in the low
energy sector is prohibitive for a straightforward deduction of hadronic 
properties from it. Therefore, models with a simplified
dynamics have been employed over the last 
decades which aim at describing hadrons 
with degrees of freedom borrowed from QCD (like quarks) or from
the observable particle spectrum (like pions) that
might be better suited in the treatment of hadrons.

The aim of the present thesis is to study a relativistic framework 
for the description of baryons employing quark and diquark degrees 
of freedom. Actual calculations are performed with
rather simple assumptions about these, but it is hoped that
further progress in knowledge about QCD's quark propagator
and the structure of 2-quark correlations will clarify
whether the reduction to quarks and diquarks in the treatment 
of baryons is compatible with QCD.

We refrain from discussing QCD here.
The relevance of QCD and its (approximate) symmetries
for hadronic physics is discussed in many good textbooks
on quantum field theory, see {\em e.g.} 
refs.~\cite{Cheng:1985bj,Peskin:1995ev}. 

In the remaining part of this Introduction we will give
a short description of the basic ideas underlying three classes of 
hadronic models which have become popular (and have remained so) 
over the last years. This brief survey is by no means
meant to be exhausting, and serves only as a motivation to consider
a fully covariant approach like the diquark-quark model.
A short overview about the topics which are covered in the
subsequent chapters is given in the last subsection of this chapter. 

\section{Modelling Baryons}

\subsection{Constituent quark models}

In the early sixties, it has become recognized that the eight
spin-1/2 baryons today referred to as ``octet baryons'' belong
to an eight-dimensional tensor representation of the unitary
group $SU(3)$. Obviously this unitary symmetry is
broken, as can be seen from the mass differences within the
baryon octet, nevertheless a couple of useful relations 
among {\em e.g.} the octet masses and magnetic moments
could be deduced from unitary
symmetry \cite{Gell-Mann:1964} that appeared to be in 
fair agreement with experiment.

Quarks with their denominations $up$, $down$ and $strange$
as building blocks of hadrons have been introduced
somewhat later as the fundamental representation of the flavor group
$SU(3)$. According to this hypothesis, the baryon octet is
an irreducible representation which arises from the decomposition
of the tensor representation formed by three quarks. Another
irreducible representation, the decuplet of spin-3/2 baryons, 
has been identified
in experiments in subsequent years.

To match the baryon quantum numbers, the quarks had to be assigned 
spin 1/2.  Spin and
flavor may be combined to the larger group $SU(6)$ and then octet
and decuplet baryons can be interpreted as a 56-dimensional 
representation of this group, with a symmetric spin-flavor
wave function. However, the overall wave function needs to be
antisymmetric, and if one is not willing to sacrifice  
the {\em spatial} symmetry of a baryon ground state wave function,
one is led to the hypothesis that all quarks appear in three
{\em colors}. Baryons are then fully antisymmetric {\em color singlets}.

Still, these group-theoretical considerations do not encompass
a dynamical model for baryons in terms of quarks. To explain 
the plethora of baryon resonances, one of the  very first models
\cite{Faiman:1968js} aims at describing baryons by solutions
of a 3-particle Schr{\"o}dinger equation using the non-relativistic 
Hamiltonian
\begin{eqnarray}
  H&=&H_{\rm kin} + H_{\rm conf}\; ,  \\
 & & H_{\rm kin}=\sum_{i=1}^3 \frac{\vect p_i^2}{2m_q}\; , \qquad
  H_{\rm conf}=\frac{1}{2}m_q \omega^2\sum_{\substack{i,j=1\\i<j}}^3
  (\vect r_i-\vect r_j)^2 \; .
\end{eqnarray}
Massive quarks with $m_q \sim$ 0.3 GeV move in a harmonic oscillator
potential which ($i$) confines the quarks, ($ii$) allows 
for the separation of variables easily and ($iii$) leads to a rich
spectrum of excitations, even far more than observed. 
However, the splitting between octet and decuplet has to be provided
by some sort of spin-dependent interaction, from the theory of atomic
spectra usually known as {\em hyperfine} interaction. As meanwhile
QCD became favored by many theorists to be the theory accounting
for the strong interactions, a tensorial and a 
spin-spin interaction term in the 
Hamiltonian could be motivated by a non-relativistic truncation
of one-gluon exchange between quarks 
\cite{DeRujula:1975ge,Isgur:1978xj},
\begin{eqnarray}
  H_{\rm spin}&=&\sum_{\substack{i,j=1\\i<j}}^3 \frac{2\alpha_s}{3m_q^2}
  \left[\frac{8\pi}{3}\delta^3(\vect r_i-\vect r_j)\,\vect S_i
  \cdt \vect S_j+ \right. \nonumber \\ 
   & & \left. \qquad \frac{1}{|\vect r_i-\vect r_j|^3}
  (3\,\vect S_i \cdt (\vect r_i-\vect r_j) \,\vect S_j \cdt (\vect r_i-\vect r_j)
   -\vect S_i \cdt \vect S_j) \right]\; .
\end{eqnarray}
Here, $\alpha_s$ is identified with the QCD fine structure constant
and $\vect S_i$ is the spin vector for quark $i$.
Spin-orbit interactions which also result from the non-relativistic
expansion of the one-gluon exchange are usually neglected using
more or less convincing arguments as they lead to practical complications
({\em e.g.} three-body interactions) and spoil the calculated 
hadronic spectrum \cite{Isgur:1978xj}.

We have introduced the basic elements 
($H_{\rm kin},H_{\rm conf}$ and $H_{\rm spin}$)
of nearly all quark potential
models that have been devised since the first models have been
developed some thirty years ago. 
The conceptual problems associated with this approach are manifest:
lack of covariance and not even an approximate chiral symmetry
as the quarks appear with masses of around 0.3 GeV in the Hamiltonian.
We shortly discuss  approaches that have been taken to remedy this.

In order to include relativistic effects, one certainly has to modify
the kinetic energy operator, 
\begin{equation}
  H_{\rm kin} \rightarrow \sum_{i=1}^3 \sqrt{\vect p_i^2+m_q^2} \; .
\end{equation}
Fortunately, it is in this case still possible to separate
overall bound state variables (total momentum and spin) from relative
variables such that the following holds: the hereby constructed
mass operator is only a function of the relative variables and
the overall and relative variables (considered as operators)
fulfill the commutation relations for the Poincar\'e algebra separately
\cite{Bakamjian:1953kh}. However, Poincar\'e invariance does not
guarantee a covariant description within Hamiltonian dynamics.
Rather, demanding covariance leads to fairly complicated
spin-dependent constraints on the Hamiltonian. They have been explicitly
formulated for light-cone variables in ref.~\cite{Leutwyler:1978vy}
but are also present for the above discussed conventional use 
of Galilean time as the Minkowski time variable in relativity.
As these constraints are usually overridden in phenomenological
applications, covariance is definitely lost in semi-relativistic potential models.

Chiral symmetry and its dynamical breaking lead to massless Goldstone
bo\-sons. They are identified with the pseudoscalar mesons which
do have mass, but this being comparatively small. Thus, chiral
symmetry appears to be explicitly broken only by small parameters
which in QCD are the current quark masses. 
Within the last years, it has become popular to replace 
potentials motivated through one-gluon exchange by 
non-relativistic approximations to one-meson exchange between
quarks \cite{Glozman:1997fs}, to pay credit to the importance of 
pions. This does not render the Hamiltonian chirally symmetric, though,
but solves some problems of the level ordering in the 
baryon spectrum. 

To conclude this brief summary of quark potential models,
we remark that the practical impossibility of a covariant description
limits these models to an effective parametrization of the
spectrum and of some static observables. In a non-covariant description,
the calculation of dynamical observables, with the simplest
being form factors, introduce too much of a frame dependence 
into the results as that they could be trusted as model predictions.

\subsection{Soliton models}

We now turn to a short discussion about another kind of baryon model
which in its original form also predates QCD and even the first
quark models. In our
exposition, we follow closely ref.~\cite{Holzwarth:1986rb}.
The central idea is to regard baryons as solitons of an effective,
non-linear meson theory. To illustrate this, let us start with a 
chirally symmetric theory describing massless
pions and its chiral partner, the $\sigma$ meson. The so-called
$\sigma$ model is defined by the Lagrangian density
\begin{equation}
 {\cal L}=\half\left( \partial_\mu \vect \pi \cdt \partial^\mu \vect \pi
   + \partial_\mu \sigma \partial^\mu\sigma\right) + V(\pi \cdt \pi +
  \sigma^2) \; .
\end{equation}
The $\sigma$ field may be regarded as the fourth component of the pion 
field, and we see that this Lagrangian density is invariant under
rotations in the ($\vect \pi, \sigma$) space leaving 
$\vect \pi^2+\sigma^2$ constant. Thus, the Lagrangian is invariant
under transformations of the group $O(4)$ which in turn shares 
the same Lie algebra with the chiral group $SU(2)\times SU(2)$. 
Depending on the functional form for $V$, the model may exhibit also
the feature of spontaneous symmetry breaking.
The non-linear version of this model is defined by demanding
\begin{equation}
  \sigma^2+\vect \pi^2=c^2={\rm const}\; .
  \label{chici}
\end{equation}
Introducing
$U=\exp(i\vect \tau \cdt \vect \phi)$, the Lagrangian simplifies
to
\begin{equation}
 \label{l2}
 {\cal L}^{(2)}= \frac{c^2}{4}{\rm Tr}\; \partial_\mu U \partial^\mu U
 ^\dagger \; ,
\end{equation}
since the potential $V$ reduces to a constant and may be omitted.
Now we look for static, finite energy configurations with a finite extension in
space which minimize the energy -- {\em solitons}. Due to the constraint
in eq.~(\ref{chici}), $\vect \pi$ is a variable which labels the points
of the three-dimensional unit sphere. A finite energy configuration only
exists if $U$ approaches a constant value for $|\vect x|\rightarrow
\infty$, with $\vect x$ being the position variable. Evidently, all
points $x$ on the sphere with radius $r$ may be identified in 
the limit $r \rightarrow \infty$, and a finite energy 
configuration is determined by the map
\begin{equation}
 \vect x \rightarrow \vect \pi( \vect x)\; : \; 
  S^3 \rightarrow S^3 \; , 
\end{equation}
which is labelled by its topological invariants, the
{\em winding numbers}. These considerations can be easily extended
to time-dependent fields $\vect \pi(\vect x,t)$ since the same
boundary conditions as for the static configurations 
must hold at infinity. Thus a topological
current may be defined,
\begin{equation}
  B^\mu=\frac{1}{24 \pi^2}\epsilon^{\mu\nu\rho\lambda}
    {\rm Tr}\;\left[ (U^\dagger\partial_\nu U) (U^\dagger\partial_\rho U)
       (U^\dagger\partial_\lambda U) \right] \; ,
\end{equation}
which conserves the winding number as the
corresponding charge. It has been the conjecture of Skyrme
\cite{Skyrme:1961vr} to identify the {\em topological} current $B^\mu$
with the {\em baryon} current and the winding number with
the baryon number. Unfortunately, one finds
by scaling $U(\vect r,t)\rightarrow U(\lambda \vect r,t)$
that a configuration of minimal energy is only obtained
for $\lambda \rightarrow \infty$, {\em i.e.} the soliton is not stable 
and collapses to size zero. This already indicates the way to resolve
the problem, by adding a term to the Lagrangian  (\ref{l2})
with more derivatives (but no more than two time-derivatives in
view of later quantization)
which stabilizes the soliton,
\begin{equation} \label{l4}
 {\cal L}^{(4)}=\frac{1}{32e^2} {\rm Tr}\,
  \left(\left[(U^\dagger\partial_\mu U),(U^\dagger\partial_\nu U)\right]\,
    \left[(U^\dagger\partial^\mu U),(U^\dagger\partial^\nu U)\right]
  \right) \; .
\end{equation}
Then a static soliton configuration is given by the ingenious
{\em hedgehog ansatz},
\begin{equation}
  U_0=\exp(i\vect \tau \cdt \vect {\hat x}\, F(r))\; .
\end{equation}
We see that the isovector pion field is simply proportional to the
position vector. The strength of the pion field is determined by
$F(r)$, the {\em chiral angle}. The boundary conditions for finite
energy configuration may be translated into the requirements
$F(0)=n\pi$ and $F(\infty)=0$ for a configuration with baryon number
$n$. Thus we have found a classical description
of baryons within the context of a simple non-linear meson theory.

There arises the immediate question of how the baryons obtain their half-integer
spin and isospin, since the fundamental pion field
possesses spin zero and isospin one. To answer this question,
the soliton needs to be quantized and to this end, time-dependent
soliton solutions are needed. We do not want to go into detail
here and just line out the basic ideas. Full time dependent solutions  
to the theory described by the Lagrangian 
${\cal L}={\cal L}^{(2)}+{\cal L}^{(4)}$ are hard to come by. Nevertheless
one observes that global rotations in isospace, caused
by a time-dependent matrix $A(t)\in SU(2)$,
\begin{equation}
 U(\vect x, t)=A(t)\,U_0(\vect x)\, A^\dagger(t) \; ,
\end{equation}
do not change the ``potential'' energy of the hedgehog,
{\em i.e.} the part in the Lagrangian depending only on spatial
derivatives. Therefore they might be a suitable {\em ansatz}
for a time-dependent soliton solution.\footnote{The static soliton 
violates isospin or, equivalently, rotational symmetry. Therefore
it needs to be projected onto good spin/isospin states. The introduction
of the collective rotation  in isospace exactly does this job.}
The time derivatives in
the Lagrangian create terms which represent the rotational energy.
Upon using the hedgehog properties and a suitable 
definition of angular velocities as canonical variables one finds
\cite{Holzwarth:1986rb}:
\begin{itemize}
  \item The absolute values of spin and isospin are equal,
      $\vect J^2=\vect T^2$.
  \item The Hamiltonian of the problem reduces to the one
   of the spherical top, $H=\frac{1}{2\theta}\vect J^2+M$.
   Here, $\theta$ is a moment of inertia and $M$ is the static
   energy (the mass) of the soliton.
\end{itemize}
Thus, upon quantizing the top, $J$ and $T$ as the quantum numbers
for spin and isospin may be well assigned half-integer values,
though there is no compelling reason to do this.

Alas, one would like to extend the theory to three flavors and
to chiral $SU(3)\times SU(3)$ symmetry in  order to describe
the baryon octet and decuplet. A detailed review on this subject 
is provided by ref.~\cite{Weigel:1996cz}. Let the Lagrangian
${\cal L}={\cal L}^{(2)}+{\cal L}^{(4)}$ remain unchanged but
define the chiral field as an $SU(3)$ field using
the mesons $\phi^a$, $a=1,\dots ,8$, built by
pions, kaons and the octet component of the $\eta$,
\begin{equation}
 U=\exp \left( i\sum_a \phi^a\lambda^a \right)\; .
\end{equation}
One may wish to add some flavor symmetry breaking terms
to the Lagrangian and can then proceed as described above,
{\em i.e.} by introducing the collective rotations and quantizing
the soliton.
At some point one would like to relate the effective meson theory
to QCD, and one could start by comparing symmetries. It turns out
that this model possesses an extra discrete symmetry that
is not shared by QCD. Under parity transformations ${\bf P}$ 
the pseudoscalar meson fields as described by QCD should obey
\begin{equation}
 {\bf P}: \; \vect \phi^a(\vect x, t) \rightarrow -\phi^a(-\vect x, t)\;.
\end{equation}
In the meson theory, one therefore defines ${\bf P}$ as
\begin{equation}
 {\bf P}: \; \vect x \rightarrow -\vect x, \; U \rightarrow U^\dagger \; .
\end{equation}
The Lagrangian is however invariant under $\vect x \rightarrow -\vect x$
and $U \rightarrow U^\dagger$ separately. To break this unwanted
symmetry, one needs to add some extra term to the meson action,
the famous {\em Wess-Zumino-Witten (WZW) term} \cite{Witten:1983tw}.
We state the consequences which come about by adding this term:
\begin{itemize}
\item If photons are added in a gauge  invariant manner to the meson
 theory, the WZW term generates a vertex for the anomalous decay
 $\pi^0 \rightarrow \gamma\gamma$. This decay is determined in QCD
 by the triangle graph which is proportional to the number of colors,
 $N_C$. Thus, the strength of the WZW term can be related to $N_C$.
\item If one considers an adiabatic $2\pi$ rotation of the soliton
 during an infinite time interval, the WZW term gives a contribution
 $N_C\pi$ to the action while all other contributions vanish. Thus
 the soliton acquires a phase $(-1)^{N_C}$ for such a rotation. 
 For $N_C$ odd,
 the soliton must therefore be quantized as a fermion, in accordance
 with $N_C=3$ in QCD.
\end{itemize}
We find that incorporating the QCD symmetries (for three flavors)
already generates a good argument to identify baryons with solitons.

There is one more connection of the soliton picture to QCD.
Witten has shown in ref.~\cite{Witten:1979kh} that in the limit
$N_C \rightarrow \infty$ QCD transforms into a theory of weakly 
interacting mesons and baryons emerge as solitons in this theory.
However, an exact mapping has not been found in this limit.

Instead, one can consider a Nambu-Jona-Lasinio (NJL) model involving
explicit quark degrees of freedom, defined by the 
Lagrangian \cite{Alkofer:1995mv}
\begin{equation}
 {\cal L}_{\rm NJL}=\bar q(i\Slash{\partial}-m^0)q - G 
  \left(\bar q { \frac{\Lambda^a_C}{2}}\gamma_\mu q\right)
  \left(\bar q { \frac{\Lambda^a_C}{2}}\gamma^\mu q\right) \; .
\end{equation}
Here, quarks ($q$) are assumed to come in 3 flavors and $N_C$ colors,
$m^0$ denotes a current quark mass matrix which is assumed to contain
small diagonal terms and $\Lambda^a_C$ are the generators of the
color group $SU(N)$. The model is an approximation to QCD if
one assumes the gluon propagator to be diagonal in color space and
$\sim \delta^4(x)$. 
Additionally, in the limit $m^0=0$ 
it conserves the chiral $SU(3)\times SU(3)$ symmetry
of QCD (with massless current quarks) which is of importance
for meson and baryon properties.
Upon Fierz transforming the NJL Lagrangian
into attractive  channels, one can show that
in the limit $N_C \rightarrow \infty$ only the meson channels survive
indeed.

Furthermore, this specific model and other quark
models of the NJL type admit solitonic solutions which
are reviewed in refs.~\cite{Alkofer:1996ph,Christov:1995vm}. 
In contrast to the non-linear $\sigma$ models discussed before,
these solitons are non-topological. In minimizing the 
energy of a field configuration, one determines the quark energy levels
in a background meson field of the hedgehog type, and a unit baryon number
is still provided by a configuration of $N_C$ quarks. It is an advantageous
feature to retain quark degrees of freedom in the solitonic picture,
if one considers phenomenological applications such as
calculations of structure functions and parton distributions
which have been performed in the last years. 

We will stop at this point and summarize some important features
of soliton models. They possess a deep conceptual appeal
as with the requirement of chiral symmetry imposed onto some effective
meson theory the baryons arise as lumps of these without further
ado. The possible
transformation of QCD in the limit $N_C \rightarrow \infty$
into such a nonlinear meson theory creates the link to the fundamental
quarks. However, in practical calculations one has to abstract from
quarks entirely or must resort to NJL models as crude approximations to
QCD. For both strategies it turns out that 
in leading order of $N_C$ (quasiclassical soliton configuration
with quantized collective variables) static baryon observables
are only accurate on the 20 \dots 30 \% level, as known from
nearly all hadronic models.
Corrections due to 
the finite number of colors (or, phrased otherwise, due to 
mesonic fluctuations)  
yield substantial contributions, and their technically involved 
calculation tends to hide the original conceptual beauty.
Another point of weakness is the non-covariant formulation
which will hamper the possible application of this model
to hadronic processes in the intermediate energy regime.

\subsection{Bag models}

Soon after QCD surfaced in theory circles, it became popular
to associate hadrons with a different phase of the theory
as compared to the vacuum phase. This picture motivates
the original M.I.T. bag model, where the vacuum phase is assumed
to prohibit the propagation of quarks and gluons but creates bubbles
of hadronic size in which quarks and gluons may propagate ordinarily.
The model is reviewed in ref.~\cite{Hasenfratz:1978dt}. It is
formalized by the Lagrangian
\begin{equation}
 {\cal L}_{\rm bag}=\left( {\cal L}_{\rm QCD} -B\right)\,\theta_V
   -\frac{1}{2}\bar q q \Delta_S \; . 
\end{equation}
Inside the bag, to be specified, $\theta_V=1$, and outside,
$\theta_V=0$. The surface function $\Delta_S$ is defined by
$\dpart{\theta_V}{x_\mu}=n^\mu\Delta_S$ and $n^\mu=(n^0,\vect n)$ 
is a normalized
unit vector perpendicular to the surface. According to this setup,
no quark or gluon current can escape the bag, due to the boundary 
conditions (for a static bag)
\begin{equation}
  \vect n \times \vect B_C =0\; , \quad 
  \vect n \cdt \vect E_C=0 \; , \quad
  \bar q \Slash{n}q =0 \; ,
\end{equation} 
where $\vect B_C$ and $\vect E_C$ denote the color 
magnetic and electric field, respectively. The bag constant $B$ has been
introduced to balance the surface pressure of the confined color
fields.

The usual procedure in finding the ground state 
is to assume a spherical, static bag and evaluate 
the quark wave functions and energy levels perturbatively.
To this end, one solves the free Dirac equation with the 
spherical boundary conditions, and the lowest-lying energy level
for a massless quark is approximately 400 MeV for a bag radius
of 1 fm, slightly above the constituent quark masses used
in the potential models. The splitting between nucleon (spin 1/2) and
$\Delta$ (spin 3/2) is evaluated by a one-loop gluon correction
to the ``bare'' quark levels. In the end one finds that the bag radius
usually comes out too large, above 1 fm, and for the QCD fine structure
constant $\alpha_s > 2$ to account for the $N-\Delta$ mass
difference. Hence the validity of a perturbative treatment is highly
questionable. Furthermore, pionic bags give just the wrong pion mass.

This led to the idea to couple $\sigma$ mesons and 
pions to the bag in order to regain
chiral invariance. These models come under different guises as
\begin{itemize}
 \item a ``hybrid'' bag where quarks and $\sigma-\vect \pi$ couple on the 
    bag surface and the latter are also allowed to propagate inside the 
    bag,
 \item a ``little'' bag where in contrast to the  hybrid bag
    pions are excluded from the interior of the bag,
 \item a ``cloudy'' bag where the mesons are constrained to the
    chiral circle  and are allowed inside the bag, and
 \item a ``chiral'' bag where the constrained mesons are described
    outside the bag by the Skyrme Lagrangian 
    ${\cal L} = {\cal L}^{(2)}+ {\cal L}^{(4)}$ which was
    discussed in the previous subsection,
    {\em cf.} eqs.~(\ref{l2},\ref{l4}).
\end{itemize}
All these facets of a softened bag are described on an introductory level
in ref.~\cite{Bhaduri:1988gc}. Usually all these models improve
on the shortcomings mentioned above. The bag radius can be chosen 
considerably smaller and $\alpha_S$ needs not be as large since 
the $N-\Delta$ splitting arises partially from the pions. Additionally,
they support the popular picture of a nucleon as consisting of
a valence quark core surrounded by a pionic cloud. As it will
be of interest later on, the non-zero electric form factor of the
neutron is frequently attributed to ``the'' pion cloud around the neutron
but such a statement has to be interpreted with care as it is
only valid in a model context such as a bag model, where valence quarks
alone yield zero for the form factor indeed.

Especially the chiral bag mentioned last in the above list
aims at combining the soliton picture of baryons 
with the valence quark picture. 
At first sight, it is appealing to think of the bag surface as
a borderline which indicates where to switch from 
a description with quark degrees of freedom to a description
with mesons. Then observables should  depend very little
on the size of the bag {\em (Cheshire Cat Principle)} and this
has been found indeed.  Unfortunately the static properties of baryons
are not improved as compared to the pure soliton, and come out
even worse than for solitons which are formulated with
pseudoscalar and vector mesons \cite{Weigel:1996cz,Hosaka:1996ee}.
Non-covariance of the formulation and possible large modifications
of observables by quantum fluctuations also remain present 
as problems to be solved.

\subsection{The diquark-quark picture: baryons as relativistic bound 
states}

Hopefully, the short review of the basic principles of three
established classes of baryon models  has illustrated the 
conceptual variance that exists among the models. Yet, all
of them describe static and a few dynamic observables fairly
well on the proverbial twenty-percent-level of accuracy, and
their degrees of freedom may be motivated from underlying QCD
and its symmetries. Nevertheless we find that none of these
is formulated in a manifestly covariant manner, and 
a model which deliberately cares  about covariance might
bridge the gap between soft and high-energy physics. Experimental
progress in this intermediate energy regime has been considerable
within the last years, and therefore understanding the corresponding 
results within a hadronic model with substructure 
constitutes a good motivation.

In the next chapter, the framework of a diquark-quark model
will be described in detail. In this model, we identify
baryons with poles in the 3-quark correlation function. These
poles may arise  by summing over infinitely many interaction graphs,
and it is well known that this summation leads to bound state 
equations such as the Bethe-Salpeter equation which is
usually formulated for two particles but can be extended to $n$ particles
as well. We will reduce the 3-quark problem to an effective two-particle
problem by introducing quasi-particles (diquarks) as separable 2-quark
correlations. The corresponding Bethe-Salpeter equation
can be solved {\em without} any non- or semi-relativistic reduction.
For the model calculations, we employ in a first attempt
massive constituent quarks with the corresponding free-fermion
propagator. It is certainly reasonable to assume
that a constituent quark mass is generated in QCD, and
this mass generation is reflected in the quark propagator 
\cite{Alkofer:2000},
but we acknowledge that the full structure of the quark propagator
in QCD is certainly described inadequately by just a free,
massive propagator.

In chapter \ref{em-chap}, we calculate the nucleon electromagnetic 
form factors maintaining covariance and gauge invariance which 
demonstrates the feasibility of a fully relativistic treatment also for 
dynamic observables. The electric nucleon properties can be described
very well, we emphasize here in anticipation of the results 
that the behavior of the 
neutron electric form factor can be understood without any explicit
reference to pions. The magnetic moments 
turn out to be underestimated, at this stage.

The results  for the  pion-nucleon form factor and the isovector
axial form factor, presented in chapter \ref{sa-chap}, 
are, for the latter and away from the soft point, in accordance 
with the available data.
For the pion-nucleon form factor, our predictions
converge with the predictions  of other baryon models.  
However, the results reveal
an overestimate for the pion-nucleon and the weak coupling constant
which are obtained in the soft limit.
Chiral symmetry demands here a further extension of the model to include
another diquark correlation, but this proved to be beyond the scope
of this work.

In contrast to the baryon models described above, confinement
is not included by construction in the diquark-quark model.
In chapter \ref{conf-chap} we investigate the possibility
of incorporating confinement by a suitable modification
of quark and diquark propagators. Thereby we are in the position
to calculate the mass spectrum of octet and decuplet baryons, and
also the nucleon magnetic moments receive some improvement. 
The justification of these modifications is discussed thoroughly.   

Chapter \ref{sal-chap} is concerned with a critical comparison to a 
semi-relativistic solution of the model which is shown to be inadequate.
Finally, we summarize the obtained results in chapter \ref{con-chap}
and comment upon further perspectives in the description of baryons 
within the context of the diquark-quark model.

Most of the  results presented here have been published previously,
{\em cf.} 
refs.~\cite{Oettel:1998bk,Oettel:1999gc,Oettel:1999bu,Oettel:2000uw,Oettel:2000jj}.
Also,  the author of this thesis has benefited a lot
from the collaboration with the authors of ref.~\cite{Hellstern:1998}
and ref.~\cite{Smekal:1998}.

Throughout the text we work in Euclidean space, if not indicated 
otherwise. The conventions relevant in this connection are summarized in
appendix \ref{conv-app}.

 \chapter{The Covariant Diquark-Quark Model}
 \label{dq-q-chap}

\section{Reduction of the relativistic three-quark problem}
\label{3qreduce}

The three-body problem has attracted a great deal of attention over 
the last decades and the formal framework presents itself to us as
being quite established, for an introduction see 
refs.~\cite{Thomas:1977,Gloeckle:1983}. Most work in the field
has been restricted to the treatment of quantum mechanical systems,
though, where certain theorems about existence and uniqueness
of solutions could be proved, see {\em e.g.} ref.~\cite{Faddeev:1965}
for the case of three-particle scattering with just two-body
potentials. For relativistic systems the few-body problem
seems to be somehow ill-posed as formal requirements like covariance
and locality have led to the formulation of quantum field theories    
which employ infinitely many degrees of freedom (that come about
by the possible creation and annihilation of particles).
Nevertheless, the spectrum of mesons and baryons, their
static properties and  transitions  indicate 
that only the degrees of freedom of a fixed number of 
(constituent) quarks are relevant. Therefore 
let us introduce the notion of baryon wave functions as matrix elements of
three quark operators that interpolate between the vacuum
and a bound state with observable quantum numbers
(see below eq.~(\ref{3qwave})).
Within QCD, the non-perturbative vacuum is a non-trivial condensate
and therefore these wave  functions would contain {\em per
definitionem} sea quark and gluonic parts. However,
it seems to be justified phenomenologically to
regard wave functions which are matrix elements of three
effective quark operators between the perturbative vacuum
and a bound state as the dominant ones compared to other operator  
matrix elements
that involve an arbitrary number of particle creation and annihilation
operators.
Then the quantum mechanical formulation
of the three-body problem 
can be taken over to field theory in the formulation of
Green's functions.

\subsection{Distinguishable quarks}

First, let us consider three distinguishable quarks. Later we
will generalize the obtained results to indistinguishable
particles. 

To save on notation, we will drop
Dirac, flavor and color indices. They are assumed
to be contained in the single particle labels.
In our definitions of Green's functions or bound state
matrix elements we always take out
one $\delta$-function representing conservation of the
total momentum, {\em e.g.} we will
denote a (dressed) single quark propagator by
$S_i$ with
\begin{eqnarray}
 (2\pi)^4  \delta^4 (k_i-p_i) \; S_i(k_i;p_i)&=&
    \int d^4 x_{k_i} \int d^4 y_{p_i} \exp\left[ i(k_i\cdt x_{k_i}-
        p_i \cdt y_{p_i}) \right]\; \nonumber \\
    &&\times\langle 0 | T q_i ({x_{k_i}}) \bar q_i ({y_{p_i}})|0 \rangle  \\
    &=:& \text{F.T.}\;
     \langle 0 | T q_i ({x_{k_i}}) \bar q_i ({y_{p_i}})|0 \rangle \; .
\end{eqnarray}
We will use the abbreviation
`F.T. $(x)$' for the Fourier transform of the subsequent expression
$(x)$ in the following. The annihilation operator for a quark of kind $i$
is denoted by $q_i$.
In this way,  the disconnected (free) three-quark propagator
is found to be
\begin{eqnarray}
  G_0(k_i;p_i) &= & \delta^4 (k_1-p_1) \delta^4 (k_2-p_2)
   \prod S_i(k_i;p_i)  
\end{eqnarray}
Sum and product run over $i=1,2,3$. 
The full quark 6-point function (or the 3-quark correlation function)
$G$ is accordingly 
\begin{eqnarray}
  (2\pi)^4  \delta^4 \left(\sum (k_i-p_i) \right) \;G(k_i;p_i)
   &=& \text{F.T.} \;
     \langle 0 | T \prod q_i(x_{k_i}) \bar q_i(y_{p_i})|0 \rangle \; .\;
\end{eqnarray}
It obeys Dyson's equation, symbolically written as
\begin{equation}
  G =G_0 + G_0 \;K \;G \; .
 \label{Dyson}
\end{equation}
$K$ is the three-quark  scattering kernel that contains
all two- and three-particle irreducible graphs and 
the whole equation involving 
the symbolic products reads explicitly
\begin{eqnarray}
 \label{symbint}
 G (k_i;p_i)& = &G_0 (k_i;p_i) + \\
            &  & \fourint{l_1} \fourint{l_2}
      \fourint{m_1} \fourint{m_2}\; G_0(k_i;l_i) \;K(l_i;m_i)
      \; G(m_i;p_i) \; . \nonumber
\end{eqnarray}
Besides the four-dimensional integrations, sums
over Dirac indices, color and flavor labels are implicitly understood
in the symbolic multiplication.

We define
a three-particle wave function $\psi$ to be 
the transition matrix element between the vacuum
and a state with an on-shell four-momentum $P$ and appropriate 
discrete quantum numbers (spin, isospin, \dots),
\begin{equation}
 (2\pi)^4 \delta^4 \left(\sum p_i-P\right) \psi(p_1,p_2,p_3) =
 \text{F.T.} \; \langle 0|q_1(x_1)q_2(x_2)q_3(x_3)| P\rangle \; .
 \label{3qwave}
\end{equation}
A bound state of mass $M$ with wave function 
$\psi$  will show up as a pole in the 6-point
function, {\em i.e.} in the vicinity of the pole $G$ becomes
\begin{equation}
 G \sim \frac{\psi(k_1,k_2,k_3) \;\bar \psi(p_1,p_2,p_3)}{P^2+M^2} \; ,
 \label{bspole}
\end{equation}
where $P=p_1+p_2+p_3$. Inserting this into Dyson's equation (\ref{Dyson})
and comparing residues, one finds the homogeneous bound state equation
\begin{equation}
 \psi= G_0 \; K \; \psi \quad \leftrightarrow \quad G^{-1}\; \psi=0 \; .
 \label{3bpsi}
\end{equation}
This is our starting point. As it stands, the
equation is far too complicated as 
neither the fully dressed quark propagator nor an 
expression for all two- and three-particle irreducible graphs
are known. One has to resort to some sort of approximation in which
the problem is tractable and judge the quality of the chosen 
approximation by the results for the bound state properties.

\begin{figure}
 \begin{center}
   \epsfig{file=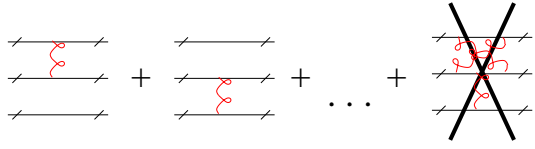,width=7cm}
 \end{center}
 \caption{Examples for admitted and excluded graphs in the 3-quark
          interaction kernel $K$ for the Faddeev problem.}
 \label{gluekern}
\end{figure}

The problem is greatly simplified when one discards three-particle
irreducible graphs from the interaction kernel $K$. We will
call this approximation the Faddeev (bound state) problem in the following.
The kernel then consists of three terms,
\begin{equation} 
 K\,=\,K_1+K_2+K_3 \; .
 \label{k2pi}
\end{equation}
The {$K_i$}, $i= 1,2,3 $,
describe the interactions of quark pairs $(jk)$, 
{\em i.e.} with {quark ($i$)} as a
spectator. $(ijk)$ is here  
a cyclic permutation of (123). In figure \ref{gluekern}, examples 
for admitted and excluded graphs in the kernel $K$ for the
Faddeev problem are shown.
The two-quark propagators $g_i$ fulfill their own Dyson's equation
with respect to the kernel $K_i$,
\begin{equation}
 \label{g_i}
 g_i= G_0 +G_0 \; K_i \; g_i \; .
\end{equation}
The objects $g_i$  and $K_i$ are defined in three-quark space.  
The former
contain a factor $S_i$, the propagator of the spectator
quark, and the latter contain a factor of $S_i^{-1}$
(although the spectator quark is not involved in the interactions
described by $K_i$). 
The matrix $\hat t_i$ is defined 
by amputating all incoming and outgoing quark legs 
from the connected part of $g_i$,
\begin{equation}
 g_i= G_0 + G_0 \; \hat t_i \; G_0 \; .
 \label{t1_def}
\end{equation}
Combining the two previous equations yields an integral equation
for $\hat t_i$,
\begin{equation}
 \hat t_i = K_i + K_i \; G_0 \; \hat t_i \; .
 \label{hatt_i}
\end{equation}
Still this matrix contains remnants of the spectator quark, this time
$S_i^{-1}$. Therefore the two-quark $t$ matrix of the interacting quarks $(jk)$
is obtained by
\begin{equation}
 t_i=\hat t_i\; S_i \; ,
 \label{t2_def}
\end{equation}
and contains just a trivial unit matrix in the subspace of the 
spectator quark.

\begin{figure}
 \begin{center}
   \epsfig{file=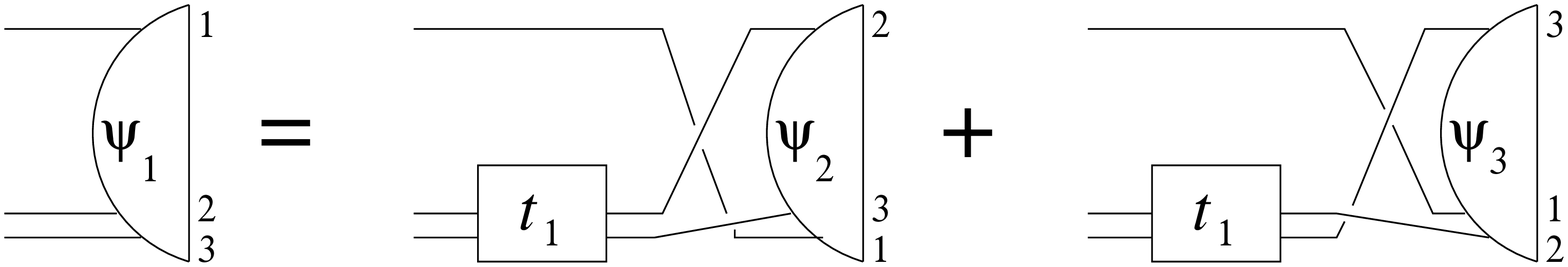,width=10cm}
 \end{center}
 \caption{The Faddeev bound state equation for the component
          $\psi_1$. The  equations for $\psi_2$ and  $\psi_3$
          follow by cyclic permutation of the particle indices.}
 \label{faddeev_fig}
\end{figure}

We need one more definition before we can tackle the bound state
problem. The so-called Faddeev components $\psi_i$ are introduced
by
\begin{equation}
 \psi_i = G_0\; K_i \; \psi \; ,
\end{equation} 
and by virtue of eqs.~(\ref{3bpsi},\ref{k2pi}) $\psi=\sum \psi_i$.
Solving eq.~(\ref{g_i}) for $G_0 K_i$ and inserting the result
into eq.~(\ref{3bpsi}) yields
\begin{equation}
 \psi =g_i \; (K_j+K_k)\;  \psi \; .
\end{equation}
Using the definition of $\hat t_i$ and $t_i$ in 
eqs.~(\ref{t1_def},\ref{t2_def}) the equation can be rewritten
as
\begin{equation}
 \psi_i = G_0 \; \hat t_i \; (\psi_j+\psi_k) = (S_j S_k) \; 
     t_i\; (\psi_j+\psi_k) \;.
 \label{Faddeev}
\end{equation}
These are the famous Faddeev bound state equations relating
the Faddeev component $\psi_i$ to $\psi_j,\psi_k$ using
the full two-quark correlation function $t_i$ (instead of
the kernel $K$). In turn the $t_i$ would have to be determined
in a full solution to the Faddeev problem by solving
eq.~(\ref{hatt_i}) which involves the kernel components $K_i$. 
The relativistic Faddeev equations, depicted in figure \ref{faddeev_fig},
are a set of coupled
4-dimensional integral equations and are thus a considerable
simplification of the original 8-dimensional integral equation problem
defined in eq.~(\ref{3bpsi}). Still, the wave function
components $\psi_i$ depend on the two relative momenta 
between the three quarks and expanding these components
in Dirac space \cite{Carimalo:1992ia} leads to an enormous number of coupled
integral equations.

\begin{figure}[b]
 \begin{center}
   \epsfig{file=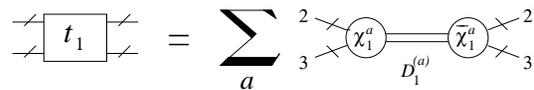,width=7cm}
 \end{center}
 \caption{The separable matrix $t_1$.}
 \label{tsep_fig}
\end{figure}

Therefore, we aim at further simplification of the bound state problem.
This will be achieved by  approximating the 
two-quark correlation function $t_i$ in terms of a sum
over separable correlations,
\begin{equation}
t_i (k_1,k_2;p_1,p_2) = \sum_a \chi_i^a(k_1,k_2)\; 
   D^{(a)}_i(k_1+k_2) \; \bar\chi_i^a(p_1,p_2)\; ,
 \label{sep_ass}
\end{equation}
pictorially shown in figure \ref{tsep_fig}.
Separability refers to the assumed property of $t_i$ that
it does not depend on any of the scalar products
$k_i \cdt p_j$ ($i,j=1,2$). We call these separable correlations
``diquarks''. The function $\chi^a$ 
is the vertex function of two quarks
with a diquark (labelled with $a$), $\bar \chi^a$
correspondingly denotes the conjugate vertex function.\footnote{
We refer here to the conjugate of a Bethe-Salpeter vertex
function. For the scalar and axialvector diquarks which will be of 
interest, it is introduced in eqs.~(\ref{dscon},\ref{dacon}).}
 Different $a$'s refer
to different representations in flavor and Dirac space, and
we name $D^{(a)}$ the ``diquark propagator'' for the
respective diquark.  
The assumption of separability in the two-quark correlations
is the starting point in the description of baryons within the
diquark-quark model. In the following section we will
introduce scalar and axialvector diquarks as the supposedly
most important correlations and explore the consequences
of quark antisymmetrization for their parametrization.

\begin{figure}[t]
 \begin{center}
   \epsfig{file=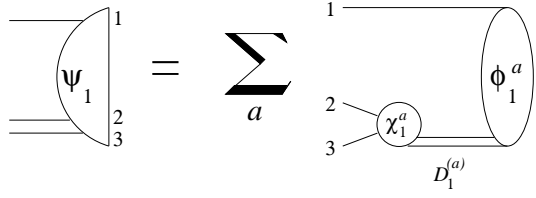,width=7cm}
 \end{center}
 \caption{The {\em ansatz} for $\psi_1$ using effective
    baryon-quark-diquark vertex functions $\phi^a_1$.}
 \label{psisep_fig}
\end{figure}

A suitable {\em ansatz} for the Faddeev components $\psi_i$, see
figure \ref{psisep_fig},
takes the form
\begin{equation}
 \psi_i (p_i,p_j,p_k) = \sum_a G_0 \; \chi^a_i(p_j,p_k)\;
    D^{(a)}_i(p_j+p_k)\; \phi^a_i (p_i,p_j+p_k) \; ,
 \label{ansatz1}
\end{equation}
where we introduced an effective vertex function of the baryon with
quark and diquark, $\phi^a_i$.
It only depends on the relative momentum
between the momentum of the spectator quark, $p_i$, and
the momentum of the diquark quasiparticle, $p_j+p_k$.
A detailed derivation of this property of $\phi^a_i$
can be found in refs. \cite{Thomas:1977,Ishii:1995bu}.

Inserting this {\em ansatz} into the Faddeev equations
(\ref{Faddeev}) yields coupled integral equations for the
effective vertex functions,
\begin{equation}
 \phi^a_i = \bar \chi^a_i \; (S_j S_k) \;
          \sum_b( \chi^b_j D^{(b)}_j \phi^b_j +
             \chi^b_k D^{(b)}_k \phi^b_k ) \; .
 \label{BS1}
\end{equation}
We will rewrite these equations into a form that will be familiar
from the theory of Bethe-Salpeter equations 
\cite{Itzykson:1980rh, Nakanishi:1969ph}. To this end
we define a relative momentum between quark $i$
and the diquark made of the quarks with label $(jk)$,
\begin{equation}
  p^{\rm rel}_i= (1-\eta) p_i -\eta (p_j+p_k) \; ,
  \label{relmom}
\end{equation}
and the total momentum
\begin{equation}
 P=p_1+p_2+p_3 \; .
\end{equation}
In contrast to the separation of variables in a non-relativistic
two-body problem, we have no unique definition of the relative momentum
between quark and diquark. Therefore we have introduced the parameter
$\eta$ which distributes relative and total momentum between quark and 
diquark.\footnote{Let $x_q$ and $x_d$ be the coordinates of quark and diquark
in configuration space. Translational invariance tells us that
$\phi^a_i(x_q+h,x_d+h)=\exp(-iP\cdt h)\phi^a_i(x_q,x_d)$. Therefore one
introduces a relative coordinate $x^{\rm rel}=x_q-x_d$ and an overall
configuration variable $X=\eta x_q + (1-\eta)x_d$. The momenta
$p^{\rm rel}$ and $P$ are the conjugate variables to $x^{\rm rel}$ and $X$.
Then $\phi^a_i(x_q+h,x_d+h)=\phi^a_i(x^{\rm rel},X+a)$ and, after
a Fourier transformation, translational invariance is seen to be fulfilled
with this choice of coordinates, leaving the freedom to choose
$\eta$ arbitrarily.}
Physical observables, like the mass of the bound states, form factors
{\em etc.} should of course not depend on this parameter.

Using the last two definitions for total and relative momentum and
taking into account all momentum conservation conditions
in eq.~(\ref{BS1}), the bound state equations read
\begin{equation}
 \phi^a_i (p^{\rm rel}_i,P) = \sum_b \sum_{j=1}^3 \fourint{k^{\rm rel}_j}
   \; K^{\rm BS}_{ij,ab} (p^{\rm rel}_i,k^{\rm rel}_j,P )\; 
   G_{0\,(b)}^{\rm q-dq}(k^{\rm rel}_j,P)\;
   \phi^b_j(k^{\rm rel}_j,P) \; .
   \label{BS2}
\end{equation}
$G_{0\,(b)}^{\rm q-dq}$ describes the disconnected (``free'') quark-diquark
propagator in the diquark channel $b$,
\begin{equation}
 G_{0\,(b)}^{\rm q-dq}(k^{\rm rel}_j,P) = S_j (\eta P+k^{\rm rel}_j)\; 
           D^{(b)}_j((1-\eta)P-k^{\rm rel}_j) \; ,
\end{equation}
and $K^{\rm BS}_{ij,ab}$ is the quark-diquark interaction kernel
given by
\begin{eqnarray}
 K^{\rm BS}_{ij,ab} (p^{\rm rel}_i,k^{\rm rel}_j,P )& =&
  \sum_{k=1}^3 \bar \delta_{ijk} \; \bar \chi^a_i (k_j,q) \;
   S_k(q) \; \chi^b_j(q,p_i) \; , \\
    k_j &=& \eta P + k^{\rm rel}_j \; , \\
    p_i &=& \eta P + p^{\rm rel}_i \; , \\
     q  &=& -p^{\rm rel}_j-k^{\rm rel}_i+(1-2\eta) P \; .
\end{eqnarray}
The symbol $\bar \delta_{ijk}$ is equal to 1 if $i,j,k$ are
pairwise distinct and 0 otherwise. 

Clearly this Bethe-Salpeter interaction kernel $K^{\rm BS}_{ij,ab}$
describes the quark exchange between the two-particle 
configurations of quark $i$/diquark $a$ and \mbox{quark $j$/} diquark $b$.
To arrive at an equation for physical baryons,
we have to project these configurations onto the respective baryon
quantum numbers. This will be done for the nucleons and the $\Delta$
resonances in section \ref{BSE-sec}.
 As it will turn out,
after the projection of the vertex functions 
the quark exchange generates
the attractive interaction that binds quarks and diquarks
to baryons.

\subsection{Identical quarks}

For identical quarks antisymmetrization is required.
Nevertheless,
 the Bethe-Salpeter equation (\ref{BS2})
will take the same algebraic form regardless 
of the particle index $i$.  
Single particle indices $i$ on the quark propagators $S_i$ and
the diquark propagators $D^{(a)}_i$ can be omitted now
as well as on the vertices $\chi^a_i$ as their functional form does 
not depend on them. 
However, some care is needed
since
the {\em particle} indices $i,j,k$ in eq.~({\ref{BS2}) specified
the {\em summation order} over color, flavor and Dirac indices.
Therefore we will introduce for each term in the 
Bethe-Salpeter equation for identical quarks
small Greek multi-indices $_{\alpha,\beta,\dots}$ for quarks that
contain all of the indices mentioned above.
Consequently, antisymmetry demands that the
diquark-quark vertices $\twosp{\chi^a}$ be antisymmetric
in their indices $_{\beta,\gamma}$.

\begin{figure}
 \begin{center}
   \epsfig{file=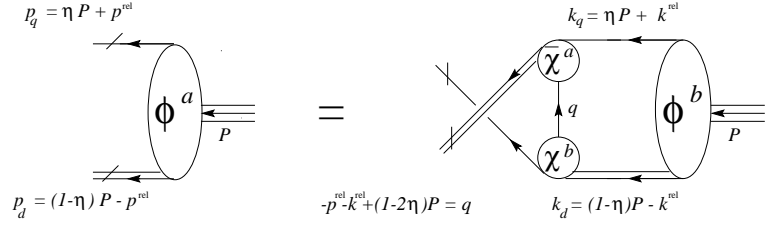,width=10cm}
 \end{center}
 \caption{The coupled set of Bethe-Salpeter equations
          for the effective vertex functions $\phi^a$.}
 \label{bse_fig}
\end{figure}

The Bethe-Salpeter equation for $\phi^a_i \equiv \phi^a$
in the case of identical quarks takes the form
\begin{eqnarray}
 \phi^a_\alpha (p^{\rm rel},P) &=& \sum_b  \fourint{k^{\rm rel}}
   K^{\rm BS}_{\alpha\gamma'} (k^{\rm rel}, p^{\rm rel}, P) \times 
  \label{BS3} \\
  \nonumber    & &  \qquad
   G^{\rm q-dq}_{0,\gamma'\gamma}(k^{\rm rel},P)\;
   \phi^b_\gamma(k^{\rm rel},P) \;, \\
   G^{\rm q-dq}_{0,\gamma'\gamma}(k^{\rm rel},P) &=&
      S_{\gamma'\gamma} (k_q)\; 
           2 D^{(b)}(k_d) \; , \\
   K^{\rm BS}_{\alpha\gamma'} (k^{\rm rel}, p^{\rm rel}, P)
      &=& \bar\chi^a_{\beta'\gamma'}(k_q,q) \;
          S_{\beta'\beta}(q) \; \chi^b_{\alpha\beta} (q, p_q) \; .
\end{eqnarray}
Repeated indices are summed over. Please note that, to save on
notation, we neglected
the indices for the involved diquark channels
in the symbols $G^{\rm q-dq}_{0}$ and $K^{\rm BS}$.
Due to the summation over
$\bar \delta_{ijk}$ an overall  factor of two emerged
that we have to absorb in the definition of the diquark
propagator in the case of identical quarks \cite{Ishii:1995bu}.
Spectator quark momenta $k_q[p_q]$, diquark momenta
$k_d[p_d]$ and the exchange quark momentum $q$ are given by
\begin{eqnarray}
 k_q[p_q] & = & \eta P + k^{\rm rel} [p^{\rm rel}] \; , \label{qm}\\
 k_d[p_d] & = & (1-\eta) P - k^{\rm rel} [p^{\rm rel}] \; , \label{dm}\\
   q      & = & -k^{\rm rel} -p^{\rm rel}+(1-2\eta)P  \; . \label{exm}
\end{eqnarray}
The set of Bethe-Salpeter equations is pictorially shown in figure
\ref{bse_fig}.

To conclude this section, we will write down
the completely antisymmetric Faddeev amplitude $\Gamma$, obtained
by cutting off the three external quark legs from
the wave function $\psi$ ({\em cf.} eq.~(\ref{ansatz1})),
\begin{equation}
  \Gamma_{\alpha_1\alpha_2\alpha_3}(p_1,p_2,p_3) =
  \sum_a 
    \sum^{\substack{ {\rm even} \\ {\rm perms}}}_{(ijk)}
     \chi^a_{\alpha_j\alpha_k}(p_j,p_k) \;
     D^{(a)}((1-\eta)-p^{\rm rel}_i) \; \phi^a_{\alpha_i} (p^{\rm rel}_i) \; . 
  \label{symm_amp}
\end{equation}
The relative momentum $p^{\rm rel}_i$ is defined as in 
eq.~(\ref{relmom}). Due to the antisymmetry of the diquark-quark 
vertices $\chi^a$ the summation over even permutations
of $(ijk)$ already generates the fully antisymmetric
Faddeev amplitude. 

Summarizing the results of this section, we have shown
that the determination of a 3-quark wave function defined
by eq.~(\ref{3qwave}) 
can be simplified enormously by the following approximations:
\begin{enumerate}
 \item Neglect all three-particle irreducible contributions
       to the full three-quark correlation function $G$.
 \item Assume that the truncated, connected  two-quark
       correlations (the $t$ matrices) can be represented
       as a sum over separable terms that are identified
       with diquark-quark vertex functions, see eq.~(\ref{sep_ass}).
\end{enumerate}
The bound state equation (\ref{3bpsi}) reduces to
a coupled set of Bethe-Salpeter equations for effective
baryon-quark-diquark vertex functions $\phi^a$, where $a$
labels the diquark types that are summed over in the expression
for the separable $t$ matrix, eq.~(\ref{sep_ass}).
These Bethe-Salpeter equations (\ref{BS3}) are the main result 
of this section and
its only ingredients are
\begin{itemize}
 \item the quark propagator $S$,
 \item the diquark-quark vertices $\chi^a$ and
 \item the diquark propagators $D^{(a)}$.
\end{itemize}
The aim of this work is to explore the consequences of choosing rather
simple forms for $S$, $\chi^a$ and $D^{(a)}$ that 
in a more elaborate approach should be determined by a suitable
effective theory of QCD. In the diquark sectors of  Nambu-Jona-Lasinio (NJL)
type of models \cite{Vogl:1991qt}, 
the corresponding bound state equations have been formulated
first in ref.~\cite{Reinhardt:1990rw} and solutions of the 
full Bethe-Salpeter equation have been obtained in 
ref.~\cite{Ishii:1995bu}. However, the framework
of an NJL model is too tight: quark and diquark are not confined and
additionally the diquark is pointlike ($\chi^a\not =\chi^a(p_1,p_2)$
where $p_{1[2]}$ are the quark momenta). 
Furthermore, the framework presented here is well-suited for
the calculation of dynamical observables. In the case of electromagnetic
observables, gauge invariance is also maintained in numerical solutions
to a high degree of precision. This is somewhat harder to achieve
in NJL model calculations employing sharp Euclidean cut-offs.
Partly due to this limitation,
 no dynamical observables of baryons have been calculated 
in the diquark sector of the NJL model so far, using fully four-dimensional solutions of the
Bethe-Salpeter equation.

\section{Diquark correlations}
\label{diquark-sec}

As explained in the previous section, ``diquarks'' or
``diquark correlations'' refer to the use of a separable two-quark
correlation function, the $t$ matrix. 
As the diquark-quark model presented here is inspired by the NJL
model, we will describe shortly how separable $t$ matrices
arise in it and discuss previous approaches to the diquark-quark picture
of baryons. The following main part  of this section 
introduces our parametrization of the $t$ matrix
including the explicit form for the diquark-quark vertices and the 
diquark propagators.

Before we embark upon these points, let us discuss heuristically
some properties of diquarks. Any two quarks within a baryon have 
to be in a color antisymmetric state as required by the Pauli 
principle and the assumed color singlet nature of baryons.
If we consider for the moment diquarks as asymptotic states
(in an unphysical subspace of the whole Hilbert space),
their spatial wave functions (assumed to be separated
from their Dirac structure) will be dominated by a 
symmetric ground state. This requires the combination 
of flavor and Dirac structure to be symmetric. 
Their mass spectrum and therefore their importance as virtual states
in baryons can be guessed by looking at the meson spectrum:
pseudoscalar mesons correspond to scalar diquarks, vector mesons
to axialvector diquarks since the intrinsic parity
of ($q\bar q$) is opposite to the intrinsic parity of ($qq$).
Although the pseudoscalar mesons are especially light
due to their suspected nature as Goldstone bosons, the 
scalar diquarks corresponding to them should be still lighter 
than axialvector diquarks. This point gets support from lattice 
calculations \cite{Hess:1998sd}. In quenched QCD employing Landau gauge 
the scalar diquark correlation function corresponds to
a massive state of around 700 MeV whereas the axialvectors
are heavier by around 100 MeV. Vector and pseudoscalar diquark masses
are then expected to lie around 1 GeV and therefore assumed to be 
less important for nucleons. This is indeed the case in a 
calculation within the Global Color Model \cite{Praschifka:1989fd}.

Having motivated the importance of scalar and axialvector
diquarks, let us return to their description within the 2-quark
correlations.
As is well-known, separable
$t$ matrices arise from separable 4-quark scattering kernels
$K \equiv K_i S_i$ in the following way\footnote{The kernels $K_i$
introduced in eq.~(\ref{k2pi}) contain a factor $S_i^{-1}$,
the inverse propagator of the spectator quark.}. Let 
\begin{equation}
 K^{{\rm sep},a}(k_1,k_2;p_1,p_2)= 
  \chi^a(k_i)\; \Lambda^{(a)}(k_1+k_2)\; \bar\chi^a(p_i)
 \; ,
\end{equation}
be the scattering kernel in the particular channel $a$. Then the 
corresponding $t$ matrix in this channel 
can be inferred 
from eq.~(\ref{hatt_i}) which for identical particles
is modified to 
\begin{equation}
 t = K + \frac{1}{2}\;K\;G_0^{(2)}\;  t \; .
\end{equation}
We denote $G_0^{(2)}$ as the free two-quark propagator, as for
distinguishable particles it corresponds to just
the product of the single propagators. 
Due to the antisymmetry of $K$ in the labels $(jk)$,
$t$ is also antisymmetric in these labels.
Its solution for the separable kernel $K^{{\rm sep},a}$ reads 
\begin{equation}
 t^a (k_i;p_i) = \chi^a(k_i) \;D^{(a)}(k_1+k_2)\; \bar\chi^a(p_i)\; .  
\end{equation} 
The inverse of the propagator $D^{(a)}$ is determined by
\begin{equation}
 \left( D^{(a)} \right)^{-1} = \left(\Lambda^{(a)}\right)^{-1} - 
   \frac{1}{2}\;\bar\chi^a \; G_0^{(2)} \chi^a \; .
   \label{dqsep}
\end{equation}
We use the symbolic product notation as in the previous chapter.
The last term 
in eq.~(\ref{dqsep}) describes a quark loop with
insertions of the vertex functions $\chi^a$ and $\bar\chi^a$.

In QCD the simplest term in the perturbative kernel, the gluon
exchange, has already a non-separable form. However,
when approximating the gluon propagator by a $\delta$-function
in configuration space, the resulting model is
a Nambu-Jona-Lasinio (NJL) model with pointlike four-quark
interactions that can be viewed as 
color octet--flavor singlet quark current interactions.
This local current-current interaction can be rewritten
into attractive interactions in color singlet meson and 
color triplet diquark channels \cite{Alkofer:1995mv}.
Let us pick the quark-quark scattering kernel 
in the scalar channel $s$ which we write in the form
\begin{eqnarray}
  (K^s)_{\alpha\beta,\gamma\delta} &=& 
      4 G_s (\chi^s)_{\alpha\beta}\; (\bar\chi^s)_{\gamma\delta} \\ 
   & = &4 G_s \; (\gamma^5 C\: \tau^2\: \lambda^k)_{\alpha\beta} \;
               (C^T \gamma^5\: \tau^2\: \lambda^k)_{\gamma\delta}\; ,
  \label{sd-njl}
\end{eqnarray} 
where we consider just an isospin doublet of two quarks that
are in a flavor antisymmetric  ($\tau^2$) and a color
antitriplet state ($\lambda^k$, $k=2,5,7$). The $\tau^i$ are
Pauli matrices and the $\lambda^k$ are Gell-Mann matrices \cite{Griffiths:1987tj}.
The coupling constant $G_s$ regulates the strength in this interaction
channel and $C$ denotes the charge conjugation matrix.
The scalar diquark propagator is by virtue of eq.~(\ref{dqsep})
\begin{equation}
 \left(D^{(s)}\right)^{-1}(k^2)= \frac{1}{4G_s}-
   2\: {\rm Tr}_D\; \fourint{q} (C^T\gamma^5) \: S(q+k/2)\: (\gamma^5 C)
   \: S^T(k/2-q)  \; .
\end{equation}
After suitable regularization of the divergent integral it turns
out that the inverse propagator has zeros for certain values of 
$k^2$, therefore bound scalar diquarks exist in the NJL model and
the propagator can be viewed as a scalar propagator which contains
the effect of the dressing quark loop.

In analogy to the meson spectrum where
pseudoscalar mesons are the lightest ones, followed by vector mesons,
axialvector diquarks are expected to be important in the diquark
channel. In fact, 
the axialvector diquark propagator acquires
poles \cite{Weiss:1993kv}, but their appearance depends
on the ratio of the coupling constant in the diquark channel
to the coupling constant in the pseudoscalar meson channel.

Having obtained separable $t$ matrices for the scalar and the axialvector
correlations, the spectrum of spin-1/2 and spin-3/2 baryons is
in principle calculable from the Bethe-Salpeter equation 
(\ref{BS3}). This was done in refs. \cite{Hanhart:1995tc,Buck:1995ch}, 
however, for
the quark exchange kernel a static approximation was used (which
amounts to neglecting all momentum dependence in the propagator
of the exchange quark). In doing this, covariance
is lost, and the rich relativistic structure of the effective 
nucleon-quark-diquark vertex functions, to be described in
section \ref{partialwaves}, has been furthermore approximated by the
dominating non-relativistic components, the $s$ waves.  
The full solution of the Bethe-Salpeter equations, {\em i.e.}
with scalar and axialvector diquarks, for nucleon and $\Delta$
was obtained in ref.~\cite{Ishii:1995bu}.
Subsequent calculations of static nucleon observables
employed unfortunately only the solutions in the scalar diquark sector
\cite{Asami:1995xq,Ishii:2000jm}. 
Especially the magnetic moments
turned out to be deficient, thus calling for including
the axialvector diquark channel.

The approximation of the two-quark $t$ matrix by a finite
sum of isolated poles at timelike momenta $k^2$ seems to be
supported by these calculations. Solutions for 
on-shell diquarks were
also obtained in the ladder-approximated Global Color model
\cite{Praschifka:1989fd}.
The Global Color Model employs a non-local current-current interaction,
mediated by an effective gluon propagator and therefore
avoids the regularization problems present in the NJL model.
Parametrizing the $t$ matrix with only the hereby obtained
scalar diquark pole the nucleon Bethe-Salpeter equation 
was solved in ref.~\cite{Burden:1988dt}. 

However, when
going beyond ladder approximation, {\em i.e.} including
higher order perturbation graphs in the quark-quark scattering kernel,
the diquark poles disappear in the NJL model \cite{Hellstern:1997nv}
as well as in the Munczek-Nemirowsky model \cite{Bender:1996bb}.
The latter employs for the gluon propagator
a $\delta$-function in momentum space and can be regarded
as quite complementary to the NJL model, indicating that
in higher-order calculations with a more realistic
gluon propagator diquark poles may disappear from
the quark-quark scattering amplitude too.

Nevertheless,
the phenomenological efficacy of using a diquark correlator 
to parametrize the 
2-quark correlations does not rely on the existence of
asymptotic diquark states. Rather, the diquark correlator may be devoid of
singularities for timelike momenta, which may be interpreted as one possible
realization of diquark confinement. In principle, one may appeal to models
employing a general, separable diquark correlator which need not have any
simple analytic structure, in which case no particle interpretation for the
diquark would be possible. 
Indeed, the fact that the interaction mechanism in the quark-diquark
picture (the quark exchange) is attractive, turns out to be independent of
the details of the employed diquark correlations. It can be traced 
back to the antisymmetry of the diquark and baryon vertices and
arises thus from the color and flavor factors in the Bethe-Salpeter equation,
see the next section. Last but not least, the usefulness 
of this picture and of employing
specific diquark correlators has to be judged by comparison
to the experiment. Calculations of observables have
to go beyond nucleon and $\Delta$ mass or the
octet and decuplet spectrum  as these are rather
insensitive to the level of sophistication in the treatment.

As the results for the $t$ matrix have such a strong model
dependence, we will not resort to a specific effective quark model
for calculating it in the following. 
Rather, we will employ 
an admittedly simple {\em ansatz} for the $t$ matrix,
 the 
pole approximation in the scalar and axialvector channel.
We will assume an isospin-doublet
of $u$ and $d$ quarks as  we are mainly interested in
nucleon properties. Later on, in chapter \ref{conf-chap}, 
we will generalize 
our treatment to the case of broken $SU(3)$ flavor symmetry.
We will use vertex functions with a finite extension in momentum
space, first because if strong diquark correlations exist
their size  is  most likely of the order of the proton
radius, and secondly, to avoid artificial regularizations
in solving the bound state equations.
The {\em ansatz} for the 2-quark $t$ matrix reads,
\begin{eqnarray} \nonumber   
 t(k_\alpha,k_\beta;p_\alpha,p_\beta) \equiv t(k,p,P) &=&
 \chi^5_{\alpha\beta}(k,P) \;D(P)\;\bar 
  \chi^5_{\gamma\delta}(p,P) \; +\; \\
  & & 
\chi_{\alpha\beta}^\mu(k,P) \;D^{\mu\nu}(P)\;
    \bar  \chi_{\gamma\delta}^\nu(p,P)  \; . 
  \label{tsep}
\end{eqnarray}
The relative momenta are defined as
\begin{equation} 
 k[p]= \sigma\, k_\alpha[p_\alpha]- (1-\sigma)\, k_\beta[p_\beta],
    \qquad \sigma \in [0,1]\; .  
\end{equation}
The diquark propagators used in the scalar and the axialvector channel
are taken to be 
\begin{eqnarray}
D(P) &=& -\frac{1}{P^2+m_{sc}^2}\, C(P^2,m=m_{sc}) \; , 
 \label{Ds} \\
D^{\mu\nu}(P) &=& -\frac{1}
   {P^2+m_{ax}^2} \left( \delta^{\mu\nu}+  (1-\xi) \frac{P^\mu P^\nu}{m_{ax}^2}
\right)\, C(P^2,m=m_{ax})  \; . \quad 
 \label{Da}
\end{eqnarray} 
The choice $C(P^2,m)=1$ and $\xi=0$ corresponds to 
free propagators of a spin-0 and  a spin-1
particle, to be employed here. 
Most generally, the dressing function $C$ is not necessarily the
same for scalar and axialvector diquarks. 
However, when investigating nontrivial forms for $C$ 
(which mimic confinement) 
in chapter \ref{conf-chap} we will 
assume the same $C$ in both channels
for simplicity. 

For the trivial form,
the diquark-quark vertices $\chi$ and
$\chi^\mu$ correspond on-shell \mbox{$(P^2=-m^2_{sc[ax]})$}
to diquark Bethe-Salpeter vertex functions.
With quark legs attached, they are Fourier transforms
of the transition matrix element
\begin{eqnarray}
 S(\sigma P+p)\,\chi^{5[\mu]}_{\alpha\beta}(p,P)
  \,S^T((1-\sigma) P-p) = \\
 \nonumber 
 \fourinta{X} \! \fourinta{x} \;\exp(iP\cdt X+ip\cdt x) \nonumber 
 \langle 0 | T q_\alpha(x_\alpha) q_\beta(x_\beta) | P, 
  {\rm sc\;[ax]} \rangle \; , 
\end{eqnarray}
with the relative coordinate $x=x_\alpha-x_\beta$ and
$X=(1-\sigma)x_\alpha+ \sigma x_\beta$.
The Dirac part of the conjugate vertex functions is obtained by 
charge conjugation, 
\begin{eqnarray}
 \bar \chi^5 (p,P)& =& C\; \left( \chi^5 (-p,-P) \right)^T
      \; C^T \;, \label{dscon} \\
 \bar \chi^{\mu} (p,P)& =& -C\; \left( \chi^{\mu} (-p,-P) \right)^T
      \; C^T \;. \label{dacon}
\end{eqnarray}

They have to be antisymmetric with respect to the interchange of the two
quarks and this entails
\begin{equation}
  \chi^{5[\mu]}_{\alpha\beta}(p,P) = \left.
  -\chi^{5[\mu]}_{\beta\alpha}(-p,P)\right|_{\sigma \leftrightarrow
         (1-\sigma)} \; .
  \label{antisymm1}
\end{equation}
As any two quarks in a baryon have to belong to the color
antitriplet representation, the diquark-quark vertices  are
proportional to the antisymmetric tensor
$\epsilon_{ABD}$ with color indices $A,B$ for the quarks,
and with $D$ labelling the color of the diquark.
Furthermore, the
scalar diquark is an antisymmetric flavor singlet $|0\rangle$
represented by
$(\tau_2)_{ab}$, and the
axialvector diquark is a symmetric flavor triplet $|1,k\rangle$
which can be represented
by $(\tau_2\tau_k)_{ab}$. Here, $a$ and $b$ label the quark flavors
and $k$ the flavor of the axialvector diquark.
For their structure in Dirac space, we maintain only the dominant
components, the antisymmetric matrix $(\gamma^5 C)$ for the
scalar diquark and the symmetric matrices $(\gamma^\mu C)$
for the axialvector diquark.
 Thus, with all these indices 
made explicit, the vertices 
read\footnote{Symbolically denoting the totality of quark indices by 
the same Greek letters that are used as their Dirac indices should 
not cause
confusion. The particular flavor structures are tied to the Dirac
decomposition of the diquarks, color-$\bar 3$ is fixed.}   
\begin{eqnarray}
 \chi^5_{\alpha\beta}(p)&=&g_s (\gamma^5 C)_{\alpha\beta}\; V(p^2,p\cdt P) \;\;
    \frac{(\tau_2)_{ab}} {\sqrt{2}}\,\frac{\epsilon_{ABD}}{\sqrt{2}} \;,
   \label{dqvertex_s} \\
 \chi_{\alpha\beta}^{\mu}(p)&=&g_a (\gamma^\mu C)_{\alpha\beta}\; V(p^2,p\cdt P) \;\;
    \frac{(\tau_2\tau_k)_{ab}} {\sqrt{2}}\,\frac{\epsilon_{ABD}}{\sqrt{2}}\; .
   \label{dqvertex_a}
\end{eqnarray}
Normalizations of the flavor and color part have been chosen such
that for the matrix elements in the respective spaces the following holds,
\begin{eqnarray}
  \langle \bar 3, I | \bar 3, J \rangle_{\rm color} &=& \half 
   \epsilon_{IMN} \epsilon_{JMN} = \delta^{IJ} \label{n1} \; ,\\
  \langle 0 | 0 \rangle_{\rm flavor} &=& \half {\rm Tr}
   \; \tau_2^\dagger\tau_2=1 \label{n2} \; ,\\
 \langle 1, i | 1, j \rangle_{\rm flavor} &= &
   \half {\rm Tr}\; (\tau_2\tau_i)^\dagger(\tau_2\tau_j) = \delta^{ij}
 \label{n3} \; .
\end{eqnarray}

The antisymmetry condition (\ref{antisymm1}) translates
into a symmetry condition for the scalar function $V$,
\begin{equation}
 V(p^2,p\cdt P) = \left. V(p^2,-p\cdt P) \right|_{\sigma \leftrightarrow
       (1-\sigma)} \;.
 \label{sym_P}
\end{equation}
For $ \sigma \not= 1/2 $ and thus for $\bar p := p
\big|_{\sigma \leftrightarrow (1-\sigma)} \not= p$, it is not
possible to neglect the $p\cdt P$ dependence in the vertex
without violating the quark-exchange antisymmetry. 
To maintain the correct quark-exchange antisymmetry, we assume
instead that the vertex depends on both scalars, $p^2$ and $p\cdt P$
in a specific way.  In particular, we assume the diquark-quark vertex is
given by a function that depends only on the scalar 
\begin{eqnarray} 
  x :=  
       (1 - 2\sigma) p\cdt P + p^2 
 =  -(1 - 2\sigma) \bar p\cdt P + \bar p^2 \label{x_def}
\end{eqnarray} 
with $\bar p  =  (1-\sigma) p_\alpha - \sigma p_\beta $ and 
$p_{\{\alpha , \beta\}}$ as given above. 
The two scalars that may be constructed from the available momenta $p$ and
$P$ which have definite symmetries
under quark exchange are given by the two independent combinations
$p_\alpha\cdot p_\beta$ (which is essentially the same as above $x$) and 
$p_\alpha^2 - p_\beta^2 $. 
The latter may only appear in odd powers which are associated with
higher moments of the Bethe-Salpeter vertex
in a polynomial expansion in $p_\alpha^2 - p_\beta^2 $.  
Hence, these are neglected by setting 
\begin{eqnarray} 
 V(p^2, p\cdt P) \rightarrow V(x)  
\end{eqnarray}  
which obviously satisfies the symmetry constraint given by
eq.~(\ref{sym_P}) $\forall \sigma \in [0,1]$.
For the actual calculations we will restrict ourselves to equal momentum 
partitioning between the quarks in the
diquark correlation,  $\sigma = 1/2$. The scalar $x$ then reduces to 
the square of the relative momentum, $x = p^2$. 

The overall strength of the diquark correlations given in
eqs.~(\ref{dqvertex_s},\ref{dqvertex_a}) is hidden
in the ``diquark-quark coupling constants'' $g_s$ and $g_a$.
To estimate their values, we can resort to the canonical
normalization condition for Bethe-Salpeter vertex functions
\cite{Itzykson:1980rh} which reads in our context,
\begin{eqnarray}
 \frac{1}{2}\fourint{p} \; \bar \chi^5_{\alpha\beta} \; P^\mu \dpartial{P^\mu} 
   (G^{(2)}_0)_{\alpha\gamma,\beta\delta} \; \chi^5_{\gamma\delta} & 
   \stackrel{!}{=}& 2 m_{sc}^2, \label{normsc} \\
  \frac{1}{2}\fourint{p} \; \left( \bar \chi^{\nu}_{\alpha\beta} \right)_T \;
   P^\mu \dpartial{P^\mu}  
   (G^{(2)}_0)_{\alpha\gamma,\beta\delta} \; 
   \left( \chi^{\nu}_{\gamma\delta} \right)_T 
    &\stackrel{!}{=}& 
                           6 m_{ax}^2. \label{normax}
\end{eqnarray}
Here, 
$\left(\chi^{\nu}\right)_T=\chi^\nu - \hat P^\nu (\hat P^\mu \chi^\mu)$
is the transverse part of the vertex $\chi^{\nu}$. Note
that the pole contribution of the axialvector diquark is transverse to its
total momentum, and the sum over the three polarization states provides 
an extra factor of 3 on the r.h.s. of eq.\ (\ref{normax}) as compared to
eq.~(\ref{normsc}). 
With normalizations as chosen in eqs.\
(\ref{dqvertex_s},\ref{dqvertex_a}) the traces over 
the color and flavor parts yield no additional factors,
see eqs.~(\ref{n1}--\ref{n3}). 

In the following, we will solve the baryon bound state
problem with the coupling constants $g_s$ and $g_a$ fixed in
the indicated manner, keeping in mind that the Bethe-Salpeter normalization
condition determines these values only on-shell.
When calculating diquark contributions to the nucleon
electric form factors, we will see that the Ward identity
provides for an indirect off-shell constraint on 
$g_s$ and $g_a$. However, the ratio between
the constants fixed either on- or off-shell will be of the
order of unity.

In the model sector with scalar diquarks only, we will
explore the ramifications of several {\em ans\"atze}
for the scalar functions $V(x)$,
\begin{eqnarray}
 V_{n{\rm -pole}} & =& \left(\frac{\lambda_n^2}{\lambda_n^2 + x}
     \right)^n \; ,\label{npole}\\
 V_{\rm exp} &=& \exp( -x/\lambda_{\rm exp}^2) \; ,\label{expo}\\
 V_{\rm gau} &=& \exp( -(x-x_0)^2/\lambda^4_{\rm gau}) \, . \label{gauss} 
\end{eqnarray}
The last form, the Gaussian form for the diquark vertices,
was suggested as a result of a variational calculation
of an approximate diquark Bethe-Salpeter equation in ref.~\cite{Praschifka:1988ry}.
The nucleon calculations of ref.~\cite{Burden:1988dt}
used this form {\em without} fixing the strength $g_s$ by
a condition such as eq.~(\ref{normsc}). It turns out that
the necessary value for $\lambda_{\rm gau}$ to obtain
a reasonable nucleon mass is about an order of magnitude smaller than
the value given in ref.~\cite{Burden:1988dt}.
Furthermore, we will show in sect. \ref{solI} that the results
for the electric form factors in the scalar diquark sector 
clearly rule out the Gaussian and the exponential form
of the diquark vertices and favor the dipole {\em ansatz}.   
\section{The diquark-quark Bethe-Salpeter equation}
\label{BSE-sec}

Equipped with the separable form of the two-quark
correlations, eq.~(\ref{tsep}), and the functional form
of the scalar and axialvector diquark correlations in \linebreak
eqs.~(\ref{dqvertex_s},\ref{dqvertex_a}), we will have a closer
look on the effective Bethe-Salpeter equation derived in 
section \ref{3qreduce} and set it up for the nucleon
and the $\Delta$ in turn.\footnote{We will sum up the four
nearly mass-degenerate states 
$\Delta^{++},  \Delta^{+} , \Delta^{0} ,\Delta^{-}$
under the label $\Delta$ in the following.} 
To complete the model definition, we have to specify the 
functional form of the quark propagator. We take it to be
a free fermion propagator with a constituent mass $m_q$,
\begin{equation}
 S(p)= \frac{i\Slash{p} -m_q}{p^2+m_q^2}\, C(p^2,m=m_q) \;
  \qquad  {\rm with}\;\; C(p^2,m=m_q)=1.
 \label{qprop}
\end{equation}
This incorporates from the beginning an effective quark mass
in the order of magnitude of several hundred MeV which is believed
to be dynamically generated in QCD. 
The property of confinement is neither manifest in eq.~(\ref{qprop})
nor in the form of the diquark propagators, eqs.~(\ref{Ds},\ref{Da}).
Effective parametrizations of confinement in the model propagators
and its influence on results will be investigated in chapter 
\ref{conf-chap}.

\subsection{Nucleon}
\label{nuc-subsec}

The Faddeev amplitude of eq.~(\ref{symm_amp}) takes
the following form,
\begin{eqnarray}
  \trisp{\Gamma} = \twosp{\chi^5}\; D \; \left( \Phi^5
   u\right)_\alpha +   
                  \twosp{\chi^\mu} \; D^{\mu\nu} \;
    \left( \Phi^\nu u \right)_\alpha \; . 
  \label{Faddeev_N}
\end{eqnarray}
The necessary antisymmetrization of $\trisp{\Gamma}$,
provided by the summation over even permutations of the
multi-indices $\trisp{\;}$ as in eq.~(\ref{symm_amp}), will
be understood implicitly from now on. The spinorial quantities
\begin{equation}
 \Phi(p,P)\, u(P) := \bp \Phi^5(p,P) \\ \Phi^\mu(p,P) \ep \, u(P)
\end{equation}
describe the effective nucleon-quark-diquark vertex functions.
The quantity $u(P)$ is a positive-energy spinor
for a spin-1/2 particle with momentum $P$.

We define Bethe-Salpeter wave functions by attaching quark and 
diquark legs to the vertex functions,
\begin{eqnarray}
 \tilde D (p_d) &:=& \bp D(p_d) & 0 \\ 0 & D^{\mu\nu}(p_d) \ep \label{tildeD} \; ,\\
 \Psi(p,P)&:=&\bp \Psi^5 \\ \Psi^\mu \ep (p,P)= S(p_q) \;
    \tilde D (p_d)
 \bp \Phi^5 \\ \Phi^\nu \ep (p,P) \, .
 \label{wavenucdef}
\end{eqnarray}
Vertex and wave function $\Phi$ and $\Psi$ are Dirac matrices
to be constructed from the quark-diquark relative momentum
$p$, the nucleon momentum $P$ and the set of $\gamma$ matrices.
The task of their decomposition will be taken up in the next section,
showing that there are altogether eight components that encompass
all possible  spin-orbit couplings of quark and diquark 
to the nucleon spin.

Inserting all these definitions into the Bethe-Salpeter equation
(\ref{BS3}), we find the equation
\begin{equation}
   \fourint{k} \left(G^{\rm q-dq}\right)^{-1} (p,k,P)
  \begin{pmatrix}\Psi^5 \\ \Psi^{\mu'}\end{pmatrix}(k,P) =0 \;, 
  \label{BS4}
\end{equation}
in which $\left(G^{\rm q-dq}\right)^{-1}(p,k,P)$ is the inverse of the full quark-diquark 4-point
function. It is the sum of the disconnected part and 
the interaction kernel,
\begin{eqnarray}
 \left(G^{\rm q-dq}\right)^{-1} (p,k,P) &=& (G_0^{\rm q-dq})^{-1}(p,k,P) -K^{\rm BS} \; ,
  \label{Gqdq}  \\
  (G_0^{\rm q-dq})^{-1}(p,k,P)  &=&
    (2\pi)^4 \;\delta^4(p-k)\; S^{-1}(p_q)\;
         \tilde D^{-1} (p_d) \; , \\
  K^{\rm BS}
    &= & 
  \frac{1}{2}
  \begin{pmatrix} - \chi^5{\Sc (p_1^2) } \; S^T{\Sc (q) }\; \bar\chi^5{\Sc
  (p_2^2) } &  
     \sqrt{3}\; \chi^{\mu'}{\Sc (p_1^2) }\; S^T{\Sc (q) }\;\bar\chi^5 {\Sc
  (p_2^2) } \\ 
    \sqrt{3}\;\chi^5{\Sc (p_1^2) }\; S^T{\Sc (q) }\;\bar\chi^{\mu}{\Sc (p_2^2) }
    &  \chi^{\mu'}{\Sc (p_1^2) }\; S^T{\Sc (q) }\;\bar\chi^{\mu}{\Sc (p_2^2) }
     \end{pmatrix} \; . \nonumber \\
   & &  
 \label{kbsdef}
\end{eqnarray}
For completeness we list the definitions of quark, diquark and
exchange quark momentum, see eqs.~(\ref{qm}--\ref{exm}),
\begin{eqnarray}
 p_q & =& \eta P +p \; , \\
 p_d & =& (1-\eta) P-p \; , \\
  q  & =& -p-k+(1-2\eta)P \; .
\end{eqnarray}
The diquark vertices and their conjugates depend on the
relative momenta between a spectator quark and the
exchange quark,
\begin{eqnarray}
 p_1 &=& p+ k/2 -(1-3\eta)P/2 \; , \\
 p_2 &=& -k -p/2 + (1-3\eta)P/2 \; .
\end{eqnarray}
Physical quantities should not depend on the momentum
partitioning parameter $\eta$ which expresses
just the invariance under reparametrizations of the relative momentum.
As we explained earlier, this follows from relativistic translation 
invariance for the solutions of the Bethe-Salpeter equation.
This implies that for
every solution  $\Psi(p,P;\eta_1)$ of the Bethe-Salpeter equation there exists 
a family of solutions of the form $\Psi(p+(\eta_2-\eta_1)P,P;\eta_2)$.
The manifestation of this invariance in actual numerical
calculations will be investigated in section \ref{solI}.

Color and flavor factors have already been worked out
in eqs.~(\ref{BS4},\ref{Gqdq}).
Their derivation has been relocated to appendix \ref{cfn}.

\subsection{Delta}

The $\Delta (1232)$ resonance is a spin-3/2, isospin-3/2
state. Therefore only axialvector diquark correlations can be present
in the 2-quark $t$ matrix since the flavor state of the $\Delta$ 
has to be fully symmetric. Accordingly its Faddeev amplitude 
reads
\begin{equation}
  \trisp{\Gamma^\Delta} = \twosp{\chi^\mu} \; D^{\mu\nu} \;
     (\Phi^{\nu\rho} u^\rho)_\alpha \; .
 \label{Faddeev_D}
\end{equation} 
The effective $\Delta$-quark-diquark vertices are now 
quantities which transform as vector-spinors:
$\Phi^{\mu\nu}(p,P)u^\nu(P)$, where $u^\nu(P)$ is a Rarita-Schwinger
spinor describing a free spin-3/2 particle with momentum $P$.
We may construct Bethe-Salpeter wave functions by attaching legs,
\begin{equation}
 \Psi^{\mu\nu}(p,P) = S(p_q)\; D^{\mu\mu'}(p_d)\; \Phi^{\mu'\nu}(p,P) \; .
 \label{wavedeldef}
\end{equation}
Both vertex and wave function $\Phi^{\mu\nu}$ and $\Psi^{\mu\nu}$
are Dirac matrices with two uncontracted Lorentz indices, respectively.
Their complete decomposition in Dirac space will
be shown in the next section.

The Bethe-Salpeter equation for the $\Delta$ is in compact notation
\begin{eqnarray}
  \fourint{k} \left(G_\Delta^{\rm q-dq}\right)^{-1} (p,k,P) 
  \Psi_\Delta^{\mu'\nu}(k,P) =0 \; ,
  \label{BS5}
\end{eqnarray}
where the inverse quark-diquark propagator $G^{-1}_\Delta$ in the
$\Delta$-channel is given by
\begin{eqnarray}
  \left(G_\Delta^{\rm q-dq}\right)^{-1}(p,k,P) &=&  (2\pi)^4 \delta^4(p-k)\; S^{-1} (p_q) \;
        (D^{\mu\mu'})^{-1} (p_d) + \nonumber \\ 
    & &  \chi^{\mu'}(p_1^2)\; S^T(q)\;\bar\chi^\mu(p_2^2) \; .
  \label{Gqdq_D}
\end{eqnarray}
The necessary color algebra that enters the explicit form
of $(G_\Delta^{\rm q-dq})^{-1}$ does not differ from the nucleon 
case as all diquarks are in a color antitriplet state. 
The computation of the flavor factor is given in appendix
\ref{cfd}.
\section{Decomposition of the Faddeev amplitudes}
\label{partialwaves}

At this point we will solve
the problem of the Dirac space decomposition of nucleon and $\Delta$
vertex functions with quark and diquark.
The exercise seems to be clearly straightforward
in the sense that, using a finite basis of Dirac matrices,
tensorial ($\Delta$) or vector and scalar matrices (nucleon)
have to be found that yield proper spin and describe positive
parity and energy solutions. We will go a step further, however,
and show that representations can be found which correspond
to a partial wave decomposition in the baryon rest frame,
{\em i.e.} a decomposition into eigenfunctions of spin
and angular momentum. Thereby we recover all non-relativistic
basis elements that would survive a corresponding reduction.
In addition, virtual components due to the off-shell
axialvector diquark and relativistic ``lower components'' will
be found. The results obtained in this section are thus an explicit 
illustration of the difference between relativistic
and non-relativistic baryon wave functions.

\goodbreak
\bigskip 

\noindent
{\it\large Nucleon}
\medskip

\noindent
In eq.~(\ref{Faddeev_N}) we have defined the nucleon Faddeev amplitudes
in terms of the effective spinors $\Phi^5(p,P)\;u(P)$ and 
$\Phi^\mu (p,P)\;u(P)$ that describe the scalar and the axialvector
correlations in the nucleon. 
In the following we will describe the complete decomposition of
the vertex function $\Phi$ 
in Dirac space such that the Faddeev amplitudes describe a spin-1/2 
particle with positive energy and have
positive parity.
 
First let us define the projectors onto positive energy, $\Lambda^+$, and
for negative energy, $\Lambda^-$, by 
\begin{equation}
 \Lambda^+(P)=\frac{1}{2} \left( 1 + \frac{\Slash{P}}{iM_n} \right)
 \; , \quad
 \Lambda^-(P)=\frac{1}{2} \left( 1 - \frac{\Slash{P}}{iM_n} \right)
 \; .
\end{equation} 
The nucleon bound state mass is denoted by $M_n$, and 
in the rest frame the nucleon momentum is given by
$P=(\vect 0, iM_n)$.
The projectors defined above form a complete set ($\Lambda^+ + \Lambda^- =1$) and 
thus the effective spinors can be written as
\begin{equation}
 \Phi\, u(P) = \Phi (\Lambda^+ + \Lambda^-)\, u(P) =  \Phi\, \Lambda^+ \,u(P).
\end{equation}
In the following, we will substitute the projected vertex function
for the vertex function itself,
$\Phi \Lambda^+ \rightarrow \Phi$. Clearly it is then 
an eigenfunction of the positive-energy projector,
\begin{equation}
  \Phi \Lambda^+ =\Phi \; .
  \label{positive}
\end{equation}

Next we consider the transformation properties of
the Faddeev amplitudes under parity. To describe
nucleons with positive parity the Faddeev amplitude
must fulfill
\begin{eqnarray} 
 {\mathbf P}\trisp{\Gamma}(q,p,P)&=& 
           \twosp{( \gamma^4 \chi^5(\tilde q) \gamma^4)}   \;
       D(\tilde p, \tilde P)\;
      ( \gamma^4 \Phi^5(\tilde p, \tilde P) u(\tilde P))_\alpha 
      + \nonumber \\
          & & 
           \twosp{( \gamma^4 \chi^\nu(\tilde q) \gamma^4)} \;
        D^{\nu\mu}(\tilde p, \tilde P) \;
        ( \gamma^4 \Phi^\mu(\tilde p, \tilde P) u(\tilde P))_\alpha \;
          \nonumber \\
          & \stackrel{!}{=}& \trisp{\Gamma}(q,p,P) \;,
 \label{parity1}
\end{eqnarray}
with $\tilde p=\Lambda_{\mathbf P}p, \tilde P=\Lambda_{\mathbf P}P$ and
\begin{equation}
 \Lambda_{\mathbf P}^{\mu\nu}= \text{diag}_4\; \{ -1,-1,-1,1\}\; .
\end{equation}
Using the transformation properties of the diquark-quark vertices 
and a Dirac spinor,
\begin{eqnarray}
  \gamma^4 \chi^5 \gamma^4 &=& \chi^5 \; , \\
  \gamma^4 \chi^\mu \gamma^4 &=& - \Lambda_{\mathbf P}^{\mu\nu} \chi^\nu
           \; , \\
  \gamma^4 u(P) &=& u(\tilde P )\; ,
\end{eqnarray}
one can deduce from eq.~(\ref{parity1})
a condition for the vertex function $\Phi$:
\begin{eqnarray}
 {\mathbf P} \bp \Phi^5(p,P) \\ \Phi^\mu(p,P) \ep &=&
   \bp \gamma^4 \Phi^5(\tilde p, \tilde P) \gamma^4 \\ 
             \gamma^4 (\Lambda_{\mathbf P}^{\mu\nu}
             \Phi^\nu (\tilde p, \tilde P)  )\gamma^4 \ep
   \stackrel{!}{=}
   \bp \Phi^5(p,P) \\ - \Phi^\nu(p,P) \ep \; . \label{parity2}
\end{eqnarray}
We see that apart from the Dirac indices $\Phi^5$ transforms
like a scalar and $\Phi^\mu$ like a pseudovector.

The conditions (\ref{positive}) and (\ref{parity2}) greatly
restrict the number of independent components in the
vertex function. The scalar correlations $\Phi^5$ are described
by two components and the axialvector correlations by six
components,
\begin{equation} 
 \label{vex_N}
 \bp \Phi^5 (p,P)  \\ \Phi^\mu(p,P) \ep  =
 \bp
   \sum \limits_{i=1}^2 S_i(p^2,p\cdt P) \; {\mathcal S}_i(p,P)  \\[4mm]
   \sum \limits_{i=1}^6 A_i(p^2,p\cdt P) \; \gamma_5 {\mathcal A}_i^\mu(p,P) 
 \ep \; .
\end{equation}
The scalar functions $S_i$ and $A_i$  depend on the
two possible scalars ($p^2$ and $p \cdt P$) that can be formed out of the
relative momentum $p$ and the nucleon momentum $P$. 
The Dirac components describing the scalar correlations
may be built out of
 $\Lambda^+$ and $\Slash{p} \Lambda^+$
and the Dirac part of the axialvector correlations 
can be constructed using the matrices
 $P^\mu \Lambda^+$, $P^\mu \Slash{p} \Lambda^+$, 
 $\gamma^\mu \Lambda^+$,   $\gamma^\mu \Slash{p} \Lambda^+$, 
 $p^\mu \Lambda^+$ and $p^\mu \Slash{p} \Lambda^+$.

In table \ref{components1}
the Dirac components ${\mathcal S}_i$ and ${\mathcal A}_i^\mu$
are given as 
 certain linear combinations
of these matrices. First, they obey a useful trace orthogonality
condition. Defining the adjoint vertex function in terms of 
conjugate Dirac components,
\begin{equation}
 \bp \bar \Phi^5 (p,P) \\ \bar \Phi^\mu (p,P) \ep  := 
 \bp \sum \limits_{i=1}^2 {\mathcal {\bar S}}_i (p,P)\; S_i (p^2,p\cdt P) \\
     \sum \limits_{i=1}^6 {\mathcal {\bar A}}_i^\mu (p,P) 
     \gamma_5 \; A_i (p^2,p\cdt P) \ep \; ,
\end{equation}
we find for them 
\begin{equation}
 \bp {\mathcal {\bar S}}_i (p,P) \\ {\mathcal {\bar A}}_i^\mu (p,P) \ep 
  =
 \bp C\; ( {\mathcal S}_i (-p,-P))^T\;C^{-1} \\
     -C\; ( {\mathcal A}_i^\mu (-p,-P))^T\;C^{-1} \ep \; .
\end{equation}
This identification of the conjugated components 
follows by charge-conjugating the
Faddeev amplitude and using the definition
of the conjugate diquark-quark vertices in eqs.~(\ref{dscon},\ref{dacon}).
The trace orthogonality property takes the following form,
\begin{eqnarray}
 \label{ortho_N1}
 \text{Tr}\; {\mathcal {\bar S}}_i {\mathcal S}_j &=& 
            2(-1)^{j+1}\delta_{ij} \\
 \label{ortho_N2}
 \text{Tr}\; {\mathcal {\bar A}}_i^\mu {\mathcal A}_j^\mu &=& 
            2(-1)^{j+1} \delta_{ij} \\
 \label{ortho_N3}
 \text{Tr}\; {\mathcal {\bar S}}_i \gamma_5 {\mathcal A}_j^\mu  =
  \text{Tr}\;{\mathcal {\bar A}}_j^\mu \gamma_5 {\mathcal S}_i &=&0
\end{eqnarray}

\begin{table}
 \begin{center}
 \begin{math}
  \begin{array}{crc} \hline \hline\\
    \phantom{++}{\mathcal S}_1\phantom{++} & \Lambda^+ & \\[2mm] 
    {\mathcal S}_2 & -i \,\hat \Slash{p}_T\, \Lambda^+  \\[2mm] \hline
    \\[-4mm]
    {\mathcal A}_1^\mu & \hat P^\mu\,\Lambda^+ \\[2mm]
    {\mathcal A}_2^\mu & -i \,\hat P^\mu\,\hat \Slash{p}_T \,\Lambda^+\\[2mm] 
    {\mathcal A}_3^\mu & \frac{1}{\sqrt{3}}\,\gamma^\mu_T\, \Lambda^+ \\[2mm]
    {\mathcal A}_4^\mu & \frac{i}{\sqrt{3}} \,\gamma^\mu_T\, 
        \hat \Slash{p}_T \,\Lambda^+ \\[2mm]
    {\mathcal A}_5^\mu & \sqrt{\frac{3}{2}}\left(\hat p^\mu_T \,\hat 
        \Slash{p}_T - \frac{1}{3}\, \gamma^\mu_T \right) \Lambda^+ \\[2mm]
    {\mathcal A}_6^\mu & i \sqrt{\frac{3}{2}} \left( \hat p^\mu_T - 
        \frac{1}{3} \,\gamma^\mu_T \,\hat \Slash{p}_T \right) \Lambda^+ \\
    \\ \hline \hline
  \end{array}
 \end{math}
 \end{center}
 \caption{Basic Dirac components of the nucleon vertex function.
   With a hat we denote normalized 4-vectors, $\hat p=\frac{p}{|p|}$.
   In the case of the complex on-shell nucleon momentum, we define
   $\hat P=\frac{P}{iM}$. The subscript $_T$ denotes the transversal
   component of a vector with respect to the nucleon momentum $P$, {\em e.g.}
   $p_T=p-(p\cdt \hat P) \hat P$.}
 \label{components1} 
\end{table}

As a second attribute, the Faddeev amplitude written with 
these Dirac components  is given as a complete
decomposition into partial waves, {\em i.e.} into components
that are eigenstates of the spin  as well as
of the angular momentum operator in the nucleon rest frame. This will
be shown in section \ref{partial_sec}.

\bigskip
\noindent
{\it\large Delta}
\medskip

\noindent
The Faddeev amplitudes for the $\Delta$ are described by an
effective Rarita-Schwinger spinor $\Phi^{\mu\nu} u^\nu$, see 
eq.~(\ref{Faddeev_D}).
As in the nucleon case we only need to consider the projected
tensor vertex function $\Phi^{\mu\rho} {\mathbb P}^{\rho\nu}
\rightarrow \Phi^{\mu\nu}$. ${\mathbb P}^{\rho\nu}$ is the 
Rarita-Schwinger projector \cite{Lurie:1968} 
onto positive-energy, spin-3/2 spinors,
\begin{equation} \label{RSP}
{\mathbb P}^{\mu\nu}:=\Lambda^+\left(\delta^{\mu\nu}
-\frac{1}{3}\gamma^\mu\gamma^\nu+
\frac{2}{3} \frac{P^\mu P^\nu}{M_\Delta^2} - 
\frac{i}{3} \frac{P^\mu\gamma^\nu-P^\nu\gamma^\mu}{M_\Delta} \right) 
= :\Lambda^+ \Lambda^{\mu \nu} \;.
\end{equation}  
It obeys the constraints
\begin{equation}
 P^\mu \,{\mathbb P}^{\mu\nu} = \gamma^\mu \,{\mathbb P}^{\mu\nu} =0 \; .
 \label{RSPcon}
\end{equation}
The tensor vertex function must be an eigenfunction of the
projector, therefore
\begin{equation}
  \Phi^{\mu\nu}=T^{\mu\rho}\, {\mathbb P}^{\rho\nu} \; .
\end{equation}
Due
to the constraints (\ref{RSPcon}), $T^{\mu\rho}$ must be either
proportional to $\delta^{\mu\rho}$ or the transverse relative
momentum $p_T=p-\hat P (p\cdt \hat P)$ 
may contract with one index of the Rarita-Schwinger 
projector and therefore 
\begin{equation}
  T^{\mu\rho} = {\mathcal D}' \delta^{\mu\rho} +
                {\mathcal E'} ^\mu \hat p_T^\rho \; ,
 \label{ED}
\end{equation}
with ${\mathcal D}'$ and ${\mathcal E'}^\mu$ being Dirac matrices
yet to be determined.

To further constrain the vertex function, let us consider the effect of
the parity transformation onto
the Faddeev amplitude,
\begin{eqnarray} 
 {\mathbf P}\trisp{\Gamma^\Delta}(q,p,P)&=& 
           \twosp{( \gamma^4 \chi^\rho(\tilde q) \gamma^4)} \;
       D^{\rho\mu}(\tilde p, \tilde P)\;
    ( \gamma^4 \Phi^{\mu\nu}(\tilde p, \tilde P) u^\nu(\tilde P))_\alpha 
      \nonumber \\
          & \stackrel{!}{=}& \trisp{\Gamma^\Delta}(q,p,P) \;.
 \label{parity3}
\end{eqnarray}
We need to evaluate the effect of  a parity transformation on the 
Rarita-Schwinger spinor.
The latter may be built as follows \cite{Nozawa:1990gt},
\begin{equation}
 u^\mu(P,s)= \sum_{m,s'}   C_{1m,\frac{1}{2}s'}^{\frac{3}{2}s}\;
              \epsilon^\mu_m(P) \;u(P,s') \;.
 \label{RSS}
\end{equation}
The $\epsilon^\mu_m$ are spin-1 polarization vectors 
$(m=0, \pm 1)$
that can be written 
in terms of a 3-dimensional spin-1 polarization vector basis
$\hat \vect{\epsilon}_m$ as
\begin{equation}
 \epsilon^\mu_m (P)= \bp \hat \vect{\epsilon}_m + 
           \frac{\hat \vect{\epsilon}_m \cdot \vect{P}}{\Dc M_\Delta}
           \frac{\vect{P}}{\Dc |P^4|+M_\Delta} \\
              i \frac{\hat \vect{\epsilon}_m\cdot \vect{P}}{\Dc M_\Delta} \ep\;.
\end{equation}
In eq.~(\ref{RSS}), these covariant polarization vectors
are combined with a Dirac spinor $u(P)$ of helicity $s'$ and 
appropriate Clebsch-Gordan
coefficients $C_{1m,\frac{1}{2}s'}^{\frac{3}{2}s}$
to give a spin-3/2 object with helicity $s$.
\begin{table}
 \begin{center}
 \begin{math}
  \begin{array}{crc} \hline \hline \\
    \phantom{++}{\mathcal D}_1^{\mu\nu}\phantom{++} & \delta^{\mu\rho}\, {\mathbb P}^{\rho\nu} &  \\[2mm]
    {\mathcal D}_2^{\mu\nu} & -i\,\frac{3}{5}\sqrt{5} 
          \left( \hat \Slash{p}_T \,\delta^{\mu\rho}
        - i\,\frac{2}{3}\,\gamma^\mu_T \,\hat p_T^\rho \right)
          {\mathbb P}^{\rho\nu}  \\[2mm]
    {\mathcal D}_3^{\mu\nu} & \sqrt{3}\,
          \hat \Slash{p}_T \, \hat P^\mu \, \hat p_T^\rho\,
          {\mathbb P}^{\rho\nu} \\[2mm]
    {\mathcal D}_4^{\mu\nu} & i\sqrt{3} \,\hat P^\mu \,\hat p_T^\rho\,
          {\mathbb P}^{\rho\nu} \\[2mm]
    {\mathcal D}_5^{\mu\nu} & \gamma^\mu_T \,\hat \Slash{p}_T\, p_T^\rho\,
          {\mathbb P}^{\rho\nu} \\[2mm]
    {\mathcal D}_6^{\mu\nu} & -i \,\gamma^\mu_T \,p_T^\rho\,
          {\mathbb P}^{\rho\nu} \\[2mm]
    {\mathcal D}_7^{\mu\nu} &  3 \left( p_T^\mu\, p_T^\rho - \frac{1}{3}
          \left[ \delta^{\mu\rho} + \gamma^\mu_T \, \hat \Slash{p}_T \,p_T^\rho
          \right]\right) {\mathbb P}^{\rho\nu} \\[2mm]
    {\mathcal D}_8^{\mu\nu} & -i \sqrt{5} 
          \left( \hat \Slash{p}_T\, p_T^\mu \, p_T^\rho
          -\frac{1}{5}\left[  \hat \Slash{p}_T\, \delta^{\mu\rho}
          +  \gamma^\mu_T \,p_T^\rho \right] \right)
          {\mathbb P}^{\rho\nu} \\
  \\ \hline \hline
  \end{array}
 \end{math}
 \end{center}
 \caption{Basic Dirac components of the $\Delta$  vertex function.}
 \label{components3} 
\end{table}
From these definitions it can be seen that
under parity the $u^\mu$ transform as follows,
\begin{equation}
  \gamma^4 \Lambda_{\mathbf P}^{\mu\nu}
         u^\nu(\tilde P) = - u^\mu(P) \; ,
\end{equation}
and therefore the vertex function must obey
\begin{equation}
 \mathbf P \Phi^{\mu\nu}(p,P) = \gamma^4 
   \;(\Lambda_{\mathbf P}^{\mu\rho}\Phi^{\rho\lambda}(\tilde p,\tilde P) 
      \Lambda_{\mathbf P}^{\lambda\nu} )\; 
      \gamma^4 = \Phi^{\mu\nu}(p,P)
\end{equation}
which translates into conditions for the unknown Dirac matrices appearing
in eq.~(\ref{ED}),
\begin{equation}
  \gamma^4 \bp {\mathcal D'}(\tilde p, \tilde P) \\ \Lambda_{\mathbf P}^{\mu\nu}
               {\mathcal E'}^\nu(\tilde p, \tilde P) \ep
  \gamma^4 = \bp {\mathcal D'}( p,  P) \\ 
               {\mathcal E'}^\mu( p, P) \ep \; .
\end{equation}
These transformation properties suggest that ${\mathcal D'}$ is
a linear combination of the ${\mathcal S_i}$ and ${\mathcal E'}^\mu$
can be composed using the ${\mathcal A}_i^\mu$ that were found for
the nucleon. Again we will reshuffle the matrices to find
a decomposition that is orthogonal and yields  partial wave components:
\begin{equation}
 \Phi^{\mu\nu} (p,P) = \sum_{i=1}^8 D_i(p^2,p\cdt P)\;
     {\mathcal D}_i^{\mu\nu}(p,P) \; .
    \label{vex_D}
\end{equation}
The  eight components ${\mathcal D}_i^{\mu\nu}$ are given in 
table \ref{components3}. To state the orthogonality
condition we write the adjoint vertex function as
\begin{equation}
 \bar \Phi^{\mu\nu} (p,P) =
     \sum_{i=1}^8 {\mathcal{\bar D}}_i^{\mu\nu}(p,P) 
       D_i(p^2,p\cdt P) \; .
\end{equation}
The definition of the adjoint Dirac components
is inferred from the charge-con\-ju\-ga\-ted Faddeev amplitude
and the corresponding relations for the axialvector
diquark vertex and the Rarita-Schwinger spinor,
\begin{equation}
 {\mathcal{\bar  D}}_i^{\mu\nu}(p,P) = C \;( {\mathcal D}_i^{\nu\mu}
          (-p,-P))^T \;C^{-1} \; .
\end{equation}
Orthogonality of the basic components can now be expressed as
\begin{equation}
 \label{ortho_D}
  \text{Tr} \; {\mathcal{\bar  D}}_i^{\mu\rho}
                {\mathcal D}_j^{\rho\mu} = 4(-1)^{j+1} \delta^{ij} \; . 
\end{equation}

\subsection{Partial wave decomposition}

\label{partial_sec}
In a general moving frame,  all the components of the 
 Faddeev amplitude $\trisp{\Gamma^{[\Delta]}}$
possess just the total angular momentum, 1/2 for
the nucleon and 3/2 for the $\Delta$, as a good quantum number.
A further interpretation of the components is not obvious. 
In the rest frame of the bound state, however, the 
trispinor $\trisp{\Gamma^{[\Delta]}}$ can be written as a sum of
trispinor components each possessing definite orbital angular
momentum and spin, thus allowing a direct interpretation
in terms of partial waves.

First we set up the general formalism and apply it 
to nucleon and $\Delta$ in turn.
In the rest frame the Pauli-Lubanski operator for an
arbitrary tri-spinor 
$\trisp{\psi}$ is given by
\begin{equation}
 W^i=\frac{1}{2}\epsilon_{ijk}{\mathcal L}^{jk} \; ,
\end{equation}
whose square characterizes the total angular momentum,
\begin{equation}
 {\mathbf W}^2\trisp{\psi}=J(J+1) \trisp{\psi} \; .
\end{equation}
The tensor ${\mathcal L}$ is the sum of an orbital part, $L$, and
a spin part, $S$, which read
\begin{eqnarray}
 L^{jk}&=&\sum_{a=1}^3 (-i)\left(p_a^j\frac{\partial}{\partial p_a^k}-
         p_a^k\frac{\partial}{\partial p_a^j}\right) \; , \\
 2(S^{jk})_{\alpha\alpha',\beta\beta',\gamma\gamma'}&=&
 (\sigma^{jk})_{\alpha\alpha'}\otimes{\delta}_{\beta\beta'}\otimes{\delta}_{\gamma\gamma'}+
 {\delta}_{\alpha\alpha'}\otimes (\sigma^{jk})_{\beta\beta'}
 \otimes{\delta}_{\gamma\gamma'}+ \nonumber \\
  & & {\delta}_{\alpha\alpha'}\otimes{\delta}_{\beta\beta'}\otimes(\sigma^{jk})_{\gamma\gamma'} \; ,  
\end{eqnarray}        
such that ${\cal L}=L+S$. 
The tensor $L$ is proportional to the unit matrix in Dirac space. 
The definition $\sigma^{\mu\nu}:=-\frac{i}{2}[\gamma^\mu,\gamma^\nu]$ differs
by a minus sign from its Minkowski counterpart.
The tensors $L$ and $S$ are written
as a sum over the respective tensors for each of the three constituent quarks
which are labelled $a=1,2,3$ and with respective Dirac indices 
${\Sc \alpha\alpha',
\beta\beta',\gamma\gamma'}$.

With the definition
of the spin matrix $\Sigma^i=\frac{1}{2}\epsilon_{ijk}\sigma^{jk}$ the Pauli-Lubanski
operator reads
\begin{eqnarray}
(W^i)_{\alpha\alpha',\beta\beta',\gamma\gamma'} &=& L^i\;
  {\delta}_{\alpha\alpha'}\otimes{\delta}_{\beta\beta'}\otimes\delta_{\gamma\gamma'}
  +(S^i)_{\alpha\alpha',\beta\beta',\gamma\gamma'}\; ,  \\
L^i    &=&(-i)\epsilon_{ijk} \left[ p^j\frac{\partial}{\partial p^k}
                   + q^j \dpartial{q^k} \right] \; , \label{Ldef} \\
(S^i)_{\alpha\alpha',\beta\beta',\gamma\gamma'}   &=& \frac{1}{2} \left( 
   (\Sigma^{i})_{\alpha\alpha'}\otimes{\delta}_{\beta\beta'}\otimes
    {\delta}_{\gamma\gamma'}+
 {\delta}_{\alpha\alpha'}\otimes (\Sigma^{i})_{\beta\beta'}\otimes
 {\delta}_{\gamma\gamma'}+ \right. \nonumber \\
    & & \left.{\delta}_{\alpha\alpha'}\otimes{\delta}_{\beta\beta'}\otimes
 (\Sigma^{i})_{\gamma\gamma'}
 \right),
\end{eqnarray}
where we have already introduced the relative momentum $p$ between quark and diquark
and the relative momentum $q$ within the diquark
via a canonical transformation:
\begin{equation}
P=p^1+p^2+p^3,\quad p=\eta (p^1+p^2)-(1-\eta)p^3, \quad q=\frac{1}{2}(p^1-p^2) \; .
\end{equation}
In the diquark-quark model, the quark ($\alpha$) and diquark 
(${\beta\gamma}$)
Dirac indices
of $\trisp{\psi}$ do separate.
In  the parametrization of  the diquark-quark vertices  
$\twosp{\chi^5}(q)$ and $\twosp{\chi^\mu}(q)$,
{\em cf.} eqs.~(\ref{dqvertex_s},\ref{dqvertex_a}), 
we retained only the dominant Dirac component along
with a scalar function of $q$. Therefore no orbital angular momentum
is carried by the diquarks,
\begin{equation}
 {\mathbf L}^2 \twosp{\chi^5}(q)= {\mathbf L}^2 \twosp{\chi^\mu}(q) =0 \; ,
\end{equation}
and the second term in the definition of $L^i$ in eq.~(\ref{Ldef})
depending on $q$ may safely be dropped.
The operator ${\bf W}^2$ now takes the form
\begin{eqnarray}
 {\bf W}^2&=& {\bf L}^2+2 \vect{L} \cdt \vect{S} + {\bf S}^2 \; ,   \\
{\bf L}^2 &=& \left(2p^i\frac{\partial}{\partial p^i}-\vect{p}^{2}\Delta_{p}+p^ip^j\frac{\partial}{\partial p^i}
          \frac{\partial}{\partial p^j}\right) \;, \label{L2} \\
 2 ({\bf L}\cdt{\bf S})_{\alpha\alpha',\beta\beta',\gamma\gamma'}&=& 
   -\epsilon_{ijk}p^j\frac{\partial}{\partial p^k} \left(
    (\Sigma^{i})_{\alpha\alpha'}\otimes{\delta}_{\beta\beta'}
     \otimes{\delta}_{\gamma\gamma'}+\right. \nonumber \\
  & &\left.{\delta}_{\alpha\alpha'}\otimes \left[(\Sigma^{i})_{\beta\beta'}
     \otimes{\delta}_{\gamma\gamma'}+
     {\delta}_{\beta\beta'}\otimes(\Sigma^{i})_{\gamma\gamma'}\right]
      \right) \; ,    \\
  ({\bf S}^2)_{\alpha\alpha',\beta\beta',\gamma\gamma'}  &= &
   \quarter\left(9\,{\delta}_{\alpha\alpha'}\otimes{\delta}_{\beta\beta'}
  \otimes{\delta}_{\gamma\gamma'} + \right. \nonumber \\
  & &\left. \quad 2\,{\Sigma^i}_{\alpha\alpha'}\otimes \left[(\Sigma^{i})_{\beta\beta'}\otimes{\delta}_{\gamma\gamma'}+
 {\delta}_{\beta\beta'}\otimes(\Sigma^{i})_{\gamma\gamma'}\right]\right. + \nonumber \\
    & & \left. \quad 2\,{\delta}_{\alpha\alpha'}\otimes(\Sigma^{i})_{\beta\beta'}\otimes 
 (\Sigma^{i})_{\gamma\gamma'}\right) \; . \label{S2}        
\end{eqnarray} 

For evaluating the action of the spin operator $\Sigma^i$
on the diquark-quark vertices $\chi^5$ and $\chi^\mu$ 
the following identities proved to be useful,
\begin{eqnarray}
\Sigma^j(\gamma^5 C)+(\gamma^5 C)(\Sigma^j)^T&=&0 \; ,\\
 & & \nonumber \\
\Sigma^j\bp (\gamma^4 C) \\ (\gamma^i C) \ep +\bp (\gamma^4 C) \\ (\gamma^i C) \ep
   (\Sigma^j)^T &=& \bp 0 \\ 2i\epsilon_{mji} (\gamma^m C)\ep \; , \\
   & & \nonumber \\
\Sigma^j (\gamma^5 C) (\Sigma^j)^T &=& -3 (\gamma^5 C) \; , \\
& & \nonumber \\
\Sigma^j \bp (\gamma^4 C) \\ (\gamma^i C) \ep (\Sigma^j)^T &=& 
   \bp -3(\gamma^4 C) \\ (\gamma^i C) \ep \; .
\end{eqnarray}

\begin{table}
 \begin{center}
 \begin{math}
  \begin{array}{cccc}  \hline \hline \\
\text{Dirac} & \text{Faddeev amplitude} & \text{EV} & \text{EV} \\ 
\text{com-} & \text{component}  & l(l+1) & s(s+1) \\ 
 \text{ponent} & & \text{of ${\bf L}^2$} & \text{of ${\bf S}^2$}\\ \\ \hline
  & & & \\
{\cal S}_1 &  \bp \varsigma \\  0 \ep_\alpha  \twosp{(\gamma_5 C)} & 0 & \frac{3}{4} \\[2mm]
{\cal S}_2 & \bp 0\\  (\vect{\sigma}\hat \vect{p})\varsigma \ep_\alpha \twosp{(\gamma_5 C)} & 2 & \frac{3}{4} \\[2mm]
 & & & \\
{\cal A}^\mu_{1} & \hat P^4\bp 0\\ \varsigma \ep_\alpha \twosp{(\gamma^4 C)} & 0 & \frac{3}{4} \\[2mm]
{\cal A}^\mu_{2} & \hat P^4\bp (\vect{\sigma}\hat \vect{p})\varsigma\\  0\ep_\alpha \twosp{(\gamma^4 C)} & 2 &\frac{3}{4}\\[2mm]
 & & & \\
{\cal A}^\mu_{3} & \sqrt{\third} \bp i\sigma^i\varsigma \\  0 \ep_\alpha \twosp{(\gamma^i C)} & 0 & \frac{3}{4} \\
{\cal A}^\mu_{4} & \sqrt{\third} \bp 0\\ \sigma^i(\vect{\sigma}\hat\vect{p})\varsigma\ep_\alpha \twosp{(\gamma^i C)} & 2 & \frac{3}{4}\\ 
 & & & \\
{\cal A}^\mu_{5} & \sqrt{\frac{3}{2}}\bp i\left(\hat p^i(\vect{\sigma}\hat\vect{p})-\frac{1}{3}\sigma^i\right)\varsigma\\  0\ep_\alpha \twosp{(\gamma^i C)} & 6 & \frac{15}{4} \\ 
{\cal A}^\mu_{6} & \sqrt{\frac{3}{2}} \bp 0\\  i\left(\hat p^i-\frac{1}{3}\sigma^i(\vect{\sigma}\hat \vect{p})\right)\varsigma\ep_\alpha \twosp{(\gamma^i C)} & 2 & \frac{15}{4} \\ 
  \\ \hline \hline            
  \end{array}
 \end{math}
 \end{center}
\caption{Classification of the components of the nucleon Faddeev amplitude
in terms of eigenfunctions of ${\bf L}^2$ and ${\bf S}^2$ in the rest frame of the
bound state. `EV' abbreviates `eigenvalue' and $\varsigma$ denotes an arbitrary Pauli two-component 
spinor. Each Faddeev component is to be multiplied with the diquark
width function $V(x=q^2)$. Normalized 4-vectors are $\hat p=p/|p|$ and
normalized 3-vectors are correspondingly $\hat\vect{p}=\vect{p}/|\vect{p}|$.}
\label{partial1}
\end{table}

\goodbreak
\bigskip
\noindent
{\it\large Nucleon}
\medskip

\noindent
We will make use of the Bethe-Salpeter
wave function $\Psi$ defined in eq.~(\ref{wavenucdef}). 
The Faddeev takes the following form, if
the wave function instead of the vertex function is used,
\begin{eqnarray}
 S_{\alpha\alpha'}(\eta P+p) \Gamma_{\alpha'\beta\gamma} (q,p,P)&=& 
      ( \Psi^5(p,P) u(P))_\alpha\; 
      \twosp{\chi^5(q)}   + \nonumber \\
          & & 
      ( \Psi^\mu (p,P) u(P))_\alpha \;
           \twosp{\chi^\mu(q) } \; . 
 \label{fadn_to_deco}
\end{eqnarray}
We will seek eigenfunctions of the spin and
orbital angular momentum operators ${\bf L}^2$ and ${\bf S}^2$,
where the eigenfunctions shall be of the general form as
indicated above. Thus they represent
components of the Faddeev amplitude
with the spectator quark leg attached, $\trisp{(S\Gamma)}$.
We note that 
for the wave function $\Psi$ an analogous decomposition as
for the vertex function holds, see eq.~(\ref{vex_N}),
\begin{equation} 
 \label{wex_N}
 \bp \Psi^5 (p,P)  \\ \Psi^\mu(p,P) \ep  =
 \bp
   \sum \limits_{i=1}^2 \hat S_i(p^2,p\cdt P) \; {\mathcal S}_i(p,P)  \\[4mm]
   \sum \limits_{i=1}^6 \hat A_i(p^2,p\cdt P) \; \gamma_5 {\mathcal A}_i^\mu (p,P)
 \ep \; .
\end{equation}
A new set of scalar functions, $\hat S_i$ and $\hat A_i$,  has been introduced
whereas the basic Dirac components are the ones from table
\ref{components1}. The reason why we choose the wave function for further
treatment is that the Faddeev amplitude written with the
vertex function depends explicitly on the axialvector
diquark propagator $D^{\mu\nu}$ which for the Proca form 
induces mixing between space components of the vertex function and
the (virtual) time component of the axialvector diquark (and 
{\em vice versa}) in which case no partial wave decomposition can be found.
This can be circumvented by using the wave function as 
in virtue of its definition the diquark propagator has become
absorbed in $\Psi$.

In table \ref{partial1} we have listed all eight components together
with the respective eigenvalues of ${\mathbf L}^2$ and ${\mathbf S}^2$.
The Faddeev components that are given in the rest frame of the nucleon
have been built using the basic Dirac components for
the wave function shown in table \ref{components1}.

There is one $s$ wave associated with the scalar diquark, described
by ${\mathcal S}_1$, and two $s$ waves associated with the axialvector
diquark, one connected with its  virtual time component, 
${\mathcal A}_1$, and the other described by ${\mathcal A}_3$.
In a non-relativistic limit where diquark off-shell components
are neglected, only the $s$ waves ${\mathcal S}_1$, ${\mathcal A}_3$
and the $d$ wave ${\mathcal A}_5$ would survive. It is remarkable
that the relativistic description leads to accompanying four $p$ waves,
the ``lower components'', which are expected to give substantial
contributions to the fraction of the nucleon spin carried
by orbital angular momentum. These $p$ waves would not be present in a 
non-relativistic model.

\bigskip
\goodbreak
\noindent
{\it\large Delta}
\medskip

\noindent
The Faddeev amplitude written in terms of
the tensor wave function $\Psi^{\mu\nu}$ defined in 
eq.~(\ref{wavedeldef}) reads
\begin{equation}
 S_{\alpha\alpha'}(\eta P+p)\Gamma^\Delta_{\alpha'\beta\gamma} (q,p,P)= 
      ( \Psi^{\mu\nu} (p,P) u^\nu(P))_\alpha \;
           \twosp{\chi^\mu(q) } \; . 
\end{equation}
Again, the wave function decomposition proceeds in the same
manner as done for the $\Delta$ vertex function
\begin{equation}
 \label{wex_D}
 \Psi^{\mu\nu} (p,P) = \sum_{i=1}^8 \hat D_i(p^2,p \cdt P)\;
     {\mathcal D}_i^{\mu\nu}(p,P) \; .
\end{equation}
Applying the operators ${\bf L}^2$ and ${\bf S}^2$, 
eqs.~(\ref{L2},\ref{S2}), to the Faddeev amplitude 
$\trisp{(S\Gamma^\Delta)}$, this decomposition is seen to
yield eigenfunctions of the spin and orbital angular
momentum operators. Their explicit form in the rest frame 
is shown in table \ref{partial2}. The strength of the 
partial waves is determined by the scalar functions 
$\hat D_i(p^2,p \cdt P)$. 

In contrast to the nucleon, only one $s$ wave (described
by ${\mathcal D}_1$) is found. Two $d$ waves that survive
the non-relativistic limit are given by ${\mathcal D}_5$
and ${\mathcal D}_7$, and one $d$ wave can be attributed
to the virtual time component of the axialvector diquark, 
${\mathcal D}_3$. All even partial waves are accompanied
by relativistic ``lower'' components that could be even more important 
as in the nucleon case for only one dominant $s$ wave is present
compared to three within the nucleon.

\begin{table}
 \begin{center}
 \begin{math}
  \begin{array}{cccc}  \hline \hline \\
\text{Dirac} & \text{Faddeev amplitude} & \text{EV} & \text{EV} \\ 
\text{com-} & \text{component}  & l(l+1) & s(s+1) \\ 
 \text{ponent} & & \text{of ${\bf L}^2$} & \text{of ${\bf S}^2$} \\ \\ \hline
  & & & \\
  {\mathcal D}_1^{\mu\nu} & \bp \varsigma^i \\ 0 \ep_\alpha \twosp{(\gamma^i C)} & 0 & \frac{15}{4} \\  
  {\mathcal D}_2^{\mu\nu} & \frac{3}{5}\sqrt{5}\bp 0\\ \left((\vect{\sigma}\hat\vect{p})\varsigma^i 
-\frac{2}{3}\sigma^i(\hat\vect{p}\vect{\varsigma})\right)\ep_\alpha \twosp{(\gamma^i C)} & 2 & \frac{15}{4} \\ 
  & & & \\
  {\mathcal D}_3^{\mu\nu} & i\sqrt{3}\hat P^4\bp 0\\ (\vect{\sigma}\hat\vect{p})(\hat\vect{p}\vect{\varsigma}) \ep_\alpha
\twosp{(\gamma^4 C)} & 6 & \frac{3}{4} \\
  {\mathcal D}_4^{\mu\nu} & i\sqrt{3}\hat P^4\bp (\hat \vect{p}\vect{\varsigma})\\ 0\ep_\alpha \twosp{(\gamma^4 C)} & 2 &\frac{3}{4}\\
  & & & \\
  {\mathcal D}_5^{\mu\nu} & \bp \sigma^i(\vect{\sigma}\hat\vect{p})(\hat\vect{p}\vect{\varsigma}) \\
 0 \ep_\alpha \twosp{(\gamma^i C)} & 6 & \frac{3}{4} \\
  {\mathcal D}_6^{\mu\nu} & \bp 0\\ \sigma^i(\hat \vect{p}\vect{\varsigma})\ep_\alpha
  \twosp{(\gamma^i C)} & 2 & \frac{3}{4} \\ 
  & & & \\
  {\mathcal D}_7^{\mu\nu} & 3\bp \hat p^i(\hat\vect{p}\vect{\varsigma})-\frac{1}{3}[\varsigma^i
   +\sigma^i (\vect{\sigma}\hat\vect{p})(\hat\vect{p}\vect{\varsigma})]   \\
         0\ep_\alpha \twosp{(\gamma^i C)} & 6 & \frac{15}{4} \\ 
  {\mathcal D}_8^{\mu\nu} & \sqrt{5} \bp 0\\ \left(\hat p^i(\vect{\sigma}\hat\vect{p})(\hat\vect{p}\vect{\varsigma})
  -\frac{1}{5}[\sigma^i(\hat \vect{p}\vect{\varsigma})+
   (\vect{\sigma}\hat \vect{p})\varsigma^i]\right) \ep_\alpha
               \twosp{(\gamma^i C)} & 12 & \frac{15}{4} \\ 
  & & & \\  \hline \hline 
  \end{array}
 \end{math}
 \end{center}
\caption{Classification of the components of the $\Delta$ 
Faddeev amplitude
in terms of eigenfunctions of ${\bf L}^2$ and ${\bf S}^2$ in the rest frame of the
bound state. $\varsigma^i$ denotes an arbitrary  
2$_{\text{Pauli}}\times$3$_{\text{vector}}$-component 
spinor that represents the non-vanishing components of $u^\mu$ in the
$\Delta$ rest frame.
Each Faddeev component is to be multiplied with the diquark
width function $V(x=q^2)$.}
\label{partial2}
\end{table}

To summarize, the relativistic decomposition of nucleon
and $\Delta$ quark-diquark wave functions yields a rich
structure in terms of partial waves. Well-known problems
from certain non-relativistic quark model descriptions
are avoided from the beginning in a relativistic treatment.
First, photoinduced $N-\Delta$ transitions that are
impossible in spherically symmetric non-relativistic
nucleon ground states (described by only one $s$ wave) will occur
in our model through overlaps in the axialvector part
of the respective wave functions. Additionally, 
photoinduced transitions
from scalar to axialvector diquarks can take place, thus creating an 
overlap of the nucleon scalar diquark correlations with
the $\Delta$ axialvector diquark correlations. 
Secondly, although the total baryon spin will be mainly made
of the quark spin in the $s$ waves,
the orbital momentum of the relativistic $p$ waves
contributes as well which could not happen in a non-relativistic
description since the $p$ waves are absent there. 

\pagebreak

\section{Numerical solutions}
\label{num-sol}

After having described the numerical algorithm 
that solves the Bethe-Salpeter equations for nucleon and $\Delta$,
we will first present solutions for the nucleon
with scalar diquark correlations only. These results, published
in ref.~\cite{Oettel:1999gc}, will provide us with constraints on the
scalar functions which enter the diquark-quark vertices. 
Especially results on the neutron electric charge radius and
the proton electric form factor
signal the superiority of using a dipole form for the
scalar function $P$ appearing in eqs.~(\ref{dqvertex_s},\ref{dqvertex_a}).
Although presenting form factors at this point is an early anticipation
of results which are not derived until chapter \ref{em-chap},
it motivates the choice of the specific diquark-quark vertex  
in the full scalar-axialvector calculations of section \ref{solII}.

\subsection{Numerical method}
\label{num-meth}

We will solve the Bethe-Salpeter equations for nucleons, eq.~(\ref{BS4}),
and for the $\Delta$, eq.~(\ref{BS5}), in the baryon rest frame 
as a system of equations for vertex
and wave function.
 In a concise notation the system
of equations reads
\begin{eqnarray}
 \psi(p,P) &=& G_0^{\rm q-dq}(p,P) \; \phi(p,P) \label{BS6} \\ 
 \phi(p,P) &=& \fourint{k} \; K^{\rm BS} \, \psi(k,P) \; . \label{BS6a} 
\end{eqnarray}
The product of quark and diquark propagators is given by
$G_0^{\rm q-dq}$ and $K^{\rm BS}$ is the quark exchange kernel. 

The procedure to solve the equations can be divided into the 
following steps:
\begin{itemize}
 \item We will project out the unknown scalar functions 
       describing the partial waves from
       the expansions for wave and vertex function by contraction
       with appropriate Dirac components.
 \item As the scalar functions depend on {\em two} Lorentz
       scalars, $p^2$ and $p \cdt P=iM|p|\,\hat p \cdt \hat P$, 
       we shift the dependence
       on $\hat p \cdt \hat P$ into an expansion in Chebyshev polynomials
       and project the equations onto the expansion coefficients,
       the Chebyshev moments. We will see that this expansion
       is close to a hyperspherical expansion.
 \item The interaction kernel can be {\em approximated} by a finite
       number of Chebyshev polynomials. Employing
       appropriate orthogonality relations, the equations
       are reduced to a coupled system of one-dimensional
       integral equations for the Chebyshev moments.       
       The numerical effort will be considerably reduced compared
       to the brute force method which puts wave and vertex function
       on a two-dimensional grid, since only a few 
       Chebyshev polynomials will be needed.
\end{itemize}       
These steps are described in  more detail in the following.

The wave function $\psi$ and the vertex function $\phi$ 
have been expanded in Dirac space according
to eqs.~(\ref{vex_N},\ref{wex_N}) in the nucleon case and 
for the $\Delta$ according 
to eqs.~(\ref{vex_D},\ref{wex_D}).
We will use the generic labels $Y_i$ ($i=1,\dots, 8$)
for the scalar functions describing the partial waves in
the vertex function and likewise $\hat Y_i$ for the
wave function. The scalar functions depend in turn on the
scalars $p^2$, $\hat p \cdt \hat P$, and the latter scalar reduces
in the baryon rest frame to $z=\cos \Theta$ where
$\Theta$ is the angle between $p$ and the 4-axis,
see appendix \ref{conv-app} for our conventions for
hyperspherical coordinates.
For numerical convenience, we redefine the scalar functions
associated with the ``lower'' components,
\begin{equation}
 Y_i[\hat Y_i] \rightarrow \frac{1}{\sqrt{1-z^2}}\;Y_i[\hat Y_i]
 ,\qquad (i \;\;{\rm odd})\; .
\end{equation}
The set of scalar functions is now expanded into
Chebyshev polynomials of the first kind \cite{Abramowitz:1965},
\begin{eqnarray}
 \label{cheby-v}
  Y_i(p^2,z) &=& \sum_{n=0}^\infty i^n Y_i^n(p^2)\;T_n(z) \; ,\\
  \hat Y_i(p^2,z) &=& \sum_{n=0}^\infty i^n
      \hat Y_i^n(p^2)\;T_n(z)\; ,  \label{cheby-w}
\end{eqnarray}
thereby defining the Chebyshev moments $Y_i^n$ and $\hat Y_i^n$.
Here, we employ a convenient (albeit non-standard) normalization
for the zeroth Chebyshev moment by setting $T_0=1/\sqrt{2}$.

In the first step, we project the left hand sides of eqs.~(\ref{BS6},\ref{BS6a})
onto the scalar functions $Y_i$ and $\hat Y_i$ using
the orthogonality relations (\ref{ortho_N1}--\ref{ortho_N3})
and (\ref{ortho_D}),
\begin{eqnarray} 
  \hat Y_i(p^2,z) &=& (g'_0)^{ij}(p^2,z) \;Y_j(p^2,z)   \\
  Y_i (p^2,z)     &=& \fourint{k} \;(H')^{ij}(p^2,k^2,y',z,z') \;
                      \hat Y_i(k^2,z') \; .
\end{eqnarray}
The new propagator matrix $(g'_0)^{ij}$ and the modified kernel
$(H')^{ij}$ are the result of taking Dirac traces after
multiplication with the respective adjoint Dirac components. 
As a consequence, the kernel $(H')^{ij}$ depends on
the possible scalar products between the vectors $k,p,P$
which are can be expressed as
$k^2$, $p^2$, $z'=\hat k \cdt \hat P$, $z=\hat p \cdt \hat P$
and $y'=\hat \vect k \cdt  \hat \vect p$.

Using the orthogonality relation of the Chebyshev polynomials,
\begin{equation}
 \int_{-1}^{1} \frac{T_n(z) T_m(z)}{\sqrt{1-z^2}}\;dz=\frac{\pi}{2}
   \delta_{nm} \; ,
  \label{ortho_T}
\end{equation}
the system of equations for the scalar quantities $Y_i$ and $\hat Y_i$
is projected onto their Chebyshev moments,
\begin{eqnarray}
 \hat Y_i^n(p^2)& = &\int_{-1}^{1}dz\, \frac{2}{\pi} (g'_0)^{ij}(p^2,z)
   \; \sum_{m=0}^{m_{\rm max}}i^{m-n} Y_j^m(p^2) \,
    \frac{T_m(z)T_n(z)}{\sqrt{1-z^2}} \quad
     \label{BS7}\\
 Y_i^m(p^2)   &=& \int_{-1}^{1}dz\, \frac{2}{\pi}  \fourint{k} \;
   (H')^{ij}(p^2,k^2,z,z',y') \;\times \nonumber \\
    && \sum_{n=0}^{n_{\rm max}}i^{n-m} \hat Y_j^n(k^2) 
       \,\frac{T_n(z')T_m(z)}{\sqrt{1-z^2}} \; . \label{BS7a}
\end{eqnarray}
Of course, in the practical calculation the Chebyshev expansion
can be taken into account only up to a $m_{\rm max}^{th}$ moment
for the $Y_i^m$ and up to a $n_{\rm max}^{th}$ moment
for the $\hat Y_i^n$. The propagator matrix $(g'_0)^{ij}$
can be expanded into Chebyshev polynomials as well and
thereby results an expression from eq.~(\ref{BS7})
that involves an integral over three Chebyshev polynomials.
The addition theorem 
\begin{eqnarray}
 T_n(z)\,T_m(z)&=& \half ( T_{|n-m|}(z) + T_{n+m}(z)) 
      \quad(n,m \not = 0) \; , \\
 T_0(z)\,T_m(z)&=& {\T \frac{1}{\sqrt{2}} } T_m(z)
\end{eqnarray}
simplifies this expression to a form where a subsequent
application of  the orthogonality
relation (\ref{ortho_T}) reduces it to a matrix equation,
\begin{equation}
 \hat Y_i^n(p^2) = (g_0)^{ij,nm} (p^2)\; Y_i^m(p^2) \; . 
 \label{eq1}
\end{equation}
The elements of the propagator matrix  
$(g_0)^{ij,nm}$ are all real due to the explicit phase
factor $i^n$ in the Chebyshev expansions.

Let us now focus attention on eq.~(\ref{BS7a}), the
4-dimensional integral over the Bethe-Salpeter kernel, 
\begin{eqnarray}
 Y_i^m(p^2) &=& \int_{0}^\infty \frac{k^3}{4\pi^4}d|k| \int_{-1}^{1} 
  \sqrt{1-z^{\prime 2}} dz' \int_{-1}^{1} dz \int_{-1}^{1} dy'\,
  (H')^{ij}(p^2,k^2,y',z,z') \nonumber \\
  & &\times\;\sum_{n=0}^{n_{\rm max}}i^{n-m} \hat Y_j^n(k^2)
       \,\frac{T_n(z')T_m(z)}{\sqrt{1-z^2}} \; .
  \label{BS7b}
\end{eqnarray}
We will take care of the angular integrations
over $z$ and $z'$ next. To this end 
we {\em approximate} the Bethe-Salpeter kernel with Chebyshev
polynomials in the following manner,
\begin{equation}
 (H'')^{ij} = (1-z^{\prime 2})(H')^{ij} = \sum_{s=1}^{m_{\rm max}+1}
    \sum_{t=1}^{n_{\rm max}+1} c^{ij,st} \; T_{s-1}(z) T_{t-1}(z') \; .
 \label{cap1}
\end{equation}
The coefficients $c^{ij,st}$ are {\em not} the Chebyshev
moments of $(H'')^{ij}$ but rather
\begin{eqnarray}
 c^{ij,st} &=& 
    { \frac{2}{(m_{\rm max}+1)} \frac{2}{(n_{\rm max}+1)}} 
   \sum_{u=1}^{m_{\rm max}+1} \sum_{v=1}^{n_{\rm max}+1} 
   \, T_{s-1}(z_u)T_{t-1}(z'_v) \nonumber \\
 & &   \times \; (H'')^{ij}(p^2,k^2,y',z_u,z'_v) \; .
 \label{cap2}
\end{eqnarray}
The $z_u\,[z'_v]$ are the {\em zeros} of the Chebyshev
polynomial $T_{m_{\rm max}+1}(z)\; [T_{n_{\rm max}+1}(z')]$.
Eqs.~(\ref{cap1},\ref{cap2}) are an example of the so-called Chebyshev
approximation of a function which we have applied to both
the arguments $z$ and $z'$.
As is known from the literature \cite{Press:89}, the 
Chebyshev approximation of a function is very close to the 
approximation by the {\em minimax polynomial}, which
(among all polynomials of the same degree) has
the smallest maximum deviation from the true function.

When inserting the kernel approximation back into
eq.~(\ref{BS7b}), we see that the orthogonality 
relation (\ref{ortho_T}) will take care of the integrations
over $z$ and $z'$. We find
\begin{eqnarray}
 Y^m_i(p^2)&=& \frac{1}{(m_{\rm max}+1)(n_{\rm max}+1)}
  \int_{0}^\infty \frac{k^3}{4\pi^2} d|k| \int_{-1}^{1} dy'
 \sum_{u=1}^{m_{\rm max}+1} \sum_{v=1}^{n_{\rm max}+1}\;i^{n-m}
 \nonumber \\
  & & \times \;T_{m}(z_u)T_{n}(z'_v) \; (H'')^{ij}(p^2,k^2,y',z_u,z'_v) 
   \; \hat Y^n_j (k^2)
 \; .
\end{eqnarray}
Here the sum runs also over the label $j$ and the Chebyshev
moment label $n$.

Finally we have succeeded to transform the original 4-dimensional
integral equation into a system of coupled one-dimensional
equations. In summary, the system reads,
\begin{eqnarray}
 \label{BS8}
 \hat Y_i^n(p^2) &=& (g_0)^{ij,nm} (p^2)\; Y_i^m(p^2) \; , \\
 \label{BS8a}
  Y_i^m(p^2)     &=&  \int_{0}^\infty d|k| \;H^{ij,mn}(k^2,p^2)\;
                       \hat Y_i^n(k^2) \; ,
\end{eqnarray}
with the definition
\begin{eqnarray}
  H^{ij,mn}(k^2,p^2)&=& \frac{1}{(m_{\rm max}+1)(n_{\rm max}+1)}
    \frac{k^3}{4\pi^2} \int_{-1}^{1} dy'  \sum_{u=1}^{m_{\rm max}+1} 
       \sum_{v=1}^{n_{\rm max}+1}\;i^{n-m} \nonumber \\
    \label{kern_final}
    & &   \times \;    T_{m}(z_u)T_{n}(z'_v) \; (H'')^{ij}(p^2,k^2,y',z_u,z'_v)\; .
\end{eqnarray}
The  momentum $k$ is discretized on a mesh with $n_{|k|}$ points,
with typically
$n_{|k|}= 20,\dots, 50$, and the integration is performed
as a Gaussian quadrature. Then the problem is equivalent
to tuning the lowest eigenvalue of a matrix equation
to one by readjusting the parameters of the model.

We have introduced the numerical method from a merely technical point
of view. Combining the Chebyshev expansion of the scalar functions
describing the partial waves with the Chebyshev approximation of 
the quark exchange kernel replaces the angular integrations
over $z=\cos \Theta$ and $z'=\cos \Theta'$ by sums that
are truncated by $m_{\rm max}$ and $n_{\rm max}$, the highest
Chebyshev moments of the vertex and wave function partial waves, 
respectively. The question is whether there is another,
more intuitive argument that supports
a quick convergence of the solution 
with increasing $m_{\rm max}[n_{\rm max}]$.

There exists a motivation using this kind of expansion
that is related to symmetries in a simpler dynamical system.
The numerical solution of the ladder approximated Bethe-Salpeter equation in the 
massless and massive Wick-Cutkosky model has been studied 
extensively in the literature, for a review see \cite{Nakanishi:1988hp}. 
In this model, two massive
mesons interact by exchanging a third (massless or massive)
meson. In the case of a massless exchange meson, the
Bethe-Salpeter equation exhibits an $O(4)$ symmetry \cite{Cutkosky:54}.
By expanding the wave functions into hyperspherical
harmonics ${\mathcal Y}_{nlm}$ ({\em cf.} appendix \ref{conv-app}), 
the solutions can be classified using the
quantum numbers $n$ (four-dimensional angular
momentum) and $l$, $m$ (three-dimensional orbital angular momentum
and its third component). The zero orbital angular momentum states of the
Wick-Cutkosky model ($l=0$) are essentially given by
\begin{equation}
  \psi^{\rm Wick}_{l=0} \sim Y^{\rm Wick}_n(p^2)\: U_n(z) \; , 
\end{equation}
with the $U_n(z)$ being Chebyshev polynomials of the
{\em second} kind. They are related to the $T_n(z)$ by
\begin{equation}
  T_n(z)=U_n(z)-z U_{n-1}(z) \; , \qquad (n>0) \; .
\end{equation}
Thus an expansion of wave functions into the $T_n$'s is closely
related to the expansion into hyperspherical harmonics.
Furthermore, the authors of ref.~\cite{Nieuwenhuis:1996qx}
have shown that in the case of  a massive exchange meson,
where the $O(4)$ symmetry is broken, the expansion into
hyperspherical harmonics converges amazingly quickly although
$n$ ceases to be a good quantum number. Typically
$n_{\rm max}<6$ was sufficient to obtain stable solutions,
except for the very weak binding regime. 

After this short digression let us return to the Bethe-Salpeter
equations in the diquark-quark model.
The Bethe-Salpeter equations treated here are ladder equations
as well, though complicated by the mixing of components
with differing angular momentum. Nevertheless,
the expansion into {\em spinor} hyperspherical harmonics
in the model sector with pointlike scalar diquarks only
has been shown to work well in ref.~\cite{Kusaka:1997vm}.
This finding has been confirmed and extended to 
pointlike axialvector diquarks in 
refs.~\cite{Hellstern:1997pg,Oettel:1998bk}. Thus
we can conclude that the $O(4)$ symmetry is also in the
diquark-quark model approximately valid.

Let us point out 
the kinematical situations in which there might arise problems.
Due to the singularities in the propagator denominators
of $G_0^{\rm q-dq}$, the allowed range of the momentum partitioning
parameter $\eta$ in the nucleon Bethe-Salpeter equation is restricted to
\begin{equation}
 \eta \in [ 1- m_{sc}/M_n, \, m_q/M_n ] \; ,
  \label{ebound1}
\end{equation}
if $m_{ax} \ge m_{sc} > m_q$ is assumed. Singularities in the
exchange quark propagator and the diquark vertices employing
$n$-pole scalar functions $V$, see eq.~(\ref{npole}),
impose the additional bounds
\begin{eqnarray}
  \label{ebound2}
 \eta & \in & \left[\half(1-m_q/M_n), \, \half(1+m_q/M_n)\right] \; , \\ 
  \label{ebound3}
 \eta & \in & \left[\third(1-2\lambda_n/M_n), \, \third(1+2\lambda_n/M_n\right] 
 \; .
\end{eqnarray}
Note that these bounds on the value of $\eta$ arise in the 
practical calculations when performed as outlined above.
In principle, $\eta$ could be chosen arbitrarily between 0 and 1, 
but beyond these bounds the connection of the respective Bethe-Salpeter equation
in Minkowski space and the present one in Euclidean space
is no longer given by a simple Wick rotation, but the Euclidean space
equation picks up additional residue terms.\footnote{We will encounter
the problem of relating Minkowski and Euclidean vertex and
wave functions again in section \ref{num_em} when calculating
current matrix elements.}

If $\eta$ is chosen to be close to  one of the boundaries
given in eqs.~(\ref{ebound1}--\ref{ebound3}) then
the propagators vary strongly due to the vicinity of the poles
and their Chebyshev expansion, employed to derive eq.~(\ref{eq1}),
will converge slowly. Thus, the number of Chebyshev moments
for the partial waves in the wave function $\psi$, 
$n_{\rm max}$, needs to be increased in order to achieve 
sufficient accuracy for the vertex function. We will
also see in the numerical solutions that the wave function 
expansion itself
will converge more slowly under these circumstances.

\subsection{Solutions I: The scalar diquark sector}
\label{solI}

In this subsection, we will present results in the model
sector with scalar diquarks only, summarizing the work
described in ref.~\cite{Oettel:1999gc}.
We recall the Faddeev amplitude for the nucleon,
{\em cf.} eq.~(\ref{fadn_to_deco}) and table \ref{partial1},
that has been decomposed into partial waves.
Using scalar diquarks only, it reads in the rest frame
\begin{equation}
 \trisp{(S\Gamma)}= \bp \hat S_1\; \varsigma\\ \hat S_2
       (\vect{\sigma}\hat \vect{p})\; \varsigma \ep_\alpha \;
       (g_s V)\,\twosp{(\gamma_5 C)} \; . 
\end{equation} 
The two scalar functions $\hat S_1$ and $\hat S_2$
(related to the wave function)
 describe upper
and lower component of the effective nucleon spinor.
Likewise only the two functions $S_1$ and $S_2$ enter
the decomposition of the vertex function.


\begin{table}
\begin{center}

\begin{tabular}{lcccc} \hline \hline \\
\multicolumn{5}{l}{Fixed width $S_1(p^2)\vert_{|p| = 0.2M_n} = 1/2$}
  \\[2mm] 
 $V(x)$ & $m_s\; [M_n]$ & $m_q\; [M_n]$  &  $\lambda \; [M_n]$ & 
                          $ g_s $ \\[2mm] \hline 
  $n= 1$ &  0.7  & 0.685  & 0.162 &    117.1\phantom{0}\\
  $n= 2$ &  0.7  & 0.620  & 0.294 &    91.79\\
  $n= 4$ &  0.7  & 0.605  & 0.458 &    85.47\\
  $n= 6$ &  0.7  & 0.600  & 0.574 &    84.37\\
  $n= 8$ &  0.7  & 0.598  & 0.671 &    83.76\\
   exp   &  0.7  & 0.593  & 0.246 &    82.16\\
   gau   &  0.7  & 0.572  & 0.238 &    71.47\\[2mm]  
\multicolumn{5}{l}{Fixed masses} \\[2mm] \hline
  $n= 1$ &  0.7  & 0.62   & 0.113 &    155.8\phantom{0}\\
  $n= 2$ &  0.7  & 0.62   & 0.294 &    91.79\\
  $n= 4$ &  0.7  & 0.62   & 0.495 &    81.08\\
  $n= 6$ &  0.7  & 0.62   & 0.637 &    78.61\\
   exp   &  0.7  & 0.62   & 0.283 &    74.71 \\ \\ \hline \hline
\end{tabular}

\end{center}
\caption{Summary of parameters used for the various scalar functions
  $V(x)$ in the diquark-quark vertices, {\it cf.}
 eqs.~(\protect\ref{npole}--\protect\ref{gauss}). \label{FWHM_table}}  
\end{table}

 We will start out
with investigations of the various forms of the scalar
function $V(x)$ that parametrizes the diquark-quark vertex,
see eqs.~(\ref{npole}--\ref{gauss}). Furthermore we pick $\sigma=1/2$
for the momentum distributing
parameter in the diquark-quark vertex.

We have chosen to consider two cases: First we fix the
the ratios of quark and scalar diquark mass to the
nucleon mass ($m_q/M_n$ and $m_{sc}/M_n$) and tune
the width parameter $\lambda$ in the scalar function
$V(x)$ until the normalization $g_s$, calculated
using eq.~(\ref{normsc}), provides enough binding to reach
the physical nucleon mass. This corresponds to
the constraint that iterating the system of equations 
(\ref{BS8},\ref{BS8a})
yields an eigenvalue 1. The results
for the ratios $m_q/M_n=0.62$ and $m_{sc}/M_n=0.7$ are given
in the upper half of table \ref{FWHM_table}. With this choice and
the choice of a dipole form for $V(x)$ we find a good 
description of the nucleon electric form factors, see
below. In the second case, we have kept fixed the diquark mass
($m_{sc}/M_n=0.7$ as before)
and vary the quark mass and the width parameter $\lambda$
until a solution of the Bethe-Salpeter equation was found with
the property
\begin{equation}
 S_1^0(p^2)\vert_{|p| = 0.2M_n} = 1/2 \; .
\end{equation}
This condition fixes the width of the zeroth Chebyshev moment of the
$s$ wave amplitude $S_1$ which we  normalize by setting
$S_1^0(0)=1$. We can regard this as a condition on the
extension of the whole nucleon-quark-diquark vertex in momentum space
since the zeroth moment of $S_1$ dominates in the expansion
of the nucleon vertex function. The resulting parameter sets
that solve the Bethe-Salpeter equation are displayed in the lower half
of table \ref{FWHM_table}.

\begin{table}
 \begin{center}
   \begin{tabular}{lllllll} \hline \hline \\
     & & \multicolumn{5}{c}{ $\eta=0.4$ }\\[2mm] \hline
      & $n_{\rm max}$ & 0 & 2 & 4 & 6 & 8  \\   
 $m_{\rm max}$ &      & & & & &   \\ \hline
    0  &  & 0.7005 & 0.9551 & 0.9731 & 0.9738 &  \\ 
    2  &  & 0.7806 & 0.9797 & 0.9967 & 0.9979 &  \\
    4  &  & 0.7844 & 0.9922 & 0.9993 & 0.9999 &  \\
    6  &  & 0.7846 & 0.9928 & 0.9999 & 1.0000 &  \\[2mm]
     & & \multicolumn{5}{c}{ $\eta=0.31$} \\[2mm] \hline
  0    & &  0.6517 & 0.8379 & 0.9285 & 0.9431 & 0.9449 \\
  2    & &  0.8755 & 0.8988 & 0.9743 & 0.9940 & 0.9965 \\
  4    & &  0.9326 & 0.9614 & 0.9904 & 0.9977 & 0.9996 \\
  6    & &  0.9406 & 0.9714 & 0.9978 & 0.9992 & 0.9998 \\
  8    & &  0.9416 & 0.9726 & 0.9992 & 0.9998 & 0.9999 \\ \\ \hline \hline
   \end{tabular} 
 \caption{Convergence of the ground state eigenvalue of the 
     Bethe-Salpeter equation in terms of $m_{\rm max}$ and $n_{\rm max}$,
     shown for two values of the momentum distributing parameter
     $\eta$. For the kernel integration the Gauss quadrature
     with $n_{y'}=20$ grid points was used, 
     {\em cf.} eq.~(\ref{kern_final}). The momentum integration
     in the iteration was performed with $n_{|k|}=20$ grid points,
     {\em cf.} eq.~(\ref{BS8a}).
     Higher accuracy in these integrations does not affect the
     eigenvalue in the precision given here. 
     \label{conv_ev} }
 \end{center}
\end{table}

All of the parameter sets suffer from a general defect that results
from the restriction to scalar diquarks only. 
The constituent quark mass
$m_q$ has to be chosen rather large (corresponding
to a strongly bound scalar diquark)
such that the 
diquark normalization $g_s$ provides enough binding energy 
for the nucleon. Furthermore, the diquark-quark vertices
are rather narrow in momentum space as can be seen from the
values for $\lambda$. In configuration space\footnote
{We refer here to the Fourier transform of the relative momentum
between the quarks within the diquark that corresponds 
to the distance between the two quarks.},
the extension of the scalar diquark (measured
by the FWHM value for the scalar function $V$)
is roughly 1 fm and thus the diquark is as broad as the nucleon appears
to be when probed with electrons or pions \cite{Povh:1991fy}! 
The inclusion of the axialvector diquark provides another
attractive channel, therefore the scalar diquark normalization $g_s$
could be adjusted to smaller values and consequently the constituent
quark mass will drop. The results to be shown in section \ref{solII}
indeed confirm this argument.

\begin{figure}
 \begin{center}
  \epsfig{file=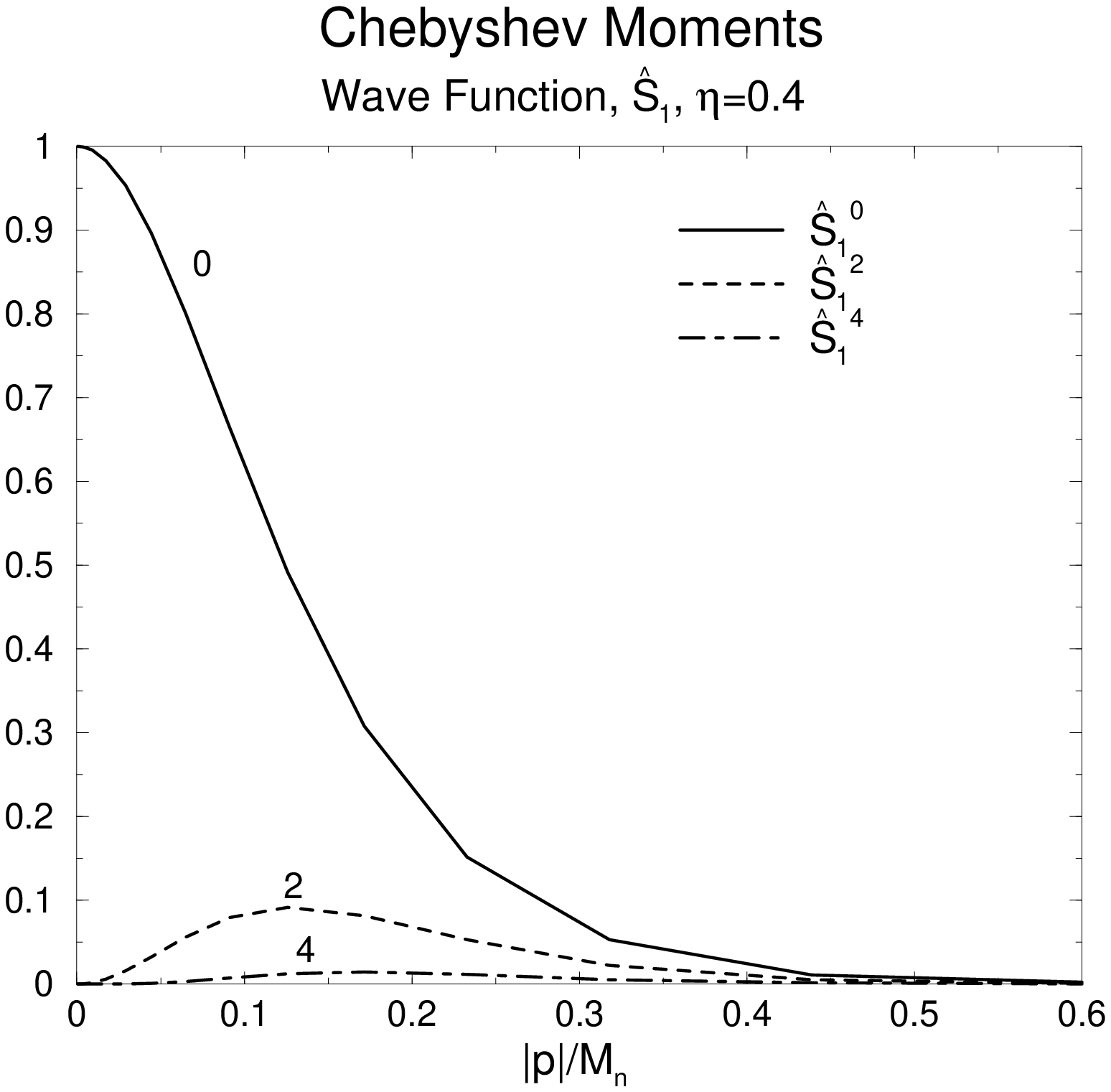,width=\figurewidth}
  \epsfig{file=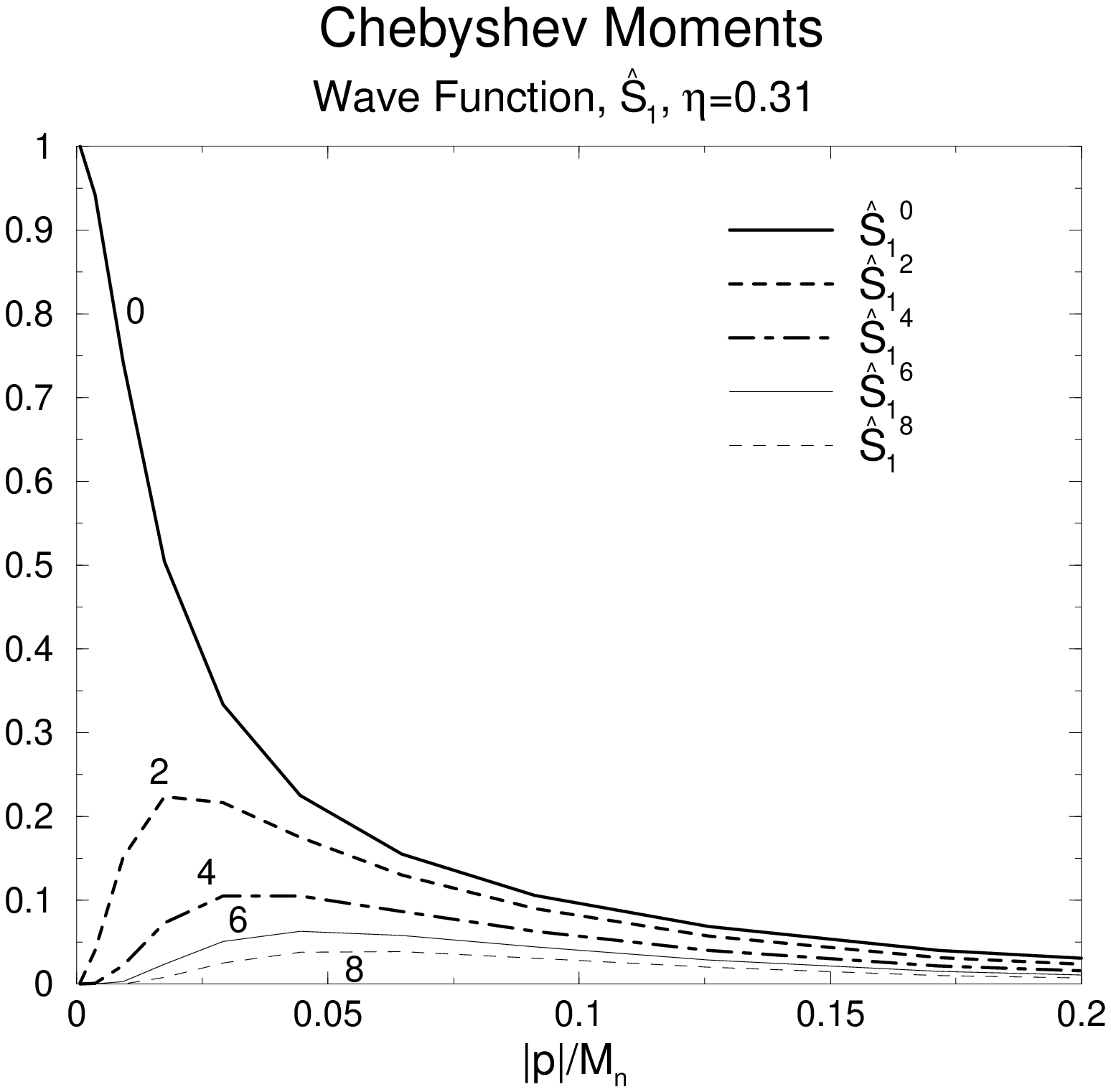,width=\figurewidth}
 \end{center}
 \caption{The Chebyshev moments of the dominating scalar
   function $\hat S_1$ in the expansion of the wave function.
   In the left panel, the moments for the choice $\eta=0.4$
   are shown. For the respective moments depicted in the right panel,
   the employed value of $\eta=0.31$ is close
   to the diquark propagator poles.}
 \label{wave1}
\end{figure}

We turn now to the convergence properties of the Chebyshev 
expansion, picking as an example the set of parameters
employing a dipole form ($n=2$) for the scalar function $V(x)$.
Here the pole in the diquark propagator and the poles
appearing in $V(x)$ put the restriction $\eta \in [0.3,0.52]$,
{\em cf.} eqs.~(\ref{ebound1},\ref{ebound3}). We exemplify
our results with the choice $\eta=0.4$ (far away from any poles)
and $\eta=0.31$ (close to the diquark propagator pole).
First we investigate the convergence of the eigenvalue (that
is obtained in the numerical iteration of the Bethe-Salpeter equation)
in terms of the highest Chebyshev moment $m_{\rm max}$
employed in the expansion of the vertex function and
$n_{\rm max}$ for the wave function. The results of
table \ref{conv_ev} show that for the ``safe'' value of $\eta=0.4$
seven Chebyshev moments for both wave and vertex function 
are sufficient to determine the eigenvalue to a precision
below $10^{-4}$. For the choice of $\eta$ near the diquark propagator
pole two more Chebyshev moments are needed to arrive at the same 
precision. Furthermore the results demonstrate that in order to
obtain an $\eta$-independent eigenvalue, the {\em full} dependence
of the scalar functions $S_i, \hat S_i$ 
on the angle $z=\cos \Theta$ has to be taken 
into account, {\em i.e.} $\eta$-invariance is not guaranteed 
for an arbitrary truncation in the Chebyshev expansion.

In the determination of observables the explicit 
solutions for wave or vertex function enter. We will
therefore have a look at their convergence properties as well.
The Chebyshev moments of the scalar functions $S_1$ and
$S_2$ building the {\em vertex function} show a very similar
convergence pattern for the different choices of $\eta$.
Each subsequent moment drops in its magnitude by a factor of
7\dots 10 as compared to the magnitude of the previous one.
The situation is completely different for the Chebyshev moments
of $\hat S_1$ and $\hat S_2$, making up the {\em wave function}.
This is depicted in figure \ref{wave1}. Whereas the convergence
for the choice of $\eta=0.4$ is still reasonably rapid (a factor
of 10 between the subsequent even Chebyshev momenta),
the moments of $\hat S_1$ for the choice $\eta=0.31$ become
squeezed in momentum space and only converge slowly with increasing
Chebyshev order. 

In the calculation of form factors we will need {\em both}
wave and vertex function, depending on the diagrams to compute.
The slow convergence of the wave function expansion will
put certain restrictions on the numerical accuracy
whenever the respective scalar functions are employed.

\begin{table}
 \begin{center}
  \begin{tabular}{llrrrr} \hline \hline \\
    & Form of diquark & $r_p$ (fm)   & $r_n^2$ (fm$^2$) &
                       $\mu_p$ & $\mu_n$ \\
    & vertex $V(x)$   &  &   & & \\ 
    \hline 
    &&& \\[-7pt]
    fixed $S_{1,0}$-width : 
    & $n=1$       & 0.78 & $-$0.17 & 0.95& $-$0.80\\ 
    & $n=2$       & 0.82 & $-$0.14 & 1.09& $-$0.93\\
    & $n=4$       & 0.84 & $-$0.12 & 1.13& $-$0.97\\
    & exp         & 0.83 & $-$0.04 & 1.16& $-$1.00\\
    & gauss       & 0.92 &  0.01   & 1.22& $-$1.07\\[2pt]
    \hline 
    &&& \\[-7pt]
    fixed masses:
    &$n=1$& 0.97 & $-$0.24& 0.98 & $-$0.86 \\
    &$n=2$& 0.82 & $-$0.14& 1.09 & $-$0.93\\
    &$n=4$& 0.75 & $-$0.03& 1.12 & $-$0.95\\
    &exp  & 0.73 & $-$0.01& 1.13 & $-$0.96\\[2pt] \hline
    &&& \\[-7pt]
    experiment & & 0.84 & $-$0.11 & 2.79 &$-$1.91 \\ \\ \hline \hline  
  \end{tabular}
 \end{center}
 \caption{Electric radii of proton/neutron and the nucleon 
   magnetic moments for the parameter sets
   having either the $S_1^0$-width or the quark mass fixed.}
 \label{radii1}
\end{table} 

\bigskip

\noindent
{\it\large Electromagnetic Properties}
\medskip

\noindent
In the following we will quote from ref.~\cite{Oettel:1999gc}
some results for electric form factors,
electric radii and magnetic moments which have been  obtained
with the parameter sets of table \ref{FWHM_table}. 
The definitions for the current operator, the form factors 
and radii 
are collected in chapter \ref{em-chap}.
The necessary formalism for calculating the appropriate 
current matrix elements
is also presented there, along with the proof of current 
conservation. Presenting the results at this point
will exhibit the shortcomings of the scalar diquark sector
and furthermore motivates the choice of the dipole form
for $V(x)$ in the calculations with scalar and axialvector diquarks.

 In table \ref{radii1} we have compiled the results for the
proton and neutron charge radii as well as for the nucleon
magnetic moments. A first glance at the proton charge radius
confirms that fixing the width of the dominant
scalar function $S_1^0$ of the nucleon-quark-diquark vertex function
fixes roughly this radius as well.  
But whereas all parameter sets show for
the proton charge radius a maximal deviation of about 15 \% to
the experimental value, the
results for the neutron charge radius differ drastically. The
dipole and the quadrupole form for $V(x)$ are here closest to
the observed radius. 

 We have plotted the nucleon electric form factors up to
a momentum transfer of $Q^2=2.25$ GeV$^2$ in figure \ref{gefig1}.
The proton form factor (left panel) has been normalized
to the phenomenological dipole fit
\begin{equation}
 G_E^{\rm phen} = \frac{1}{\left(1 + Q^2/(0.71\: {\rm GeV}^2)\right)^2}\;,
\end{equation} 
which represents for momentum transfer below 1 GeV$^2$
a quite accurate description of the experimentally measured
$G_E$ \cite{Hoehler:1976}. Here the results for 
the parameter set employing
the dipole shape for $V(x)$ follows the phenomenological dipole
closest as compared to the other curves. Therefore we
employ the dipole in the diquark vertex for further
calculations. Regarding the overall shape of the neutron form
factor, only the Gaussian parametrization for $V(x)$ produces
results grossly deviating from the scarce experimental information
one has about this quantity. A discussion of the available 
experimental results is postponed until section \ref{results1}.

\begin{figure}
 \begin{center}
  \epsfig{file=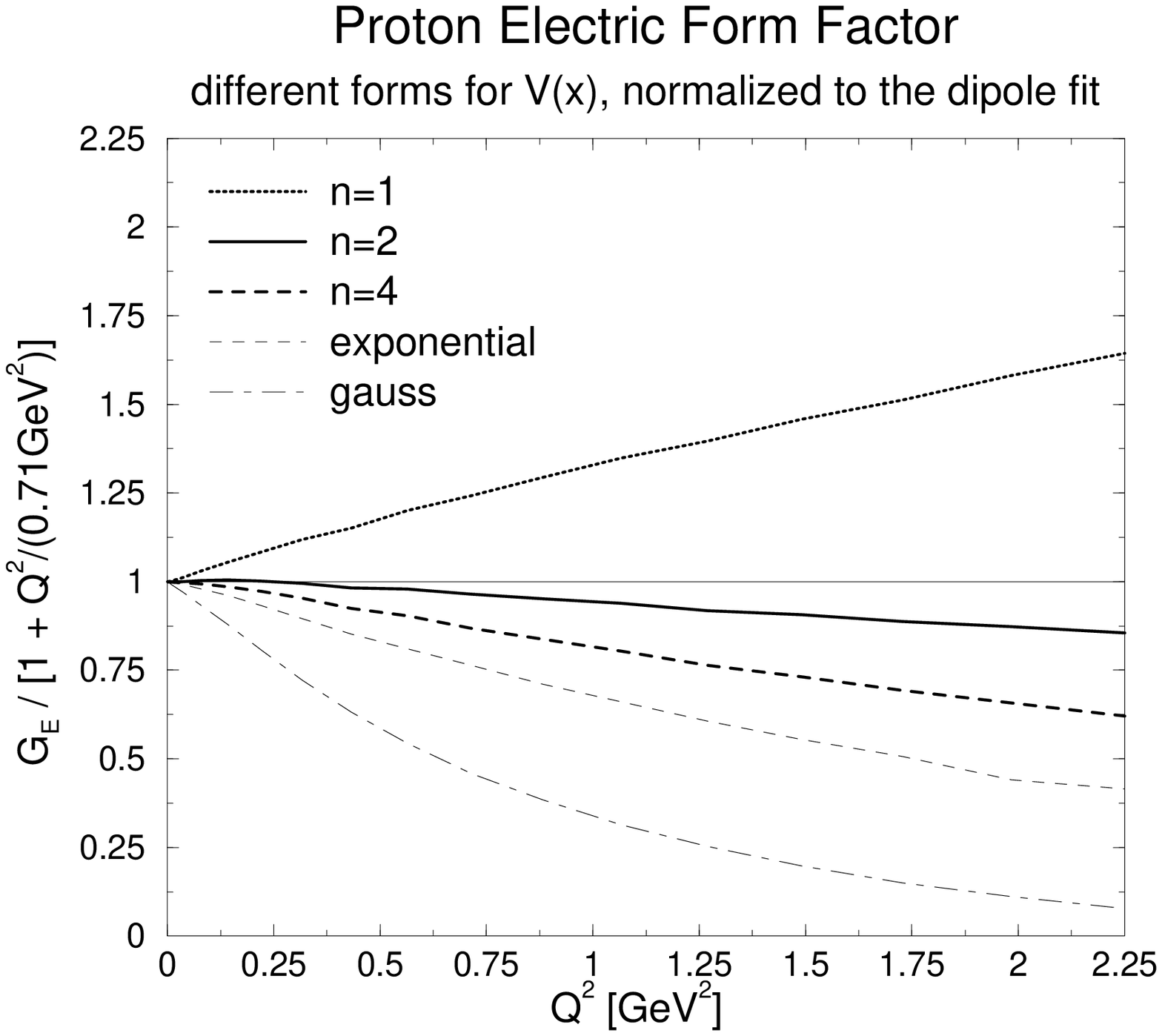,width=\figurewidth}
  \epsfig{file=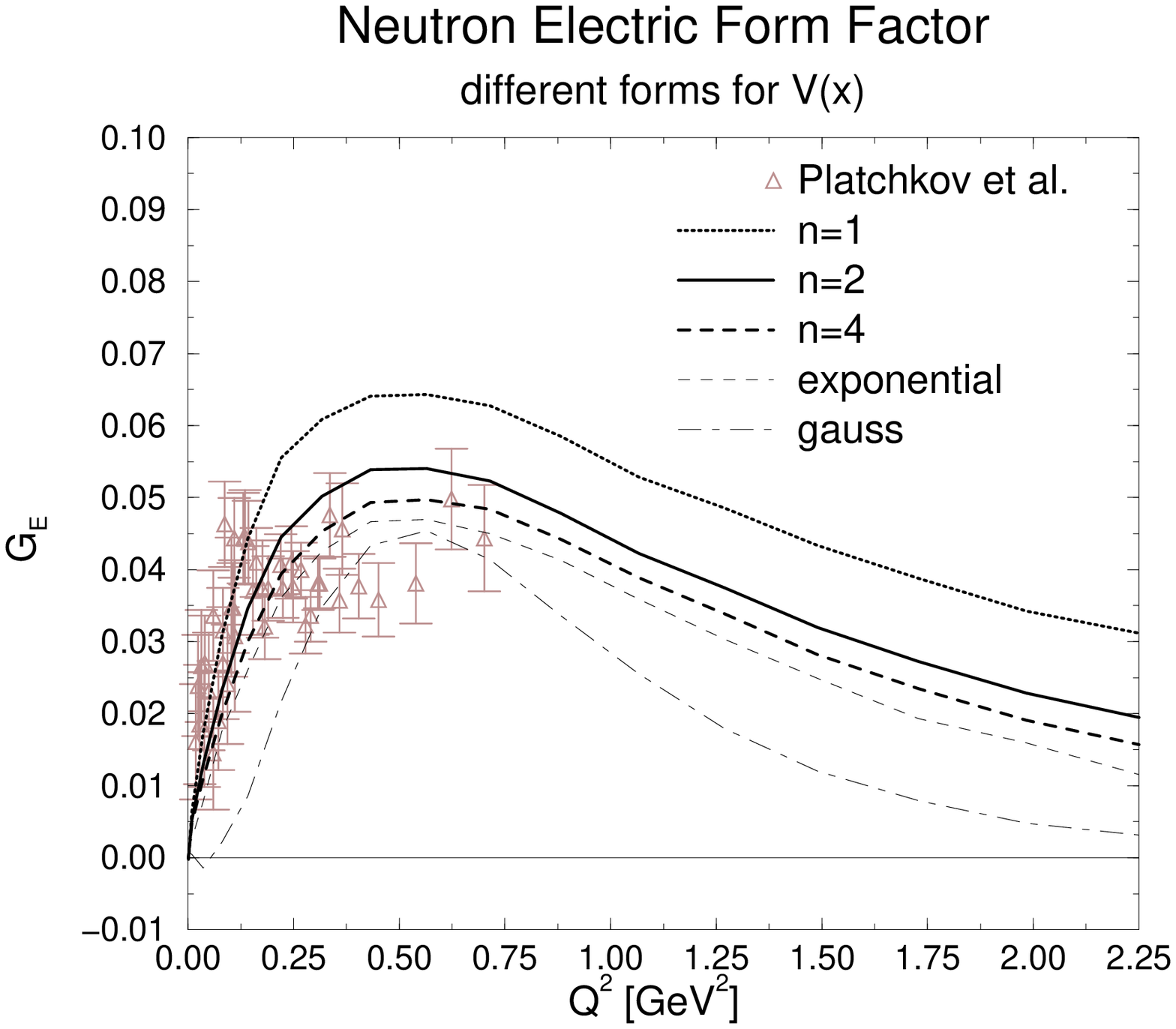,width=\figurewidth}
 \end{center}
 \caption{Proton (left panel) and neutron (right panel) electric
  form factors for the various parameter sets with fixed
  width of the zeroth Chebyshev moment $S_1$. The experimental data
  for the neutron is taken form ref.~\cite{Platchkov:1990ch}.}
 \label{gefig1}
\end{figure}

Returning to the discussion of the results in table \ref{radii1},
we see that the proton and neutron magnetic  moments are ($i$) much too 
small and
($ii$) their ratio is far from the experimentally measured
value. Both defects are known from exploratory studies
within other forms of quark-scalar diquark models \cite{Keiner:1996bu}
and can be attributed to the oversimplification of the
two-quark correlations to just the propagation of a spinless
scalar diquark that  cannot exhibit a spin flip  when struck by a 
photon. 
The problem is furthermore enhanced by the comparatively large
values of the constituent quark mass in all parameter sets.
Remember that the main indication for a quark mass scale
of around 0.3 GeV comes from a non-relativistic estimate
of the nucleon magnetic moments. According to this argument,
the nucleon magnetic moments are proportional to $M_n/m_q$
and thus quark masses being larger than 0.5 GeV are expected
to yield magnetic moments that are far too small.

Although the results on the electromagnetic properties
have been reported at this stage without derivation, we 
regard  the previously unspecified form of the diquark vertex
as being fixed now by the dipole form.
With the ingredients of the model being complete, we present in the
next subsection solutions for nucleon and $\Delta$ 
using the scalar and the axialvector diquark channel.

\subsection{Solutions II: Scalar and axialvector diquarks}
\label{solII}

Including axialvector diquarks, the Bethe-Salpeter equations can 
now be solved for nucleon and $\Delta$. The full decomposition
of wave and vertex functions, shown in table \ref{partial1}
for the nucleon and in table \ref{partial2}  for the $\Delta$,
has to be taken into account and the equations are solved
for the eight scalar functions that describe vertex and wave
function respectively. Altogether there are four parameters:
the quark mass $m_q$, the diquark masses $m_{sc}$ and $m_{ax}$,
and the width $\lambda$ of the dipole-shaped diquark-quark vertex.

The solutions in the scalar diquark sector suffered from
the large constituent quark masses that set a wrong scale
within the nucleon, as seen by the results for the magnetic moments.
The inclusion of the attractive axialvector diquark channel
now allows for solutions with smaller quark masses and smaller
binding energy for the scalar diquark. The 
diquark normalization (or coupling) constant $g_s$ drops for this reason
and therefore the nucleon binding energy becomes smaller in the
scalar channel, but
the additional binding provided by the axialvector diquark
compensates this effect to yield the physical nucleon mass.

In the calculations presented here and in the following chapters
 we shall illustrate the consequences of
 the model assumptions with two different parameter 
sets which emphasize slightly different aspects.
For Set I, we
employ a constituent quark mass of $m_q=0.36$ GeV which is
close to the values commonly used by non- or semi-relativistic
constituent quark models. Due to the free-particle poles in the
bare quark and diquark propagators, the
axialvector diquark mass is below 0.72 GeV and the delta mass
below 1.08 GeV. On the other hand, nucleon {\em and} delta masses 
are fitted by Set II, {\em i.e.} the parameter space 
is constrained by these two masses.
In particular, this implies $m_q>0.41$
GeV. Both parameter sets together with the corresponding values resulting 
for the effective diquark couplings and baryon masses are given in
table~\ref{pars}.  

\begin{table}
 \begin{center}
  \begin{tabular}{ccccccccc} \hline \hline \\
  Set 
  & $m_q$& $m_{sc}$& $m_{ax}$ & $\lambda$ & $g_s$ & $g_a$ & $M_\Delta$ & $M_n$ \\
  & [GeV]& [GeV]&   [GeV]&   [GeV]&              &       & [GeV] & [GeV]  
  \\[2mm] \hline
   I  & 0.36 & 0.625   & 0.684  &   0.95 &     9.29  & 6.97  & 1.007  & 0.939\\
   II & 0.425& 0.598   & 0.831  &   0.53 &    22.10  & 6.37  & 1.232  & 0.939
 \\ \\ \hline \hline
  \end{tabular}
 \end{center}
 \caption{The two parameter sets of the full model 
together with the values of diquark normalizations
(couplings) and the bound masses $M_n$, $M_\Delta$ 
that arise for these sets.}
\label{pars}
\end{table}

Two differences between the two sets
are important in the following: The strength of the axialvector correlations
within the nucleon is rather weak for Set II, as the ratio of axialvector
to scalar diquark normalization (coupling), $g_a/g_s=0.29$,
is much smaller than for Set I where $g_a/g_s=0.75$.
We will quantify the correlation strength more precisely
when assessing the contributions of scalar and axialvector
correlations to the normalization of the nucleon, {\em cf.} 
section \ref{results1}. Secondly, the 15 \% difference between the
constituent quark masses of the two sets will have 
an influence on the results for magnetic moments that is even stronger
than could be expected from the non-relativistic argument. 
  
Although the wave and the vertex function are no physical observables
they do enter observable matrix elements via Mandelstam's formalism 
\cite{Mandelstam:1955} (to be described in the next chapter) and 
therefore
the strengths of the single components give a hint on their
effect on observables. We have plotted the leading Chebyshev moments
of the scalar functions describing the nucleon $s$ waves in
figure \ref{s_wfig}. These are $\hat S_1^0$ and $\hat A_3^0$,
describing the $s$ waves for scalar and axialvector diquark
and $\hat A_1^0$ that is connected with the virtual time component
of the latter. We see for both sets that the functions
$\hat A_1^0$ are suppressed by a factor of $10^3$ compared to the
dominating $\hat S_1^0$. The strength of the other $s$ wave associated
with the axialvector diquark is roughly proportional to the
ratio $g_a/g_s$ for the respective parameter set. 

\begin{figure}
 \begin{center}
   \epsfig{file=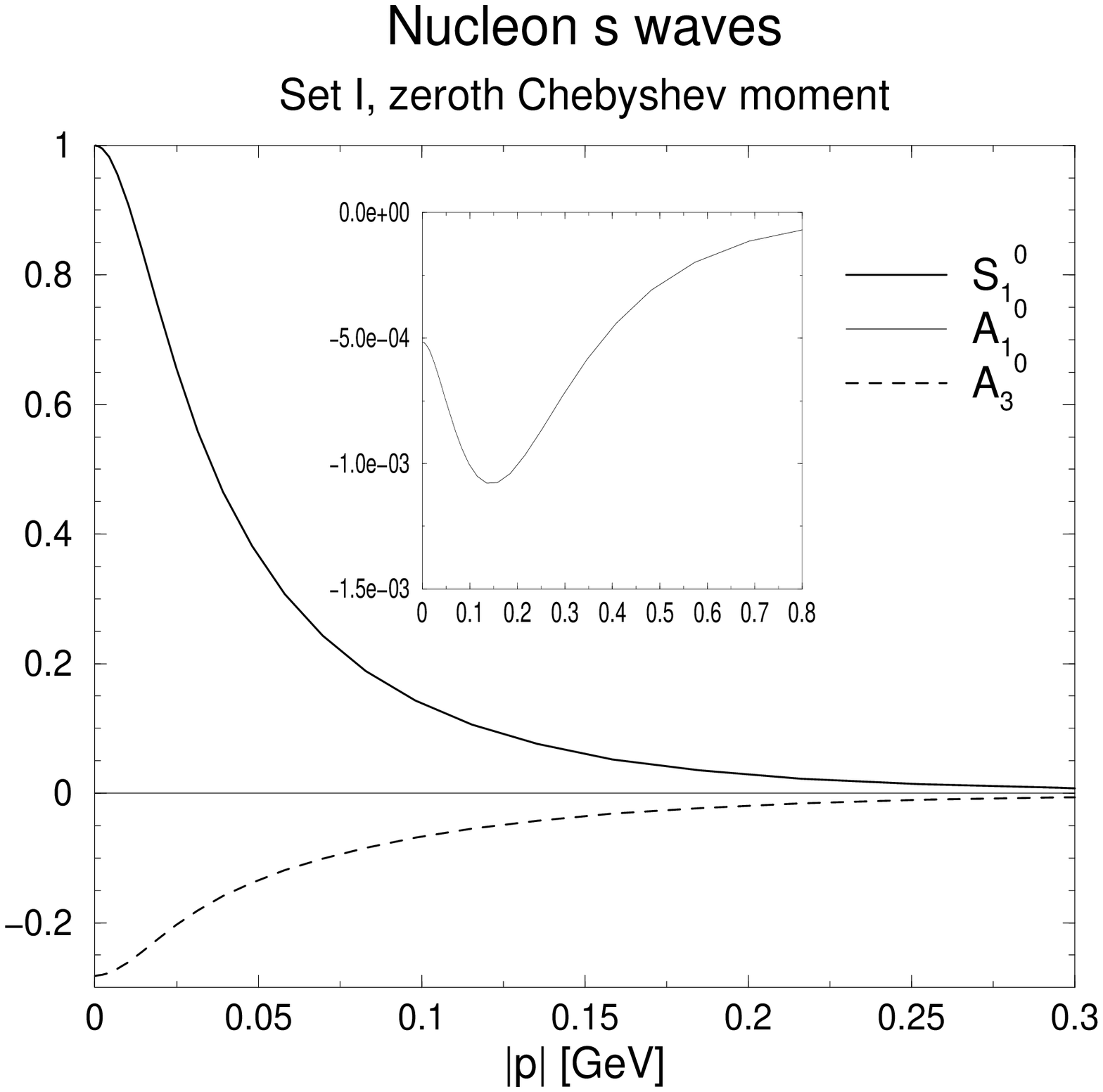,width=\figurewidth}
   \epsfig{file=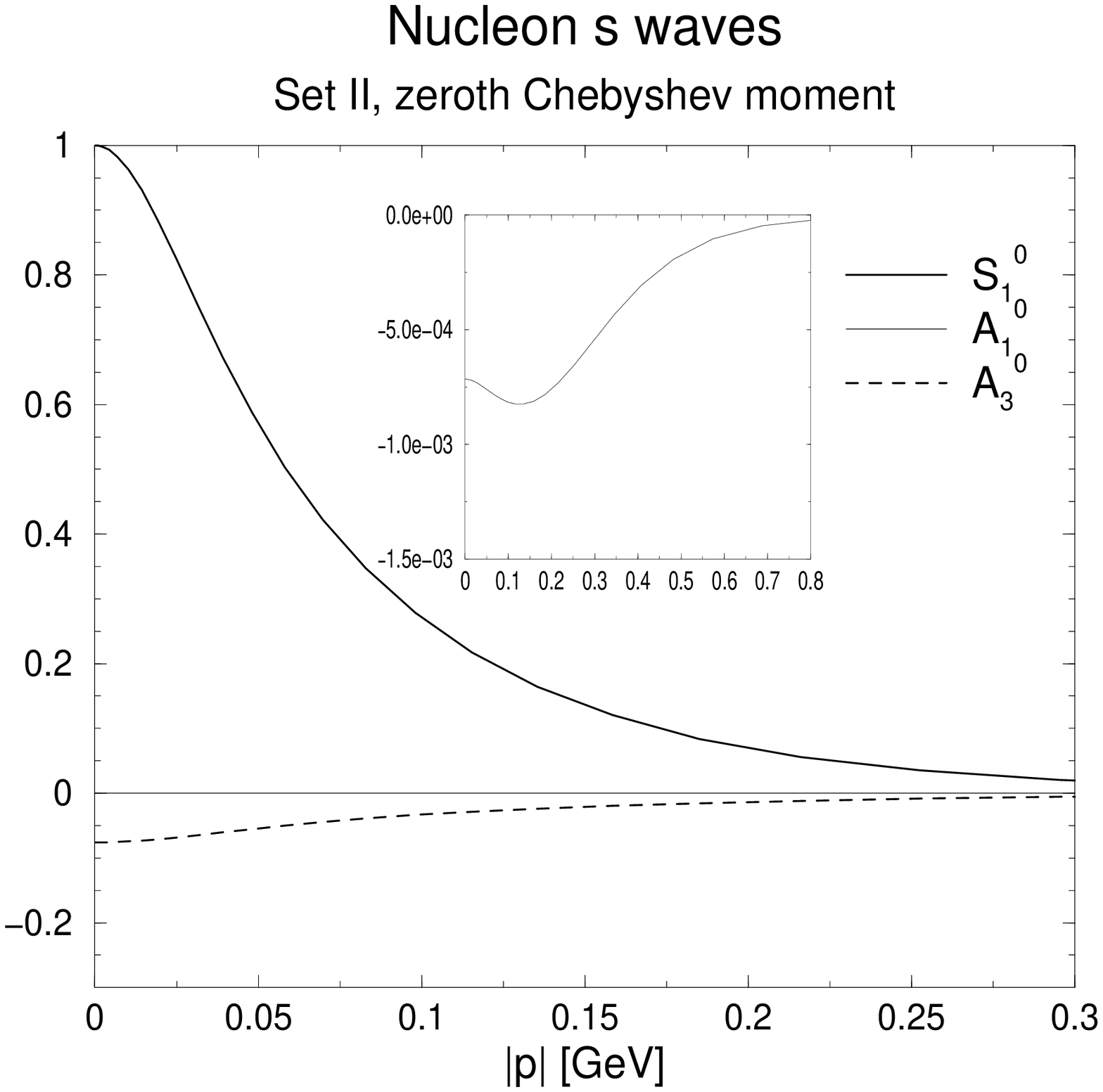,width=\figurewidth}
 \end{center}
 \caption{The leading Chebyshev moments of the
   functions $\hat S_1$, $\hat A_1$ and $\hat A_3$ related to the 
   nucleon $s$ waves. All functions are normalized by the condition
   $\hat S_1^0=1$.}
 \label{s_wfig} 
\end{figure}

In   appendix \ref{wave-app}
the nucleon partial waves for Set I and the $\Delta$ partial
waves for Set II (here the $\Delta$ bound state mass corresponds
to the physical one) are shown  and discussed. In summary,
all the remarks from the previous subsection 
on the numerical accuracy and the convergence
properties for wave and vertex functions apply also in the
case of the full model. Especially for Set II, the $\Delta$ 
is  an extremely weakly bound state and thus  the Chebyshev
expansion for the wave function converges quite slowly.

Before we turn to the detailed discussion of the 
nucleon electromagnetic properties
we will sum up the results obtained so far. From a 
physical point of view, the parameters of the model
could be chosen more intuitively after having included
the axialvector diquark sector of the model. This especially
applies to the value of the constituent quark mass which appeared
to be rather large in the scalar diquark sector. Still, the
inclusion of the $\Delta$ resonance leads to rather large
quark and axialvector diquark masses and consequently small
axialvector correlations in the nucleon. This is 
an artefact of employing free quark and diquark propagators,
therefore we will discuss appropriate modifications which
mimic confinement  later in chapter \ref{conf-chap}.

From a technical point of view, we have succeeded to transform the
full 4-dimensional problem into a manageable form. The chosen
expansion in Dirac space into partial waves has been done covariantly
and the results for the partial wave strengths identify
the leading and subleading  components. The Chebyshev expansion for
the scalar functions which describe the partial wave strengths
has proved to be quite efficient and accurate, furthermore 
the expansion has been done for a Lorentz invariant variable,
$z=\hat p \cdt \hat P$, with the consequence that the boost of 
the wave and vertex function solutions is rendered feasible.
A similar method for treating mesons in a Bethe-Salpeter
approach has been put forward in 
refs.~\cite{Maris:1997tm,Maris:1999nt} where the efficacy
of this technique has been demonstrated for the solution 
of the Bethe-Salpeter equations as well as in calculations of 
mesonic observables.

As a conclusion, there is no technical reason why one 
should stick to some kind of non-relativistic approximation
in solving a ladder Bethe-Salpeter equation. Such an approximation
could only be justified in a {\em very} weak binding situation 
and results for physical observables should depend in a 
controlled manner on the non-relativistic truncation.
We investigate this issue in chapter \ref{sal-chap} for the case
of the diquark-quark model where we compare results
for vertex function solutions and observables obtained
($i$) with the exact treatment and ($ii$) in the
Salpeter approximation \cite{Salpeter:1952}, a popular semi-relativistic
approximation of the Bethe-Salpeter equation. 
We find that the Salpeter approximation strongly violates
the approximate $O(4)$ symmetry present in the solutions of
the full Bethe-Salpeter equation and observables obtained
with both methods deviate in a rather unpredictable manner.

 \chapter{Electromagnetic Form Factors}
 \label{em-chap}
  
Electromagnetic form factors describe one of the most 
basic dynamic observables of a composite system: the elastic scattering
of an electron on the composite target. Electron and
target interact with each other by exchanging quanta of the
electromagnetic field. For the target being a nucleon, the scattering 
is from a classical point of view a function
of the charge and magnetization distributions within the target
which interact with charge and spin of the electron, respectively.
The Sachs form factors $G_E$ and $G_M$, to be introduced in 
eqs.~(\ref{gedef},\ref{gmdef}), describe in the Breit system exactly the
Fourier transforms of these distributions 
\cite{Sachs:1962}. 
In the usual field theoretical treatment it is justified to assume that the
scattering is mediated by one virtual photon (due
to the smallness of the coupling constant in quantum electrodynamics), 
thereby the
information on the electromagnetic structure of the probe
is contained in the following matrix element, defined in momentum space,
\begin{equation}
 \langle J^\mu \rangle(Q) =
 \langle {\rm target;}\;P_f |\;J^\mu \; |{\rm target;}\;P_i
 \rangle (Q)\; ,
\end{equation}  
where initial and final target state possess on-shell four-momenta
$P_i=P-Q/2$ and $P_f=P+Q/2$, respectively. The current operator $J^\mu$
is obtained from a suitable model Lagrangian where it
is the coefficient of the linear term in $A^\mu$, the electromagnetic 
vector potential. For nucleon targets the matrix element 
is written in the most general form as 
\begin{equation}
\langle J^\mu \rangle =\bar{u}_{\sigma_f}(P_f) 
   \left[-i\gamma^\mu {\mathcal F}_1
  +\frac{i \kappa {\mathcal F}_2}{2M_n} \sigma^{\mu\nu} Q^\nu \right]
  u_{\sigma_i}(P_i) \;.
\end{equation}
Here,  ${\mathcal F}_1$ and ${\mathcal F}_2$ are the Dirac charge and
the Pauli anomalous magnetic form factor, respectively.
The constant $\kappa$ describes the anomalous magnetic moment.
No further independent structures exist if the current is assumed
to transform like a vector under parity transformations and if
current conservation,
\begin{equation}
 Q^\mu \; \langle J^\mu  \rangle = 0 \; , 
 \label{cc}
\end{equation}
is to hold \cite{Aitchison:1989bs}. The Sachs form factors 
$G_E$ and $G_M$ are introduced as
\begin{eqnarray}
 \label{gedef}
 G_E &=& {\mathcal F}_1 - \frac{Q^2}{4M^2}\kappa {\mathcal F}_2 \; , \\
 \label{gmdef}
 G_M &=& {\mathcal F}_1 + \kappa {\mathcal F}_2 \; .
\end{eqnarray}
By definition, the values of $G_E$ and $G_M$ at zero momentum
transfer are the nucleon charge and magnetic moments. 

It is convenient in the following to introduce 
matrix elements which are themselves matrix valued due to
summations over initial and final spin,
\begin{equation}
\langle P_f| \;  J^\mu\; |P_i \rangle  := \sum_{\sigma_f,\sigma_i}
   u_{\sigma_f}(P_f)\;
      \langle J^\mu \rangle\;
     \bar{u}_{\sigma_i}(P_i) \;  , \label{m-el}
\end{equation}
to remove the nucleon spinors in the parametrization of $\langle J^\mu \rangle$. Using the Gordon identities
\cite{Aitchison:1989bs}, this matrix element is written in terms
of $G_E$ and $G_M$ as
\begin{equation}
 \langle P_f| \;  J^\mu\; |P_i \rangle =
 -i \Lambda^+(P_f) \left[ 	 \gamma^\mu G_M +
  iM \frac{P^\mu}{P^2} (G_E-G_M) \right] \Lambda^+(P_i) \; .
 \label{m-el2}
\end{equation}
The form factors can be extracted from this expression by taking
traces as follows,
\begin{eqnarray}
 \label{getrace}
G_E(Q^2) &=& \frac{M_n}{2P^2}\; {\rm Tr}\;  
    \langle P_f| \;  J^\mu\; |P_i \rangle  P^\mu \; ,\\
 \label{gmtrace}
G_M(Q^2) &=& \frac{iM_n^2}{Q^2}\; {\rm Tr}\;\langle P_f| \;  J^\mu\; |P_i \rangle  
   (\gamma^\mu)_T \; ,\qquad 
   \left( (\gamma^\mu)_T=\gamma^\mu -\hat P^\mu \hat\Slash{P}\right)\; . 
  \quad
\end{eqnarray}

This little introductory exercise provided us with the
necessary formulae to obtain the form factors from the
nucleon current matrix element. The more for\-mi\-dable task is
to construct a suitable current operator for the diquark-quark model
and to give a prescription how to obtain nucleon matrix elements
from the current operator and the solutions of the Bethe-Salpeter
equation. The current operator should respect gauge invariance  
and consequently the condition (\ref{cc}), {\em i.e.} current 
conservation, should hold.

The following section aims at constructing the current operator.
This amounts to clarify how the photon couples to the
full quark-diquark propagator $G^{\rm q-dq}$ given in eq.~(\ref{Gqdq}),
the 4-point function for quark and diquark in the channel
with nucleon quantum numbers. The resulting 
5-point function has to fulfill a Ward-Takahashi identity which in turn
guarantees that for the current matrix elements current conservation
holds. Thereby we find 
non-trivial irreducible couplings of the photon to the diquark-quark 
vertices $\chi^5$ and $\chi^\mu$, the  {\em seagull vertices},
besides the expected photon couplings to quark and diquark. 
In total we will find that the nucleon current operator 
consists of
\begin{itemize}
 \item the photon coupling to quark and diquark
       ({\em impulse approximation}), 
 \item the photon-mediated
       transitions from scalar to axialvector diquarks and 
       the reverse transition
       ({\em extended impulse approximation}) and
 \item contributions from the quark exchange kernel
       which
       contains the couplings to the exchange quark and
       the seagull couplings to the diquark-quark vertices.
\end{itemize}
Unknown constants like the anomalous magnetic moment
of the axialvector diquark and the strength of the 
scalar-axialvector transitions will be calculated by taking into account 
the diquark substructure in a first approximation which
does not violate gauge invariance for the nucleon current.

The construction of the current operator by using 
Ward-Takahashi identities provides us furthermore with a proof
of the equivalence of the canonical normalization condition
for Bethe-Salpeter wave function with the normalization to correct
bound state charge. The canonical normalization amounts to fixing
the residue of the bound state pole in the full propagator
$G^{\rm q-dq}$.

Equipped with the explicit form of the current operator
we will then proceed by calculating the form factors for the 
two parameter sets
introduced in section \ref{solII}. We find that 
current conservation is reflected in the numerical solutions very 
accurately. The electric form factors of proton and neutron
are described very well by our solutions whereas the magnetic
form factors fall short by some 15 \%, due to our simplified
assumptions for the quark and diquark propagators.
An interesting point is that the recent experimental data on
the ratio $G_E/G_M$ for the proton \cite{Jones:1999rz} 
strongly constrains the axialvector correlations in the nucleon 
to be rather moderate. We will conclude this chapter by 
discussing possible further improvements.

\section{The nucleon current operator in the diquark-quark model}
\label{curop-sec}

We shall make use of the ``gauging of equations'' formalism
outlined in refs.~\cite{Haberzettl:1997jg,Kvinikhidze:1998xn,Kvinikhidze:1999xp}.
To simplify the treatment, we will derive the current operator
by assuming that the diquark quasiparticle could in 
principle be represented by a diquark creation operator
$d^\dagger=\left( d_s^\dagger,d^{\mu\dagger}_a\right)$ which creates
scalar and axialvector diquarks. The resulting expressions 
are equivalent to the results when constructing
the current operator by starting from the 3-quark problem,
provided a Ward identity for the diquark propagator holds
({\em cf.} refs.~\cite{Kvinikhidze:1999xp,Blankleider:1999xp}).

We employ a notation for Green's functions as in section
\ref{3qreduce}. 
We remind that their multiplication is to be understood
symbolically and involves summation over Dirac indices
and integration over the relative momenta as in eq.~(\ref{symbint}).
We assume that, in contrast to eq.~(\ref{Gqdq}),
 the quark-diquark propagator in the nucleon channel
\begin{eqnarray}
 (2\pi)^4 \delta^4(k_q+k_d-p_q-p_d)\; G^{\rm q-dq}(k_q,k_d;p_q,p_d)=\\
 {\rm F.T.}\; \langle 0|\, T q(x_{k_q}) d(y_{k_d})\, \bar q(x_{p_q}) 
    d^\dagger(y_{p_d})\, | 0 \rangle  \nonumber
\end{eqnarray}
is given 
as a matrix that describes the propagation of charge
eigenstate combinations of 
quarks $u,d$ and scalar diquark $(ud)$ / axialvector
diquarks $[uu],[ud],[dd]$,
{\em cf.} appendix \ref{cfn}. The part described by
the free quark and diquark propagators is of course
diagonal in the charge eigenstates whereas the 
possible (multiple) quark exchange 
describes transitions between them.  
Likewise the 5-point function is defined to be
\begin{eqnarray}
 G^\mu(k_q,k_d;p_q,p_d) = {\rm F.T.}\;
 \langle 0|\, T q(x_{k_q}) d(y_{k_d})\, \bar q(x_{p_q})
    d^\dagger(y_{p_d}) J^\mu(0)\, | 0 \rangle \; . 
\end{eqnarray}
We fix momentum conservation by $k_q+k_d=p_q+p_d+Q$, where
$Q$ denotes the photon momentum. It is evident from the
application of Wick's theorem to this equation that $G^\mu$
contains (at least) all diagrams that can be constructed
from $G$ by attaching a photon to all propagators 
(like $\langle 0|\,Tq\bar q\,|0 \rangle$) and vertices
(like the diquark-quark vertices $\langle 0|\,T d\,\bar q \bar q
\, |0 \rangle$). An equivalent definition of
$G^\mu$ expresses the procedure of attaching photons on
propagators and vertices by a functional derivative,
\begin{eqnarray} \nonumber
 G^\mu(k_q,k_d;p_q,p_d) = -\frac{\delta}{\delta A^\mu(0)}
  \hskip 3cm \\ \left.
  {\rm F.T.}\; \langle 0|\, T q(x_{k_q}) d(y_{k_d})\, \bar q(x_{p_q})
 d^\dagger(y_{p_d}) \exp\left(-{\T \int}d^4 x\, J^\mu(x)A^\mu(x)\right) \, | 0 \rangle \right|_{A^\mu=0}\; .\quad 
\end{eqnarray}
``Gauging'' is an operation which achieves the same result by the
following simple rules \cite{Haberzettl:1997jg}:
\begin{enumerate}
  \item Notation: $F^\mu=\left.-\frac{\Dc \delta}{\Dc \delta A^\mu(0)} \,F_A\right|_{A^\mu=0}$
      where \\[1mm] \hspace*{1cm} $F={\rm F.T.}\; 
      \langle 0|\,q(x_1)\dots\bar q(y_1)\dots \dots | 0 \rangle $
      and \\ \hspace*{1cm}
      $F_A={\rm F.T.}\; \langle 0|\,q(x_1)\dots \bar q (y_1) \dots 
   \exp\left(-{\T \int}d^4 x\, J^\mu(x)A^\mu(x)\right)\, |0\rangle$.
  \item It is a linear operation, and for a constant $c$
    its action yields zero, $c^\mu=0$.
  \item It fulfills Leibniz' rule, $(F_1F_2)^\mu=F_1^\mu F_2+F_1F_2^\mu$.
  \item Gauging the inverse, $F^{-1}$, is accomplished by applying
    rule 2 and 3, {\em i.e.} $(F^{-1}F)^\mu=0 \rightarrow 
       (F^{-1})^\mu = - F^{-1} F^\mu F^{-1}$.
\end{enumerate}

Besides the above mentioned contributions to $G^\mu$ from
attaching photons to propagators and diquark-quark vertices
(which are obtained by gauging $G^{\rm q-dq}$),
there exist other contributions that cannot be obtained this way:
Although a transition from scalar to axialvector diquark
cannot occur, $\langle 0|\, T d_a^\nu\, d_s^\dagger\,| 0 \rangle=0$,  
this transition is possible if mediated by a photon,
$\langle 0|\,T d_a^\nu\, d_s^\dagger J^\mu\, | 0 \rangle_{\rm irred}
\not = 0$. The special feature of this scalar-to-axialvector transition 
is that it satisfies gauge invariance on its own, {\em i.e.}
its contribution vanishes upon contraction with $Q^\mu$.

We know from gauge invariance that
the five-point function $G^\mu$ must obey a Ward-Takahashi identity,
\begin{eqnarray}
 Q^\mu G^\mu(k_q,k_d;p_q,p_d) = \hskip 4cm \nonumber \\
    q_{q_f} G^{\rm q-dq} (k_q-Q,k_d;p_q,p_d) + 
    q_{d_f} G^{\rm q-dq}(k_q,k_d-Q;p_q,p_d)- \nonumber \\
    q_{q_i} G^{\rm q-dq}(k_q,k_d;p_q+Q,p_d) - q_{d_i}
   G^{\rm q-dq}(k_q,k_d;p_q,p_d+Q) \; .
  \label{WTG}
\end{eqnarray}
The charges $q_{q_f[q_i]}$ refer to the quark charges in the final
[initial] state, likewise $q_{d_f[d_i]}$ refer to the respective 
diquark charges. 
The notation is somewhat sloppy, since $G^{\rm q-dq}$ is
a matrix that describes the propagation between the different
charge eigenstates of quark and diquark. Therefore, eq.~(\ref{WTG})
is to be understood {\em for each matrix element} of 
$G^{\rm q-dq}$.

We will show that gauging $G^{\rm q-dq}$ leads
to a $G^\mu$ that fulfills the Ward-Takahashi identity
(\ref{WTG}). First,
we fix some notation for the basic quark and diquark vertices. 
Defining $-(S^{-1})^\mu=\Gamma^\mu_q$ to be the quark-photon vertex,
we find for the gauged quark propagator,
\begin{equation}
   S^\mu(k_q;p_q):= S(k_q)\;\Gamma^\mu_q\;S(p_q) \; . \label{Sgauge}
\end{equation}
The Ward-Takahashi identity for it reads
\begin{equation}
 Q^\mu S^\mu(k_q;p_q) =  q_q\;(S(k_q-Q)-S(p_q+Q)) \label{WTS}\; . 
\end{equation}
In the same manner we gauge the diquark propagator, given by 
eq.~(\ref{tildeD}), and write its Ward-Takahashi identity as
\begin{eqnarray}
 \tilde D^\mu (k_d;p_d)&:=& \tilde D(k_d)\;\Gamma^\mu_{dq}\;
        \tilde D(p_d) \\
      &=& \bp D(k_d) & 0 \\ 0 & D^{\alpha\alpha'}(k_d) \ep
      \bp \Gamma^\mu_{sc} & 0 \\ 0 & \Gamma^{\mu,\alpha'\beta'}_{ax}\ep
       \bp D(p_d) & 0 \\ 0 & D^{\beta'\beta}(p_d) \ep \nonumber\\
  Q^\mu \tilde D^\mu (k_d;p_d) &=& \bp q_{sc} & 0 \\ 0 & q_{ax} \ep
                  (\tilde D(k_d-Q) - \tilde D(p_d+Q)) \; ,
   \label{WTD}
\end{eqnarray}
where we have introduced the photon vertex with the
scalar diquark, $\Gamma^\mu_{sc}$ and the vertex with the
axialvector diquark, $\Gamma^{\mu,\alpha\beta}_{ax}$.
For the gauged free quark-diquark propagator we obtain 
after applying Leibniz' rule,
\begin{eqnarray}
 (G_0^{\rm q-dq})^\mu & :=& G_0^{\rm q-dq}\; \Gamma_0^\mu\; 
       G_0^{\rm q-dq} \; , \\ 
  \Gamma_0^\mu &=& \Gamma^\mu_q\;\tilde D^{-1} + S^{-1}\;
                  \Gamma^\mu_{dq} \; . 
 \label{G0mu}
\end{eqnarray}

The rules 2 and 3 can now be applied to the {\em inhomogeneous}
quark-diquark Bethe-Salpeter equation in order to find 
$G^\mu$,\footnote{As for $G^{\rm q-dq}$, the
kernel $K^{\rm BS}$ has not exactly the form as given
in eq.~(\ref{kbsdef})  but is rather written as a larger matrix
describing the quark exchange between all charge eigenstate
combinations of quark and diquark.}
\begin{equation}
 G^{\rm q-dq}=G_0^{\rm q-dq} + G_0^{\rm q-dq}\;K^{\rm BS}\;G^{\rm q-dq}
  \; .
\end{equation}
After simple algebraic manipulations using the above equation we find
\begin{eqnarray}
 G^\mu &=& (G_0^{\rm q-dq})^\mu + \left( G_0^{\rm q-dq}\; K^{\rm BS}
             \;G^{\rm q-dq} \right)^\mu \; , \\
       &=& G^{\rm q-dq}\;\left( \Gamma_0^\mu + (K^{\rm BS})^\mu \right)\; 
           G^{\rm q-dq} \; . \label{gmudef}
\end{eqnarray}
Since the Ward-Takahashi identities for the free diquark and 
quark vertices hold, eqs.~(\ref{WTS},\ref{WTD}), a Ward-Takahashi
identity for $\Gamma_0^\mu$ follows trivially. Now 
we consider: {\em if}  the following identity for $(K^{\rm BS})^\mu$
holds,
\begin{eqnarray}
 Q^\mu (K^{\rm BS})^\mu(k_q,k_d;p_q,p_d) = \hskip 3cm\nonumber \\
    q_{q_f} K^{\rm BS} (k_q-Q,k_d;p_q,p_d) + 
    q_{d_f} K^{\rm BS}(k_q,k_d-Q;p_q,p_d)- \nonumber \\
    q_{q_i} K^{\rm BS}(k_q,k_d;p_q+Q,p_d) - q_{d_i}
   K^{\rm BS}(k_q,k_d;p_q,p_d+Q) \; ,
 \label{WTK}
\end{eqnarray} 
{\em then} the identity
\begin{eqnarray}
 Q^\mu  \left( \Gamma_0^\mu + (K^{\rm BS})^\mu \right) = \hskip 3cm \nonumber \\
   q_{q_i} (G^{\rm q-dq})^{-1}(k_q,k_d;p_q+Q,p_d) +
   q_{d_i} (G^{\rm q-dq})^{-1}(k_q,k_d;p_q,p_d+Q) - \nonumber \\
   q_{q_f} (G^{\rm q-dq})^{-1}(k_q-Q,k_d;p_q,p_d) -
   q_{d_f} (G^{\rm q-dq})^{-1}(k_q,k_d-Q;p_q,p_d) 
  \label{WTGamma}
\end{eqnarray}
is valid, as can be inferred from eq.~(\ref{gmudef}) and
$(G^{\rm q-dq})^{-1}= (G_0^{\rm q-dq})^{-1} - K^{\rm BS}$, 
eq.~(\ref{Gqdq}). The Ward-Takahashi identity for $G^\mu$
itself, eq.~(\ref{WTG}), follows immediately after left and
right multiplication with $G^{\rm q-dq}$. 
Indeed, eq.~(\ref{WTK}) which we will call the Ward-Takahashi
identity
for the quark exchange kernel is derived from
a Ward-Takahashi identity\footnote{Any perturbative graph 
with external legs and an external photon can be shown 
to fulfill a corresponding Ward-Takahashi identity.} for 
$(G_0^{\rm q-dq} K^{\rm BS} G_0^{\rm q-dq})^\mu$
and eqs.~(\ref{WTS},\ref{WTD}).
Therefore we have established  eq.~(\ref{WTG}).

Matrix elements of the current between bound states are 
obtained from the expression for $G^\mu$, eq.~(\ref{gmudef}).
The full quark-diquark propagator, which sandwiches
the full vertex $\Gamma_0^\mu+(K^{\rm BS})^\mu$,
 can be approximated
at the nucleon bound state pole by
\begin{equation}
 G^{\rm q-dq} \sim \frac{\Psi \; (i\Slash{P}-M_n) \; \bar\Psi}
                        {P^2+M_n^2}+ {\rm regular\; terms} \;
\end{equation}
as will be shown in the beginning of section \ref{normsec}.
Inserted into eq.~(\ref{gmudef}), we find for the current matrix 
element (the residue term of the leading pole) 
\begin{equation}
 \langle P_f| \;  J^\mu\; |P_i \rangle =
  \bar \Psi_f \; \left( \Gamma_0^\mu + (K^{\rm BS})^\mu \right)\;
  \Psi_i \; .
  \label{nuccurdef}
\end{equation}
This expression describes how to obtain bound state matrix elements
from its Bethe-Salpeter wave function and is usually referred
to as {\em Mandelstam's formalism} \cite{Mandelstam:1955} although in
the original paper only contributions in impulse approximation have been
discussed.

When contracting eq.~(\ref{nuccurdef}) with $Q^\mu$, we can use the 
Ward-Takahashi identity (\ref{WTGamma}). Current conservation,
$Q^\mu \langle P_f| \;  J^\mu\; |P_i \rangle=0$, follows
immediately upon using the Bethe-Salpeter equations
\begin{equation}
  (G^{\rm q-dq})^{-1}\; \Psi = \bar \Psi \;(G^{\rm q-dq})^{-1} =0\; .
\end{equation} 
We have arrived at an important point in this chapter.
The nucleon current defined in eq.~(\ref{nuccurdef}) above
is conserved.
To guarantee gauge invariance, we must include in the
definition of the current operator the gauged
Bethe-Salpeter kernel $(K^{\rm BS})^\mu$ besides the
photon coupling to the free propagators $\Gamma_0^\mu$.
The latter contribution is commonly referred to as
the {\em impulse approximation}. Incidentally, the necessity of including
current contributions from the Bethe-Salpeter kernel
for a gauge invariant matrix element was first discussed
in ref.~\cite{Gross:1987bu} in a deuteron bound state equation
where the binding is provided by meson exchange.

We turn now to the construction of explicit expressions for 
the current operator in impulse approximation,
$\Gamma_0^\mu$, and
the gauged kernel $(K^{\rm BS})^\mu$.

\subsection{Impulse approximation}
\label{impulse-sec}

\begin{figure}[t]
\begin{center}
 \epsfig{file=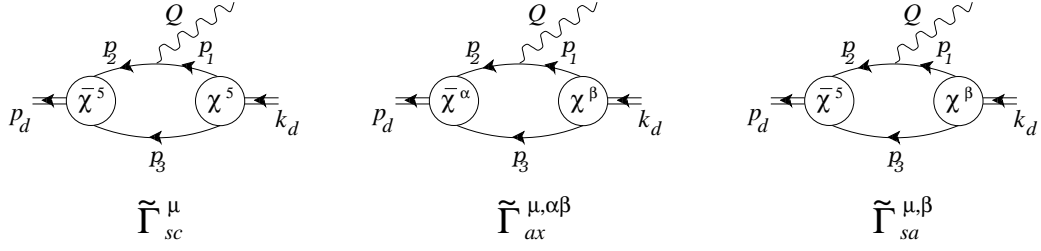, width=\columnwidth}
\end{center}
\caption{Resolved vertices: photon-scalar diquark, photon-axialvector diquark
  and anomalous scalar-axialvector diquark transition.}
\label{emresolve}
\end{figure}

Let us  specify the electromagnetic vertices of the basic constituents,
quark and diquark. For the quark-photon vertex, we employ the
perturbative expression as the model propagator
in eq.~(\ref{qprop}) is the tree-level one,
\begin{equation}
  \Gamma^\mu_q=-i q_q \gamma^\mu\; . \label{vertq}
\end{equation}
Of course it satisfies the corresponding Ward identity (\ref{WTS}).

The diquark propagators are the ones of a free spin-0 and spin-1
particle, therefore we also employ the corresponding free vertices
\begin{eqnarray}
  \Gamma^\mu_{dq} &=& \bp \Gamma^\mu_{sc}& 0 \\ 0 & 
     \Gamma^{\mu,\alpha\beta}_{ax}  \ep \; , \\
     \Gamma^\mu_{sc} &=& -q_{sc}\; (k_d+p_d)^\mu \; , \label{vertsc} \\
     \Gamma^{\mu,\alpha\beta}_{ax} &=& q_{ax} \left(
         -(p_d+k_d)^\mu\; \delta^{\alpha\beta} +
      (p_d^\alpha- \xi\;k_d^\alpha)\;\delta^{\mu\beta} +
      ( k_d^\beta - \xi\;p_d^\beta)\;\delta^{\mu\alpha} +
         \right. \nonumber \\
        & &  \left.\qquad \kappa\; (Q^\beta \; \delta^{\mu\alpha} -Q^\alpha\; 
      \delta^{\mu\beta} ) \right) \; . \label{vertax}
\end{eqnarray} 
As the spin-1 vertex is not necessarily
textbook material, we refer to ref.~\cite{Lee:1962vm} for its derivation.
The constant $\kappa$ which appears in eq.~(\ref{vertax}) 
denotes a possible anomalous magnetic moment
of the axialvector diquark.
We obtain its value from a calculation for vanishing momentum transfer
($Q^2=0$) in which the quark substructure of the diquarks is
resolved, {\em i.e.} in which a (soft) photon couples to the quarks within the
diquarks, {\em cf.} the second graph in figure \ref{emresolve}.
The calculation of   $\kappa$
is provided in appendix~\ref{dqres1}.
Although the parameters presented in table \ref{pars}
differ for the two sets, especially the diquark width
parameter $\lambda$, we obtain in both cases $\kappa = 1.0$ 
({\em cf.} table \ref{cc_1} in appendix \ref{dqres1}). This might seem
understandable from nonrelativistic intuition: the magnetic moments of two
quarks with charges $q_1$ and $q_2$ add up to $(q_1+q_2)/m_q$, the magnetic
moment of the axialvector diquark is $(1+\kappa)(q_1+q_2)/m_{ax}$ and if the
axialvector diquark is weakly bound, $m_{ax}\backsimeq 2 m_q$, then $\kappa
\backsimeq 1$. 

As already mentioned, there is one contribution to the current
which cannot be obtained by the gauging method: photon-induced 
anomalous\footnote{If the product of the parity eigenvalues
$\pi_i$
of the three particles at some vertex that describes their interaction
is $(-1)$ we call the vertex {\em anomalous}. For the
scalar-axialvector transitions, we have $\pi_\gamma=\pi_{sc}=
-\pi_{ax}=1$, thus the corresponding vertex is anomalous.}
transitions from a scalar to axialvector diquark and {\em vice versa}.
The corresponding vertex describing the transition from axialvector (with index $\beta$) to
scalar diquark  must have the form
 \begin{equation}
 \hat\Gamma^{\mu\beta}_{sa}=-i\frac{\kappa_{sa}}{2M_n}\, 
   \epsilon^{\mu\beta\rho\lambda}
                 (p_d+k_d)^\rho Q^\lambda \;,
 \label{sa_vert}
\end{equation}
and that for the reverse transition from an scalar to axialvector (index $\alpha
$) is given by,
\begin{equation}
 \hat\Gamma^{\mu\alpha}_{as}=i\frac{\kappa_{sa}}{2M_n}\, 
   \epsilon^{\mu\alpha\rho\lambda}
                 (p_d+k_d)^\rho Q^\lambda \;.
 \label{as_vert}
\end{equation}
The tensor structure of these anomalous diagrams  is derived
in appendix \ref{dqres1} by resolving the diquarks in a way as
represented by the right diagram in figure~\ref{emresolve}. The  explicit
factor $1/M_n$ was
introduced to isolate a dimensionless constant $\kappa_{sa}$. Its value is
obtained roughly as $\kappa_{sa}\simeq 2.1$ (with the next digit depending on
the parameter set, {\em cf.} table \ref{cc_1}).
Since the only possible transition
for nucleons involves the scalar $(ud)$ diquark and the axialvector
$[ud]$ diquark, there is only a constant charge (or flavor) factor
in eqs.~(\ref{sa_vert},\ref{as_vert}). It is $(q_d-q_u)/2=-1/2$.
We see immediately that  nucleon current contributions
using these vertices are transversal and thus do not affect current conservation.

The current operator associated with the anomalous transitions
reads
\begin{eqnarray}
 \bar\Psi_f\:\Gamma^\mu_{\rm sc-ax}\:\Psi_i 
   &=& w_{[ud]}\; \left( \langle J^\mu_{sa} \rangle^{\rm sc-ax} + 
       \langle J^\mu_{as} \rangle^{\rm ax-sc} \right )\\
  &=& w_{[ud]}\;
 \begin{array}[t]({cc}) \bar\Psi^5 & \bar \Psi^\alpha \end{array}
 \begin{array}[t]({cc})
   0  & S^{-1} \hat \Gamma^{\mu\beta}_{sa} \\  S^{-1} 
    \hat \Gamma^{\mu\alpha}_{as} & 0
 \end{array} 
 \begin{array}[t]({c}) \Psi^5 \\ \Psi^\alpha \end{array}\; . \quad
 \label{Gsamu}
\end{eqnarray}
As for nucleons only the transition between scalar $(ud)$ diquark
and axialvector $[ud]$ diquark is possible, the flavor weight factor $w_{[ud]}$
for the strength of the $[ud]$ diquark correlations
in the total axialvector correlations is included,
{\em cf.} eqs.~(\ref{pce},\ref{nce}). 

\begin{figure}[t]
\begin{center}
 \epsfig{file=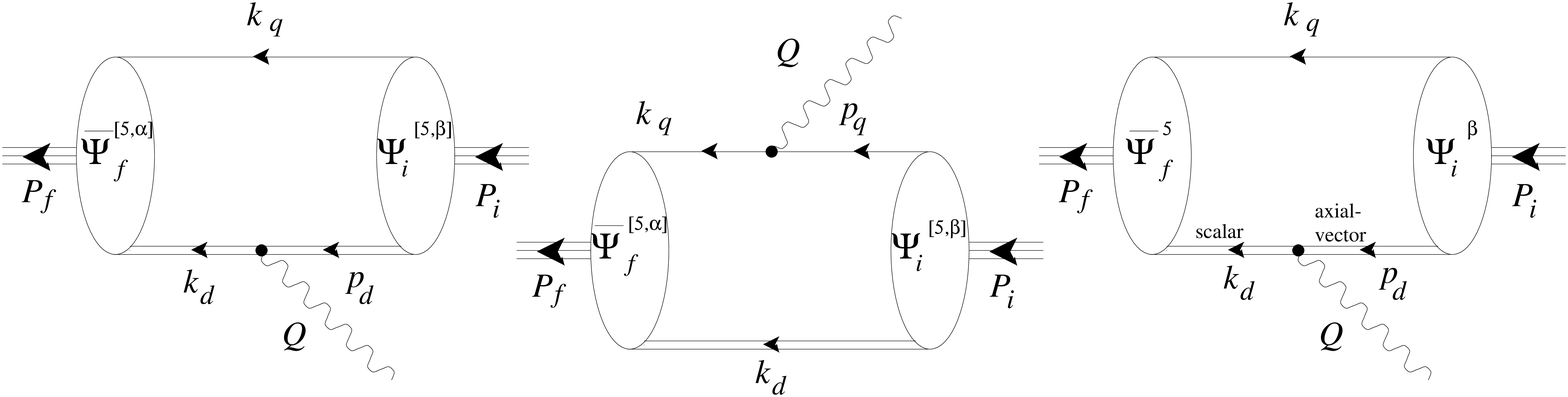,width=\columnwidth}
\end{center}
\caption{Impulse approximate contributions to the electromagnetic
current. For the scalar-axialvector transition, a diagram analogous to
the third one (with initial and final nucleon states interchanged)
has to be computed.}
\label{impulse}
\end{figure}

To conclude this subsection, we write down the explicit expressions
for the proton and neutron current matrix elements 
in the extended impulse approximation,
\begin{eqnarray}
  _{\rm p}\langle P_f|\; J^\mu\;|P_i \rangle^{\rm e-imp}_{\rm p} &=&
   \frac{2}{3} \langle J^\mu_q \rangle^{\rm sc-sc} +
   \frac{1}{3} \langle J^\mu_{sc} \rangle^{\rm sc-sc} +
   \langle J^\mu_{ax} \rangle^{\rm ax-ax} + \nonumber \\
   & &\frac{\sqrt{3}}{3} \left( \langle J^\mu_{sa} \rangle^{\rm sc-ax} +
         \langle J^\mu_{as} \rangle^{\rm ax-sc} \right) \; , \label{jimp}\\
 _{\rm n}\langle P_f|\; J^\mu \;|P_i \rangle^{\rm e-imp}_{\rm n} &=&
   -\frac{1}{3}\left( \langle J^\mu_q \rangle^{\rm sc-sc} -
    \langle J^\mu_q \rangle^{\rm ax-ax} -
    \langle J^\mu_{sc} \rangle^{\rm sc-sc} + \right. \nonumber \\
   &&\left. \langle J^\mu_{ax} \rangle^{\rm ax-ax} \right)-
   \frac{\sqrt{3}}{3} \left( \langle J^\mu_{sa} \rangle^{\rm sc-ax} +
         \langle J^\mu_{as} \rangle^{\rm ax-sc} \right) .\;\;\label{jimn}
\end{eqnarray}
The superscript `sc-sc' indicates that
the current operator is to be sandwiched between scalar
nucleon amplitudes for both the final and the initial state.
Likewise `sc-ax' denotes current operators that are sandwiched
between scalar amplitudes in the final and axialvector amplitudes in
the initial state, {\em etc.}  As an example we give
\begin{eqnarray}
 \langle J^\mu_q \rangle^{\rm ax-ax} &=& \bar\Psi^\alpha \; 
   (D^{-1})^{\alpha\beta}\, \hat\Gamma^\mu_q \; \Psi^\beta  \qquad 
    (\hat\Gamma^\mu_q = \Gamma^\mu_q/q_q) \label{sme}\\
   &=& \fourint{p} \bar\Psi^\alpha(P_f,p+(1-\eta)Q)\;
   (D^{-1})^{\alpha\beta}(k_d)\, \hat\Gamma^\mu_q \; \Psi^\beta(P_i,p) \; .
  \label{imp-example} \qquad\;
\end{eqnarray}
The weight factors multiplying the single matrix elements
arise from flavor algebra: we have chosen to define the matrix elements
to be independent of the charge of the struck particle, 
{\em cf.} eq.~(\ref{sme}).
The nucleon is decomposed into its charge eigenstates
of quark and diquark as in eqs.~(\ref{pce},\ref{nce}). 
Then the weight factors for the matrix elements are composed of the 
weight factors of the charge eigenstates and the charge of the
struck quark or diquark.

We note that the
axialvector amplitudes contribute to the proton current
only in combination with diquark current couplings.

For visualization, the diagrams of the extended impulse approximation
are shown in figure \ref{impulse}.

\subsection{Seagulls}
\label{sg-sec}

\begin{figure}
  \epsfig{file=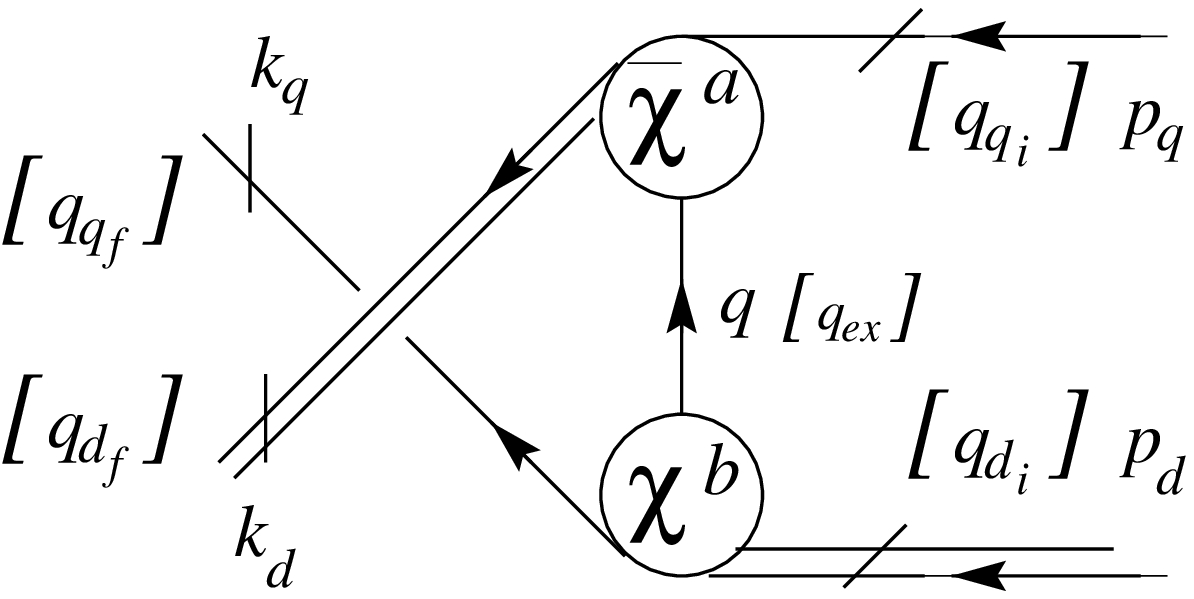,width=5cm} \hspace{3cm}
  \epsfig{file=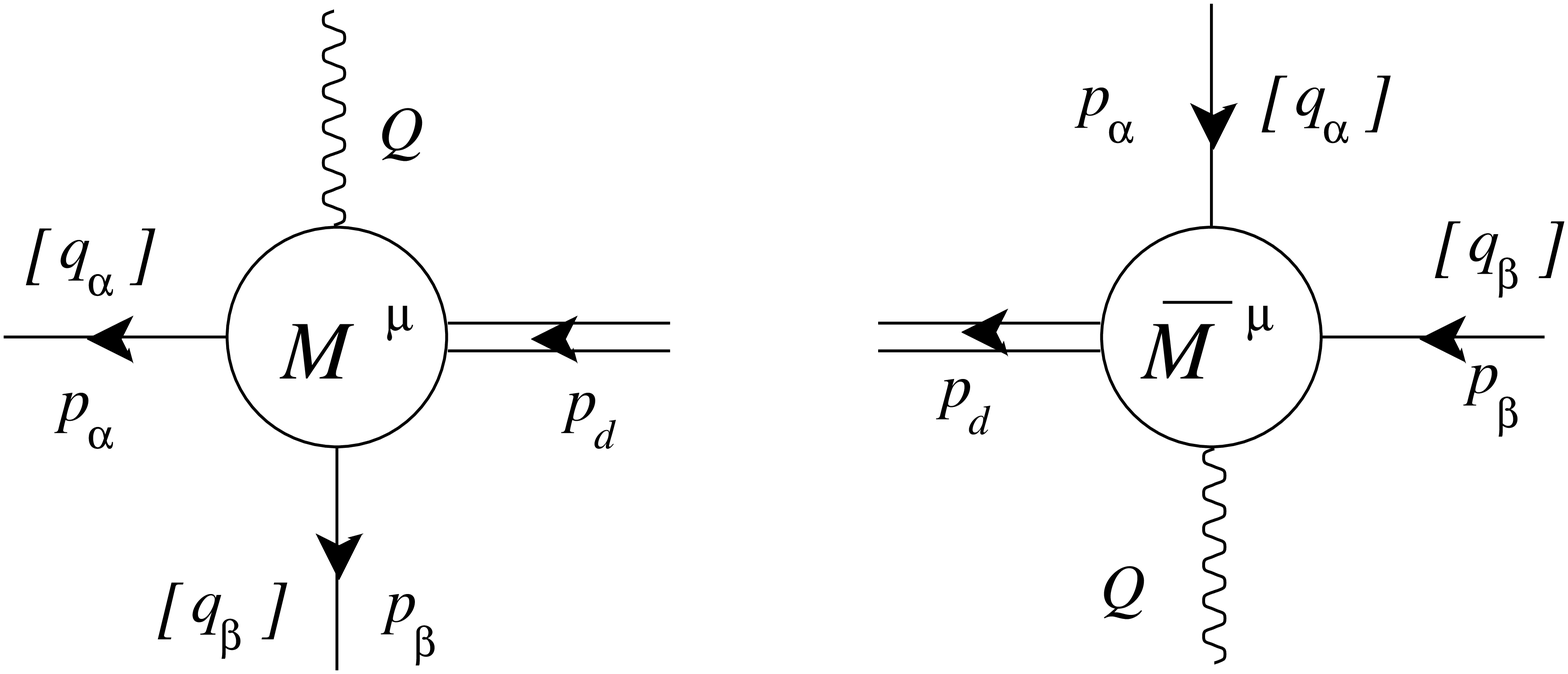,width=5cm}
  \caption{Left panel: The quark exchange kernel.
           Right panel: The seagull vertices.
           The charges belonging to the incoming and outgoing
           particles are enclosed in brackets, the other labels
           denote their momenta.}
  \label{kernfig1}
\end{figure}

A matrix element of the quark exchange kernel (between charge
eigenstates) is given by
\begin{equation}
  K^{\rm BS}_{ba} = -f_{ba}\; \chi^b(k_q,q) \;S^T(q)\;\bar\chi^a(q,p_q) \,
\end{equation} 
{\em cf.} figure \ref{kernfig1}. The corresponding flavor factor
which is unimportant in the following is given by $f_{ba}$. 
As in the actual solutions of the nucleon Bethe-Salpeter equation
we have assumed that the diquark-quark vertices depend on just
the relative momentum between the two ``participating'' quarks,
$\chi^a(k,p)\rightarrow \chi^a((k-p)/2)$.
Gauging the kernel leads to three terms,
\begin{equation}
  (K^{\rm BS}_{ba})^\mu = -f_{ba} \left[
   (M^b)^\mu \;S^T\; \bar\chi^a + \chi^b\; (S^\mu)^T\; \bar\chi^a +
    \chi^b \; S^T \;(\bar M^a)^\mu \right]\; .
   \label{Kgauge}
\end{equation} 
The gauged diquark-quark vertices $\chi^a$ and $\bar \chi^a$
are called $(M^a)^\mu$ and $(\bar M^a)^\mu$, respectively.
These are the seagull vertices mentioned in the beginning of the 
chapter. 

The Ward-Takahashi identity (\ref{WTK}) which leads to a conserved
nucleon current constrains the longitudinal part of the
seagull vertices. Contracting (\ref{Kgauge}) with $Q^\mu$,
inserting on the left hand side eq.~(\ref{WTK})
and using
the Ward-Takahashi identity (\ref{WTS}) for the gauged exchanged
quark on the right hand side, we find
\begin{eqnarray}
 Q^\mu (M^a)^\mu(p_\alpha,p_\beta) & \stackrel{!}{=} & q_\alpha\; \left[
   \chi^a(p_\alpha-Q,p_\beta) - \chi^a(p_\alpha,p_\beta) \right] 
    + \nonumber \label{co1}\\
   & &q_\beta\; \left[ \chi^a(p_\alpha,p_\beta-Q)- \chi^a(p_\alpha,p_\beta)
 \right] \; \\ 
   && \left( {\rm with}\quad p_d+Q=p_\alpha+p_\beta \right) \; , \nonumber \\
 Q^\mu (\bar M^a)^\mu(p_\alpha,p_\beta) & \stackrel{!}{=} &
  - q_\alpha\; \left[
  \bar\chi^a(p_\alpha+Q,p_\beta) - \bar\chi^a(p_\alpha,p_\beta) \right]  
   - \nonumber \label{co2}\\
   & & \;\; q_\beta\; \left[ \bar\chi^a(p_\alpha,p_\beta+Q)- \bar\chi^a(p_\alpha,p_\beta) \right] \; \\
   && \left( {\rm with}\quad p_d-Q=p_\alpha+p_\beta \right) \; . \nonumber 
\end{eqnarray}  
The notation is explained in the right panel of
figure \ref{kernfig1}.
Of course, such constraints do not fix the seagull vertices
completely. We choose a form which solves for this constraint
and is regular in the limit $Q\rightarrow 0$. With
$p=(p_\alpha-p_\beta)/2$ the vertices read
\begin{eqnarray}
  (M^a)^\mu(p,Q;q_\alpha,q_\beta)&=& q_\alpha \;\frac{(4p-Q)^\mu}{4p\cdt Q-Q^2}
    \;\left[ \chi^a(p-Q/2) -\chi^a(p) \right] + \nonumber \\
  & &  q_\beta\; \frac{(4p+Q)^\mu}{4p\cdt Q+Q^2}
     \;\left[ \chi^a(p+Q/2) -\chi^a(p) \right] \; , \label{seag1} \\
  (\bar M^a)^\mu(p,Q;q_\alpha,q_\beta)&=& -q_\alpha\; \frac{(4p-Q)^\mu}{4p\cdt Q-Q^2}
    \;\left[ \chi^a(p+Q/2) -\chi^a(p) \right] + \nonumber \\
  & &  -q_\beta\; \frac{(4p+Q)^\mu}{4p\cdt Q+Q^2}
     \;\left[ \chi^a(p-Q/2) -\chi^a(p) \right] \; . \label{seag2}
\end{eqnarray}
In the limit $Q\rightarrow 0$ they simplify to
\begin{eqnarray}
  (M^a)^\mu(p,0;q_\alpha,q_\beta) &=& 
   -\frac{1}{2}\,(q_\alpha-q_\beta)\;p^\mu\;
    \frac{d\,\chi^a}{d\,p^2} \; ,\label{sg0-1}\\
  (\bar M^a)^\mu(p,0;q_\alpha,q_\beta) &=&
   -\frac{1}{2}\,(q_\alpha-q_\beta)\;p^\mu\;
    \frac{d\,\bar \chi^a}{d\,p^2}  \;. \label{sg0-2}
\end{eqnarray}

We add as a side remark that these seagull terms were derived 
a little bit differently in ref.~\cite{Oettel:1999gc}.
There the Ward-Takahashi identity for the matrix element
$\langle 0|\, T qq\bar q \bar q J^\mu \, |0 \rangle $ has
been considered. Using the separable {\em ansatz} for the
two-quark correlations, eq.~(\ref{tsep}), for expressions
like $\langle 0|\, T qq\bar q \bar q \,|0\rangle $, the constraints
(\ref{co1},\ref{co2}) have been derived under the assumption
that the photon-diquark vertices contain no quark loop,
{\em i.e.} they are given by the perturbative 
expressions (\ref{vertsc},\ref{vertax}). However, in the limit
$Q\rightarrow 0$ the seagulls {\em must} have the form 
as in eqs.~(\ref{sg0-1},\ref{sg0-2}), as here
the Ward-Takahashi identities reduce to the (differential)
Ward identities which fix the vertices completely. In this kinematical
situation a dressing quark loop in the photon-diquark vertices
must yield the diquark charge (for an on-shell diquark).

Furthermore we note that seagull vertices of  a similar kind were
first considered in ref.~\cite{Ohta:1989ji}. The author
discusses extended meson-baryon vertices and derives 
meson-baryon-photon vertices from a Lagrangian point of view
by applying the minimal substitution rule.

For completeness we give the explicit expressions for the
seagull and exchange quark contributions that are obtained
when recombining the quark-diquark charge eigenstates
to the correct nucleon flavor state. The symbolic matrix 
multiplications are to be understood as accompanied
by an integration over the relative quark-diquark momenta
in the initial and final states, respectively,
and a summation over Dirac and Lorentz indices. First we
write down the current term
for the proton that corresponds to the diagram in the lower right
of figure \ref{7dim},
\begin{eqnarray}
 \label{sg}
 \langle J^\mu_{sg} \rangle^{\rm proton} &=& -\half\;
   \begin{array}[t]({cc}) \bar\Psi^5 & \bar \Psi^\alpha \end{array}
    \;
   \begin{array}[t]({cc})
    K^\mu_{sg,ss} & K^\mu_{sg,sa} \\ K^\mu_{sg,as} & K^\mu_{sg,aa} 
   \end{array} 
   \;
    \begin{array}[t]({c}) \Psi^5 \\ \Psi^\beta \end{array} 
    \begin{array}[t]{c} \; \\ \; ,\end{array} \\
     \nonumber
  K^\mu_{sg,ss} &=& 
   (M^5)^\mu(p_1',Q;q_u,q_d)\,S^T(q')\,\bar\chi^5(p_2)  \; , \\
     \nonumber
  K^\mu_{sg,sa} &=& -\frac{1}{3}\sqrt{3}\;
   (M^\beta)^\mu(p_1',Q;3q_u,2q_u+q_d) \,S^T(q')\,\bar\chi^5(p_2) \; , \\
     \nonumber
  K^\mu_{sg,as} &=& -\frac{1}{3}\sqrt{3}\;
   (M^5)^\mu(p_1',Q;q_u+2q_d,2q_u+q_d)\,S^T(q')\,\bar\chi^\alpha(p_2)\; ,\\
     \nonumber
  K^\mu_{sg,aa} &=& -\frac{1}{3}\;
   (M^\beta)^\mu(p_1',Q;q_u+2q_d,4q_u-q_d)\,S^T(q')\,\bar\chi^\alpha(p_2)
   \; ,
\end{eqnarray}
to be followed by the conjugated seagull contributions as depicted in the diagram
to the lower left in figure \ref{7dim},
\begin{eqnarray}
 \label{sgbar}
 \langle J^\mu_{\overline{sg}} \rangle^{\rm proton} &=& -\half\;
    \begin{array}[t]({cc}) \bar\Psi^5 & \bar \Psi^\alpha \end{array}
    \;
   \begin{array}[t]({cc})
    K^\mu_{\overline{sg},ss} & K^\mu_{\overline{sg},sa} 
      \\ K^\mu_{\overline{sg},as} & K^\mu_{\overline{sg},aa} 
   \end{array} 
   \;
    \begin{array}[t]({c}) \Psi^5 \\ \Psi^\beta \end{array} 
    \begin{array}[t]{c} \; \\ \; ,\end{array} \\
     \nonumber
  K^\mu_{\overline{sg},ss} &=&
    \chi^5(p_1)\,S^T(q)\,(\bar M^5)^\mu(p_2',Q;q_d,q_u) \; , \\
     \nonumber
  K^\mu_{\overline{sg},sa} &=& -\frac{1}{3}\sqrt{3}\;
    \chi^\beta(p_1)\,S^T(q)\,(\bar M^5)^\mu(p_2',Q;q_d+2q_u,q_u+2q_d) \; , \\
     \nonumber
  K^\mu_{\overline{sg},as} &=& -\frac{1}{3}\sqrt{3}\;
    \chi^5(p_1)\,S^T(q)\,(\bar M^\alpha)^\mu(p_2',Q;2q_u+q_d,3q_u) \; , \\
     \nonumber
  K^\mu_{\overline{sg},aa} &=&  -\frac{1}{3}\;
    \chi^\beta(p_1)\,S^T(q)\,(\bar M^\alpha)^\mu(p_2',Q;4q_u-q_d,q_u+2q_d) \; , 
\end{eqnarray} 
and to be completed finally by the exchange quark contribution, {\em cf.} the upper
diagram in  figure \ref{7dim}, 
\begin{eqnarray}
 \label{ex}
 \langle J^\mu_{ex} \rangle^{\rm proton} &=& -\half\;
    \begin{array}[t]({cc}) \bar\Psi^5 & \bar \Psi^\alpha \end{array}
    \;
   \begin{array}[t]({cc})
    K^\mu_{ex,ss} & K^\mu_{ex,sa}
      \\ K^\mu_{ex,as} & K^\mu_{ex,aa}
   \end{array}
   \;
    \begin{array}[t]({c}) \Psi^5 \\ \Psi^\beta \end{array}
    \begin{array}[t]{c} \; \\ \; ,\end{array} \\
     \nonumber
   K^\mu_{ex,ss} &=& q_d \; \chi^5(p_1) \, S^T(q)
       \left(\Gamma^\mu_q\right)^T S^T(q')\, \bar\chi^5(p_2) \; , \\
     \nonumber
   K^\mu_{ex,sa} &=& -\frac{1}{3}\sqrt{3}\;(2q_u+q_d) \; \chi^\beta(p_1) \, S^T(q)
       \left(\Gamma^\mu_q\right)^T S^T(q')\, \bar\chi^5(p_2) \; , \\
     \nonumber
   K^\mu_{ex,as} &=& -\frac{1}{3}\sqrt{3}\; (2q_u+q_d) \; \chi^5(p_1) \, S^T(q)
       \left(\Gamma^\mu_q\right)^T S^T(q')\, \bar\chi^\alpha(p_2) \; , \\
     \nonumber
   K^\mu_{ex,aa} &=& -\frac{1}{3}\; (4q_u-q_d)  \; \chi^\beta(p_1) \, S^T(q)
       \left(\Gamma^\mu_q\right)^T S^T(q')\, \bar\chi^\alpha(p_2) \; . 
\end{eqnarray}
The respective contributions to the neutron current are obtained
by interchanging the charges of the $u$ and $d$ quarks,
$q_u \leftrightarrow q_d$. The relative momenta in the diquark-quark
vertices and seagulls are according to figure \ref{7dim}:
$p_1=k_q-q$, $p_1'=k_q-q'$, $p_2=q'-p_q$ and $p_2'=q-p_q$.

\begin{figure}[t]
 \begin{center}
   \epsfig{file=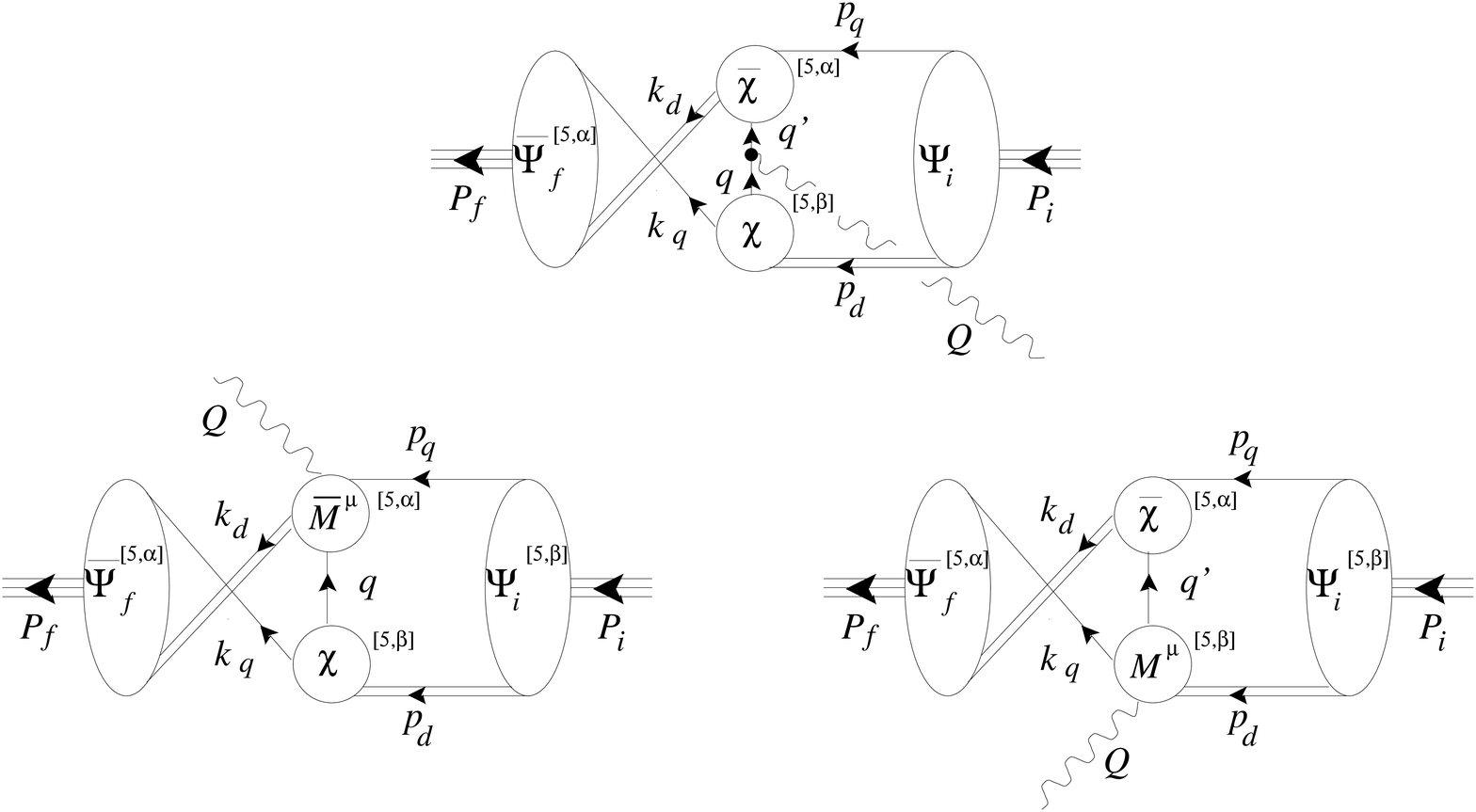,width=\columnwidth}
 \end{center}
 \caption{Exchange quark and seagull diagrams.}
 \label{7dim}
\end{figure}

The nucleon current operator is now complete. We recapitulate that
the current matrix element is written as,
\begin{equation}
  \langle P_f| \;  J^\mu\; |P_i \rangle :=
  \bar \Psi_f \;\Gamma^\mu_{\rm nuc} \; \Psi_i =
  \bar \Psi_f \; \left( \Gamma_0^\mu + \Gamma_{\rm sc-ax}^\mu +
   (K^{\rm BS})^\mu \right)\;
  \Psi_i \; \label{curtot} ,
\end{equation}
including all the discussed elements.
The impulse approximation operator $\Gamma_0^\mu$ involves the
couplings to quark and diquark, {\em cf.} 
eqs.~(\ref{G0mu},\ref{vertq}--\ref{vertax}). The extended
impulse approximation comprises additionally the anomalous transitions
between  scalar and axialvector diquarks. Their operator 
$\Gamma_{\rm sc-ax}^\mu$ is given in eq.~(\ref{Gsamu}). Finally 
we have discussed the
coupling to the exchange quark kernel, $(K^{\rm BS})^\mu$,
in the last subsection.

\subsection{Normalization of the Bethe-Salpeter wave function 
    and the nucleon charges}
\label{normsec}

A bound state Bethe-Salpeter equation is a homogeneous integral
equation, therefore its solutions need subsequent normalization.
The canonical normalization condition is derived by demanding
that the bound state contributes a pole to the full two-body
propagator with residue 1. 
As the nucleons are spin-1/2 particles, we demand 
that near the nucleon pole the full quark-diquark propagator  
behaves like a free fermion propagator,
\begin{eqnarray}
   S(P)=\frac{i\Slash{P}-M_n}{P^2+M_n^2}  &=& 
   (-2M_n)
  \sum_\sigma \frac{ u_{\sigma}\;\bar u_{\sigma} }
   {P^2+M_n^2}\; \Rightarrow \\
   (G^{\rm q-dq})_{\rm pole} &\stackrel{!}{=} &(-2M_n)
   \frac{(\Psi u_{\sigma})\;(\bar u_{\sigma}\bar \Psi) }
   {P^2+M_n^2} \; . \label{Gqdqpole}
\end{eqnarray}
Now we write the inhomogeneous Bethe-Salpeter equation 
(\ref{Gqdq}) for the full quark-diquark
propagator   in the form
\begin{equation}
  G^{\rm q-dq} = G^{\rm q-dq}\; \left( (G_0^{\rm q-dq})^{-1} -
         K^{BS} \right)  G^{\rm q-dq}
\end{equation}
and insert the pole form (\ref{Gqdqpole})
on both sides to find the condition
\begin{equation}
 \lim_{P^2\rightarrow -M_n^2} = \bar u_{\sigma}\bar \Psi\,
    \frac{(G_0^{\rm q-dq})^{-1} -  
         K^{BS}}{P^2+M_n^2}  \, \Psi u_{\sigma} 
    \stackrel{!}{=} -\frac{1}{2M_n}\; .
\end{equation}
Summing over spins, applying l'H{\^o}spital's rule and 
writing the result covariantly yields the final normalization
condition
\begin{equation}
  -\frac{1}{2M_n}\;{\rm Tr}\; \bar \Psi \left[ P^\mu \dpartial{P^\mu}
   \left( (G_0^{\rm q-dq})^{-1} -  
         K^{BS} \right) \right]_{P=P_n} \Psi \stackrel{!}{=} 1 \; ,
 \label{nucnorm}
\end{equation}
where $P_n$ denotes the on-shell nucleon momentum.
 
On the other hand, the electric form factor $G_E$ is normalized
to yield the physical nucleon charges in the soft limit, 
$Q\rightarrow 0$.  Using eq.~(\ref{getrace})
that extracts $G_E$ from the current matrix element and
eq.~(\ref{nuccurdef}) for the definition of the 
current operator\footnote{As the scalar-to-axialvector transitions
are transversal, they do not contribute to the nucleon charges.}
  we find another condition,
\begin{equation}
 -\frac{1}{2M_n}\;{\rm Tr}\; \bar \Psi
   \left( \Gamma_0^\mu + (K^{\rm BS})^\mu \right) \Psi
  \stackrel{!}{=} q_N \; .
 \label{cnorm}
\end{equation}
In fact, the two conditions (\ref{nucnorm},\ref{cnorm}) are equivalent
for the proton which has non-zero charge $q_N$. In the literature, a
proof of this statement can be found \cite{Nishijima:1967}
where the authors treat the case of 
one interaction channel and the momentum partitioning fixed
by $\eta=1/2$. Crucial for their proof is the 
differential form of the Ward-Takahashi identity for the gauged 
full 2-particle
propagator which corresponds in our case to 
the gauged quark-diquark propagator, eq.~(\ref{WTG}).
The proof is easily accomodated for our purposes and here it comes.

In our notation for the full quark-diquark propagator
indicating all four particle momenta,
$G^{\rm q-dq}(k_q,k_d;p_q,p_d)$, the momenta are not independent.
We employ therefore the equivalent notation 
$G^{\rm q-dq}(k,p;P)$ with $P=k_q+k_d=p_q+p_d$ and  $p,k$
being the relative momenta. Our usual conventions hold:
\begin{equation}
 k_q[p_q]=\eta P+k[p]\;, \quad k_d[p_d]=(1-\eta)P-k[p] \; .
\end{equation}
For the gauged propagator $G^\mu(k_q,k_d;p_q,p_d)$ the four momenta
are independent, and here we define 
\begin{equation}
 P_f=p_q+p_d \;, \quad P_i=k_q+k_d \; .
\end{equation}
The Ward-Takahashi identity (\ref{WTG}) can now be formulated
as
\begin{eqnarray} 
  \mbox{\hskip 3cm} (P_f-P_i)^\mu G^\mu &=&  \\
 & & \mbox{\hskip -7cm}
    q_{q_f} G^{\rm q-dq}(k-(1-\eta)(P_f-P_i),p;P_i) +
   q_{d_f} G^{\rm q-dq}(k+\eta(P_f-P_i),p;P_i) - \nonumber \\
 & & \mbox{\hskip -7cm}
   q_{q_i} G^{\rm q-dq}(k,p+(1-\eta)(P_f-P_i);P_f) -
   q_{d_i} G^{\rm q-dq}(k,p-\eta(P_f-P_i);P_f)  \; . \nonumber
\end{eqnarray}
As we remarked earlier, this identity holds for each matrix element
of $G^{\rm q-dq}$ that connects initial and final charge eigenstates
of the quark-diquark system.
We differentiate both sides with respect to $P_f^\mu$ and subsequently assume
the soft limit, $P_i=P_f=:P$,  to find
\begin{eqnarray} \label{hilf1} 
  G^\mu &=& -(q_{q_i}+q_{d_i}) \dpart{G^{\rm q-dq}}{P^\mu}(k,p;P)+ \\
  & &  \mbox{\hskip -1cm}
(\eta q_{d_f}- (1-\eta) q_{q_f} )\dpart{G^{\rm q-dq}}{k^\mu}
    (k,p;P) \nonumber
+(\eta q_{d_i}- (1-\eta) q_{q_i} )\dpart{G^{\rm q-dq}}{p^\mu}
  (k,p;P) \; .
\end{eqnarray}
Since $q_{q_i}+q_{d_i}=1$ (proton charge) for all
the matrix elements of $G^{\rm q-dq}$, we employ for the first
term on the right hand side 
\begin{eqnarray}
  \dpartial{P^\mu}\left[G^{\rm q-dq}(G^{\rm q-dq})^{-1}\right]&=&0
 \quad \Rightarrow \nonumber \\ 
  \dpart{G^{\rm q-dq}}{P^\mu} &=& -G^{\rm q-dq}
   \dpart{(G^{\rm q-dq})^{-1}}{P^\mu} G^{\rm q-dq} \; .
\end{eqnarray}
On the other hand, we can use eq.~(\ref{gmudef}) for $G^\mu$
on the left hand side of eq.~(\ref{hilf1}). Therefore
we obtain
\begin{eqnarray}
    \nonumber
 G^{\rm q-dq}\left( \Gamma_0^\mu + (K^{\rm BS})^\mu \right) G^{\rm q-dq}
  = -G^{\rm q-dq}
   \dpart{(G^{\rm q-dq})^{-1}}{P^\mu} G^{\rm q-dq} + \\
(\eta q_{d_f}- (1-\eta) q_{q_f} )\dpart{G^{\rm q-dq}}{k^\mu} 
+(\eta q_{d_i}- (1-\eta) q_{q_i} )\dpart{G^{\rm q-dq}}{p^\mu}
   \; . \label{hilf2}
\end{eqnarray}
Near the bound state the full quark-diquark propagator
can be written in the pole form of eq.~(\ref{Gqdqpole}).
We see  that the first terms on the left and right hand
side of eq.~(\ref{hilf2}) have a double pole at $P^2=-M_n^2$,
whereas the remaining terms can only contribute a single pole.
Fortunately these drop out when comparing residues and we are left with
\begin{equation}
  \bar u_{\sigma}\bar \Psi\,\dpart{(G^{\rm q-dq})^{-1}}{P^\mu}\,
  \Psi u_{\sigma'}  =
  \bar u_{\sigma}\bar \Psi\, \left( \Gamma_0^\mu + 
     (K^{\rm BS})^\mu \right) \,
 \Psi u_{\sigma'}  \; .
\end{equation}
Putting $\sigma=\sigma'$, performing the spin summation and contracting
with $P^\mu$, we indeed find the equality of the canonical
normalization condition (\ref{nucnorm}) and the charge normalization 
condition (\ref{cnorm}). Also the neutron has the correct charge:
Although the overall normalization cannot be fixed in this case,
eq.~(\ref{hilf1}) guarantees that its electric form factor is
zero in the soft limit since $q_{q_i}+q_{d_i}=0$ for all
quark-diquark channels. 

It is possible to prove the equivalence of 
eqs.~(\ref{nucnorm},\ref{cnorm}) without explicitly 
recurring to the Ward-Takahashi
identity (\ref{WTG}). The proof is slightly more technical and
for the scalar diquark sector of the model it is described
in ref.~\cite{Oettel:1999gc}.

\subsection{Summary}

In conclusion we want to emphasize the power of the general Ward-Takahashi 
identity for the full quark-diquark propagator, eq.~(\ref{WTG}).
Its implications allowed us to derive 
($i$) conservation of the nucleon current,
($ii$) constraints on the longitudinal part of
{\em all} contributing graphs, notably of the irreducible 
quark-diquark-photon vertices (seagulls) and finally ($iii$)
the equivalence of the canonical and the charge normalization
condition. It is interesting to note that actually current conservation
holds for {\em each single diagram} depicted in figures
\ref{impulse} and \ref{7dim}. This has been proved in 
ref.~\cite{Keiner:1996bu} by using just the transformation
properties of the respective matrix elements under
time reversal and parity. But as we know, gauge invariance
provides {\em more} information than  conserved current matrix elements
on-shell.
This additional information is manifest in the 
(off-shell) Ward-Takahashi 
identities and it is certainly wrong to restrict oneself to 
assorted current contributions since at least the 
correct nucleon charges cannot be obtained in general.
This is a shortcoming of a quark-diquark model investigated
in refs.~\cite{Bloch:1999ke,Bloch:1999rm}. The authors
employ parametrized nucleon Faddeev amplitudes and
calculate form factors in a generalized impulse
approximation which, besides containing overcounted contributions
\cite{Blankleider:1999xp}, does not respect a Ward-Takahashi identity
like eq.~(\ref{WTG}). The immediate consequence turned out to be that
the correct nucleon charges have been obtained only for a special choice
of parameters.

\section{Numerical calculations}

Before we present the results for the nucleon electric and magnetic 
form factors, we will discuss shortly the numerical method.
As we have calculated the Bethe-Salpeter wave functions $\Psi$
in the rest frame of the nucleon, we have to discuss their boosting
to a moving frame. Thereby we will find that an accurate description of
the boosted wave function is linked to the accuracy of its Chebyshev
expansion in the rest frame.
Since the {\em vertex} function $\Phi$ show better 
convergence in the Chebyshev expansion, it would be desirable
to use $\Phi$ instead of the wave function $\Psi$. We find that
this is feasible only for the diagrams of the extended impulse
approximation but that due to their singularity structure the diagrams
evaluated with the vertex function acquire additional residue 
contributions in a certain kinematical region. In this region
the (Euclidean) vertex function $\Phi$ is not connected to the
(Euclidean) wave function $\Psi$ by simply cutting quark and diquark legs    
off $\Psi$.

\subsection{Numerical method}
\label{num_em}

The numerical computation of the form factors is done in the Breit frame
where 
\begin{eqnarray} 
      Q^\mu & = &  (\vect Q,0 ) \; , \nonumber\\
      P_i^\mu & = &  (- \vect Q/2,i\omega_Q) \; , \label{BF_def}\\
      P_f^\mu & = &  (\vect Q/2,i\omega_Q) \; , \nonumber \\
      P^\mu =(P_i+P_f)/2 & = & 
             (\vect 0,i\omega_Q ) \; ,   \nonumber 
\end{eqnarray}
with $\omega_Q = \sqrt{ M_n^2 + \vect Q^2/4 }$. 

The total current operator, eq.~(\ref{curtot}), employs the wave functions
$\Psi$ in the initial and final state. These are determined by
the basic covariants and the corresponding scalar functions
$\hat S_i(p^2,z=\hat p \cdt \hat P_n)$, 
$\hat A_i(p^2,z=\hat p \cdt \hat P_n)$, {\em cf.} eq.~(\ref{wex_N}).
We recall that we absorbed the angular ($z$) dependence in the Chebyshev 
expansion, eq.~(\ref{cheby-w}). In the rest frame,
$P_n=(\vect 0,iM_n)$, $z$ is real and
$|z|\le 1$, therefore the expansion is {\em always} convergent.
In the Breit frame we have the real relative momenta in the
initial state, $p$, and the final state, $k$, which are integrated
over. Therefore the angular variables $z_k=\hat k \cdt \hat P$ and
$z_p=\hat p \cdt \hat P$ are real and not larger than 1.
The angular variables that enter the initial and final state 
wave functions as an argument
are, however,
\begin{eqnarray}
 \label{zi}
 z_i=\hat p \cdt \hat P_i &=&\frac{\omega_Q}{M_n}\,z_p - i\, \frac{1}{2}
      \frac{|\vect{Q}|}{M_n}\, \hat p\cdt \hat Q \; ,\\  
 \label{zf}
 z_f=\hat k \cdt \hat P_f &=&\frac{\omega_Q}{M_n}\,z_k + i\, \frac{1}{2}
      \frac{|\vect{Q}|}{M_n}\, \hat k\cdt \hat Q \; .  
\end{eqnarray}
Therefore, in order to use the rest frame solutions,
analytical continuation to complex values for $z_i$ and $z_f$ is necessary.
This can be justified for the bound-state Bethe-Salpeter 
wave functions $\Psi$.  These can be expressed as vacuum 
expectation values of local (quark field)
and almost local operators (diquark field) and we can resort to 
the domain of holomorphy of
such expectation values to continue the relative momenta of the 
bound-state 
Bethe-Salpeter wave function $\Psi (p,P) $ into the 4-dimensional complex Euclidean
space necessary for the computation of Breit-frame matrix elements from
rest frame nucleon wave functions. 
The necessary analyticity properties are
manifest in the expansion in terms of Chebyshev polynomials with complex
arguments.  

There is a practical obstacle in the outlined procedure.
As the maximum of the squared absolute values
$|z_{i[f]}|^2_{\rm max} = 1+Q^2/(4M_n^2)$ rises with increasing
$Q^2$, we encounter convergence problems for the current matrix elements
as the Chebyshev moments of higher order $n$ become more and more
important ($T_n$ is a polynomial of degree $n$). 
Consider the strengths of the wave function Chebyshev moments
presented in appendix \ref{wave-app} (Set I). 
The fourth Chebyshev moment of 
the dominating function $\hat S_1$ is about 7 \% of 
the zeroth moment. With $T_4(z)=8z^4-8z^2+1$ 
one can estimate the critical $Q^2$ for which contributions from $\hat S_1^4$
are comparable to the ones from $\hat S_1^0$,
\begin{equation}
 | \hat S_1^4\,T_4(z_f) |_{\rm max} \sim 
  | \hat S_1^0\,T_0(z_f) |_{\rm max}
 \quad \Rightarrow \quad Q^2 \sim 1.4\; {\rm GeV}^2\; .
\end{equation}
As a consequence, the grid size in the angular integrations has to be 
increased beyond this value of $Q^2$ because the higher Chebyshev
polynomials introduce rapid oscillations in the integrand. 
Since the Chebyshev moments of the vertex function converge  
much more rapidly ( $|S_1^4|_{\rm max} =0.01 \,|S_1^0|_{\rm max}$
for Set II),
it would be desirable to use $\Phi$ instead of $\Psi$ in
the expressions for the current, eq.~(\ref{curtot}). 
Having $\Psi=G_0^{\rm q-dq}\, \Phi$, the free quark and diquark 
propagators with the corresponding  singularities are introduced  
in the integrand. For sufficiently small
$Q^2$ these are outside the complex integration domain. For larger $Q^2$,
these singularities enter the integration domain, therefore,
a proper treatment of these singularities is required when they
come into the domain of interest.

For the impulse approximation contributions to the form factors 
we have been able to take the corresponding residues 
into account explicitly in
the integration. Here the integration is reduced to one four-dimensional
integral over either $p$ or $k$ due to the inherent $\delta$-function
of the free quark-diquark propagator $G_0^{\rm q-dq}$. 
Being technically  somewhat involved, we have placed the matter  into
appendix~\ref{residue}. 
For these contributions, one can compare the calculations
with boosted $\Psi$ and $\Phi$ and verify
numerically that they yield the same, unique results
(up to several $Q^2$).
This is demonstrated in appendix \ref{residue}.

For the current contributions from the gauged Bethe-Salpeter kernel,
{\em cf.} figure \ref{7dim}, 
two four-dimensional integrations over $p$ and $k$
are required. The complex singularity structure of the integrand
did not allow a similar treatment as described in appendix 
\ref{residue} for the diagrams of the impulse approximation.
Therefore we have resorted to employ the boosted wave functions
$\Psi$. As a result, the numerical uncertainty (for these diagrams only)
exceeded the level of a few percent beyond momentum transfers
of 2.5 GeV$^2$. Due to this limitation (which could
be avoided by increasing simply the used computer time)
the form factor results presented in section \ref{results1}
are restricted to the region
of momentum transfers below 2.5 GeV$^2$.

In addition to the aforementioned complications, there is another bound on
the value of $Q^2$ above which the exchange and seagull diagrams cannot be 
evaluated. It is due to the singularities in the diquark-quark vertices 
$\chi^{5[\mu]} (p_i)$ and in the exchange quark propagator. The rational
$n$-pole forms of the diquark-quark vertices,
$V_{n-{\rm pole}}(p^2)=(\lambda_n^2/(\lambda_n^2+p^2))^n$ 
for example yield the
following upper bound,
\begin{eqnarray}
 Q^2 < 4 \left( \frac{4\lambda_n^2}{(1-3\eta)^2} -M_n^2 \right) \; .
\end{eqnarray}
A free constituent propagator for the exchange quark gives the additional 
constraint,
\begin{eqnarray}
 Q^2 < 4 \left( \frac{m_q^2}{(1-2\eta)^2}-M_n^2 \right) \; .
\end{eqnarray}
It turns out, however, that these bounds on $Q^2$ are insignificant for the
model parameters employed in the calculations described herein.

\subsection{Results}
\label{results1}

For the scalar diquark sector, we have presented results
for the nucleon electric form factors, radii and magnetic moments
already in section \ref{solI}. There we found a good description of
both the proton and neutron electric form factor, whereas the magnetic
moments turned out to be less than  
half of the respective experimental value.
We attributed this failure to the absence of axialvector diquarks
(which carry spin and therefore contribute to the spin-flip currents)
and to the high constituent quark mass employed for the solutions.
Furthermore there is  a substantial contribution to the electric
form factors by the exchange quark and the seagull diagrams
\cite{Oettel:1999gc} with the latter being of the order of
30 \%. We attribute this finding to the relatively small
diquark width in momentum space and the quite sizeable nucleon 
binding energy (which is around 300 MeV for the parameter
sets from table \ref{FWHM_table}).

We turn now to the form factors for the two parameter sets 
of table \ref{pars} which include axialvector diquark correlations.
First we want to judge the strength of the axialvector correlations
within the nucleon by evaluating the contributions of 
$\Psi^5$ and $\Psi^\mu$ to the {\em norm} integral, eq.~(\ref{nucnorm}).
The strength of the axialvector correlations
is rather weak for Set II, since the scalar diquark
contributes 92 \% to the norm integral of eq.~(\ref{nucnorm}) while
the axialvector correlations and scalar-axialvector
transition terms together give rise only to the remaining 8 \% for this
set. Scalar diquark contributions are terms $\sim \bar\Psi^5 \Psi^5$,
axialvector contributions are $\sim \bar\Psi^\mu \Psi^\nu$ and
the transition terms account for the two cross terms.
For Set I, the fraction of the scalar correlations
is reduced to 66 \%, the axialvector correlations are therefore expected
to influence nucleon properties more strongly for Set I than for Set II.

\begin{figure}
 \begin{center}
  \epsfig{file=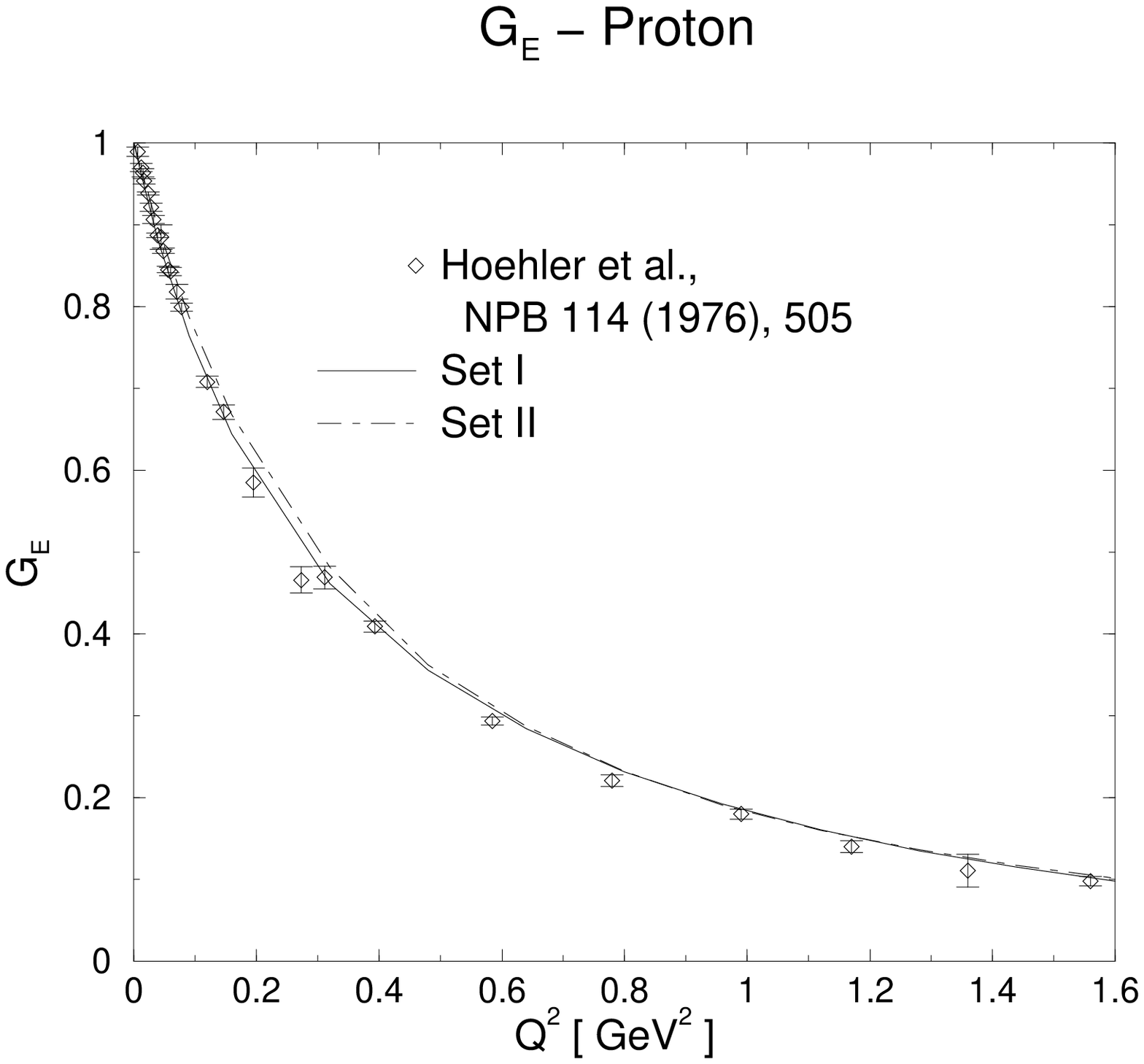,width=\figurewidth}
  \epsfig{file=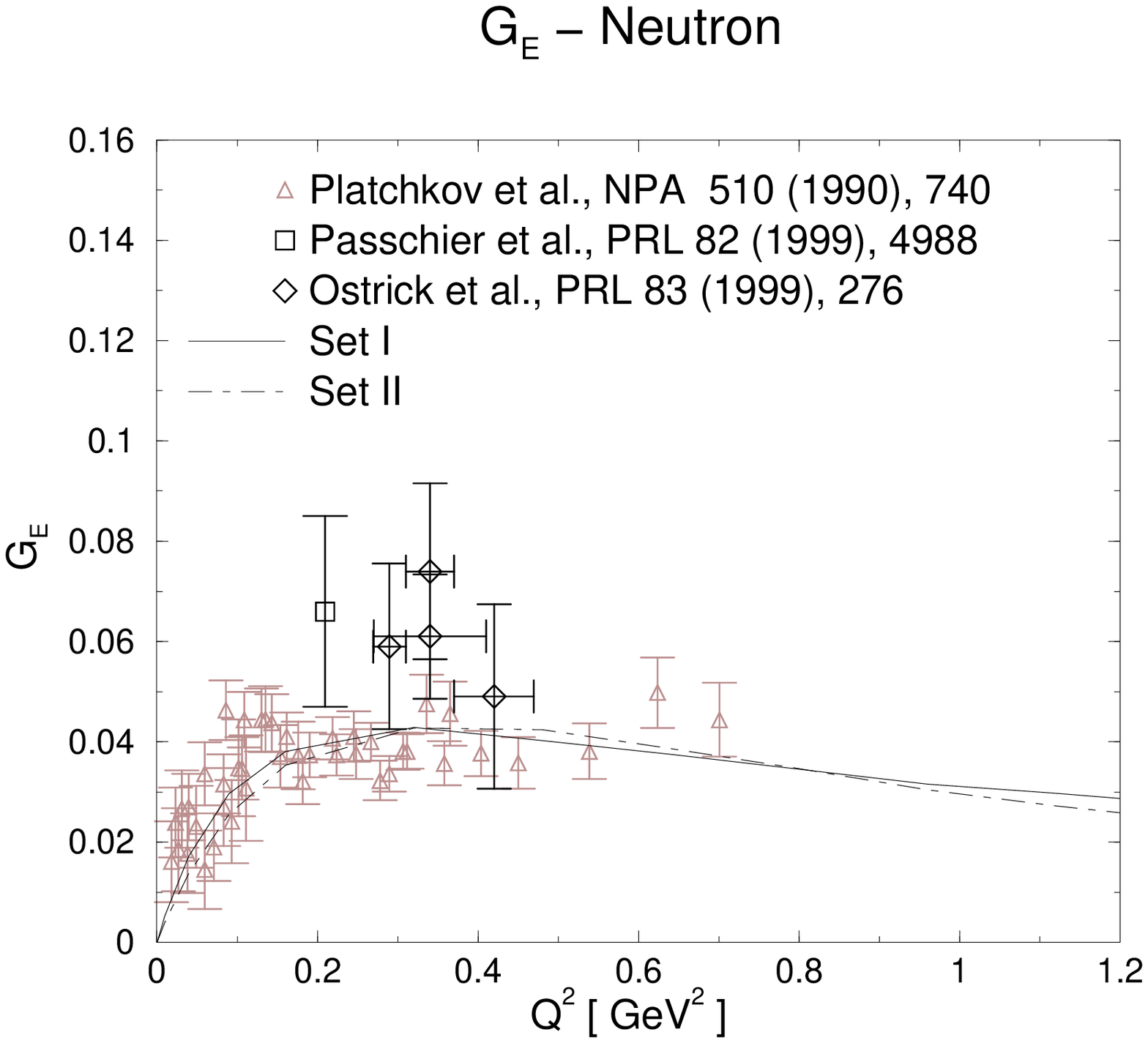,width=\figurewidth}
 \end{center}
 \caption{Electric form factors of both nucleons for the parameter
  sets of table~\ref{pars}. Experimental data
  for the proton is taken from \cite{Hoehler:1976}. The older neutron data
  analysis \cite{Platchkov:1990ch} contains more systematic 
  uncertainties (due to
  specific nucleon-nucleon potentials) than the more recent data from
  \cite{Passchier:1999cj,Ostrick:1999xa}.}
 \label{ge2fig}
\end{figure} 

\begin{table}[t]
 \begin{center}
  \begin{tabular}{llrrr} \hline \hline \\
 &  & Set I & Set II  & experiment \\
 & & & & \\ \hline
 $(r_p)_{\rm el}$ & [fm]  & 0.88 & 0.81 & 0.836 $\pm$ 0.013 \\
 $(r^2_n)_{\rm el}$ & [fm$^2$] & $-$0.12 & $-$0.10 & $-$0.113 $\pm$ 0.007 \\
 $(r_p)_{\rm mag}$  & [fm] & 0.84  & 0.83 & 0.843 $\pm$ 0.013   \\
 $(r_n)_{\rm mag}$  & [fm] & 0.84 & 0.83  & 0.840 $\pm$ 0.042 \\
 &  & & & \\ \hline \hline
  \end{tabular}
  \caption{Nucleon electric and magnetic radii for the two parameter sets
  compared to the experimental values from \cite{Kopecky:1995}
  (for $(r^2_n)_{\rm el}$) and \cite{Hoehler:1976} 
  (for the remaining radii).}
  \label{radtab}
 \end{center}
\end{table}

The results for the nucleon electric form factors are presented 
in figure \ref{ge2fig}. For the two sets we show here the sum
of {\em all} diagrams, formed by the extended impulse approximation
and the  exchange kernel graphs. 
In table \ref{radtab} we have collected the corresponding 
radii.\footnote{A radius $r_F$ corresponding to a form factor
$F$ is defined by $r_F^2=-6\frac{dF}{dQ^2}|_{Q^2=0}/F(0)$
if $F(0)$ is non-zero. For the neutron electric charge radius,
the quotient $F(0)$ is simply omitted.}
The phenomenological dipole behavior of the proton $G_E$ is well
reproduced by both parameter sets. In the neutron case, the slope 
at the soft point
and the general behavior are well accounted for although
the data points of ref.~\cite{Platchkov:1990ch} have to be interpreted
with care since here the electric form factor has been extracted 
from electron-deuteron scattering data using specific nucleon-nucleon 
potentials. Data from more recent experiments which largely
avoid this model dependence are still rare. Our results
lie here at the lower experimental bound.

\begin{figure}
 \begin{center}
  \epsfig{file=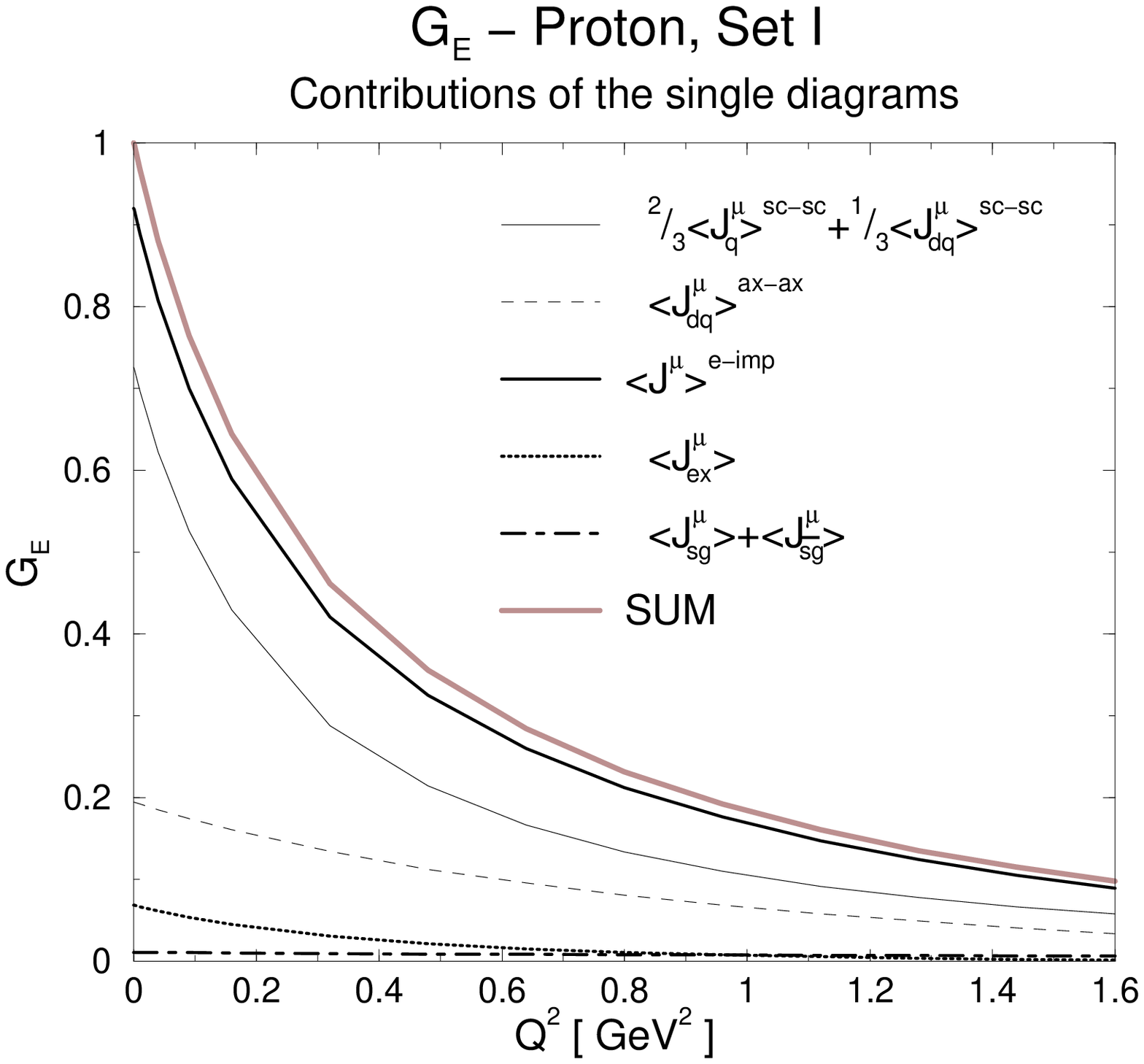,width=\figurewidth}
  \epsfig{file=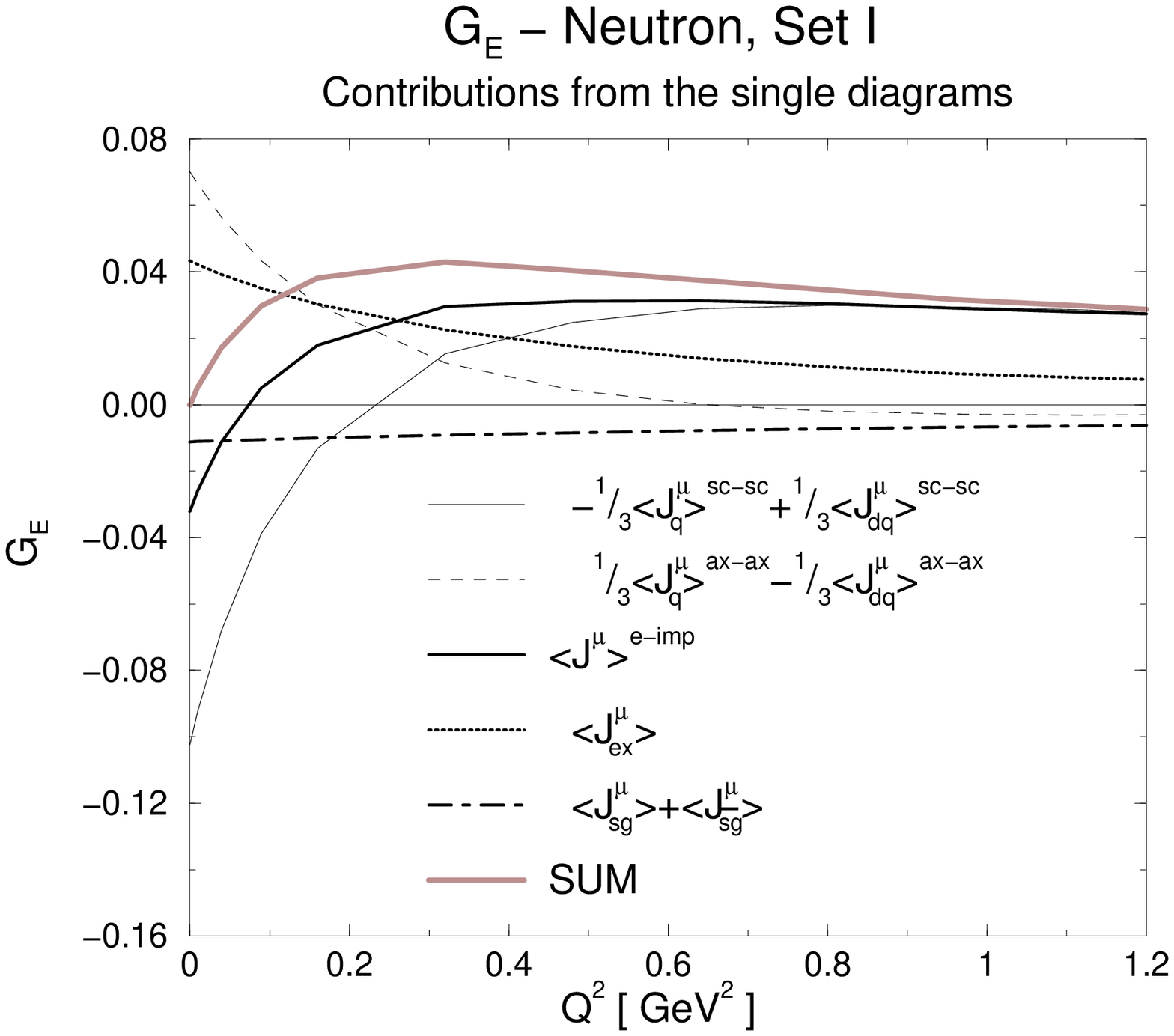,width=\figurewidth}
 \end{center}
 \caption{Contributions of the single diagrams to the nucleon electric 
  form factors for Set I. They are obtained by applying 
  eq.~(\ref{getrace}) to the current matrix elements.}
 \label{ge3fig}
\end{figure} 

In figure \ref{ge3fig} we evaluate the contributions from
the single diagrams for Set I. 
For the notation of the single current matrix 
elements we refer to eqs.~(\ref{jimp},\ref{jimn}) and
(\ref{sg}--\ref{ex}). As can be seen, the overall behavior
of the impulse approximation diagrams is close to the total form factor.
Clearly the ones sandwiched between the scalar nucleon correlations   
(straight thin line) dominate over the matrix elements
of the axialvector correlations (dashed thin line). Seagulls
and exchange quark represent just a correction below 10 \%
but are nevertheless important to guarantee the correct 
behavior for $Q\rightarrow 0$.
On the other side, the neutron electric form factor is much more
sensitive to the behavior of the single diagrams as it arises
as a result of delicate cancellations. Especially we want to point
to the contribution from the impulse approximation graphs
which are proportional to the nucleon axialvector correlations
in the initial and final states,
 {\em i.e.} to the matrix elements $\langle J^\mu_{q} \rangle^{\rm ax-ax}$
 and $\langle J^\mu_{dq} \rangle^{\rm ax-ax}$
(dashed thin line):
At $Q^2=0$, they add positively to the neutron charge but for
increasing $Q^2$ their contribution quickly drops.
As a consequence, stronger axialvector correlations tend
to suppress the peak in the form factor which the data suggest 
to be located at $Q^2=0.2\dots 0.5$ GeV$^2$. Again, the seagulls
and the exchange quark graph provide for the correct neutron charge.

\begin{figure}
 \begin{center}
  \epsfig{file=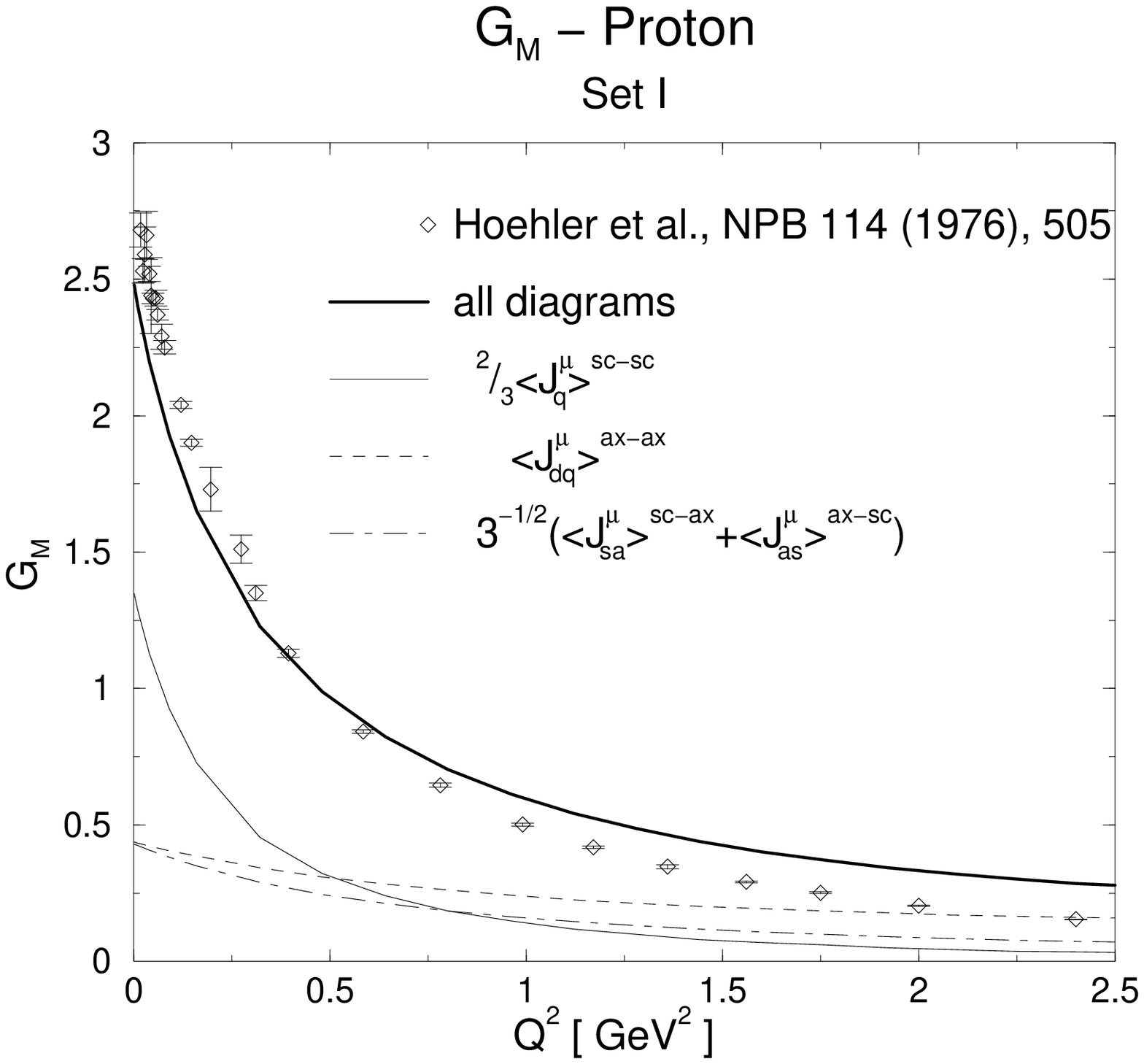,width=\figurewidth}
  \epsfig{file=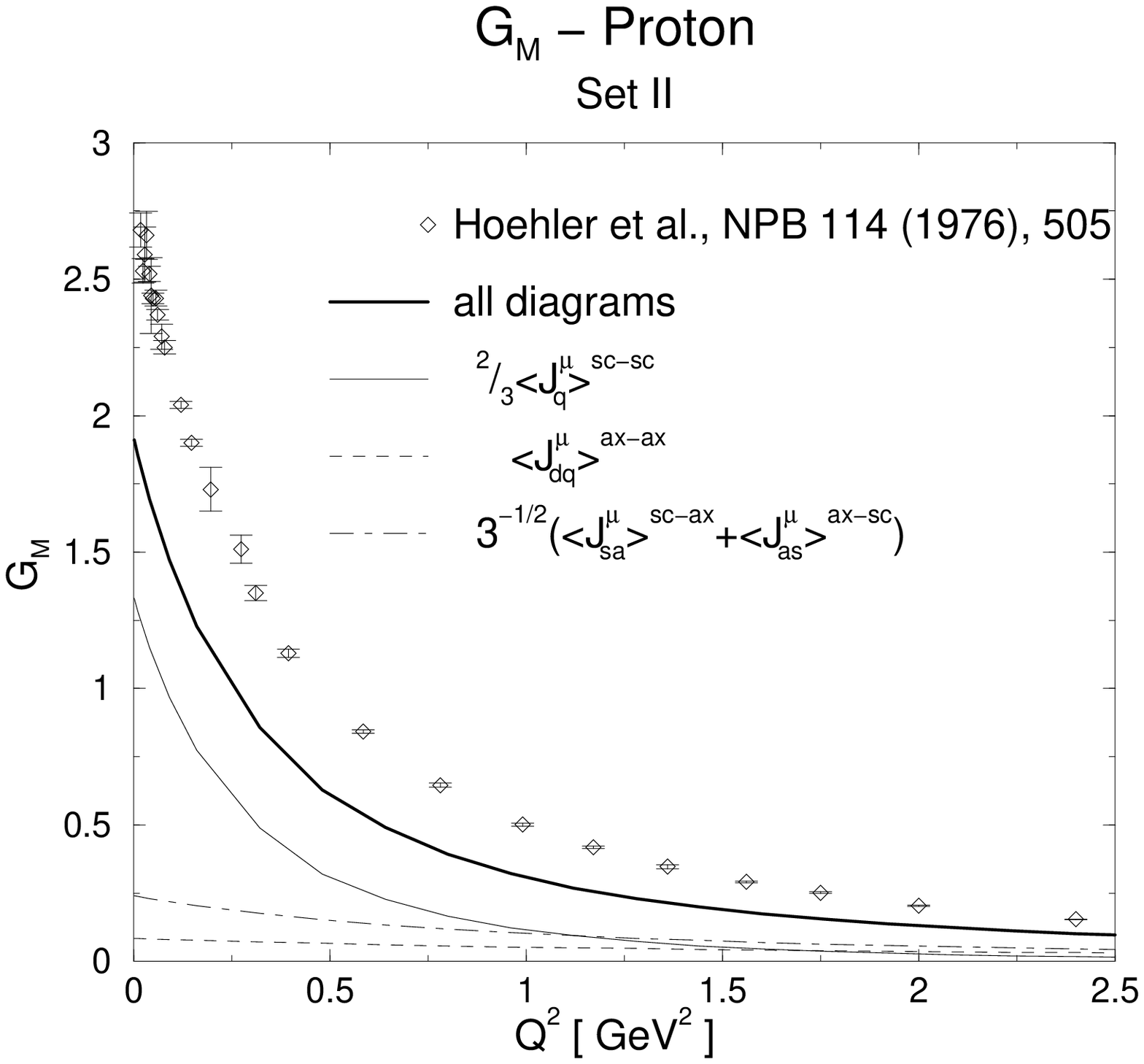,width=\figurewidth}
 \end{center}
 \caption{The proton magnetic form factor for Set I and II along
   with the dominating contributions from the impulse approximation.
   Experimental data are provided by ref.~\cite{Hoehler:1976}.}
 \label{gm1fig}
\end{figure} 

\begin{table}[t]
\begin{center}
\begin{tabular}{llrrr} \hline \hline \\
   &  & Set I & Set II & experiment \\
   &  &  &  &  \\ \hline
   & $\langle J^\mu_q \rangle^{\rm sc-sc}$ & 1.35  & 1.33 &  \\
   & $\langle J^\mu_{dq} \rangle^{\rm ax-ax}$ & 0.44 & 0.08 & \\
  $\mu_p$ & $\langle J^\mu_{sa} \rangle^{\rm sc-ax}+
            \langle J^\mu_{as} \rangle^{\rm ax-sc}$ 
 & 0.43 & 0.24 & \\
   & $\langle J^\mu_{ex} \rangle$ & 0.25 & 0.22 &\\
   & SUM  & 2.48 & 1.92 & 2.79\\ \hline
 $\mu_n$  & SUM  & $-$1.53 & $-$1.35 & $-$1.91 \\ \hline
 $\mu_p + \mu_n$   & isoscalar & 0.95  & 0.57 & 0.88 \\
 $\mu_p - \mu_n$  &  isovector & 4.01  & 3.27 & 4.70 \\ \\ \hline \hline
\end{tabular}
\end{center}
\caption{Magnetic moments of proton and neutron.
  The contributions from the indicated current matrix elements
  are to be understood as coming with the appropriate flavor
  factors, {\em cf.} eq.~(\ref{jimp}).} 
\label{magtab}
\end{table}

The results for the proton magnetic form factor are shown in
figure \ref{gm1fig}. 
Correspondingly, we have collected the single diagram 
contributions to the
magnetic moment $\mu=G_M(0)$ in table \ref{magtab}. There, the influence 
of two parameters is visible. First, 
in contrast to non-relativistic constituent models,
the dependence of the proton magnetic moment on the ratio $M_n/m_q$
is stronger than linear. As a result, the quark impulse contribution to
$\mu_p$ with the scalar diquark being spectator, which is the dominant one,
yields about the same for both sets, even though the
corresponding nucleon amplitudes of Set I contribute about 25 \% less
to the norm than those of Set II.
Secondly, the scalar-axialvector
transitions contribute equally strong (Set I) or stronger (Set II)
than the spin flip
 of the axialvector diquark itself.
While for Set II (with weaker axialvector diquark correlations)
the magnetic moments are about 30 \% too small,
the stronger diquark correlations of Set I yield an isovector contribution
which is only 15\% below and an isoscalar magnetic
moment slightly above the phenomenological value.

Stronger axialvector diquark correlations are favorable for larger values of
the magnetic moments as expected. If the isoscalar magnetic moment is taken
as an indication that those of Set I are somewhat too strong, however,
a certain mismatch with the isovector contribution remains, also with
axialvector diquarks included. Looking at the left panel
of figure \ref{gm1fig}, we note that the axialvector diquark diagram
(dashed thin line) falls very slowly with increasing $Q^2$ and eventually
causes a violation of the observed dipole shape. For Set II, this effect
is absent due to the weakness of the axialvector correlations.

We can add another twist to the story by comparing our results
to the recently measured ratio $\mu_p\, G_E/G_M$ for the 
proton \cite{Jones:1999rz},
{\em cf.} figure \ref{gemfig}. 
The ratio obtained from Set II with weak axialvector correlations
lies above the experimental data, and that for Set I below.
The experimental observation that this ratio decreases significantly with
increasing $Q^2$ (about 40 \% from $Q^2 = 0 $ to $3.5$ GeV$^2$), can be
well reproduced with axialvector diquark correlations of a certain strength
included. The reason for this is the following:
The impulse-approximated photon-diquark couplings yield contributions that
tend to fall off slower with increasing $Q^2$ than those of the quark.
This is the case for both, the electric and the magnetic form factor.
If no axialvector diquark correlations inside the nucleon are maintained,
however, the only diquark contribution to the electromagnetic current arises
from $\langle J^\mu_{sc} \rangle^{\rm sc-sc}$,
see eqs.~(\ref{jimp},\ref{jimn}). Although this term does provide 
a substantial contribution to $G_E$, its respective contribution
to $G_M$ is of the order of $10^{-3}$. This reflects the fact that
an on-shell scalar diquark would have no magnetic moment at all, and
the small contribution to $G_M$ may be interpreted as an off-shell effect.
Consequently, too large a ratio $\mu_p\, G_E/G_M$ results, if only scalar
diquarks are maintained \cite{Oettel:1999gc}. 
For Set II (with weak axialvector
correlations), this effect is still visible,
although the scalar-to-axialvector transitions
already bend the ratio towards lower values at larger $Q^2$. These transitions
almost exclusively contribute to $G_M$, and it thus follows that the stronger
axialvector correlations of Set I enhance this effect.
Just as for the isoscalar magnetic moment, the axialvector diquark
correlations of Set I tend to be somewhat too strong here again.

\begin{figure}[t]
 \begin{center}
 \epsfig{file=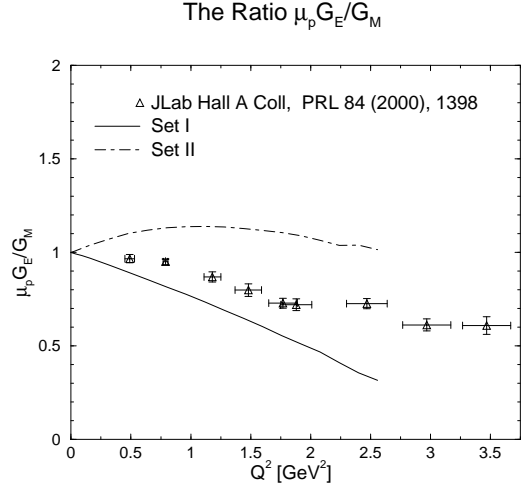,width=\figurewidth}
 \caption{The ratio $(\mu_p\;G_E)/G_M$ compared to the data
          from ref.~\cite{Jones:1999rz}.}
 \label{gemfig}
 \end{center}
\end{figure}

\subsection{Conclusions}

The numerical results for the electric form factors support
the already emphasized importance of using the Ward-Takahashi
identities for constructing a suitable current operator.
The correct value for $G_E(0)=q_N$, the nucleon charge,
has been obtained not only for the two parameter sets that have
been presented here but also for a general choice of parameters
and of the momentum partitioning $\eta$. Furthermore, the proton
and neutron electric form factors are described very well in
the diquark-quark-model.

The strength of axialvector correlations $\Psi^\mu$ within the nucleon 
turns out to be constrained by two observations:
\begin{itemize}
 \item The impulse approximation graphs involving $\Psi^\mu$
   tend to suppress the neutron electric form factor. This becomes
   effective only for stronger correlations than presented here, though. 
 \item Although larger axialvector correlations are favorable for
   the magnetic moments as expected, the ratio $\mu_p\,G_E/G_m$ provides
   tight bounds. As a result, $\Psi^\mu$ should contribute
   to the nucleon norm around 20 \%.
\end{itemize} 
As  has been mentioned,  a considerable improvement in the description
of the nucleon magnetic moments has been achieved compared to the
scalar diquark sector. The anomalous transitions between
scalar and axialvector diquark  are in this connection equally important as
the contributions of the axialvector diquark itself. We hereby
confirm the conjecture from ref.~\cite{Weiss:1993kv} where the
nucleon magnetic moments have been ascertained in a simple additive
diquark-quark picture including scalar and axialvector diquark.

Nevertheless, the isovector magnetic moments are still too small
by about 15~\%. We discuss two possibilities that might account
for the mismatch.
\begin{itemize}
 \item As we will see in the next chapter on the pionic and 
  the axial form factor of the nucleon, {\em vector} diquarks
  need to be included in the parametrization of the two-quark
  correlations on theoretical grounds. Although
  their contributions to the nucleon {\em binding energy}
  are expected to be negligible, {\em cf.} the discussion in 
  the beginning of section \ref{diquark-sec}, the {\em photon coupling}
  of the vector diquarks could be strong enough to compensate
  the underestimated magnetic moments.
 \item The tree-level form of the employed quark and diquark
  propagators is certainly too na{\"\i}ve an assumption.
  As an example, a non-trivial quark propagator would lead 
  to some effective {\em anomalous}
  magnetic moment of the (constituent) quark
  via the associated non-trivial quark-photon vertex. 
  The effect of the latter would be visible, {\em e.g.},
  in an enhanced magnetic moment contribution of the
  quark impulse approximation
  diagram. In chapter \ref{conf-chap} we will
  therefore discuss some aspects of a dressed
  quark propagator and employ 
  an especially simple parametrization 
  (which avoids the free-particle threshold)
  in calculations of the octet-decuplet spectrum and of form factors. 
\end{itemize}

 \chapter{Strong and Axial Form Factors}
 \label{sa-chap}
  
In this chapter we will investigate the pion-nucleon form factor
as an example of a strong form factor and the isovector
axial form factor which is of interest in the
description of nucleon weak interactions. In the first 
case, we are interested in the matrix elements of a pseudoscalar
current operator $J_5^a$ and in the second case it is the
pseudovector\footnote{We reserve the label {\em pseudovector}
for the external currents to distinguish them from {\em axialvector}
diquarks. Both objects have of course the same Lorentz and parity 
transformation properties.}  
current operator $J^{a,\mu}_5$ whose matrix elements are the
object under scrutiny. The superscript $a$ labels the component
of the isovector triplet which both current operators belong to.

If the squared momentum transfer is equal to the pion mass,
$Q^2=-m_\pi^2$, then the pseudoscalar matrix element is entirely
determined by the {\em pion-nucleon coupling constant} 
$g_{\pi NN}$ which
is accessible experimentally. There
appears to be no experimentally feasible scattering process
which is dominated by one-pion exchange such that the form factor
could be determined away from the pion mass shell.
Therefore corresponding theoretical results from microscopic 
nucleon models such as the diquark-quark
model only serve as a guideline for effective nucleon
potential models which incorporate one-pion exchange.

The matrix element of the pseudovector current at $Q^2=0$
is determined by the $\beta$ decay and the corresponding
{\em weak decay} or {\em axial coupling  constant} $g_A$ is known quite 
accurately\footnote{Although in 1959 the constant was assumed to be 
known precise enough, $g_A|_{1959}=1.17\pm 0.02$, 
it is quite different from the presently known value,
$g_A|_{2000}=1.267\pm 0.0035$ \cite{Mukhopadhyay:1998iw,p-data:2000}.   
}. For spacelike $Q^2$, the form factor may be determined
by neutrino-deuteron scattering \cite{Kitagaki:1990vs} but
these experiments are difficult to perform. Thus the axial
structure of the nucleon is known far less precisely than its
electromagnetic structure.

As it is well-known, these seemingly unrelated constants
$g_A$ and $g_{\pi NN}$ are linked with each other by
the famous {\em Goldberger-Treiman relation}, see below.
It can be derived from the assumption of chiral symmetry
of an underlying Lagrangian which conserves the 
pseudovector current.
Additionally it is assumed that this chiral symmetry is
spontaneously broken and pions appear as massless
Goldstone bosons. Now the physical pion is massive,
albeit its mass is small, so chiral symmetry is explicitly broken
and the pseudovector current can be  only partially conserved 
({\em PCAC hypothesis}).
However, it can be shown that the Goldberger-Treiman relation is 
still valid under the assumption that the axial form factor 
is slowly varying in the vicinity of $Q^2=0$.
Furthermore, the Goldberger-Treiman relation holds experimentally
on the five per cent accuracy level which, among
other observations, lends support to the concept
of spontaneously broken chiral symmetry and the partially conserved
pseudovector current.

We tried to show by these introductory arguments that the pion-nucleon
coupling constant  and the weak decay constant are related 
with each other in a model independent way. Thereby we imply that
if one were to start with a chirally symmetric {\em quark} 
theory one should always recover
the Goldberger-Treiman relation for the {\em nucleon}, 
similar to the case of electromagnetic
interactions where the nucleon charges should always come out
correctly. Again similar to the electromagnetic case, there are
powerful {\em chiral} Ward identities 
which constrain the pseudovector current vertices.

Unfortunately we will not be able to comply with the demand 
of a conserved pseudovector current and to actually derive 
the Goldberger-Treiman relation in the diquark-quark
model in its present form. As will become clear in section \ref{vectors},
{\em vector} diquark correlations are
inevitably needed to conserve the pseudovector current.
These are neglected in our model, for their influence on
the binding energy is assumed to be small and they introduce
six additional scalar functions in the nucleon vertex function $\Phi$,
therefore being a substantial numerical complication. 
Our calculations presented in this chapter 
aim therefore at quantitative results for the
expected discrepancy and at possible explanations for 
the measured weak coupling constant. In contrast to the
anomalous magnetic moments, the non-relativistic quark model
estimate $g_A=5/3$ is 31 \% larger than the measured value,
$g_A=1.27$, and therefore a relativistic treatment such as taken up here
might offer an explanation for this difference. 

We will begin with fixing our conventions for the 
matrix elements and re\-derive quickly the Goldberger-Treiman relation.
Section \ref{vectors} explains the necessity of vector diquarks 
for pseudovector current conservation.
In section \ref{curops} we are concerned with the construction  
of the pseudoscalar and pseudovector current operator 
for our model. Here we start with a suitable quark-pion vertex
which in conjunction with a chiral Ward identity
yields the quark-pseudovector vertex. 
The structures and strengths  of the couplings of 
the diquarks
to the pion and the pseudovector current 
are obtained from resolving
the diquarks in a way similar to their electromagnetic couplings.
The numerical results,
again obtained by employing the two parameter sets from table
\ref{pars},
 are presented in section \ref{num2}. We will
find a slight overestimation of $g_A$  and a violation of
the Goldberger-Treiman relation of around 15 \%. 
The  axial form factor results for finite $Q^2$
are consistent with the presently available
experimental results.

Let us start with a few definitions. 
The (spin-summed) matrix element of the pseudoscalar current $J_5^a$ 
is pa\-ra\-me\-trized as
\begin{equation}
 \langle P_f| \;  J^a_5\; |P_i \rangle = \Lambda^+(P_f)
  \,\tau^a \gamma^5 g_{\pi NN}(Q^2)\,
  \Lambda^+(P_i) \; ,    \label{psdens}
\end{equation}
which defines the pion-nucleon form factor $g_{\pi NN}(Q^2)$.
As before, $Q=P_f-P_i$ holds.
For numerical evaluation, we will choose the neutral component,
$J^3_5$, and sandwich between proton flavor states, indicated
by the subscript $_{\rm p}$. 
Straightforward algebra allows to extract the form factor as the
following Dirac trace,
\begin{equation}
 g_{\pi NN}(Q^2) = - \frac{2M_n^2}{Q^2} \;\text{Tr} \;\left[
              _{\rm p}\langle P_f| \;  J^3_5\; |P_i \rangle_{\rm p} 
          \;\gamma^5\right]    \;. \label{gpNNtr}
\end{equation}
We take for granted in the following that $g_{\pi NN}(0)$ is close
to the pion-nucleon coupling constant $g_{\pi NN}=g_{\pi NN}(-m_\pi^2)$
as consistent with the assumption of PCAC.

The matrix elements of the pseudovector current are para\-me\-trized
by the form factor $g_A(Q^2)$ and the induced pseudoscalar form factor
$g_P(Q^2)$,
\begin{equation}
 \langle P_f| \; J^{a,\mu}_5 \; |P_i \rangle = 
  \Lambda^+(P_f)\,\frac{\tau^a}{2}
  \left[ i\gamma^\mu \gamma^5 g_A(Q^2) +Q^\mu \gamma^5 g_P(Q^2) 
   \right]\,\Lambda^+(P_i) \;.
  \label{nucax}
\end{equation}
For $Q^2 \rightarrow 0$, the induced pseudoscalar form factor
will be dominated by the pion pole,
\begin{equation}
 \lim_{Q^2 \rightarrow 0}  g_P(Q^2) = 2 \frac{f_\pi\,
    g_{\pi NN} (0)}{Q^2} \; .
   \label{gpgpNN}
\end{equation} 
We denote the pion decay constant by $f_\pi=92.7$ MeV.
The Goldberger-Treiman relation follows from current conservation,
$Q^\mu \langle J^{a,\mu}_5 \rangle =0$, and after replacing
$g_P$ by eq.~(\ref{gpgpNN}) we find,
\begin{equation}
 g_A(0) =  f_\pi  g_{\pi NN}(0)/{M_n} \; .
\end{equation}
The regular part of the matrix element, $g_A(Q^2)$, 
and the pion pole part, $g_P(Q^2)$,
can be extracted from eq.~(\ref{nucax}) as follows:
\begin{eqnarray}
 g_A (Q^2) & = &- \frac{i}{2\left( 1+\frac{Q^2}{4 M_n^2}\right) }
             \;\text{Tr}\;\left[ _{\rm p}\langle P_f| \; J^{3,\mu}_5 \; |P_i 
                     \rangle_{\rm p} 
               \left( \gamma^5 \gamma^\mu -
             i \gamma^5 \frac{2M_n}{Q^2} Q^\mu \right)\right]  \; ,
 \label{gatrace} \nonumber \\
           & & \\
 g_P (Q^2) & = & \frac{2M_n}{Q^2} \left( g_A (Q^2) - \frac{2M_n}{Q^2}
              \;\text{Tr}\; \left[ _{\rm p}\langle P_f| \; J^{3,\mu}_5 \; |P_i 
               \rangle_{\rm p} \;
            Q^\mu \gamma^5\right] \;
               \right) \; .
 \label{gptrace}
\end{eqnarray}
Having specified our conventions, we turn now to a short discussion
about the consequences of excluding vector diquarks from our
considerations.

\section{Scalar and vector/axialvector diquarks as chiral
  multiplets}  
\label{vectors}

The necessity of including vector diquarks along
with the axialvectors was observed only very recently 
in ref.~\cite{Ishii:2000zy}
within the context of an NJL model. Remember that diquarks
as separable 2-quark correlators are in the NJL model solutions of the 
$t$ matrix integral equation, with the interaction kernel given 
in lowest order in the coupling constant. 
The separable interaction kernel
describes here the 4-quark point interaction: two quarks  
in a channel with certain diquark quantum numbers interact
locally with two antiquarks with respective antidiquark
quantum numbers, see {\em e.g.} eq.~(\ref{sd-njl}) for the scalar channel.
The statement of ref.~\cite{Ishii:2000zy}
is summarized as follows: Chiral invariance of the NJL Lagrangian
requires the inclusion of quark-quark interaction
kernels with both vector and axialvector diquark quantum numbers
and the same coupling constant. Solutions for the separable vector
diquark $t$ matrix do not have the same form as the axialvector
$t$ matrix, {\em e.g.}, a possible vector diquark mass pole will 
appear at larger invariant momentum values than an axialvector pole.
Nevertheless vector diquarks have to be carried all the way
through the nucleon Faddeev problem in order to comply
with chiral invariance. 

In our model, we have parametrized the diquark correlations
without recurring to a specific model Lagrangian. To adapt
the above argument to our case, we introduce 
interpolating diquark fields as done in section \ref{curop-sec}.
Let $d_s^\dagger$ denote scalar diquark creation operators
and $d_v^{\mu\dagger},d_a^{b\mu\dagger}$ the ones for vector and
axialvector diquarks. 
Scalar and vector diquarks are flavor singlets, the axialvector
diquark belongs to a flavor triplet, with its components 
denoted by the index $b$.
The appropriate
color indices are suppressed as they are unimportant
for the argument. The interaction
of (pointlike) diquarks with quarks can now be written as
an interaction Lagrangian,
\begin{eqnarray}
  \mathcal{L}_{\rm int}&=&   \nonumber
    g_s \; \left( q^T \,C\gamma^5\, \tau^2\, q\; \right)
              d_s^\dagger+ 
    g_v \; \left( q^T \,Ci\gamma^\mu\gamma^5\, \tau^2\, q\; \right)
              d_v^{\mu\dagger} + \\
    & &   g_a \; \left( q^T \,Ci\gamma^\mu\, \tau^2\tau^b\, q\; \right)
              d_a^{b\mu\dagger} + \quad
{\rm herm.\; conj.} \; , \label{intlag}
\end{eqnarray}
where $q$ denotes as usual the quark field. A combined infinitesimal
chiral and isospin variation of the quark field reads
\begin{equation}
  \delta q= -i \omega_b \tau^b \gamma^5\, q \; ,
\end{equation}
with $\omega_b$ denoting a set of three infinitesimal parameters.
Straightforward evaluation to first order in the variation yields
\begin{eqnarray}
  \label{v1}
  \delta \left( q^T \,C\gamma^5\, \tau^2\, q\; \right) & = & 0 \; ,\\
  \label{v2}
  \delta \left( q^T \,Ci\gamma^\mu\gamma^5\, \tau^2\, q\; \right) &=& 
  -i \omega_b \left( q^T \,Ci\gamma^\mu\, \tau^2\tau^b\, q\; \right) 
    \; , \\
  \label{v3}
  \delta \left( q^T \,Ci\gamma^\mu\, \tau^2\tau^b\, q\; \right) &=&
   -i \omega_b \left( q^T \,Ci\gamma^\mu\gamma^5\, \tau^2\, q\; \right)
   \; .  
\end{eqnarray}
We demand that the interaction Lagrangian (\ref{intlag}) be invariant
under an infinitesimal chiral variation. Substituting 
eqs.~(\ref{v1}--\ref{v3})
into eq.~(\ref{intlag}), we see that this leads to the requirements
\begin{eqnarray}
  \delta\,d_a^{b\mu\dagger} = i\omega_b d_v^{\mu\dagger}, \quad
  \delta\,d_v^{\mu\dagger} = i\omega_b d_a^{b\mu\dagger} \quad
  {\rm and} \quad g_a=g_v \; .
\end{eqnarray} 
Vector and axialvector diquarks mix under a chiral (isovector)
transformation. The scalar diquark channel is unaffected by the 
transformation as its  variation is zero. This is equivalent to the
statement that a pseudovector (isovector) current does not
couple to the scalar diquark. 
Summarizing  these findings we have found  that
the scalar diquark on the one side and vector/axialvector
diquarks on the other side belong to different chiral multiplets
\cite{Ishii:2000zy}.

We learn something else. The above argument is only valid for a {\em
local} quark-diquark interaction. If we introduced extended diquarks 
by a non-local interaction with some momentum dependent
vertex we would find additional {\em chiral seagull vertices} 
as a result of the variation. This is very similar to the electromagnetic
case. Although we derived the seagulls in section \ref{sg-sec}
by requiring gauge invariance via
a Ward-Takahashi identity for the Bethe-Salpeter interaction kernel,
they can also be obtained from an interaction Lagrangian  
with momentum-dependent vertices \cite{Ohta:1989ji} in much the same
way as indicated here.

In the following 
we will neglect both the chiral seagulls and the 
vector diquark correlations. As we remarked earlier, the expected
high mass of the vector diquarks would suppress the strength 
of the corresponding vector correlations within the nucleon. Another 
argument is concerned with the numerical feasibility. In the Dirac 
algebra, the vector correlations could be decomposed similarly to the
axialvector correlations, leading to six additional scalar functions.
This would increase the numerical effort considerably and since
the numerical calculations are already very involved at this stage,
the inclusion of vector diquarks is beyond the scope of this thesis. 
On the other hand, we have no
serious argument for the neglect of the seagulls. However,
we can estimate from the numerical results presented
in section \ref{num2} that their contribution to the
weak coupling constant should be below ten per cent.

\section{Pseudoscalar and pseudovector current operator}
\label{curops}

First we will specify the quark vertices with pseudoscalar 
and pseudovector current operators and then continue with a discussion
of the effective diquark vertices that can be obtained by coupling
the currents to an intermediate quark loop.
   
As we use a free fermion propagator for the quark, eq.~(\ref{qprop}),
the scalar functions $A,B$ in the parametrization $S^{-1}=
-i\Slash{p}A-B$ reduce to $A=1$ and $B=m_q$. This suggests that
we use for the pion-quark vertex\footnote{We drop the flavor 
indices on the vertices in the following. Each pseudoscalar vertex
comes with a factor of $\tau^a$ and each pseudovector vertex
with a factor of $\tau^a/2$ as these are widely used conventions.}
\begin{equation}
 \Gamma_{5,q}
         = -\gamma^5 \frac{B(k^2)+B(p^2)}{2f_\pi} = 
           -\gamma^5 \frac{B}{f_\pi} \; ,
 \label{vertpion}
\end{equation}
and discard the three additionally possible Dirac structures.
The momenta of outgoing and incoming quark are $k$ and $p$,
respectively. 
The reason is that in the chiral limit
eq.~(\ref{vertpion}) represents the exact pion Bethe-Salpeter amplitude
(on-shell, {\em i.e.} $Q^2=0$)
 for equal quark and
antiquark momenta, since in this limit, with constant $A$,
the Dyson-Schwinger equation for the scalar function
$B$ agrees with the Bethe-Salpeter equation for a pion of zero momentum. 
Of course, the subdominant amplitudes should in principle be included
for physical pions (with momentum $P^2 = -m_\pi^2$), when
solving  the Dyson-Schwinger equation for $A$ and $B$
and the Bethe-Salpeter equation for the pion in mutually consistent truncations
\cite{Bender:1996bb,Maris:1997hd,Maris:1997tm}. 
The generalization to pion off-shell momenta as used in 
eq.~(\ref{vertpion}) is by no means unique but certainly the most simple
one.

We use chiral symmetry constraints to construct the pseudovector-quark 
vertex $\Gamma^{\mu}_{5,q}$.
In the chiral limit, the Ward-Ta\-ka\-ha\-shi identity for this vertex reads,
\begin{equation}
 Q^\mu \Gamma^{\mu}_{5,q} = 
       S^{-1}(k) \gamma^5 + \gamma^5 S^{-1}(p) \; ,  \quad (Q=k-p)\;.
\label{pvWTI}
\end{equation}
To satisfy this constraint we use the form of the vertex proposed in
ref.~\cite{Delbourgo:1979me},
\begin{equation}
 \Gamma^{\mu}_{5,q} =
       -i \gamma^\mu \gamma^5  +  \frac{Q^\mu}{Q^2}
       2 f_\pi \Gamma_{5,q} \;.
 \label{vertpv}
\end{equation}
The physical picture behind this form of the vertex is clear. In
the spontaneously broken phase of chiral symmetry the quark 
acquires mass and at the same time the Goldstone boson pole 
must appear in the
full quark-pseudovector vertex, and the  Goldstone boson pole  
in turn is proportional to the 
generated quark mass.
The second term which contains the massless pion pole
does not contribute to $g_A$ as can be seen from eq.~(\ref{gatrace}).

The pion and the pseudovector current can couple to the diquarks
by an intermediate quark loop. As for the anomalous contributions to the
electromagnetic current, we derive the Lorentz structure of the
diquark vertices and calculate their effective strengths from this
quark substructure of the diquarks in appendix \ref{dqres2}.
For momentum definitions at these vertices we refer to 
figure \ref{emresolve}.

As already mentioned in the last section, no such couplings arise for 
the scalar diquark. This can be inferred from parity or covariance
or, simpler, one considers the flavor trace of an isovector current
between two isoscalar states (scalar diquark vertex) which yields
zero. The axialvector diquark and the pion couple by an anomalous vertex.
Its Lorentz structure is similar to that for the photon-induced
scalar-to-axialvector transition in eq.\ (\ref{sa_vert}),
\begin{eqnarray}
 \Gamma^{\alpha\beta}_{5,ax} &=& \frac{\kappa^5_{ax}}{2M_n}
                      \frac{m_q}{f_\pi}
                     \epsilon^{\alpha\beta\mu\nu}
                  (p_d+k_d)^\mu Q^\nu
                 \label{vert5ax} \;.
\end{eqnarray}
Here, $\alpha$ and $\beta$ are the Lorentz indices of outgoing
and incoming diquark, respectively. 
The factor $m_q/f_\pi$ comes from the quark-pion vertex (\ref{vertpion})
in the quark loop (see appendix \ref{dqres2}), and the nucleon mass
was introduced to isolate a dimensionless constant $\kappa^5_{ax}$.

The pseudovector current and the axialvector diquark are also coupled by
anomalous terms. As before, we denote with $\alpha$ and $\beta$ the Lorentz
indices of outgoing and incoming diquark, respectively,
and with $\mu$ the pseudovector index. Out of three possible Lorentz
structures for the regular
part of the vertex,
\begin{equation*}
  p_d^\mu \;\epsilon^{\alpha\beta\rho\lambda} p_d^\rho Q^\lambda \; , \quad
  \epsilon^{\mu\alpha\beta\rho}Q^\rho \quad {\rm and} \quad
  \epsilon^{\mu\alpha\beta\rho} (p_d+k_d)^\rho\;,
\end{equation*}
only the last term contributes to $g_A$
in the limit $Q \rightarrow 0$.
We furthermore verified numerically that the first two terms yield negligible
contributions to the form factor also for nonzero $Q$. Again, the pion pole
contributes proportionally to $Q^\mu$, and our {\em ansatz}
 for the vertex thus reads
\begin{eqnarray}
 \Gamma^{\mu\alpha\beta}_{5,ax} &=& \frac{\kappa^5_{\mu,ax}}{2}
                        \epsilon^{\mu\alpha\beta\nu} (p_d+k_d)^\nu+ 
                   \frac{Q^\mu}{Q^2}
                 2f_\pi \Gamma^{\alpha\beta}_{5,ax} \label{vert5muax}\; .
\end{eqnarray}
For both coupling strengths in the vertices 
$\Gamma^{\alpha\beta}_{5,ax}$ and $\Gamma^{\mu\alpha\beta}_{5,ax}$
we roughly obtain
$\kappa^5_{ax} \simeq \kappa^5_{\mu,ax} \simeq 4.5$ slightly dependent on the
parameter set, see table~\ref{cc_2}  in appendix \ref{dqres2}.

Scalar-to-axialvector transitions are also possible by the pion and the
pseudovector current. An effective vertex for the pion-mediated transition
has one free Lorentz index to be contracted with the axialvector
diquark. Therefore, two types of structures exist, one with the pion momentum
$Q$, and the other with any combination of the diquark momenta $p_d$ and
$k_d$. If we considered this transition as being described by an interaction
Lagrangian of scalar, axialvector and pseudoscalar fields, terms of the
latter structure would be proportional to the divergence of the axialvector field
which is a constraint that can be set to zero.
We therefore adopt the following form for the transition vertex,
\begin{equation}
 \Gamma^{\beta}_{5,sa} = -i \kappa^5_{sa}
                     \frac{m_q}{f_\pi}
                    Q^\beta \; .
 \label{vert5sa}
\end{equation}
Here the Lorentz index of the participating axialvector diquark
is given by $\beta$.
This vertex corresponds to a derivative coupling of the
pion to scalar and axialvector diquark.

The pseudovector-induced transition vertex has two Lorentz indices,
denoted by $\mu$ for the pseudovector current and $\beta$ for the axialvector
diquark.
From the momentum transfer $Q^\mu$ and one of the diquark momenta
altogether five independent tensors can be constructed,
\begin{equation*}
  \delta^{\mu\beta} \; , \quad
  Q^\mu Q^\beta \; , \quad
  Q^\mu p_d^\beta \; , \quad
  p_d^\mu Q^\beta \quad {\rm and} \quad 
  p_d^\mu p_d^\beta \; .
\end{equation*}
Note that a term constructed with the totally antisymmetric tensor
would have the wrong parity. We assume, as before, that all terms proportional to $Q^\mu$ are contained
in the pion part. They do not contribute to $g_A$ anyway.
From the
diquark loop calculation in appendix \ref{dqres2} we find that
the terms proportional to $p_d^\mu Q^\beta$ and $p_d^\mu p_d^\beta$
can again be neglected with an error on the level
of one per cent. Therefore, we use a vertex of the form,
\begin{equation}
 \Gamma^{\mu\beta}_{5,sa} = i M_n \kappa^5_{\mu,sa}\;
                            \delta^{\mu\beta}
                    + \frac{Q^\mu}{Q^2} 2 f_\pi \Gamma^{\beta}_{5,sa}.
 \label{vert5musa}
\end{equation}
For the strengths of these two transition vertices
we obtain $\kappa^5_{sa}\simeq 3.9$ and (on average)
$\kappa^5_{\mu,sa} \simeq 2.1$, {\em cf.} table~\ref{cc_2} in appendix
\ref{dqres2}. 

The vertices $\Gamma^{\beta}_{5,as},\Gamma^{\mu\beta}_{5,as}$ for 
the reverse transitions are obtained by simply reversing the sign of $Q$.

Having constructed all vertices, we can now perform the flavor algebra
and write down the pseudoscalar current matrix element in the
extended impulse approximation. As done in section \ref{impulse-sec}
we introduce quark current matrix elements by
\begin{equation}
 \langle J_{5,q} \rangle^{\rm sc-sc}= 
  \bar \Psi^5 \;D^{-1} \,\Gamma_{5,q} \; \Psi^5 \; ,
\end{equation}
where this time {\em no} flavor matrix factor comes with
$\Gamma_{5,q}$ as we compute the flavor factors separately. 
Likewise scalar/axialvector diquark and the transition matrix elements
are defined. With these definitions we find for the matrix element
of the extended impulse approximation
\begin{eqnarray} \nonumber
 _{\rm p}\langle P_f|\; J^3_5\;|P_i \rangle^{\rm e-imp}_{\rm p} 
  &=&
  \langle J_{5,q} \rangle^{\rm sc-sc} 
  -\frac{1}{3} \langle J_{5,q} \rangle^{\rm ax-ax} 
  +\frac{2}{3} \langle J_{5,dq} \rangle^{\rm ax-ax} + \\
     & &
  \frac{\sqrt{3}}{3} \left(\langle J_{5,sa} \rangle^{\rm sc-ax}+
             \langle J_{5,as} \rangle^{\rm ax-sc} \right) \; .
 \label{imp5-me}
\end{eqnarray}
As we neglect seagull contributions, the only contribution from
the Bethe-Salpeter kernel is due to the coupling to the exchange
quark and it reads explicitly
\begin{eqnarray}
 \label{ex5}
  _{\rm p}\langle P_f|\; J^3_5\;|P_i \rangle^{\rm BS-ker}_{\rm p}
 &=& \langle J_{5,ex} \rangle \\ \nonumber 
 &=& -\half\;
    \begin{array}[t]({cc}) \bar\Psi^5 & \bar \Psi^\alpha \end{array}
    \;
   \begin{array}[t]({cc})
    K^5_{ex,ss} & K^5_{ex,sa}
      \\ K^5_{ex,as} & K^5_{ex,aa}
   \end{array}
   \;
    \begin{array}[t]({c}) \Psi^5 \\ \Psi^\beta \end{array}
    \begin{array}[t]{c} \; \\ \; ,\end{array} \\
     \nonumber
   K^5_{ex,ss} &=&  \chi^5(p_1) \, S^T(q)
       \left(\Gamma_{5,q}\right)^T S^T(q')\, \bar\chi^5(p_2) \; , \\
     \nonumber
   K^5_{ex,sa} &=& \frac{1}{3}\sqrt{3}\; \chi^\beta(p_1) \, S^T(q)
       \left(\Gamma_{5,q}\right)^T S^T(q')\, \bar\chi^5(p_2) \; , \\
     \nonumber
   K^5_{ex,as} &=& \frac{1}{3}\sqrt{3} \; \chi^5(p_1) \, S^T(q)
       \left(\Gamma_{5,q}\right)^T S^T(q')\, \bar\chi^\alpha(p_2) \; , \\
     \nonumber
   K^5_{ex,aa} &=& \frac{5}{3} \; \chi^\beta(p_1) \, S^T(q)
       \left(\Gamma_{5,q}\right)^T S^T(q')\, \bar\chi^\alpha(p_2) \; .
\end{eqnarray}
The diagrams of the extended impulse approximation are equivalent
to those displayed in figure \ref{impulse} and likewise the exchange
quark diagram is similar to the first diagram in figure \ref{7dim}.
The black dot which represents the quark-photon or diquark-photon
vertex should be replaced with the appropriate quark-pseudoscalar
or diquark-pseudoscalar vertex.
The matrix elements for the pseudovector current differ from
eqs.~(\ref{imp5-me},\ref{ex5}) only by an overall factor of 1/2
after replacing the pseudoscalar vertices by their pseudovector
counterparts.

\section{Numerical results}
\label{num2}

Having defined the pseudoscalar and pseudovector current operators
in the last section and having fixed all unknown coupling constants
by appropriate quark loop calculations, {\em cf.} appendix
\ref{dqres2}, we can now compute the form factors
for the two parameter sets of table \ref{pars}.
Regarding the numerical procedure, the method and its intricacies have been
lined out in section \ref{num_em} and all the remarks made in this section
also apply here. We summarize: impulse approximation diagrams
are computed in the Breit frame employing boosted vertex functions
$\Phi$ and taking into account the residue contributions, {\em cf.}
appendix \ref{residue}. The exchange quark diagram can only be computed
 using boosted wave functions $\Psi$ and therefore the related
numerical accuracy
becomes very limited beyond momentum transfers $Q^2>2.5$ GeV$^2$.

Examining $g_{\pi NN}(0)$ which is assumed to be close to
the physical pion-nucleon coupling 
and the weak coupling constant $g_A(0)$, we find large contributions to
both arising from the scalar-axialvector transitions, {\em cf.} table
\ref{ga&gpitab}. As mentioned
in the previous section, the various diquark contributions violate
the Goldberger-Treiman relation.
Some compensations occur between the small contributions from
the axialvector diquark coupling in impulse approximation
and the comparatively large
ones from scalar-axialvector transitions which provide the dominant effect
to yield 
$g_A(0) > 1$.

Summing all these contributions, the Goldberger-Treiman discrepancy,
\begin{equation}
  \Delta_{GT} \equiv \frac{g_{\pi NN}(0)}{g_A(0)}\,\frac{f_\pi}{M_n}-1
\end{equation}
amounts to 0.14 for Set I and 0.18 for Set II. The larger discrepancy for Set
II (with weaker axialvector correlations) is due to the
larger violation of the Goldberger-Treiman relation from the
exchange quark contribution in this case. This contribution
is dominated by the scalar amplitudes,
and its Goldberger-Treiman violation should therefore
be compensated by appropriate {\em chiral seagulls} as already mentioned
in section \ref{vectors}. For Set II which is dominated
by scalar correlations within the nucleon, their contribution
to $g_A(0)$  can be 
estimated to be around 0.1 because the seagulls together
with the contributions from $\langle J^{[\mu]}_{5,q} \rangle^{\rm sc-sc}$ 
and $\langle J^{[\mu]}_{5,ex} \rangle$ should approximately
obey the Goldberger-Treiman relation.

\begin{table}[t]
\begin{center}
\begin{tabular}{lrrrr} \hline \hline \\
   &  \multicolumn{2}{c}{Set I}  &  \multicolumn{2}{c}{Set II}   \\
   &  $g_{\pi NN}(0)$ & $g_A(0)$ & $g_{\pi NN}(0)$ & $g_A(0)$  \\
   &  &  &  &  \\  \hline
 $\langle J^{[\mu]}_{5,q} \rangle^{\rm sc-sc}$  & 7.96 & 0.76 & 9.25 & 0.86  \\
 $\langle J^{[\mu]}_{5,q} \rangle^{\rm ax-ax}$ & 0.50 & 0.04 & 0.10 & 0.01 \\
 $\langle J^{[\mu]}_{5,dq} \rangle^{\rm ax-ax}$  & 1.44 & 0.18 & 0.34 & 0.04 \\
 $\langle J^{[\mu]}_{5,sa} \rangle^{\rm sc-ax}+
            \langle J^{[\mu]}_{5,as} \rangle^{\rm ax-sc}$
  & 5.66 & 0.39 & 3.79  & 0.22 \\
 $\langle J^{[\mu]}_{5,ex} \rangle$  & 1.69 & 0.12 & 2.70 & 0.22 \\
 SUM  & 17.25 & 1.49 & 16.18 & 1.35  \\ \hline
     & \multicolumn{2}{c|}{$g_{\pi NN}$} & 
       \multicolumn{2}{c}{$g_{A}$} \\
 experiment & \multicolumn{2}{r|}{13.14$\pm$ 0.07 \cite{Arndt:1994bu}} &
          \multicolumn{2}{r}{ 1.267 $\pm$ 0.0035 \cite{p-data:2000}} \\
   & \multicolumn{2}{r|}{13.38$\pm$ 0.12 \cite{Rahm:1998jt}} &  &
 \\ \\ \hline \hline
\end{tabular}
\end{center}
\caption{Various contributions to $g_{\pi NN}(0)$ and $g_A(0)$, labelled as in
   table~\ref{magtab}.} 
\label{ga&gpitab}
\end{table}

\begin{table}[b]
 \begin{center}
 \begin{tabular}{lllll} \hline \hline \\
  & & Set I & Set II & experiment \\
  & & & & \\ \hline
 $r_{\pi NN}$ & [fm] & 0.83 & 0.81 &  \\
 $r_A$ & [fm] & 0.82 & 0.81 & 0.70$\pm$0.09 \\ \\ \hline \hline
 \end{tabular}
 \end{center}
 \caption{Strong radius $r_{\Sc\pi NN}$ and weak radius
         $r_A$, the experimental value of the latter
         is taken from \cite{Kitagaki:1990vs}.}
\label{radtab_2}
\end{table}

The strong and weak radii are presented in table \ref{radtab_2} and the
corresponding form factors in figure \ref{gpifig}. Experimentally  the
axial form factor is known much less precisely than
the electromagnetic form factors. In the right panel
of figure\ \ref{gpifig}  the experimental situation is summarized
by a band of dipole parametrizations of $g_A$
that are consistent with a wide-(energy)band neutrino experiment
\cite{Kitagaki:1990vs}.
Besides the slightly too large values obtained for $Q^2 \to 0$
which are likely to be due to the PCAC violations of axialvector diquarks 
and of the missing seagulls as
discussed in section \ref{vectors},
our results yield quite
compelling agreement with the experimental bounds.
As in the case of the electromagnetic form factors we observe
a dominance of the diquark contributions for larger $Q^2$.

\begin{figure}[t]
 \begin{center}
   \epsfig{file=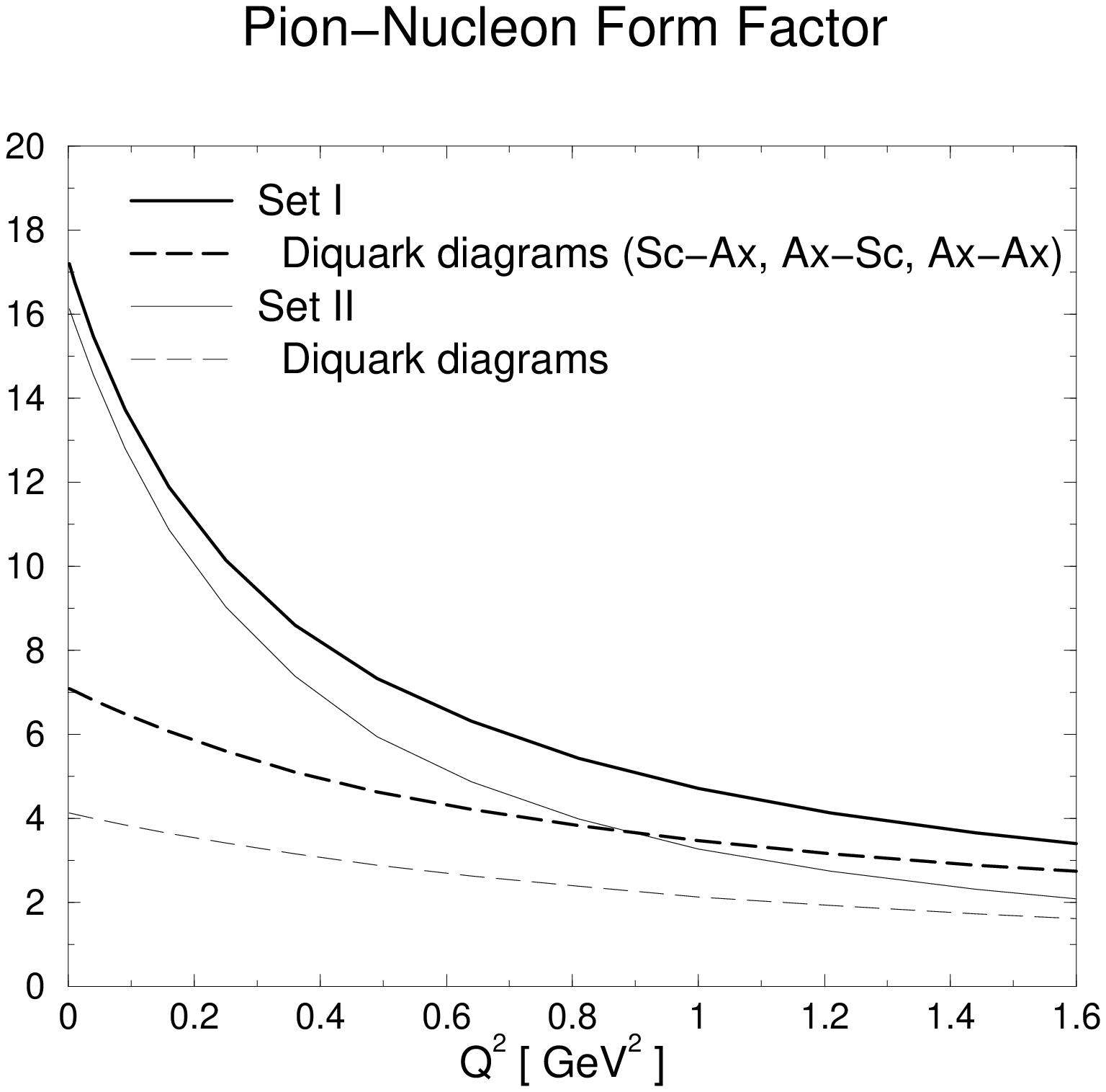,width=\figurewidth} 
   \epsfig{file=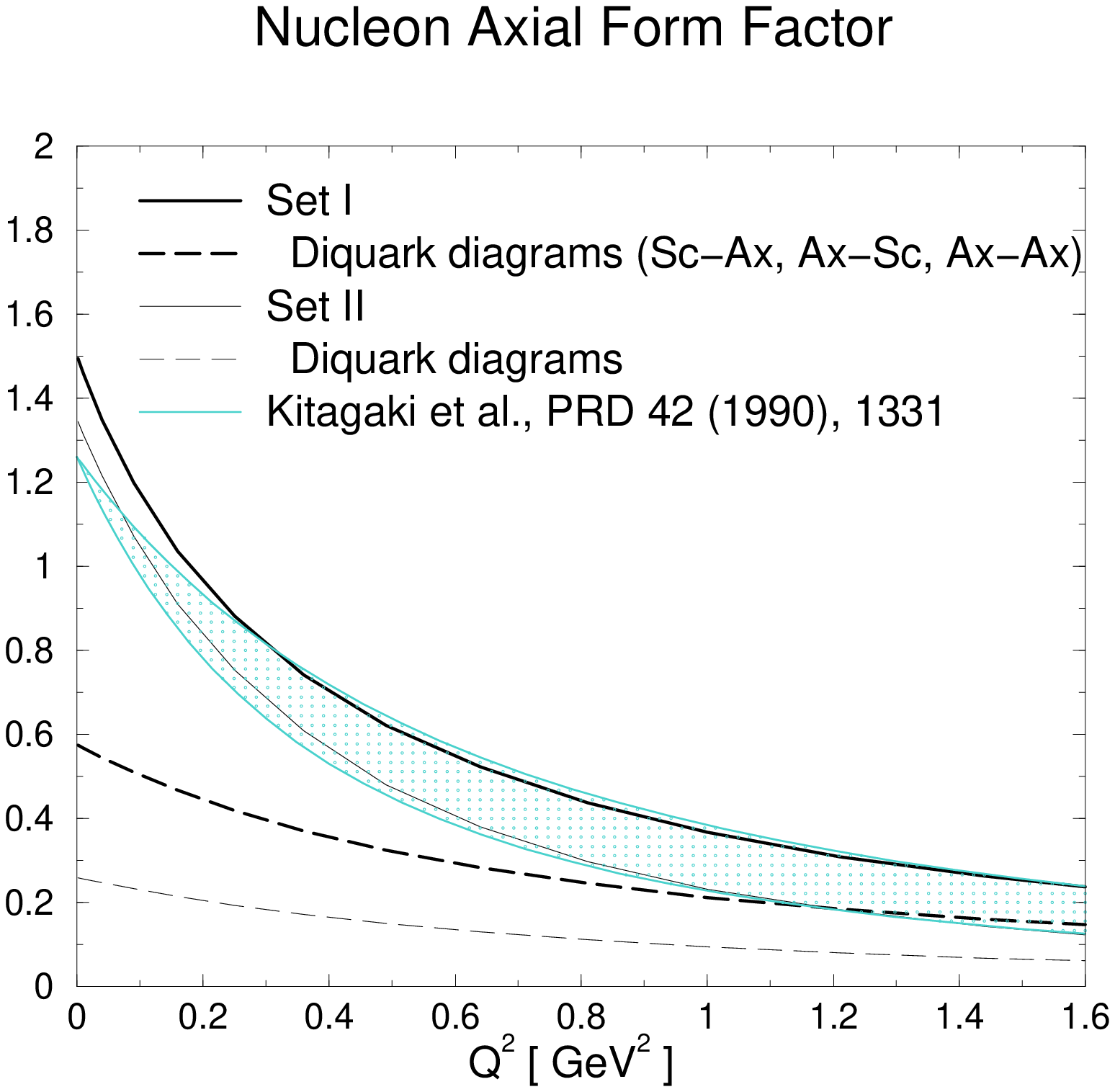,width=\figurewidth}
 \end{center}
\caption{The pion-nucleon form factor $g_{\pi NN}(Q^2)$ and
 the axial form factor $g_A(Q^2)$. 'Diquark diagrams'
 labels the sum of the impulse-approximate axialvector contributions
 and scalar-axialvector transitions. The shaded
 region in the right panel represents the uncertainty in $g_A$ as determined
 from quasi-elastic neutrino scattering when a dipole form is fitted to both,
 the vector and the axial form factor \cite{Kitagaki:1990vs}.}
 \label{gpifig}
\end{figure}

Turning to the pion-nucleon form factor we have
fitted our results to an $n$-pole,
\begin{equation}
 g_{\pi NN}(Q^2) = \frac{g_{\pi NN}(0)}{(1+Q^2/\Lambda_n^2)^n}\;. 
\end{equation} 
The form factor of Set I is close to a {\em monopole}, $n=1$,
with cut-off $\Lambda_1=0.6$ GeV, whereas the one of Set II
is better approximated by a {\em dipole}, $n=2$,
with cut-off $\Lambda_2=0.9$ GeV. The exponent $n$ is determined
by the strength of the diquark contributions whereas the cut-off
has to comply with the slope at $Q^2=0$ (the radius). It is
not a big surprise that the proton electromagnetic radii
and the pionic/axial radii are very close to each other
since the dominating impulse approximation diagrams
have a very similar behavior for all observables.

As we have mentioned in the beginning of the chapter, pion-nucleon
form factors are used in one-boson exchange potential
{\em ans{\"a}tze} to describe the $NN$ force. Most of the
potentials that fit the data employ a monopole with a large
cut-off, $\Lambda_1\ge 1.3$ GeV,
see {\em e.g.} ref.~\cite{Machleidt:1987hj}. However, other nucleon models,
lattice and QCD sum rule calculations indicate a monopole
behavior with a much smaller cut-off $\Lambda_1 \approx
0.5-0.95$ GeV, {\em cf.} ref.~\cite{Meissner:1995ra} 
and references therein.
Our calculations support this point of view.

Summarizing the results of this section, we have found moderate
violations of PCAC and the Goldberger-Treiman relation. These
violations are estimated to be partitioned equally between the
effects of neglecting chiral seagulls and vector diquarks.
The inclusion of axialvector diquarks is essential to obtain
values for the weak coupling constant $g_A>1$. Of particular importance
for this result
are the transitions between scalar and axialvector diquarks,
additionally they restrict the axialvector correlations
within the nucleon to be small. For  the pion-nucleon form factor,
our model calculations support a monopole- or dipole-like
fall-off with a cut-off well below 1 GeV.

 \chapter{Effective Confinement}
 \label{conf-chap} 
  
In this chapter we will investigate the possibility
of incorporating confinement into the diquark-quark model 
by a suitable modification of quark and diquark propagators.
As will become clear in the following,
this will enable us to calculate the mass spectrum of octet
and decuplet baryons.
 Electromagnetic form factors recomputed
with the modified propagators reveal a drastic increase  on the magnetic 
moments, being almost too much of an improvement. We have done the calculations for {\em pointlike} and
for {\em extended} diquarks  to demonstrate the necessity of
the latter for a consistent description of masses and
electromagnetic structure.
Although the results for the spectrum and the magnetic
moments are encouraging, there are serious difficulties
associated with this approach of implementing confinement, 
both motivated from
``theoretical'' and ``phenomenological'' arguments.

Remember that the propagators along with the diquark vertices are
the only ingredients of the model after the full 3-quark
problem has been reduced, {\em cf.} the summary at the end of section 
\ref{3qreduce}.
By choosing the most simple
{\em ans\"atze} for them, {\em i.e.} free spin-1/2 and
spin-0/spin-1 propagators for quark and diquark we have succeeded
to describe quite successfully various spacelike nucleon form factors.
However, shortcomings remain which might be cured
by allowing for non-trivial propagators. These shortcomings are:
\begin{itemize}
\item To incorporate the $\Delta$ into the description, rather large
 quark and  axialvector diquark masses have to be chosen, as their
 sum must be larger than the mass of the $\Delta$. This problem
 extends to the strange quark mass and strange diquarks
 when aiming at including the whole octet-decuplet spectrum.
\item The isovector part of the nucleon magnetic moments is 
 too small, even after inclusion of the axialvector diquarks.
 The difference is substantial for parameter sets which fit
 the $\Delta$ mass.
 A reason for this, besides too large a constituent quark mass, 
 might be the insufficiency of the bare
 quark-photon vertex to account for the quark contributions
 to the magnetic moments.
 Gauge invariance demands, however, that a nonperturbative
 quark-photon vertex is intimately connected  with
 a nonperturbative quark propagator.
\item Hadronic reactions where {\em timelike} momenta of the
 order of 1 GeV are deposited  onto nucleons, {\em e.g.}
 meson production processes, cannot be described 
 within the model in its present form. The free-particle poles
 of quark and diquark cause unphysical thresholds 
 in these processes.
\end{itemize}
As we see, the lack of confinement which is manifest in the poles
of the quark and diquark propagators used so far is
the main motivation to consider alterations in the model premises.
At this stage we hope that the necessary modifications in
the quark-photon vertex which are required by gauge invariance
will improve the results on the magnetic moments as well.

\section{Confinement in model propagators}

Confinement is understood as the absence of colored states in the
spectrum of observed particles. From a phenomenological point of view
(as adopted here), calculations of $S$ matrix elements using diagrams 
with internal quark loops
must not have any imaginary parts which are associated with 
the singularities of the quark propagators. So, either they are absent
or their contributions cancel in some manner
\cite{Alkofer:2000}. In this way, one is led
to consider the following possibilities, which are not necessarily
the only ones:
\begin{itemize}
\item[a.] Propagators are entire functions in the whole complex plane
 \cite{Efimov:1993zg}. If they are to be analytic, they must possess
 an essential singularity in the infinite. An example is the function
  $f(x)=e^x$.
\item[b.] Propagators have complex conjugate poles \cite{Stingl:1994nk}.
 In some sense the poles correspond to virtual excitations
 that cancel each other in physical amplitudes. Here an example
 is provided by the function $f(x)=x/(a^2+x^2)$.
\end{itemize}

We will investigate the consequences of propagators being entire
analytic functions. We insist here on analyticity in order not to loose
the relativistic repa\-ra\-me\-trization invariance related to the
choice of the momentum partitioning parameter $\eta$.\footnote{
As explained in section \ref{nuc-subsec}, $\eta$ invariance requires
the following: If $\Psi(p,P;\eta_1)$ is a solution of the 
Bethe-Salpeter equation then $\Psi(p+(\eta_2-\eta_1)P,P;\eta_2)$
with $\eta_2\not = \eta_1$ is also one. In the rest frame
of the bound state, $P=(\vect 0,iM_n)$, this can be shown to be valid
if the integration of the quark exchange kernel
$K^{\rm BS}(p,k,P)$ over the component $k^4$ can be shifted as
$k^4 \rightarrow k^4+(\eta_2-\eta_1)iM_n$. This is of course possible,
if the propagators are analytic in the complex domain
of $k^4$ given by $({\rm Re}\, k^4, {\rm Im}\,k^4) \in
((-\infty,+\infty),[0,(\eta_2-\eta_1)M_n])$.}
A word of caution is in order here.
The Wick rotation which connects Minkowski and Euclidean metric
is no longer applicable as the propagators possess the essential 
singularity at infinity. Furthermore we are aware that this propagator 
contradicts general arguments about the behavior of propagators 
in gauge theories with confinement. 
Following ref.~\cite{Oehme:1995pv}, it should
vanish for $p^2 \rightarrow \infty$ in {\em all} directions
of the complex $p^2$-plane faster than the free propagator
($p$ is the particle momentum). Furthermore
the propagators should have cuts along the {\em real} $p^2$-axis.
Unphysical colored states associated with these cuts are then to be
removed from the observable spectrum by other conditions such as
a suitable definition of a physical subspace ({\em cf.} the
long-known Gupta-Bleuler mechanism to remove scalar and longitudinal 
photons from the physical
particle spectrum in QED \cite{Bleuler:1950}). It seems to be impossible
to implement propagators which behave as indicated above and
such a projection onto physical states in actual matrix element calculations at
present, though. Therefore we believe it is instructive 
to investigate some phenomenological advantages and deficiencies 
of an ``entire'' propagator. In practical calculations, the
quark propagator will be needed only on a limited domain 
in the complex plane and the essential singularity thus poses
no practical difficulty, at least for the form factor calculations
presented here.

The second possible form for the propagator mentioned above 
suffers from only one principal deficiency, as 
there are complex conjugate pole associated with it. 
The asymptotic conditions in the complex plane  are
in this case easy to incorporate.
Actual calculations with this form are harder to perform
and are currently under investigation.

Before we turn to the propagator parametrization chosen here,
we discuss some aspects of the quark propagator that are known
so far. We regard the quark propagator to be given in its most general
form, $S^{-1}=-i\Slash{p}A-B$, with $A,B$ being scalar functions.
Information about the quark propagator may be obtained from
calculations within the framework of Dyson-Schwinger equations,
for a recent review see ref.~\cite{Alkofer:2000}.
In the near future, lattice results will provide us with  more
``empirical'' information, for a recent study in quenched QCD
see ref.~\cite{Skullerud:2000un}. Unfortunately,
the quark propagator is usually computed in both schemes 
only for positive real
$p^2$, where $p$ is the Euclidean quark momentum,
therefore the immediate benefit for phenomenological calculations
is limited. However, these calculations have established an important
generic feature.
In rainbow truncation 
of the Dyson-Schwinger equations
as well as in improved truncation schemes it could be shown
that the quark acquires an effective constituent mass, {\em
i.e.} $M(0)=B(0)/A(0)$ is non-zero for a vanishing current quark mass.
This feature of dynamical chiral symmetry breaking  
has also been observed in the already mentioned lattice calculation
\cite{Skullerud:2000un}.
For phenomenological calculations, the numerical results for $A,B$
have been fitted to entire functions which
has been successfully used in the description of (spacelike)
meson \cite{Maris:1998hc} and nucleon properties \cite{Bloch:1999ke,Bloch:1999rm}. 
The most impressive success in this direction
has been achieved in a combined rainbow/ladder truncation scheme
for quarks and mesons employing a specific  
gluon propagator which displays significant enhancement at momenta
of the order of a few hundred MeV.
Here the quark propagator has been computed
from Dyson-Schwinger equations
on {\em all} complex momentum points\footnote{This was rendered
possible through the analytical {\em ansatz} for the gluon propagator.}
where it has been needed to solve
the Bethe-Salpeter equation for mesons and to compute their electric
form factors \cite{Maris:1997tm,Maris:1999nt,Maris:2000}. Additionally
the quark-photon vertex has been obtained from the solution
of a corresponding inhomogeneous Bethe-Salpeter equation 
\cite{Maris:1999bh}. In these studies the solution for the quark
propagator exhibited no poles for the (complex) momenta
sampled by the meson calculations. These calculations support
the working hypothesis of a quark propagator having no poles,
at least in the interesting momentum regime. 

In eq.~(\ref{qprop}) we have already indicated a possible parametrization
of the quark propagator by multiplying the free one with
a single dressing function $C(p^2,m)$. We choose a function for $C$
which is entire and analytic and removes the free-particle pole
in the denominator of the free propagator,
\begin{equation}
  C(p^2,m) \equiv C^{\rm exp}(p^2,m) = 1-
      \exp\left(-d\;\frac{p^2+m^2}{m^2}\right) \; .
  \label{cexp}
\end{equation}
The essential singularity occurs here for $p^2 \rightarrow -\infty$.
The parameter $d$ regulates the modification as compared to the free
propagator. If $d$ is large, the spacelike properties remain nearly 
unchanged, whereas for timelike momenta the propagator blows up quickly.
For small $d\lesssim 1$, even the spacelike behavior is substantially
modified. 

For simplicity, we apply the same modification to the
propagators of scalar and axialvector diquarks, {\em cf.}
eqs.~(\ref{Ds},\ref{Da}). Additionally we choose $\xi=1$ in 
eq.~(\ref{Da}). This parameter is a gauge parameter when describing
the interaction of a vector field with photons \cite{Lee:1962vm}.
The photon vertex with the axialvector diquark has to be modified
correspondingly.

We note that due to the absence of poles the ``masses'' 
$m_q,m_{sc},m_{ax}$ have no physical interpretation. They
are merely width parameters of the propagator.

\section{The octet-decuplet mass spectrum}
\label{od-sec}

Equipped with propagators which effectively mimic confinement
we can calculate the mass spectrum of octet and decuplet baryons.
This investigation has been reported in ref.~\cite{Oettel:1999bu}.
In doing so, we have to extend all previous considerations to
flavor $SU(3)$ and introduce explicit symmetry breaking as visible
in the experimental spectrum. However, we will confine ourselves
to the limit of unbroken isospin and, accordingly, the {\em strange} quark mass
is the only source of symmetry breaking, $m_s \not = m_u=m_d$, where
$m_u,m_d$ denote the mass parameters of $up$ and $down$ quark.

In order to limit the number of parameters we assume the
scalar and axialvector diquark mass parameters to be equal and introduce 
a diquark mass coefficient $\zeta$ by
\begin{equation}
  m_{sc}^{ab}=m_{ax}^{ab}=\zeta(m_a+m_b) \; \quad
  (a,b \in \{u,d,s\}) \; .
\end{equation}
Thus $ab$ denotes the flavor content of the diquark. It will
be interesting to see whether the mass differences
between spin-1/2 and spin-3/2 baryons can be explained
without an explicitly larger axialvector diquark mass parameter. 

Diquark 3$\times$3 flavor matrices $\left(t_{(ab)}\right)_{jk}$
are chosen antisymmetric
for scalar diquarks,
\begin{equation}
  t_{(ud)}=-\frac{i}{\sqrt{2}}\,\lambda^2\;, \quad 
  t_{(us)}=-\frac{i}{\sqrt{2}}\,\lambda^5\;, \quad 
  t_{(ds)}=-\frac{i}{\sqrt{2}}\,\lambda^7\;.
\end{equation}
Here we use for the $\lambda$'s the standard Gell-Mann matrices 
\cite{Griffiths:1987tj}. Axialvector diquark matrices
$\left(t_{[ab]}\right)_{jk}$
 are symmetric in their indices $j,k$,
\begin{equation}
 \begin{array}{lll}   
   t_{[ud]}={\Dc \frac{1}{\sqrt{2}}}\,\lambda^1\; , & 
   t_{[us]}={\Dc \frac{1}{\sqrt{2}}}\,\lambda^4\; , &
   t_{[ds]}={\Dc \frac{1}{\sqrt{2}}}\,\lambda^6\; , \\[5mm]
   t_{[uu]}=\delta_{j1}\delta_{k1} \; , &
   t_{[dd]}=\delta_{j2}\delta_{k2} \; , &
   t_{[ss]}=\delta_{j3}\delta_{k3} \; . 
 \end{array}
\end{equation}
To derive the Bethe-Salpeter equations for the octet 
($N$, $\Lambda$, $\Sigma$, $\Xi$) and decuplet
baryons ($\Delta$, $\Sigma^\ast$, $\Xi^\ast$,
$\Omega$), one starts with their flavor wave functions prescribed
by the Eightfold Way.
The octet baryons have scalar diquark correlations, therefore
this part of the flavor wave function is mixed antisymmetric.
The axialvector diquark correlations for  octet members
are described in flavor space by a sum of two mixed antisymmetric
states. For the decuplet members with only axialvector diquarks
contributing these are either totally symmetric 
($\Delta^{++},\Delta^{-},\Omega$) 
or mixed symmetric states.  
These wave functions may contain different mass eigenstates
of quark-diquark pairs. For these mass eigenstates the system
of coupled Bethe-Salpeter equations has to be solved, where
the decomposition in the Dirac algebra according to
section \ref{partialwaves} can be used. As an example we consider
the $\Sigma^+$ hyperon with its flavor wave function
\begin{equation}
 | \Sigma^+ \rangle_{\rm flavor} =
 \bp u(su) \\ \sqrt{\third}u[us] -\sqrt{\frac{2}{3}}s[uu] \ep\; .
\end{equation}
The scalar diquark part, $u(su)$ is clearly mixed antisymmetric
and the axialvector diquark part is the normalized sum of
$(us-su)u$ and $uus-suu$, both also mixed antisymmetric.
The Dirac-flavor wave function contains three independent
terms $\Psi^5_{u(us)}$, $\Psi^\mu_{s[uu]}$ and $\Psi^\mu_{u[us]}$
with a total of 2+6+6=14 unknown scalar functions which shows
the necessary increase of computer resources.

We have listed the Bethe-Salpeter equations for all octet and
decuplet baryons in appendix \ref{od-app} and quote from the
results given there just one  remarkable fact.
Due to the symmetry breaking, there occurs mixing of the flavor singlet
wave function with the scalar diquark correlations within
the $\Lambda$. This mixing is absent in quark potential models.

\begin{table}[t]
\begin{center}
\begin{tabular}{lcccc} \hline \hline & & & & \\
 && experi-  & pointlike & extended   \\
 &&  ment     & \multicolumn{2}{c}{diquark} \\ \hline
  &   &&& \\
 &&  & {${ V(p)=1}$} &
  {$ V(p)=\left(\frac{\Dc\lambda^2}{\Dc \lambda^2+p^2}\right)^4$}\\[3mm]
 & & & & { $\lambda=$ 0.5  GeV} \\ \hline
 &&&& \\
$m_u\,$ & [GeV]    &      &0.50       &0.56  \\
$m_s$ & [GeV] &      &0.63    &0.68   \\
$\xi$ &    &      &0.73 &0.60  \\ 
$g_a/g_s$ & & &  1.37 & 0.58 \\ \hline 
& &&& \\
$\left.
  \begin{array}[c]{l}
   M_\Lambda \\ M_\Sigma \\ M_\Xi \\ M_{\Sigma^\ast} \\ M_{\Xi^\ast} \\
   M_\Omega \end{array}
 \right\}$ &
    [GeV] &
 \begin{tabular}[c]{l}
  1.116 \\ 1.193 \\ 1.315 \\ 1.384 \\ 1.530 \\ 1.672 
 \end{tabular} &
 \begin{tabular}[c]{r}
  1.133\\ 1.140\\ 1.319\\ 1.380\\ 1.516\\ 1.665
 \end{tabular} &
 \begin{tabular}[c]{r}
  1.098 \\ 1.129 \\ 1.279 \\ 1.396 \\ 1.572 \\ 1.766
 \end{tabular} \\
 &&&&\\ \hline \hline
\end{tabular}       
\end{center}
\caption{Octet and decuplet masses.}
\label{od-masses}
\end{table}

As the diquarks here can never be on-shell, we are not allowed to use
the normalization conditions (\ref{normsc},\ref{normax}) 
to fix the quark-diquark coupling strengths $g_s$ and $g_a$.
Therefore we adjust them to the masses of nucleon and $\Delta$,
$M_n=0.939$ GeV and $M_\Delta=1.232$ GeV.

In table \ref{od-masses} we show results for the masses
of the remaining baryons ($i$) for pointlike diquarks and
($ii$) for a quadrupole-like diquark vertex function.
For both cases the octet-decuplet mass difference is a result
of solely the relativistic dynamics; it is even overestimated for
extended diquarks. The parameter set using the extended
diquarks allows a considerable reduction of the ratio $g_a/g_s$
as compared to the ratio for pointlike diquarks.
Still, axialvector correlations are somewhat too strong regarding
the constraints which have been obtained in chapters
\ref{em-chap} and \ref{sa-chap}.

As can be seen from table \ref{od-masses}, the calculated baryon masses
are higher for the spin-3/2 states than the sum of the 
quark and diquark
mass parameters. Due to the absence of poles, it is possible
to smoothly cross the tree-level thresholds given by the boundaries 
for the momentum partitioning parameter $\eta$ in 
eqs.~(\ref{ebound1},\ref{ebound2}), {\em cf.} the discussion
in refs.~\cite{Hellstern:1997pg,Oettel:1998bk}.

These results have been obtained with the choice $d=1$
in the propagator modification function $C^{\rm exp}$, {\em cf.}
eq.~(\ref{cexp}), corresponding to a quite substantial modification
of the spacelike propagator properties. Since  mainly these  are important for
solving the Bethe-Salpeter equation as well as for the calculation
of form factors, we expect visible changes for the latter as well.
We will now turn to this issue.

\section{Electromagnetic form factors}

In order to use the machinery developed in chapter \ref{em-chap}
for calculating the form factors, we have to re-specify the
basic quark and diquark vertices. To satisfy gauge invariance,
the quark vertex must fulfill the Ward-Takahashi identity
\begin{equation}
  (k-p)^\mu\; \Gamma^\mu_q=q_q\;\left(S^{-1}(k)-S^{-1}(p)\right) \; ,
  \label{WTS2}
\end{equation}
with the quark propagator being dressed by the function 
$C^{\rm exp}$ of eq.~(\ref{cexp}). 
The scalar functions $A,B$ are given in terms of $C^{\rm exp}$ by
\begin{equation}
 A(p^2)=\frac{1}{C^{\rm exp}(p^2,m_q)}\; , \qquad
 B(p^2)=\frac{m_q}{C^{\rm exp}(p^2,m_q)} \; .
\end{equation}
It turns out that
the longitudinal part of the quark vertex is unambiguously 
given by the Ball-Chiu {\em ansatz} \cite{Ball:1980ay},
labelled by the superscript BC,
\begin{eqnarray}
  \Gamma^\mu_q& =& q_q\left(\Gamma^{\mu,{\rm BC}}_q + \Gamma^{\mu}_{q,T}
          \right) \; ,\\
  \Gamma^{\mu,{\rm BC}}_q & =& -i \,\frac{A(k^2)+A(p^2)}{2}\, \gamma^\mu - \\
   &&  i\,\frac{(k+p)^\mu}{k^2-p^2} \left[ (A(k^2)-A(p^2)) 
     \frac{\Slash{k}+\Slash{p}}{2} -i\,(B(k^2)-B(p^2)) \right] \; . \nonumber
\end{eqnarray}
Its exclusivity results if the following four conditions
are valid:
\begin{enumerate}
 \item It satisfies the Ward-Takahashi identity (\ref{WTS2}).
 \item It does not possess any kinematical singularities.
 \item It transforms under parity, charge conjugation and time reversal
   as the free vertex.
 \item It reduces to the free vertex for $A=1$ and $B=m_q$.
\end{enumerate}
The transverse part of the vertex, $\Gamma^{\mu}_{q,T}$, remains
undetermined. We will put it to zero in the following which
is certainly an oversimplification regarding the information
about it which has been obtained during the last years.
It has been shown in 
ref.~\cite{Curtis:1990zs} that the requirement
of multiplicative renormalizability greatly restricts the tensor structure
of the transversal vertex. For large spacelike momenta $Q$, where
no resonances are present, this leads to a phenomenologically
unimportant vertex modification. 
For the small and moderate
momentum transfers at which we investigate the form factors,
a transversal term which
contains the effect of the
$\rho$ meson pole might turn out to be of importance 
for the quark-photon vertex.
The already mentioned study in ref.~\cite{Maris:1999bh} has obtained
the {\em full} quark-photon vertex  as a solution of an inhomogeneous
Bethe-Salpeter equation. The quark-antiquark scattering kernel 
in the vector channel must enter the equation for the quark-photon
vertex as well, thereby the solution for the full 
vertex contains the vector meson  pole as the pole also appears
in the full quark-antiquark scattering amplitude. Besides having found
this vector meson pole contribution (being transversal) 
to the quark-photon vertex, the study has numerically confirmed that
the longitudinal part of the vertex is given by the Ball-Chiu recipe.
Within this model context,
the remnants of the vector meson pole have been shown to 
contribute about 30 \% to the full pion charge radius $r_\pi$, with the 
remainder being attributable to the Ball-Chiu vertex.

In order to create a link to the old and, at this time,
fairly successful picture of vector meson dominance,
it would be interesting to investigate the influence of the
vector meson pole also for the nucleon form factors. This requires,
however, a tremendous effort to use an effective 
gluon interaction like the one from ref.~\cite{Maris:1999bh}
for the quark-antiquark and quark-quark scattering kernels
to describe mesons {\em and} diquarks, to obtain the quark-photon
vertex self-consistently, and to calculate or reliably parametrize
the quark propagator in the complex plane where it is needed for
calculations of nucleon properties. 
This is beyond the scope of this work. 

Having introduced the Ball-Chiu vertex as an approximation
to the full quark-photon vertex and having discussed possible shortcomings,
let us turn to the diquark-photon vertices.
We make  use of the Ward-Takahashi identities for 
scalar and axialvector diquarks,
\begin{equation}
  (k-p)^\mu \;\Gamma^{\mu,[\alpha\beta]}_{sc[ax]}= q_{sc[ax]}\;\left(
  (D^{-1})^{[\alpha\beta]}(k)-(D^{-1})^{[\alpha\beta]}(p)\right)\; ,  
\end{equation} 
to fix the longitudinal part of the respective photon vertices.
Neglecting again transversal parts, we find for the vertices,
\begin{eqnarray}
  \label{bc-scalar}
  \Gamma^{\mu}_{sc} &=& -q_{sc}\;(k+p)^\mu\,
     \frac{\Dc \frac{k^2+m_{sc}^2}{C^{\rm exp}(k^2,m_{sc})} - 
            \frac{p^2+m_{sc}^2}{C^{\rm exp}(p^2,m_{sc})}}{k^2-p^2}\; , \\
  \Gamma^{\mu,\alpha\beta}_{ax} &=& -q_{ax}\;(k+p)^\mu\,
     \delta^{\alpha\beta}\;
     \frac{\Dc \frac{k^2+m_{ax}^2}{C^{\rm exp}(k^2,m_{ax})} -
            \frac{p^2+m_{ax}^2}{C^{\rm exp}(p^2,m_{ax})}}{k^2-p^2}+
         \nonumber \\        
       \label{bc-ax}
         & & \;\;q_{ax}\;(1+\kappa)\left( Q^\beta\,\delta^{\mu\alpha}-Q^\alpha\,
               \delta^{\mu\beta}\right) \; .
\end{eqnarray}
The last term in eq.~(\ref{bc-ax}) is transversal and therefore
no dressing can be inferred from the Ward-Takahashi identity.
For this reason, we do not modify this term.
Compared to the bare vertex, eq.~(\ref{vertax}),
the transversal term appears with a modified prefactor, $(1+\kappa)$ instead
of $\kappa$ where $\kappa$ denotes the anomalous magnetic moment.
This is a consequence of choosing the gauge parameter $\xi=1$
in the dressed propagator \cite{Lee:1962vm}.

Anomalous transitions between scalar and axialvector diquarks 
are of course possible as well. Their vertices remain unchanged,
{\em cf.} eqs.~(\ref{sa_vert},\ref{as_vert}). 

We can determine the unknown anomalous magnetic moment $\kappa$
appearing in eq.~(\ref{bc-ax}) and the strength of the
scalar-to-axialvector transitions $\kappa_{sa}$ from
eqs.~(\ref{sa_vert},\ref{as_vert}) by much the same procedure
as described in section \ref{curop-sec} and appendix \ref{dqres1}.
That is, we calculate the resolved diquark vertices
from figure \ref{emresolve}, of course by using the dressed quark 
propagator and quark-photon vertex given above, and re-adjust
the diquark normalization (coupling) constants $g_a \rightarrow 
g_a^{\rm resc}$ and $g_s \rightarrow g_s^{\rm resc}$
such that the differential Ward identity is fulfilled, 
{\em cf.} appendix \ref{dqres1}. Using these rescaled quantities
$g_a^{\rm resc}$ and $g_s^{\rm resc}$, we determine 
$\kappa$ and $\kappa_{sa}$.

\begin{table}[b]
 \begin{center}
  \begin{tabular}{lllll} \hline \hline
    & & & & \\
    & $g_s^{\rm resc}/g_s$ & $g_a^{\rm resc}/g_a$ & $\kappa$ & $\kappa_{sa}$ \\ \hline
    & & & & \\
    extended diquarks & 0.73 & 1.33  & 0.69 & 1.76 \\
     & & & & \\ \hline \hline
  \end{tabular} 
 \caption{
   Rescaled diquark normalizations
   and constants of photon-diquark couplings
   for the parameter set with extended quarks, {\em cf.}
   table \ref{od-masses}.}
  \label{cc_3}
 \end{center}
\end{table}

For pointlike diquarks as used by
the first parameter set of table \ref{od-masses} the calculation of the
resolved vertices leads to divergent integrals and therefore prohibits
their use. 
We have performed the calculation for the second parameter set from
table \ref{od-masses}, {\em i.e.} the set which employs
extended diquarks. From the results found  in table
\ref{cc_3} we see that the anomalous magnetic moment and the transition
strength are reduced compared 
to the previous calculations
with free propagators, {\em cf.} table \ref{cc_1}.

\begin{figure}
 \begin{center}
  \epsfig{file=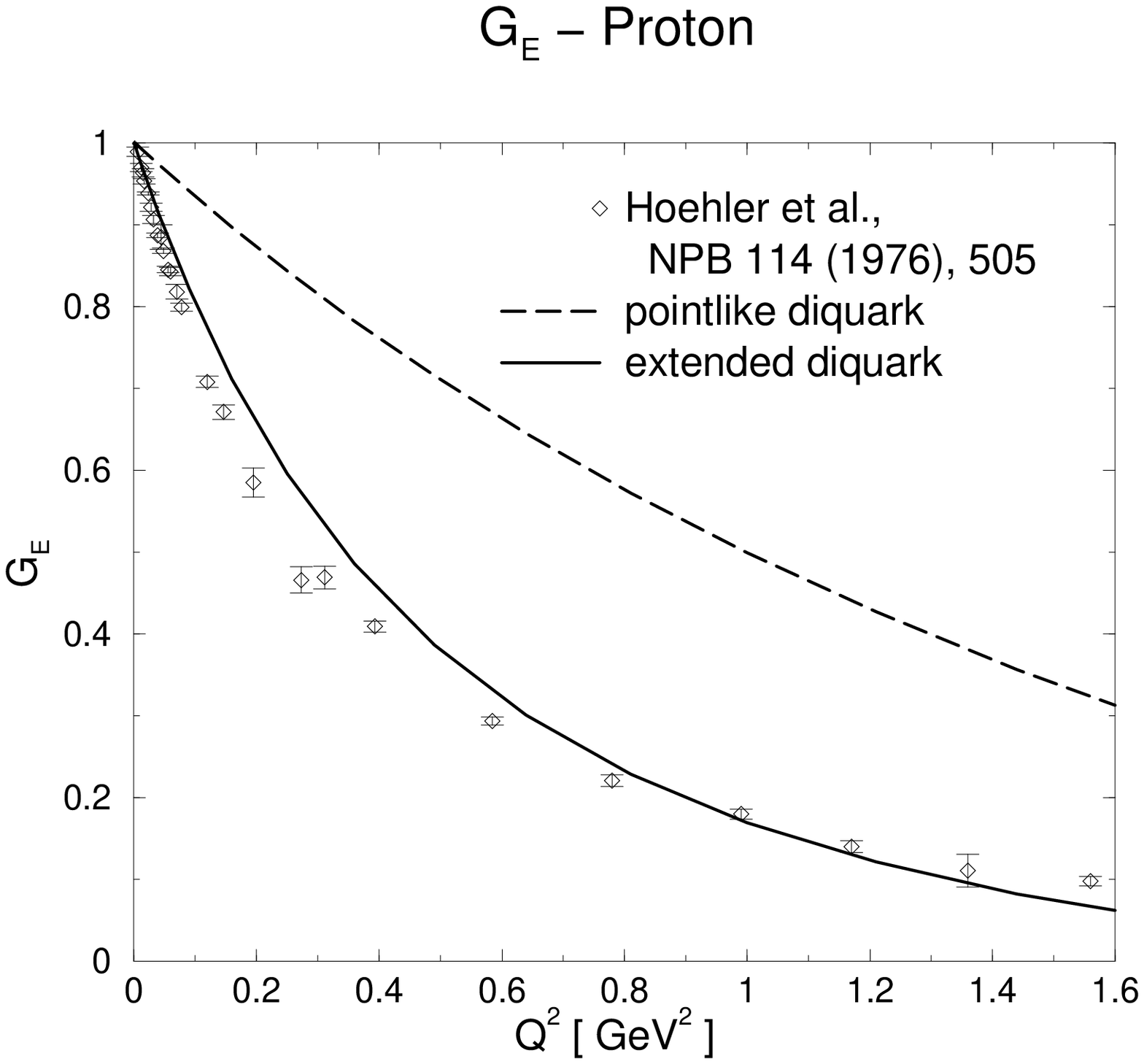,width=\figurewidth}
  \epsfig{file=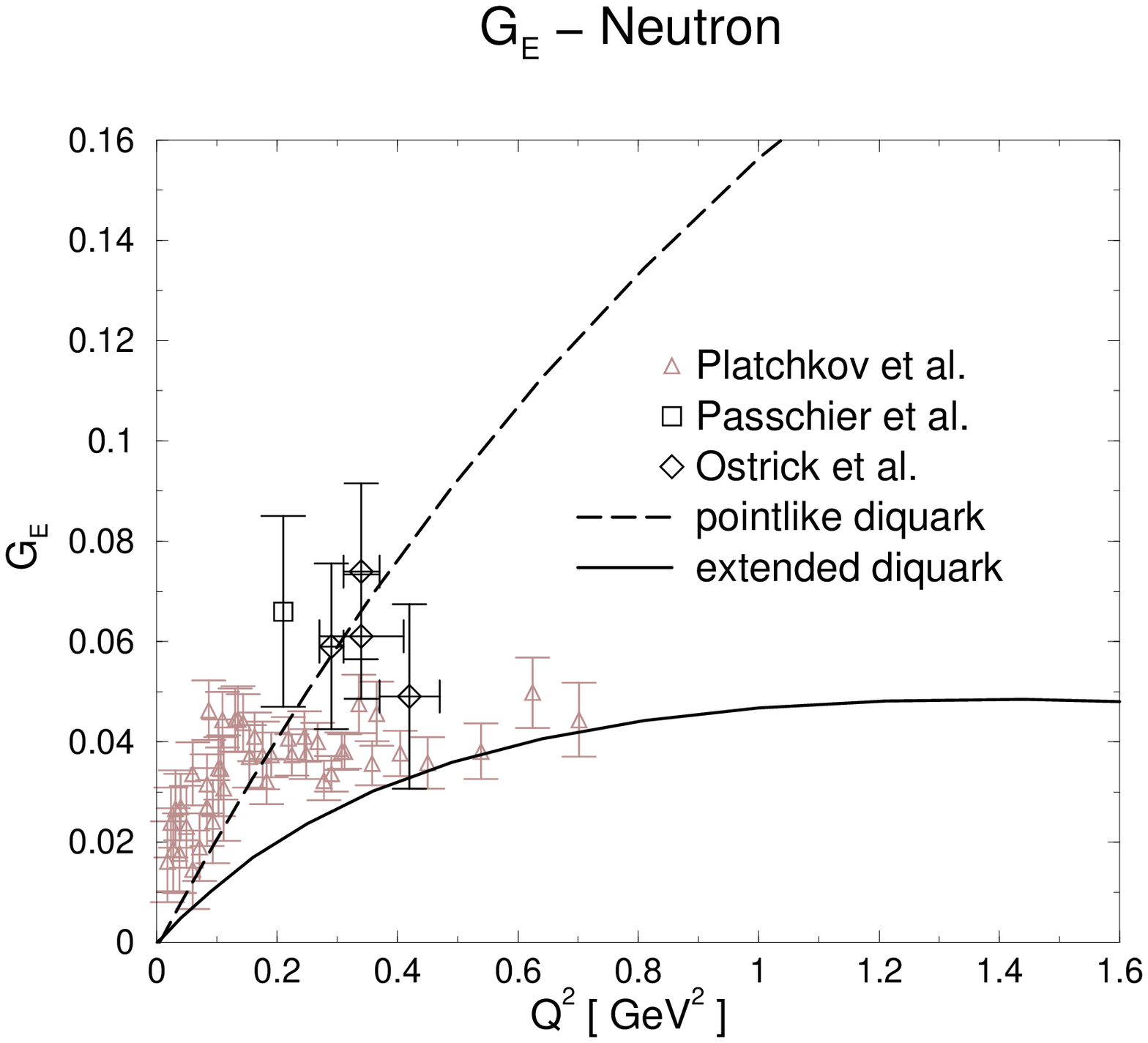,width=\figurewidth}
 \end{center}
 \caption{Electric form factors of both nucleons for the parameter
  sets of table~\ref{od-masses}.}
 \label{ge4fig}
\end{figure}

The form factors are now calculated as the sum of the extended
impulse approximation diagrams, {\em cf.} figure \ref{impulse},
and of the quark exchange kernel diagrams depicted
in figure \ref{7dim}. The relevant formulae for the impulse approximation
are given by eqs.~(\ref{jimp},\ref{jimn}) and by eqs.~(\ref{sg}--\ref{ex})
for the exchange kernel contributions. 

Turning to the results for the electric form factors, shown
in figure \ref{ge4fig}, we see the favorable influence of the
diquark-quark vertices with finite width. They lead to a basically
correct description of the proton electric form factor. The neutron 
electric form factor is also much better described by the parameter
sets which employs extended diquarks. Here we see the already mentioned
quenching effect of the axialvector correlations, though. They are quite
strong as compared to the parameter sets  from table \ref{pars},
with the scalar diquark contributions to the norm integral reduced
to 51 \% (the numbers are 66 \% for Set I and 92 \% for Set II).

Regarding the overall shape of the calculated magnetic form factors,
depicted in figure \ref{gm3fig}, we obtain the same conclusion as
before. The parameter set with extended diquarks fairly well describes 
the empirical dipole shape whereas pointlike diquarks assume the nucleons
to be far too rigid. It is interesting to observe that for the extended
diquarks the calculated magnetic moments, $\mu_p=$ 3.32 
and $\mu_n=-$2.14,
now exceed the experimental values by 19 and 12 per cent, 
respectively. We find that the magnetic moment contribution
of {\em every single} diagram weighed with its contribution to the total
nucleon charge is enhanced when comparing with the corresponding ratios
for the Sets I and II which employ bare propagator and vertices.
Clearly the dressing of propagators and vertices has a quite
substantial effect on the magnetic moments and a more refined choice
for them may provide for an exact description of both
electric and magnetic form factors.

\begin{figure}
 \begin{center}
  \epsfig{file=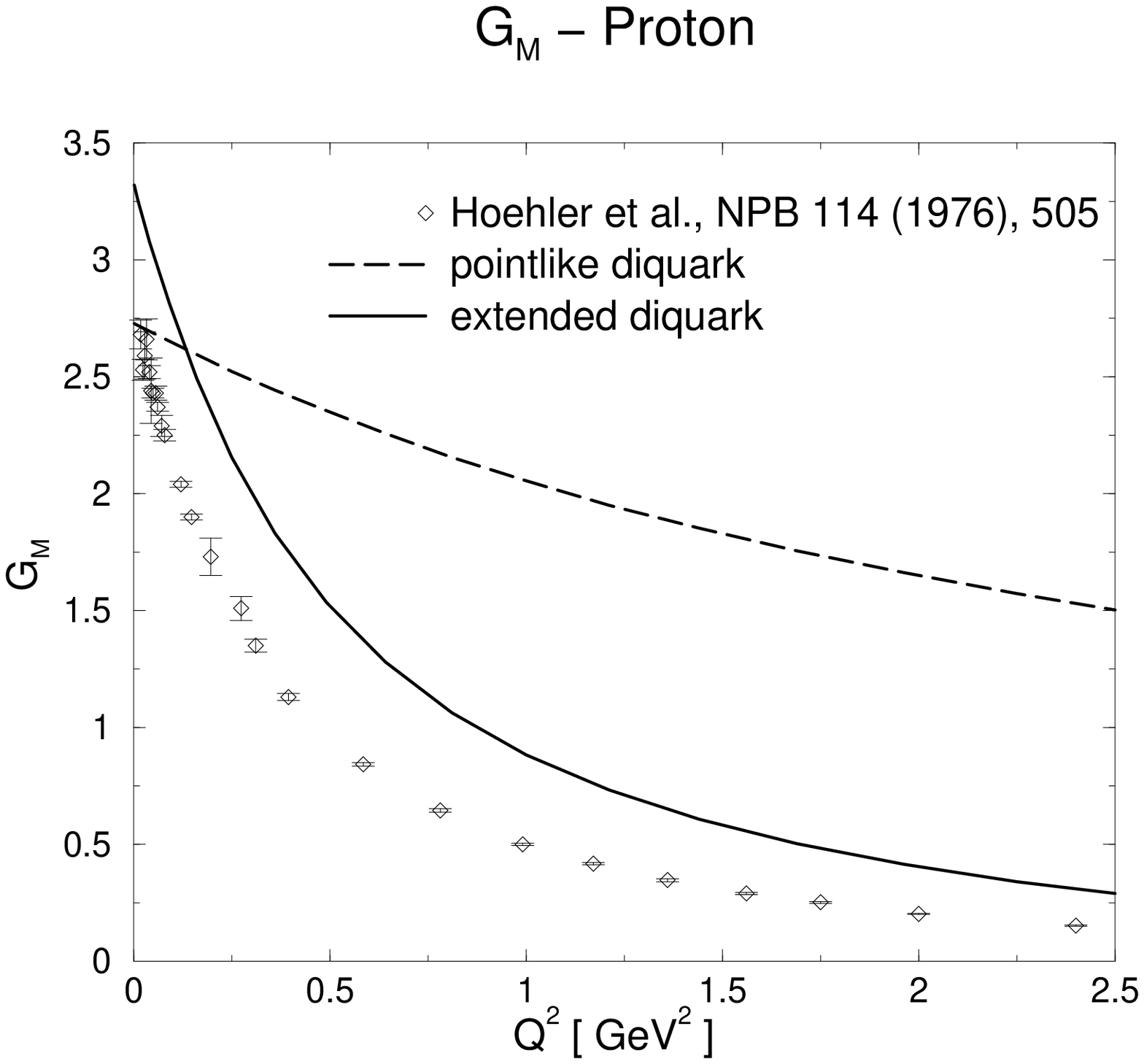,width=\figurewidth}
  \epsfig{file=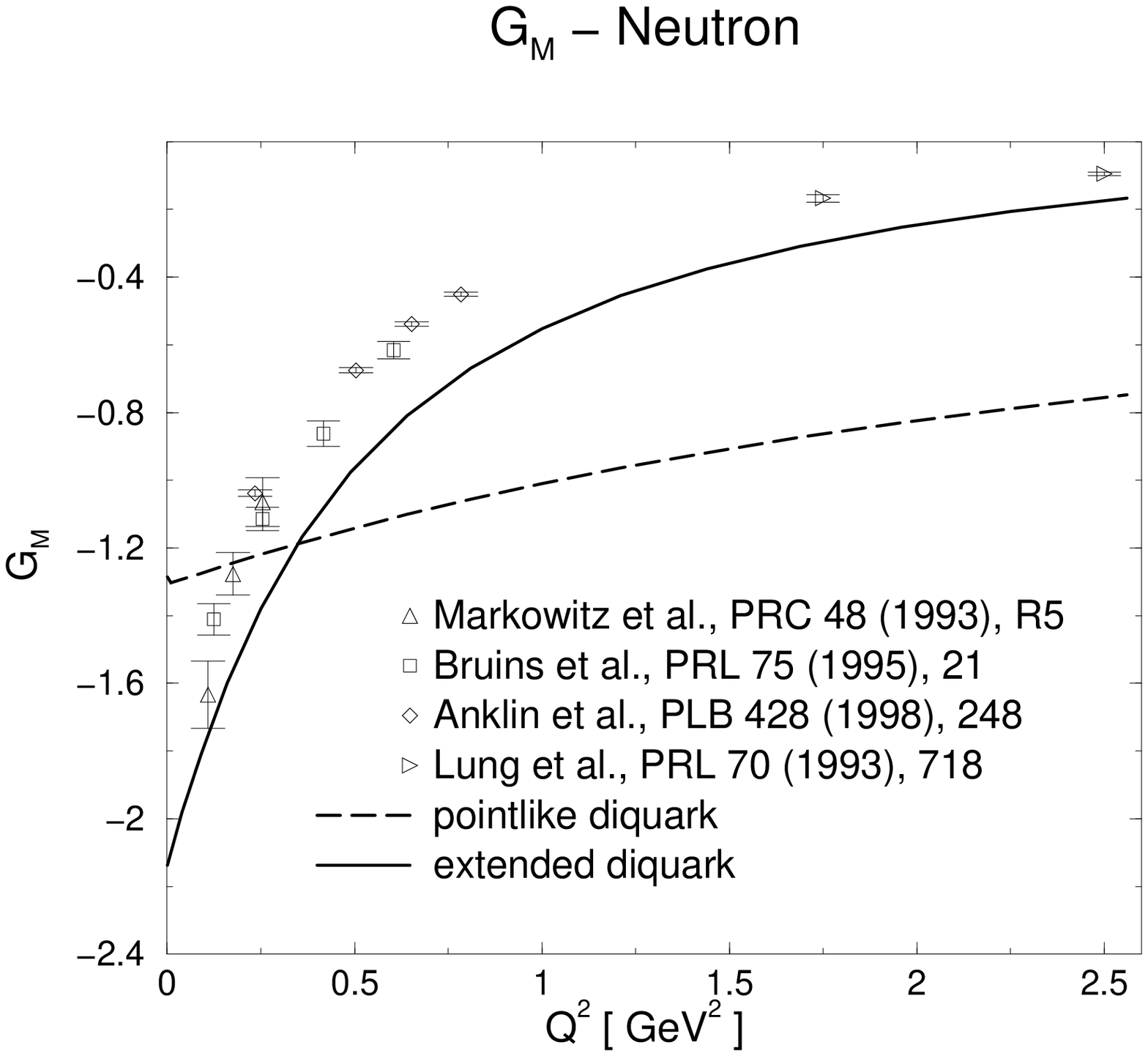,width=\figurewidth}
 \end{center}
 \caption{Magnetic form factors of both nucleons for the parameter
  sets of table~\ref{od-masses}. Experimental data
  for the neutron is taken from 
  refs.~\cite{Markowitz:1993hx,Bruins:1995ns,Anklin:1998ae,Lung:1993bu}.
  }
 \label{gm3fig}
\end{figure}

\section{Discussion}

We have employed  a simple parametrization of quark and
diquark propagators which renders them to be devoid of
poles in the complex momentum plane except for an essential 
singularity at infinity.
Thereby we could demonstrate that  an overall satisfying description of
both the electromagnetic form factors
and the mass spectrum of octet and decuplet baryons 
has been achieved. However, these results have to be taken
with a grain of salt as has been already alluded to. 
General arguments about the functional form of the quark propagator
in QCD simply prohibit such a behavior for timelike momenta as
has been assumed here. Since mass spectrum and form factor calculations
test the propagator only to a moderate extent in the timelike region
(Re$(p^2)\gtrsim -$0.3 GeV$^2$) one might argue that 
the chosen form for the propagator parametrizes the 
momentum regime needed well enough for practical calculations of physical processes
and only the continuation to large timelike momenta needs to be adapted.

On the other hand, there are processes which permit to 
test the quark propagator properties for larger timelike momenta.
Consider the photoproduction of kaons off the proton
($\gamma p \rightarrow K^+ \Lambda$), 
measured by the SAPHIR collaboration \cite{Tran:1998qw}.
The dominant contributions to this process in the diquark-quark model
are depicted in figure \ref{pgkl}. The threshold 
photon energy for this process to occur (with the proton being at rest)
is $E_\gamma=$ 0.913 GeV and a broad, dipped  maximum for the total
cross section is found for $E_\gamma=$ 1.1\dots1.5 GeV.
Now consider the left diagram in figure \ref{pgkl}. For the
momentum $p$ of the quark propagating between the points of the
photon absorption and kaon emission we find the bound
Re$(p^2)\ge -(\eta M_n+E_\gamma)^2$ \cite{Ahlig:2000}, with $M_n$ being
the nucleon mass and $\eta$ the momentum partitioning parameter
between quark and diquark. For photon energies
larger than 1 GeV a propagator dressed as given by eq.~(\ref{cexp})
will show considerable enhancement and this has indeed been found 
\cite{Alkofer:1999rn,Ahlig:2000}. The calculation of the total cross section
using the parameter set with pointlike diquarks from table 
\ref{od-masses} reveals an  upshot of the cross section beyond
$E_\gamma=$ 1.2 GeV, clearly in contrast to the data. Only with
a (non-analytical) modification function,
\begin{equation}
  C(p^2,m)= 1-\exp \left(-\frac{1}{4}\, \frac{|p^2+m_q^2|}{m_q^2}
        \right)\; ,
\end{equation}
which damps the propagator for timelike momenta,
the experimental cross section could be reproduced.
Unfortunately solutions of the Bethe-Salpeter equation and
results for observables employing such a propagator do not
exhibit invariance under shifts of the momentum partitioning parameter
$\eta$ anymore due to the non-analyticity of $C$. Thereby relativistic
translation invariance seems to be lost.

\begin{figure}
 \begin{center}
   \epsfig{file=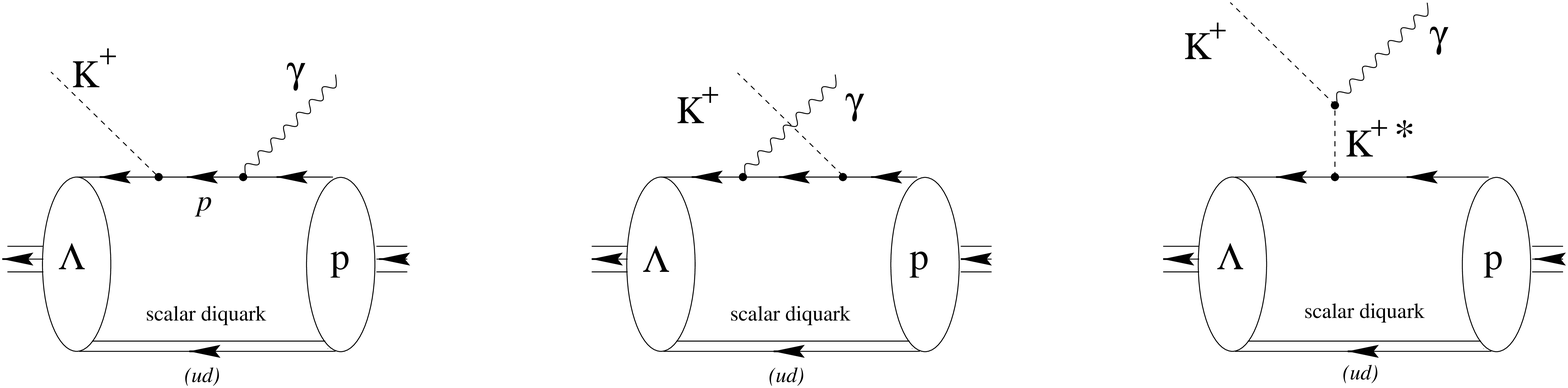,width=\textwidth}
 \end{center}
 \caption{The dominating processes in the diquark-quark model for
  kaon photoproduction, assuming small axialvector correlations.}
 \label{pgkl}
\end{figure}

Another phenomenological argument against the
pole-free form for the quark propagator comes from
the analysis of Deep Inelastic Scattering in the diquark-quark picture.
In the Bjorken limit the leading contribution to the hadronic tensor
is obtained by calculating the imaginary part of the 
``handbag diagram'' shown in figure \ref{compton-fig}. This diagram
constitutes the leading part of the Compton tensor
in forward scattering. By Cutkosky's rule the imaginary part is obtained 
by cutting the diagram as indicated in the figure and setting the
propagators on-shell at the cuts. This procedure is of course not possible
when employing pole-free propagators and in this case the imaginary part
of the Compton scattering amplitude could only be obtained by 
calculating more complicated diagrams with intermediate mesons and 
baryons. In the limit of the squared photon momentum $Q^2$ 
going to infinity,
each of these diagrams will be suppressed and the hadronic
tensor vanishes, in obvious contradiction to the measured finite 
structure functions.

Therefore we are in the peculiar situation that results obtained
by using  free
quark and diquark propagators compares favorably with experiments
when calculating quark distributions. For the quark-diquark model
with pointlike scalar diquarks this has been done in 
ref.~\cite{Kusaka:1997vm}. A simple quark-diquark spectator model
which parametrizes the nucleon vertex function in terms of the
``$s$ wave'' covariants ${\cal S}_1$ and ${\cal A}_3$,
{\em cf.} table \ref{components1}, has been used 
in ref.~\cite{Jakob:1997wg} to calculate
various distribution and fragmentation functions. These calculations 
indicate that the approximation of the plethora of intermediate (colorless)
states possible in Compton scattering by a simple (colored) quark
and a (colored) diquark works amazingly well. However, in the beginning
of the chapter we have emphasized that exactly these thresholds
are sought to avoid in other calculations such as those
related to kaon photoproduction. It therefore remains an interesting task
to resolve this puzzle in a consistent manner.

\begin{figure}[t]
 \begin{center}
   \epsfig{file=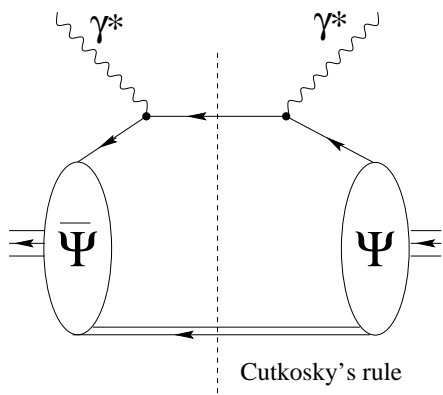,width=6cm}
 \end{center}
 \caption{Handbag diagram for Virtual Compton Scattering in the
       diquark-quark model. The imaginary part of the forward amplitude
       is obtained by Cutkosky's rule, {\em i.e.} by
        setting the intermediate quark and diquark
       propagators on-shell, indicated by the dashed line.}
 \label{compton-fig}
\end{figure}

\clearpage

 \chapter{The Salpeter Approximation}
 \label{sal-chap}
  
In this chapter we will critically compare results
for vertex functions and nucleon observables obtained
($i$) by using the full solution as described in chapters
\ref{dq-q-chap}--\ref{sa-chap} and ($ii$) in the 
{\em Salpeter approximation} \cite{Salpeter:1952} to the (ladder) Bethe-Salpeter
equation. The Salpeter approximation, to be
explained below, is a popular 3-dimensional reduction of
the original 4-dimensional problem. We mention that the here
obtained results are summarized in ref.~\cite{Oettel:2000uw}.

We can identify two motivations to study
semi-relativistic reductions to the Bethe-Salpeter
equation. First, the gain in technical simplicity is substantial
when the equation is transformed into a 3-dimensional problem
which can be solved with the tried and tested methods from
quantum mechanics. Secondly, 
excited states which have negative norm are absent
in the spectrum of a Salpeter approximated equation.
We will shortly dwell upon this point. Let us consider the spectrum of the 
Bethe-Salpeter equation in ladder approximation
within two well-studied  models, the Wick-Cutkosky model
\cite{Wick:1954} and the positronium system \cite{Ahlig:1998,Ahlig:1998qf}.
They have a class of ``normal''
solutions, {\em i.e.} solutions with a non-relativistic 
quantum mechanical analogue,
which exist already for infinitesimally small coupling constants.
On the other side, there are solutions which begin to exist
only for a finite coupling constant, and these are called 
``abnormal'' states. In the Wick-Cutkosky model, they are excitations 
in the relative time between the two constituents and clearly
in a non-relativistic description with no separation in the time 
coordinate between the two constituents corresponding states
do not exist. Furthermore, these abnormal states have partly negative norm
\cite{Nakanishi:1969ph}. Therefore, the interpretation 
of excited states within a given model employing
the Bethe-Salpeter equation for modelling bound states is seriously
hampered. Here, the Salpeter approximation 
allows a transformation of the problem into a Hamiltonian
form with the Hamiltonian being hermitian and thus no negative
norm states can arise\footnote{Abnormal states may still exist, {\em cf.}
ref.~\cite{Bijtebier:1997ir} for an investigation on the persistence of
abnormal states in 3-dimensional reductions of a two-fermion
Bethe-Salpeter equation.}. This transformation is possible as
the interaction kernel in the Bethe-Salpeter equation
is assumed to be independent of the relative time coordinate
or, equivalently, independent of the fourth component of the relative momenta
({\em equal time approximation}). Studies employing
this technique 
in solving for ground state and excited meson masses can be found in 
refs.~\cite{Metsch:1996zx,Tiemeijer:1994bj}. 
 
Appreciating the advantages of the Salpeter approximation 
in the computation of excited states
one could proceed
and calculate dynamical observables for the bound states. 
However, we will show in the following that nucleon form factors,
the pion-nucleon and weak coupling constants
calculated in the diquark-quark model differ substantially for
the full solution and the Salpeter approximation. 
Therefore, results from the latter might lead to a misinterpretation
of the model's parameters.
As there
exist studies for the diquark-quark model using the Salpeter 
approximation \cite{Keiner:1996bu,Keiner:1996at}, we will
employ model parameters as chosen there to furnish the comparison. 
 
\section{Solutions for vertex functions}

We will use the diquark-quark model as defined in chapter
\ref{dq-q-chap}. Following refs.~\cite{Keiner:1996bu,Keiner:1996at},
we employ free quark and diquark propagators, with 
the gauge parameter $\xi$ for the axialvector diquark
chosen to be 1, {\em cf.} eq.~(\ref{Da}),
\begin{eqnarray}
  S(p) &=& \frac{i\Slash{p} -m_q}{p^2+m_q^2}\; , \\
  D(p) &=& -\frac{1}{p^2+m_{sc}^2} \; , \qquad
  D^{\mu\nu}(p) = -\frac{\delta^{\mu\nu}}
   {p^2+m_{ax}^2} \; . 
\end{eqnarray}
The Dirac part of the diquark-quark vertices is given as in
eqs.~(\ref{dqvertex_s},\ref{dqvertex_a}) by
\begin{equation}
 \chi^5=g_s\;(\gamma^5 C)\,V(q^2)\; , \qquad
 \chi^\mu=g_a\;(\gamma^\mu C)\,V(q^2) \; ,
\end{equation}
with the scalar function $V$ chosen to be 
\begin{equation}
  V(q^2)=\exp (-4\lambda^2 q^2)\; .
\end{equation}
The relative momentum between the quarks is denoted by $q$. We 
recapitulate the Bethe-Salpeter equation in short notation,
\begin{eqnarray}
  \Psi(p,P) &=& S(p_q)\,\tilde D(p_d)\;\Phi(p,P) \label{BS10}\\
  \Phi(p,P) &=& \fourint{k} K^{\rm BS}(p,k,P)\;\Psi(k,P) \; .\label{BS11}
\end{eqnarray}
All terms appearing here are defined in section
\ref{nuc-subsec}. Let $q_T=q-\hat P(q\cdt \hat P)$ denote
the transversal part of any momentum $q$ with respect to the bound
state momentum $P$.
The Salpeter approximation consists
in reducing the momentum dependence of the quark exchange kernel
in the following manner,
\begin{equation}
 K^{\rm BS}(p,k,P) \rightarrow K^{\rm BS}(p_T,k_T) \; .
\end{equation}
In the rest frame of the bound state,
$K^{\rm BS}(p_T,k_T) \equiv K^{\rm BS}(\vect p, \vect k)$, and
hence the (other) names {\em equal time} or {\em instantaneous} 
approximation. It follows immediately from eq.~(\ref{BS11})
that $\Phi(p,P)\equiv \Phi(\vect p)$. 
Although covariance is lost by assuming instantaneous interactions,
the solutions for $\Phi$ are invariant against shifts in the
parameter $\eta$ which defines the distribution of momentum
between quark and diquark.

Again following ref.~\cite{Keiner:1996bu}, the reduced Bethe-Salpeter 
equation can be processed by introducing 
\begin{equation}
  \tilde \Psi(\vect p) = \gamma^4\int \frac{dp^4}{2\pi} \Psi(p,P) \; ,
\end{equation}
integrating eq.~(\ref{BS10}) over $p^4$ likewise, and thus arriving
at an equation with the formal structure,
\begin{equation}
 H \tilde \Psi = M \tilde \Psi\; .
\end{equation}
The ``Hamiltonian'' $H$ is an integral operator which
contains ``kinetic'' parts stemming from the propagators and
a ``potential'', being the instantaneous quark exchange.

Instead of solving this Schr\"odinger-type equation, we will
employ the numerical method developed in section \ref{num-meth}.
Having expanded the wave function $\Psi$ and the vertex function
$\Phi$ according to eqs.~(\ref{vex_N},\ref{wex_N}), we solve
the Bethe-Salpeter equation for the sets of scalar functions
$\{S_i,A_i\}$ and $\{\hat S_i,\hat A_i\}$. As described in section
\ref{num-meth}, each of the scalar functions depends on $p^2$
and $z=\hat p \cdt \hat P$. The dependence on $z$ will be 
absorbed by the Chebyshev expansion of 
eqs.~(\ref{cheby-v},\ref{cheby-w}). Note that in the Salpeter 
approximation the
scalar functions $S_i,A_i$ pertaining to the vertex function
depend only on the single variable $\vect p^2=p^2(1-z^2)$,
thus their Chebyshev expansions contains just even moments.

As stated above we adopt the parameters of 
refs.~\cite{Keiner:1996bu,Keiner:1996at}. 
The first study investigates one parameter set in the scalar diquark 
sector only and the second
one includes the axialvector diquark channel using another parameter set.
In these studies, the author did not
fix the diquark-quark coupling constants by a normalization condition
like in eqs.~(\ref{normsc},\ref{normax}), but rather 
adjusted $g_s$ so as to obtain the physical nucleon mass
for given values of quark and diquark mass, and of $g_a/g_s$
when including the axialvector diquarks.
This procedure is convenient as the Bethe-Salpeter equation
is an eigenvalue problem for the eigenvalue $g_s$.
In table \ref{par-sal} we have listed the two sets of parameters
with their corresponding 
eigenvalues $g_s$ obtained by us in the full calculation and the Salpeter 
approximation. The values in parentheses are the ones from 
refs.~\cite{Keiner:1996bu,Keiner:1996at}. 
Please note that due to a different
flavor normalization these values had to be multiplied by $\sqrt{2}$ to be
directly comparable to ours.

\begin{table}
\begin{center}
\begin{tabular}{clcccccl} \hline \hline \\ 
Set & & $m_q$& $m_{sc}$& $m_{ax}$ & $\lambda$&  $g_a/g_s$ &$g_s$ \\
   & & ~[GeV]~ & ~[GeV]~ & ~[GeV]~ & ~[fm]~ && \\ \hline
Ia & Salpeter~ & 0.35 & 0.65 & - & 0.18 & -&20.0 ~~(20.0) \\
  & full     & & & & & &16.5 \\ \hline
IIa & Salpeter~ & 0.35 & 0.65 & 0.65 & 0.24 &1& 12.3 ~~(11.5) \\
  & full     & & & & && \phantom{1}9.6 \\ &&&&&&& \\ \hline \hline
\end{tabular}
\end{center}
\caption{The two parameter sets of the model as taken from
   refs.~\cite{Keiner:1996bu,Keiner:1996at}.}
\label{par-sal}
\end{table} 

Although we could reproduce the eigenvalue for the model case with scalar
diquarks only, this is not the case for Set IIa. We observe that the
calculations of \cite{Keiner:1996at} 
involved only 4 instead of 6 axialvector
components of $\Phi^\mu$, namely the projected ones onto
zero orbital angular momentum and the corresponding lower components.
Still one would expect a higher eigenvalue in the reduced system.
The more striking observation is the amplification of the eigenvalue by 
about 20\dots25 \% in the Salpeter approximation although the binding energy
is small, being only 6\% of the sum of the constituent masses.
This is in contrast to results obtained in the massive Wick-Cutkosky model
\cite{Nieuwenhuis:1996mc} where the Salpeter approximation 
leads to a {\em reduction} of the 
eigenvalue. This may be attributed to the exchange of a boson instead of
a fermion as here in the diquark-quark model.

The substantial difference between the two approaches is also reflected
in the vertex function solutions themselves. 
Figure \ref{s1-sal} shows the Chebyshev moments
of the dominant scalar function $S_1$ for both methods using the parameters
of Set Ia. Only the even momenta are given, since the odd ones are zero in the
Salpeter approximation (but are present, of course, in the full calculation).
Two things are manifest: the Salpeter amplitudes have a much broader spatial
extent than the full amplitudes. Secondly, the expansion in Chebyshev 
polynomials that relies on an approximate $O(4)$ symmetry converges much more
rapidly for the full solution but is hardly convincing in the Salpeter
approximation. Again, since the scalar functions depend in the Salpeter
approximation on just one variable, $p^2(1-z^2)$, our expansion is a 
cumbersome way of visualizing the solution but makes clear that the 
Salpeter approximation does not exhibit an approximate $O(4)$-symmetry.

We remind ourselves that Salpeter in his original work \cite{Salpeter:1952}
devised the instantaneous approximation of the Bethe-Salpeter
kernel as the first term in a perturbation series for the full solution.
We clearly see that one should not stop after the first term 
as the character of the exact and the approximated solution differ so
profoundly.

\begin{figure}
 \begin{center}
  \epsfig{file=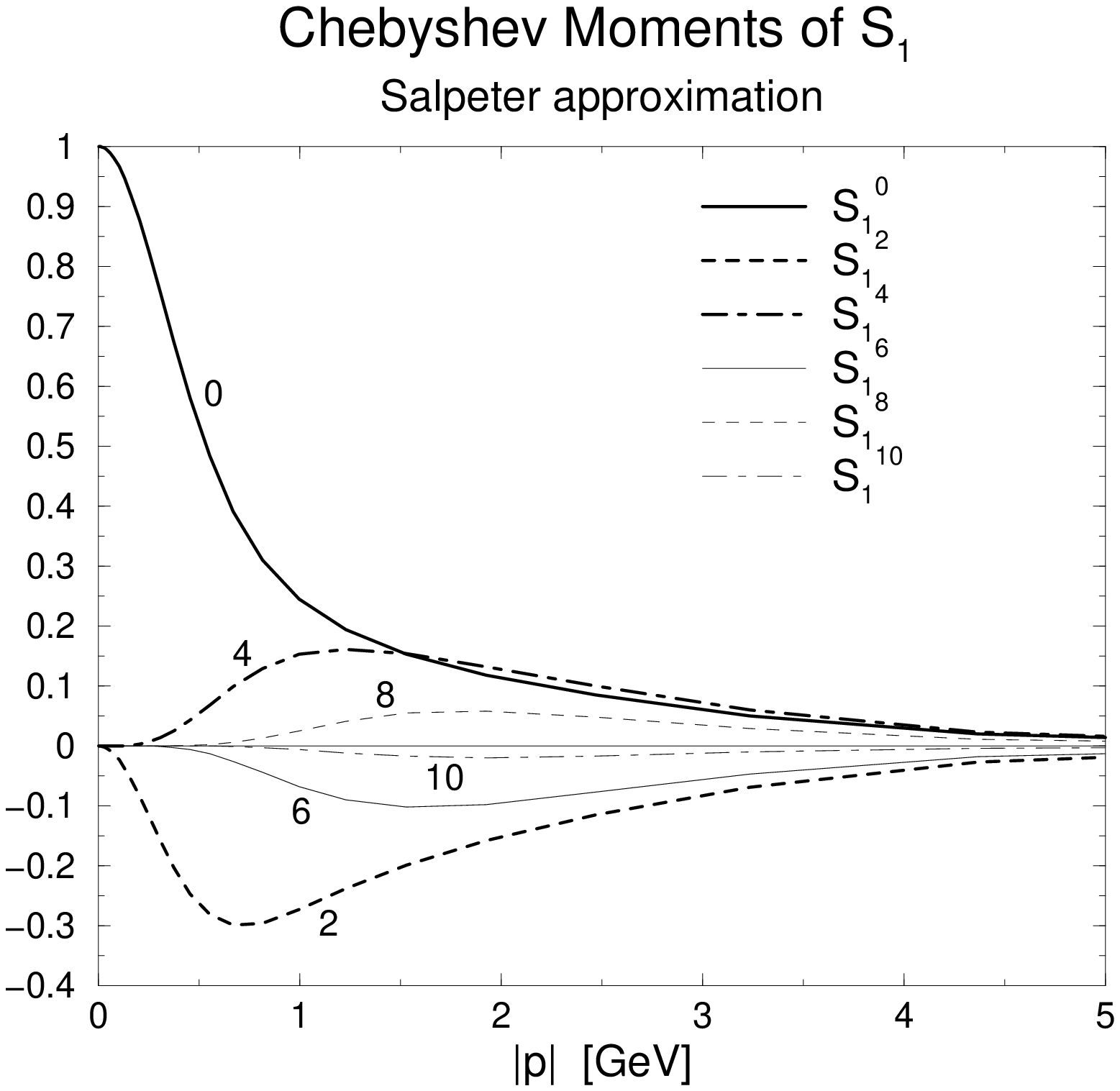,width=\figurewidth}
  \epsfig{file=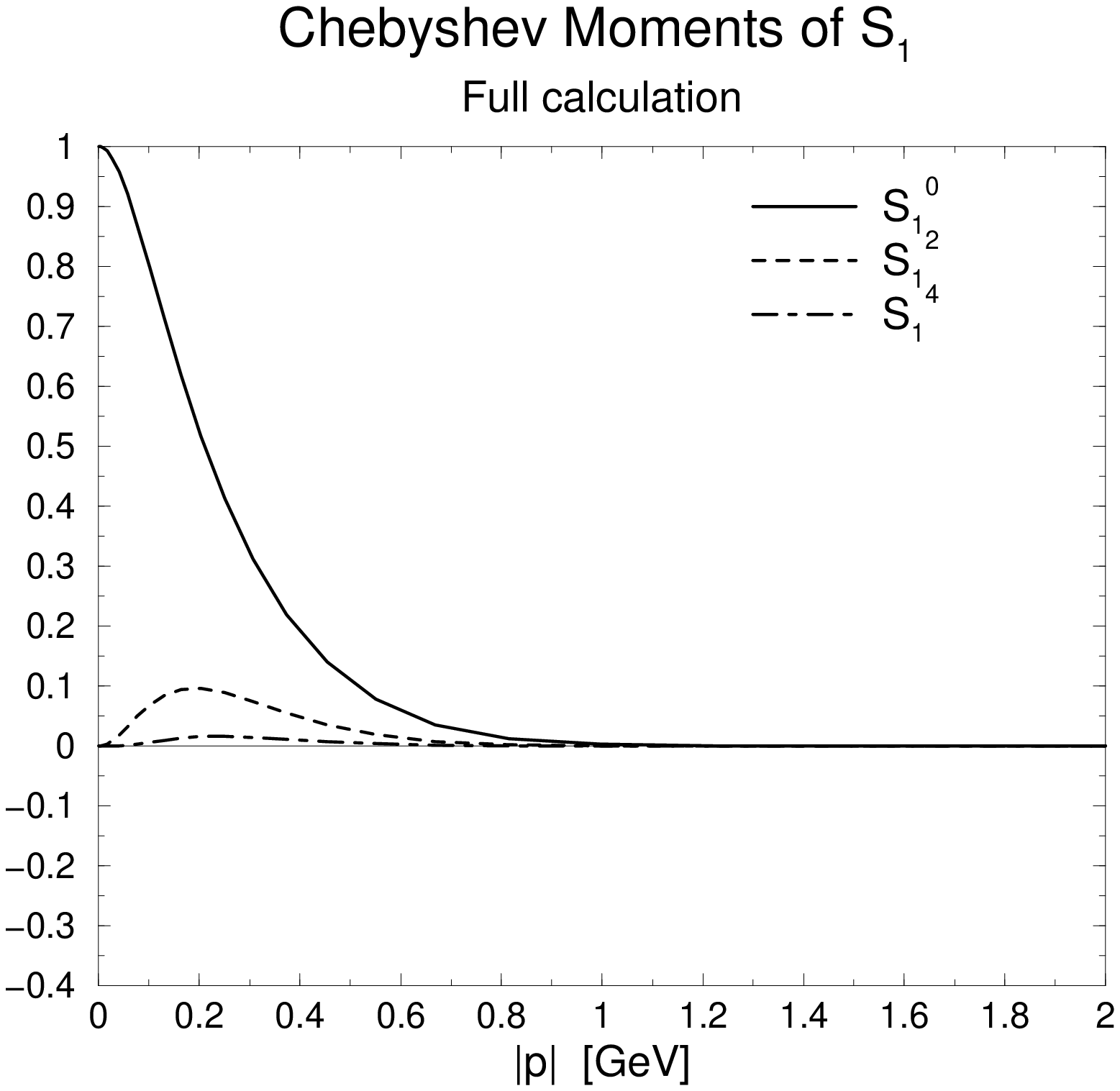,width=\figurewidth}
 \end{center} 
 \caption{The Chebyshev expansion of the dominant scalar function $S_1(p,z)$
  in the Salpeter approximation and the full calculation.}
 \label{s1-sal}
\end{figure}

\section{Results for observables}

Let us turn first to the calculation of the electromagnetic 
form factors $G_E$ and $G_M$. They can be extracted from
the current matrix element as prescribed by 
eqs.~(\ref{getrace},\ref{gmtrace}). In impulse approximation 
the current operator
for proton and neutron reads 
\begin{eqnarray}
    _{\rm p}\langle P_f|\; J^\mu\;|P_i \rangle^{\rm imp}_{\rm p} &=&
   \frac{2}{3} \langle J^\mu_q \rangle^{\rm sc-sc} +
   \frac{1}{3} \langle J^\mu_{sc} \rangle^{\rm sc-sc} +
   \langle J^\mu_{ax} \rangle^{\rm ax-ax} \; , \label{jimp-sal}\\
  _{\rm n}\langle P_f|\; J^\mu \;|P_i \rangle^{\rm imp}_{\rm n} &=&
   -\frac{1}{3}\left( \langle J^\mu_q \rangle^{\rm sc-sc} -
    \langle J^\mu_q \rangle^{\rm ax-ax} -
    \langle J^\mu_{sc} \rangle^{\rm sc-sc} + 
    \langle J^\mu_{ax} \rangle^{\rm ax-ax} \right) \; . \nonumber \\
    \label{jimn-sal}
\end{eqnarray}
Here we adopted the notation from section \ref{curop-sec}. 
Diagrammatically the single matrix elements correspond to the
left two graphs of figure \ref{impulse}. The quark and diquark vertices
which are needed in these diagrams are given by 
eqs.~(\ref{vertq}--\ref{vertax}).
These diagrams are the only ones which have been calculated in 
   refs.~\cite{Keiner:1996bu,Keiner:1996at}, with the following
justification given:
as already mentioned in the summary of section \ref{curop-sec},
the diagrams of the impulse approximation separately conserve
the current. Furthermore, peculiar identities at zero momentum
transfer hold in the Salpeter approximation,
\begin{eqnarray}
   \langle J^4_q(Q^2=0) \rangle^{\rm sc-sc}&=&
   \langle J^4_{sc}(Q^2=0) \rangle^{\rm sc-sc} \; \\
   \langle J^4_q(Q^2=0) \rangle^{\rm ax-ax} &=&
   \langle J^4_{ax}(Q^2=0) \rangle^{\rm ax-ax} \; ,
\end{eqnarray}
which are proved by straightforward calculation. These identities,
in conjunction with eqs.~(\ref{jimp-sal},\ref{jimn-sal})
and the normalization condition (\ref{nucnorm})
for the nucleon wave function, guarantee that proton 
and neutron have their correct charge, {\em i.e.} the
electric form factors satisfy
$G_{E}^{\rm proton} (Q^2=0)=1$ and  $G_{E}^{\rm neutron} (Q^2=0)=0$. 

We have seen that in the full calculation we need the current 
contributions from the quark exchange kernel to establish
the correct nucleon charges. This followed as a consequence
of the general Ward-Takahashi identity, eq.~(\ref{WTG}),
for the quark-diquark propagator. We will therefore compare
the impulse approximation for the Salpeter calculation
to the calculations with full vertex functions
with and without the exchange diagrams of figure
\ref{7dim}.

\begin{figure}[t]
 \begin{center}
  \epsfig{file=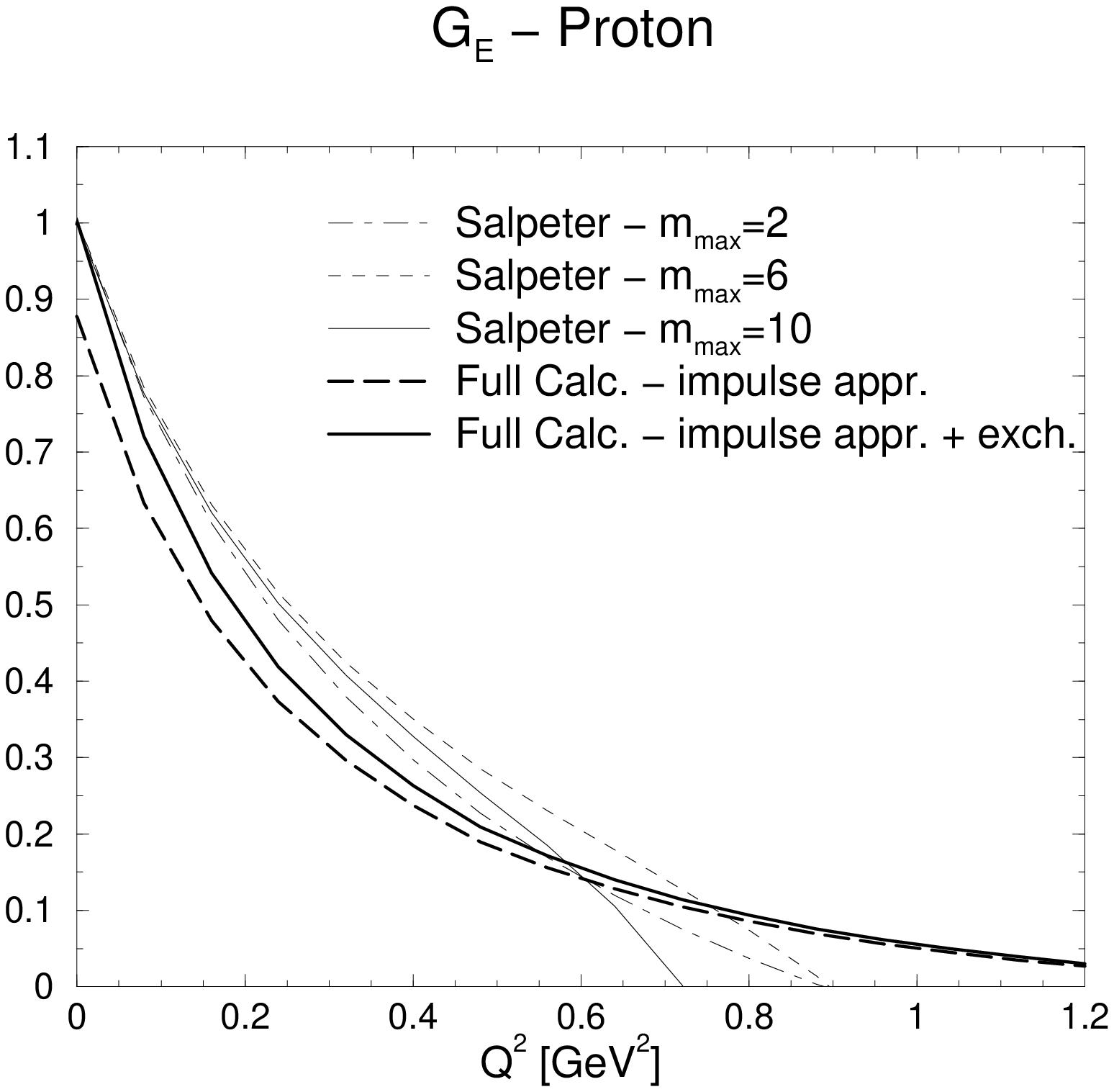,width=\figurewidth}
  \epsfig{file=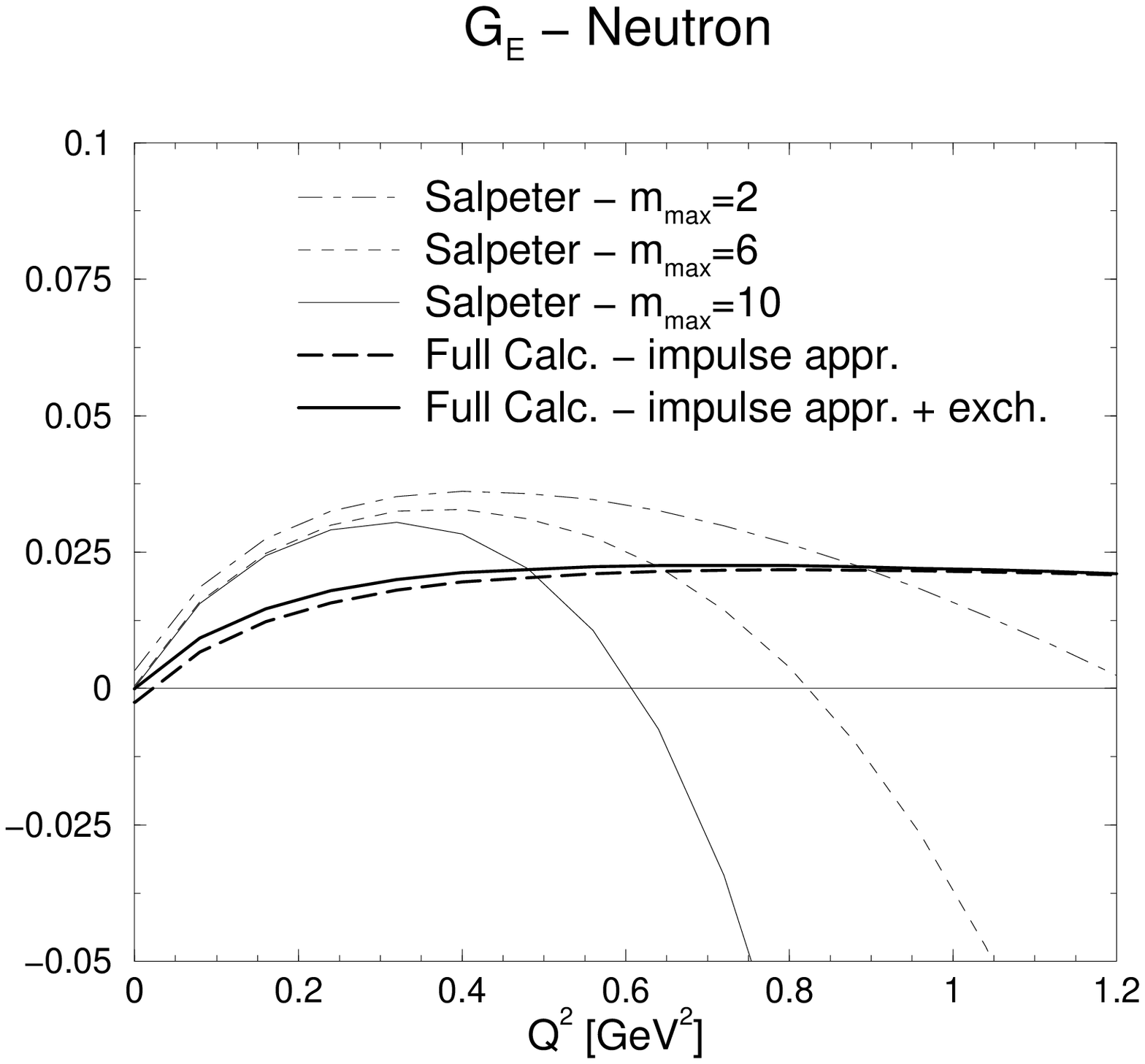,width=\figurewidth}
 \end{center} 
 \caption{Nucleon electric form factors in the Salpeter approximation
and in the full calculation. The curves for the Salpeter approximation
have been obtained by including the Chebyshev moments of the vertex function
up to order $m_{\rm max}$.} 
 \label{ge5-fig}
\end{figure}  

In figure \ref{ge5-fig} the electric form factors of proton and neutron are
displayed, using the parameters of Set IIa.  
The first observation is that in the
Salpeter approximation we could not obtain convergence with the expansion in
Chebyshev polynomials beyond $Q^2 \approx$ 0.4 GeV$^2$. 
As described in section \ref{num_em}, we computed the form
factors in the Breit frame where $Q$ is real but 
$z_i=\hat p \cdot \hat P_i$
and  $z_f=\hat k \cdot \hat P_f$, the angular variables
which enter as their arguments the Chebyshev polynomials 
for the boosted vertex functions, 
 have imaginary parts.  Their absolute
values may exceed one, except for the case of zero momentum transfer,
{\em cf.} eqs.~(\ref{zi},\ref{zf}).
So this expansion that works in the rest frame, and it does
barely so for the Salpeter approximation,  will not generally work in 
a moving frame.  On the other hand, the decrease  of the higher 
Chebyshev moments is good enough
for the full four-dimensional solution that the numerical values
for the form factors 
converge up to several GeV$^2$. However, as can be
seen from fig. \ref{ge5-fig} the Salpeter approximation badly fails above 0.5
GeV$^2$ thereby revealing its semi-relativistic nature.

The second finding concerns the electromagnetic radii. The
Salpeter approximation tends to underestimate the proton charge radius
and to overestimate the absolute value of the neutron charge radius,
see the first two result lines of table \ref{static-sal}.
The axialvector correlations tend to suppress the neutron electric form factor
much more  in the full calculation than in the Salpeter approximation.
Remember that this suppression is due to the axialvector
correlations which have been chosen to be rather strong for
Set IIa, as $g_s=g_a$.

\begin{table}
\begin{center}
\begin{tabular}{clllllll}\hline\hline \\ 
    & &\multicolumn{4}{c}{--- $\mu_p$ ---} \\&&&&& \\ 
  Set &  & $\langle J^\mu_q \rangle^{\rm sc-sc}$ & 
           $\langle J^\mu_{ax} \rangle^{\rm ax-ax}$ &
       $\langle J^\mu_{ex} \rangle$ & SUM \\ \hline
 Ia & Salpeter~~~~~   & 1.57  & -    & - & 1.58 (1.58)\\
   & full       & 2.04  & -    & 0.31& 2.38 \\
  IIa & Salpeter & 0.92 (1.08)  & 1.37 (1.7) & - & 2.29 (2.78) \\
     & full     & 1.09  & 0.98 & 0.37~~~~ & 2.45 \\ &&&&& \\
  \hline\hline 
\end{tabular}
\end{center}
\caption{The most important contributions to the proton magnetic moment 
from the single diagrams. The notation is as in table
\ref{magtab}.  In parentheses
the values of refs.~\cite{Keiner:1996bu,Keiner:1996at} are given.}
\label{magmom-sal}
\end{table} 

Turning to the magnetic moments, the contributions of the various diagrams
are tabulated in table \ref{magmom-sal} for the proton. 
Following ref.~\cite{Keiner:1996at} we ascribed to the
axialvector diquark a rather large anomalous magnetic moment of 
$\kappa=1.6$
which was needed in that study to fit the proton magnetic moment.
As already observed earlier, we could reproduce the magnetic moment
for Set Ia, however, for Set IIa, the values differ and especially the
coupling to the axialvector diquark is much weaker in our Salpeter
calculation. Rather more interesting is the comparison between  the full
calculations  using either Set Ia or Set IIa: The axialvector diquark
improves  the magnetic moment only marginally, due
to the neglect of the anomalous transitions between
scalar and axialvector diquarks. The Salpeter approximation 
tends to overestimate the contribution of 
$\langle J^\mu_{ax} \rangle^{\rm ax-ax}$ quite drastically.

\begin{table}[t]
 \begin{center}
  \begin{tabular}{llrrrcrrr} \hline\hline \\
 & Set & \multicolumn{3}{c}{Ia} && \multicolumn{3}{c}{IIa} \\
 &     & \multicolumn{1}{c}{Salpeter} & \multicolumn{2}{c}{full} 
       && \multicolumn{1}{c}{Salpeter} & \multicolumn{2}{c}{full} \\
 &     & IA\phantom{o} & IA\phantom{o} & all\phantom{o} &&  IA\phantom{o} & IA\phantom{o} & all\phantom{o} \\
 \\ \hline
   $(r_p)_{\rm el}$& [fm] & 0.89 & 1.00 & 0.99 && 0.88 & 1.01 & 1.01\\
   $(r^2_n)_{\rm el}$& [fm$^2$] &
       $-$0.28 & $-$0.21 & $-$0.23 && $-$0.06 & $-$0.04 & $-$0.04 \\
   $ \mu_p$ & & 1.58 & 2.05 & 2.58 && 2.29 & 2.07 & 2.45    \\
   $(r_p)_{\rm mag}$ & [fm] & 1.04 & 1.09 & 1.06 && 0.85 & 0.99 &0.99\\
   $\mu_n$ && $-$0.77 & $-$1.02 & $-$1.65 && $-$0.98 & $-$1.02 & $-$1.26 \\
   $(r_n)_{\rm mag}$ & [fm]&1.05&1.09&1.00&&0.94&1.06&1.05 \\ \hline
   $g_{\pi NN}$ &&8.96&11.71&15.34&&6.03&6.95&9.36 \\
   $r_{\pi NN}$ & [fm]&1.04&1.07&1.04&&1.15&1.21&1.15 \\
   $g_A$ &&0.93&1.16&1.46&&0.53&0.65&0.82\\
   $r_A$ & [fm]&0.93&0.99&1.02&&1.08&1.14&1.10 \\ 
   \\ \hline \hline 
  \end{tabular}
 \end{center}
 \caption{Some static nucleon observables. The acronym `IA'
  indicates that the results in these columns have been obtained
  in the impulse approximation whereas `all' means
  that also the diagrams of the quark exchange kernel have been
  taken into account.}
 \label{static-sal}
\end{table}

We have also calculated the pion-nucleon constant, the weak coupling 
constant and the corresponding radii
for the two sets in a first approximation, taking into account only
the diagrams where the pseudoscalar and pseudovector current
couple to the spectator or the exchanged quark. 
Please refer to table \ref{static-sal} for the numerical results.
Of course, the numbers
for $g_A$ and $g_{\pi NN}$ obtained for the parameters of Set IIa 
signal the missing contributions from the resolved axialvector diquark and 
from the scalar-to-axialvector transitions. But we want to emphasize
another point: Already in impulse approximation
the results vary strongly (on the level of 20 \%) between
the full and the Salpeter calculation with the Salpeter approximation
{\em underestimating} the couplings. But as we have seen for the magnetic 
moments for Set IIa, the Salpeter approximation {\em overestimates} 
them, so we arrive at the conclusion that observables
vary quite strongly and unpredictably when employing the
equal time approximation. In this case, it is hardly of use to relate 
parameters to the physical content which the model hopefully
describes.

Let us summarize the results obtained in this chapter.
The Bethe-Salpeter equation for the nucleon has been solved in a fully
covariant way and in the instantaneous Salpeter approximation. 
As for the model with scalar
diquarks only we have verified the results of 
ref.~\cite{Keiner:1996bu} whereas 
discrepancies remain if the axialvector diquark is included. These
differences are partly due to the fact that in 
ref.~\cite{Keiner:1996at} not all 
(ground state) axialvector components have been taken into account.
Additionally, we take our result as an indication that the calculations
presented in ref.~\cite{Keiner:1996at} might suffer from some minor error.

However, the main purpose has been  the comparison of observables calculated in
the Salpeter approximation to the ones obtained in the fully four-dimensional
scheme. The first very surprising observation is the overestimation of the Bethe-Salpeter 
eigenvalue in the Salpeter approximation. Phrased otherwise, for a given
coupling constant the binding energy would be much too small in the Salpeter 
approximation.  We have also demonstrated that the 
character of the vertex function solutions in the full treatment and for the
Salpeter approximation is grossly different, since the 
full solutions exhibit an approximate $O$(4) symmetry which is absent
in the Salpeter approximation.  This has
drastic consequences for the resulting nucleon electromagnetic form factors if
the  photon virtuality exceeds 0.4 GeV$^2$.
Although the non-convergence  of the form factors in the Salpeter
approximation in this momentum regime can be solely ascribed to the
Chebyshev expansion which is inappropriate for the instantaneous 
approximation, we find nevertheless differences for static observables.
On the one hand, different nucleon radii
differ  only mildly in these two approaches, with
the exception of the neutron charge radius. On the other hand,
 one sees very clearly that the
results (obtained in the Salpeter approximation) 
for the magnetic moments, the pion-nucleon coupling and the weak
coupling constant deviate sizeably from the full calculation.

 \chapter{Summary and Conclusions}
 \label{con-chap}
  
In this thesis we have presented a framework for
a fully covariant diquark-quark model of baryons and 
obtained solutions for bound states and certain
nucleon observables 
having employed  simple {\em ans\"atze} for the quark
and diquark propagators, and the diquark-quark vertices.
We derived the basic Bethe-Salpeter equations governing the dynamics
of the quarks and the diquark quasi-particles by a suitable 
reduction of the
relativistic three-quark problem.
This reduction consists in neglecting three-quark irreducible
interactions and assuming separable two-quark correlations, the diquarks.
Scalar and axialvector diquarks have been included into the
description as the presumably most important diquark correlations.

We solved the Bethe-Salpeter equation which sums up the
continuous quark exchange between quark and diquark in a fully
covariant manner, thereby obtaining relativistic baryon wave
functions which contain far more structure than non-relativistic
wave functions. These wave functions have been subsequently employed
to calculate electromagnetic form factors. Here, we constructed
the current operator for the electromagnetic field using
Ward-Takahashi identities and thereby maintaining gauge invariance.
Covariance and gauge invariance are manifest in the numerical results.
For the nucleon electric form factors we found good
agreement with the experimental data. The magnetic moments fall short
by approximately 15 \dots 30 \%, 
depending on whether the $\Delta$ resonance is included
in the description or not. 
A possible source of this failure  is found in the  
free constituent quark and diquark propagators  which have
been employed in these calculations. The lack of confinement forced
us to choose rather large quark and diquark masses to obtain
a bound $\Delta$, and the perturbative quark-photon vertex is probably 
insufficient to account for the observed magnetic moments.
The recently measured ratio of electric to magnetic form factor
of the proton restricts the strength of the axialvector correlations
within the nucleon to be small, on the level of 20 \%, if our results
are to match the data.

In an attempt to circumvent the problem posed by the inclusion of
the $\Delta$ resonance and the underestimated magnetic moments,
we have investigated a  possible parametrization of confinement
by modifying quark and diquark propagators to render them pole-free
in the complex momentum plane. These  modifications
allowed us to calculate the mass spectrum of octet and decuplet 
baryons in fair agreement with experiment and the nucleon magnetic moments
received additional contributions which even rendered them
to be larger by 15 \% than the experimental values. Nevertheless,
 there are serious
doubts about the justification of such a modification.
The modified propagators have an essential singularity
at timelike infinity which is not in accordance with general theorems
about the behavior of propagators in quantum field theory.
Additionally, this singularity causes a spurious enhancement
of the calculated cross section in kaon photoproduction, and
would render the calculation of the leading twist contributions to 
Deep Inelastic
Scattering impossible. Therefore, refined methods to incorporate
confinement are needed and work on this problem is currently done.

The isovector axial form factor for the nucleon and the
pion-nucleon form factor have also been calculated. 
Chiral symmetry could be  used to construct the 
pseudovector and the pseudoscalar quark current operators.
As it turns out, chiral symmetry requires also 
the inclusion of vector diquarks
besides the axialvector diquarks, since they form a chiral multiplet.
The neglect of the former leads
to a moderate violation of the Goldberger-Treiman relation, and
the pion-nucleon and the weak coupling constant 
are found to be somewhat too large compared to their experimental values.
Since the transitions between scalar and axialvector diquarks
proved to be the main source of the calculated Goldberger-Treiman
discrepancy and the large pion-nucleon coupling constant, we
again conclude that axialvector correlations within the nucleon
should be small. 

We also compared results for nucleon vertex function solutions 
and observables
between the fully covariant treatment of the diquark-quark 
Bethe-Salpeter equation and the semi-relativistic Salpeter
approximation. The vertex functions for the latter differed 
strongly from the
full solution, on a qualitative as well as on a quantitative
level. Furthermore, we found  rather erratic deviations
between the results for static observables obtained with both
methods and therefore conclude that the semi-relativistic
treatment is inadmissible within the diquark-quark model. 

Regarding the simplicity of the quark-diquark picture,
the model reproduces fairly well the 
basic nucleon observables investigated here. It should be considered
a success that maintaining covariance is feasible in calculating
the form factors and this is a good starting point to investigate
more involved hadronic processes. For kaon photoproduction,
$\gamma p \rightarrow K^+ \Lambda$, and associated
strangeness production, $pp\rightarrow pK^+\Lambda$, preliminary
results are already available, {\em cf.} 
refs.~\cite{Alkofer:1999rn,Fischer:1999}. 

To check the viability of the chosen approach within QCD,
some more information about the quark propagator and the
two-quark correlations are needed. Whereas there is justified hope
that lattice or Dyson-Schwinger calculations will provide us with
more refined results for the quark propagator in the near future,
it is not clear at all whether a non-perturbative characterization
of the two-quark correlations can be found. To proceed further in
this direction,
one probably has to resort in a next step to an approximate model
for the quark-gluon dynamics.

 \begin{appendix}
  \chapter{Varia}
   \section{Conventions}
\label{conv-app}

Throughout the text we work in Euclidean space with
the metric $g^{\mu\nu}=\delta^{\mu\nu}$. Accordingly, a hermitian
basis of Dirac matrices $\{\vect \gamma, \gamma^4\}$ is used which is 
related to the matrices $\{\gamma^0_D,\vect \gamma_D\}$ of
the Dirac representation by
\begin{equation}
  \gamma^4=\gamma^0_D\; , \quad \vect \gamma =-i \vect \gamma_D\; .
\end{equation}
 Furthermore we define $\gamma^5=-\gamma^1\gamma^2\gamma^3\gamma^4$.
The charge conjugation matrix is defined by $C=-\gamma^2 \gamma^4$.
The free quark propagator $S(k)$ is determined by the
relation, valid in the theory for a free, massive fermion,
\begin{eqnarray}
  \langle 0|T q(x)\bar q(y) | 0 \rangle &=&\fourint{k}
    \exp(ik\cdt (x-y)) S(k) \; , \\
  S(k) &=& \frac{i\Slash{k}-m_q}{k^2+m_q^2} \; .
\end{eqnarray}
We arrive at this equation, if we formulate 
the continuation between Minkowski and Euclidean space using the
well-known rules
\begin{eqnarray}
  x^0_M \rightarrow -ix^4\; , \quad \vect x_M \rightarrow \vect x\; , \\
  k^0_M \rightarrow ik^4 \; , \quad \vect k_M \rightarrow -\vect k\; 
\end{eqnarray}
for a formal transcription of position variables $x_M$ and
momentum variables $k_M$ in Minkowski space to their Euclidean
pendants $x$ and $k$. Additionally we prescribe for the Euclidean
momentum variables $k \rightarrow -k$. Furnished with an additional
transcription rule for the metric tensor in Minkowski space,
\begin{equation}
 g^{\mu\nu} \rightarrow -\delta^{\mu\nu}\; , 
\end{equation}
we find for the scalar and the Proca propagator,
\begin{equation}
D(p) = -\frac{1}{p^2+m_{sc}^2}\; , \quad
D^{\mu\nu}(p) = -\frac{\delta^{\mu\nu}+  \frac{p^\mu p^\nu}{m_{ax}^2}}
   {p^2+m_{ax}^2}  
 \; .
\end{equation}

Four-dimensional momentum integrals, appearing in the Bethe-Salpeter
equation or in the expression for current matrix elements, are
evaluated by using hyperspherical coordinates. We parametrize
the Cartesian components of an Euclidean vector $q^\mu$ in 
the following way,
\begin{equation}
 q^\mu=|q| \bp \sin\phi \sin\theta \sin\Theta \\
               \cos\phi \sin\theta \sin\Theta \\
               \cos\theta \sin\Theta \\
               \cos\Theta \ep \; . 
\end{equation}
With the abbreviations $z=\cos\Theta$ and $y=\cos\theta$ the
four-dimensional volume integral measure is
\begin{equation}
 \fourint{q}=\frac{1}{(2\pi)^4}\,\int_0^\infty |q|^3d|q|
  \int_{-1}^1\sqrt{1-z^2}\,dz \int_{-1}^1 dy \int_0^{2\pi}d\phi \; .
\end{equation}
Hyperspherical harmonics are defined by
\begin{equation}
  {\cal Y}_{nlm}(\Theta,\theta,\phi)=\sqrt{\frac{2^{2l+1}}{\pi}\frac
   {(n+1)(n-l)!(l!)^2}{(n+l+1)!}} \sin^l\Theta\, C^{1+l}_{n-l}(\cos\Theta)
   Y_{lm}(\theta,\phi),
\end{equation}
where the polynomials $C^{1+l}_{n-l}(\cos\Theta)$ are
the {\em Gegenbauer polynomials} \cite{Abramowitz:1965} and 
the $Y_{lm}(\theta,\phi)$ are the familiar spherical harmonics.
For vanishing three-dimensional angular momentum, $l=0$,
the Gegenbauer polynomials reduce to the Chebyshev polynomials
of the second kind, $C^{1}_{n}=U_n$, and the corresponding hyperspherical
harmonics are simply
\begin{equation}
  {\cal Y}_{n00}= \sqrt{ \frac{1}{2\pi^2}}\,U_n(\cos\Theta) \; .
\end{equation}
Thus, the expansion of relativistic ground state wave functions ($l=0$)
in terms of Chebyshev polynomials, as done in section \ref{num-sol} 
for the baryons, corresponds to an expansion into hyperspherical harmonics.

\section{Color and flavor factors in the 
Bethe-Salpeter equations}

\subsection{Nucleon}
\label{cfn}

For the derivation of the color factor,
let us introduce the color singlet state
of the nucleon as being proportional to a 3-dimensional unit matrix
$(\bf 3 \otimes \bar{ \bf 3} = \bf 1 \oplus 8)$
\begin{equation}
  |0\rangle_{\rm color} = \sqrt{\third} (\lambda^0)_{AB} :=
    \sqrt{\third} \delta_{AB},
  \qquad \langle 0| 0 \rangle =1 \; .
\end{equation}
Since both diquarks are in an antitriplet representation,
the color part of the Bethe-Salpeter kernel reads (see figure \ref{bsekern})
\begin{equation}
 K^{\rm BS}_{\rm color} = \half\: \epsilon_{LKA}\:\epsilon_{JNA}=-
  \half(\delta_{LN}\delta_{KJ}-\delta_{LJ}\delta_{KN}) \; .
\end{equation}
Sandwiching between the nucleon color singlet state yields 
\begin{equation}
 \langle 0| K^{\rm BS}_{\rm color} | 0 \rangle =-
   {\textstyle\frac{1}{6}}\delta_{JK}(\delta_{LN}\delta_{KJ}-
        \delta_{LJ}\delta_{KN})\delta_{LN} =-1 \; .
 \label{co}
\end{equation}

The nucleon flavor state distinguishes of course between        
scalar and axialvector diquarks. 
We rewrite the diquarks as eigenstates of their total isospin and
the projection onto the respective $z$ axis. They are easily expressed by the
combinations of $u$ and $d$ quark. $(ud)=-\sqrt{\half}(ud-du)$ 
has isospin zero (scalar diquark). $[uu]$, 
$[ud]=\sqrt{\half}(ud+du)$ and $[dd]$ have isospin one with the
respective third ($z$) component being $+1,0,-1$. 
Expressed with spherical isospin matrices $\tau_m$,
$\tau_{\pm 1}= \mp \sqrt{\half}(\tau^1 \pm i\tau^2)$ and
$\tau_0=\tau^3$, they read
\begin{eqnarray}
  (ud) & \equiv & -\frac{1}{\sqrt{2}}\:i \tau^2 \\
 {\bp {[uu]} \\ {[ud]} \\ {[dd]} \ep} & \equiv &  \frac{1}{\sqrt{2}}
 {\bp i \tau_1 \tau^2 \\ i \tau_0 \tau^2 \\ i \tau_{-1} \tau^2 \ep}
\end{eqnarray}
Remember the expression for the nucleon Faddeev amplitude,
\begin{equation}
 \trisp{\Gamma} = \twosp{\chi^5}\; D \; \left( \Phi^5
   u\right)_\alpha +
                  \twosp{\chi^\mu} \; D^{\mu\nu} \;
    \left( \Phi^\nu u \right)_\alpha \; .
\end{equation}
The two summands on the right hand side represent scalar and axialvector
correlations, respectively, and the Greek indices contain also the 
flavor indices. We will write the nucleon flavor state as a 2-vector,
with scalar and axialvector correlations as components, and keep
in mind that they add up to the Faddeev amplitude as indicated above.
Then the proton
flavor wave function can be constructed with the appropriate
Clebsch-Gordan coefficients as
\begin{eqnarray}
 | p \rangle_{\rm flavor} &=&
 \bp | p \rangle_{\rm sc}^{abc} \\ | p \rangle_{\rm ax}^{abc} \ep =
 \bp u(ud) \\ \sqrt{\third}u[ud] -\sqrt{\frac{2}{3}}d[uu] \ep \nonumber \\ 
  &=& \bp \sqrt{\half} \,{\T {1\choose 0}}_a (i\tau^2)_{bc}  \cr
     \sqrt{\frac{1}{6}}\, {\T {1\choose 0}}_a  (\tau^1)_{bc} -
      \sqrt{\third}\,{\T {0\choose 1}}_a (i\tau_{-1}\tau^2)_{bc} \ep \; .
  \label{pce}
\end{eqnarray}
The scalar and axialvector flavor components are separately normalized
to yield unity,
\begin{equation}
  _{\rm sc[ax]}^{acb}\langle p | p \rangle^{abc}_{\rm sc[ax]} 
  \stackrel{!}{=} 1 \; ,
\end{equation}
where the conjugate flavor states are obtained by Hermitian
conjugation. The absolute strength of the scalar and axialvector 
correlations is determined by the solution to the Bethe-Salpeter
equation. A redefinition of the diquark flavor matrices
appearing in eq.~(\ref{pce}) also affects the diquark normalization
conditions, eqs.~(\ref{normsc},\ref{normax}), and the 
Bethe-Salpeter kernel such that the contributions
of the scalar and axialvector correlations to observables remain
unchanged.
The neutron flavor state accordingly writes
\begin{eqnarray}
 | n \rangle_{\rm flavor} &=&
 \bp d(ud) \\ -\sqrt{\third}d[ud] +\sqrt{\frac{2}{3}}u[dd] \ep 
  \label{nce}
\end{eqnarray}
\begin{figure}
 \begin{center}
   \epsfig{file=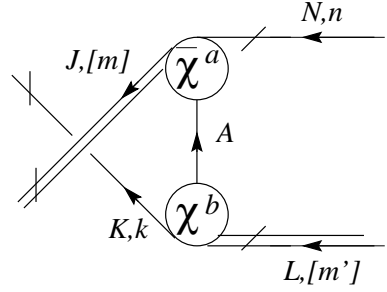,width=5cm}
 \end{center}
 \caption{Color and flavor indices appearing in the Bethe-Salpeter kernel.
  Capital letters run from 1\dots 3 and label the color state
  of quarks ($\bf 3$) and of diquarks ($\bar {\bf 3}$).
  Small letters $n,k=1,2$ label quark isospin and $m,m'=\pm 1,0$
  label the third component of the axialvector diquark isospin.  }
 \label{bsekern}
\end{figure}
The matrix representation of the Bethe-Salpeter kernel with respect to 
the diquark isospin eigenstates defined above is given by
(see figure \ref{bsekern})
\begin{eqnarray}
   & &  \qquad \qquad \qquad {\Sc I=0} \quad\qquad\qquad\qquad {\Sc I=1}
  \nonumber  \\
 K^{\rm BS}_{\rm flavor} &=& \begin{matrix} {\Sc I=0} \\[2mm] {\Sc I=1} 
        \end{matrix} \;\; \frac{1}{2} 
   \bp \left( \tau^2 \,\tau^2\right)_{kn} &  
       -\left((\tau_{m'} \tau^2)\,\tau^2\right)_{kn} \\[2mm]
       -\left(  \tau^2 \, (\tau^2 \tau_m^\dagger)\right)_{kn} & 
       \left((\tau_{m'} \tau^2)\, (\tau^2 \tau_m^\dagger)\right)_{kn} \ep \; .
\end{eqnarray}
Sandwiching between either the proton or the neutron
flavor state yields the identical flavor matrix
\begin{eqnarray}
 & & \quad\qquad\;\;  {\Sc I=0} \qquad {\Sc I=1} \nonumber \\
 \langle p| K^{\rm BS}_{\rm flavor} | p \rangle =
 \langle n| K^{\rm BS}_{\rm flavor} | n \rangle &=&
   \begin{matrix} {\Sc I=0} \\ {\Sc I=1} \end{matrix} \;\;
   \frac{1}{2} \bp +1  & -\sqrt{3} \\ -\sqrt{3} & -1 \ep \; .
 \label{fl}
\end{eqnarray}
Eqs.~(\ref{co},\ref{fl}) 
 provide for
the explicit factors in the Bethe-Salpeter kernel of eq.~(\ref{Gqdq}).

\subsection{$\Delta$}
\label{cfd}

Taking over the notations from the previous subsection, we
write the flavor states of the four nearly mass-degenerate
$\Delta$ resonances as
\begin{equation}
 \bp | \Delta^{++} \rangle \\  | \Delta^{+} \rangle \\
    | \Delta^{0} \rangle \\  | \Delta^{-} \rangle \ep
 =
 \bp  u{[uu]} \\ \sqrt{\frac{2}{3}}u{[ud]} +\sqrt{\third}d{[uu]} \\  
    \sqrt{\frac{2}{3}}d{[ud]} + \sqrt{\third}u{[dd]}  \\ d{[dd]} \ep \; .
\end{equation}
The flavor matrix of the Bethe-Salpeter kernel involves only
the axialvector diquark ($I=1$) channel,
\begin{equation}
  (K^{\rm BS}_\Delta)_{\rm flavor}  =
       \left((\tau_{m'} \tau^2)\, (\tau^2 \tau_m^\dagger)\right)_{kn} \; .
\end{equation}
This yields for all $\Delta$ states $i \in \{++,+,0,-\}$
 identical flavor factors,
\begin{equation}
 \langle \Delta^{(i)} | (K^{\rm BS}_\Delta)_{\rm flavor} | \Delta^{(i)}
 \rangle = 1 \;. 
\end{equation}

\section{Wave functions}
\label{wave-app}

On the next pages the scalar functions describing the partial waves
of nucleon and $\Delta$ are displayed. In figure \ref{i} we show
the nucleon functions $\hat S_i$, $\hat A_i$ for the parameter Set I
of table \ref{pars}. The momentum partitioning parameter $\eta=0.36$ 
has been  chosen. This value is roughly half way between
the bounds set by the constituent poles, {\em cf.} eq.~(\ref{ebound1}),
\begin{equation}
 \eta \in [0.334,0.383] \; . 
\end{equation}
The even Chebyshev momenta dominate 
the expansion of the scalar functions and therefore we show the three
leading even momenta for each function. The two dominating
$s$ waves $\hat S_1$ and $\hat A_3$ converge quickly in their expansion.
The $p$ waves $\hat S_2$, $\hat A_2$ and $\hat A_6$ which follow next 
with respect to the overall magnitude display a more moderate convergence. 
One can estimate that these $p$ waves will have an influence on observables
on the level of below 10 \%. All other functions are subdominant,
especially the $d$ wave $\hat A_5$ turns out to be extremely small.

The $\Delta$ partial waves depicted in figure \ref{ii}
have been calculated using the parameter Set II which fits the
experimental $\Delta$ mass. The binding energy is only 24 MeV and
consequently the range of allowed $\eta$ values is restricted
by eqs.~(\ref{ebound1},\ref{ebound2}) to
\begin{equation}
 \eta \in [0.328,0.345] \; .
\end{equation}
We have chosen $\eta=1/3$. The small binding energy is reflected 
in the scalar functions as follows. The only $s$ wave $\hat D_1$
clearly dominates and the $p$ waves $\hat D_2$, $\hat D_4$ and
$\hat D_6$ represent corrections of the order of one per cent.
All other partial waves are suppressed.
Although the convergence of the Chebyshev expansion
is still moderate for $\hat D_1$, the remaining functions 
possess a number of higher moments which are larger than 
one tenth of the zeroth moment. In figure \ref{ii} we have picked
the moments 0, 4 and 8 to illustrate this behaviour.

\newlength{\figurewidtha}
\setlength{\figurewidtha}{5.5cm}
\begin{figure}[t]
 \begin{center}
   \epsfig{file=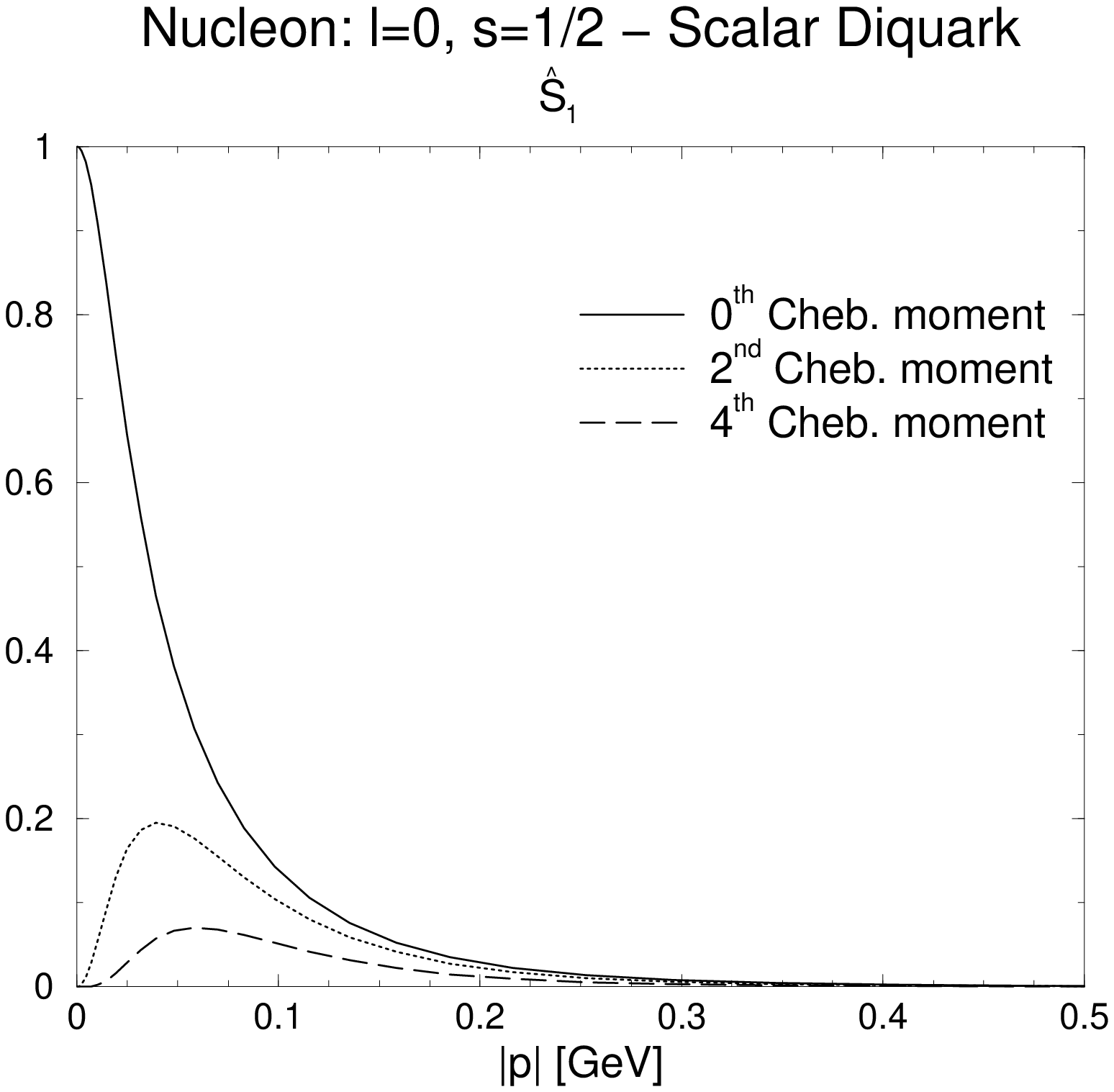,width=\figurewidtha} \hspace{5mm} 
   \epsfig{file=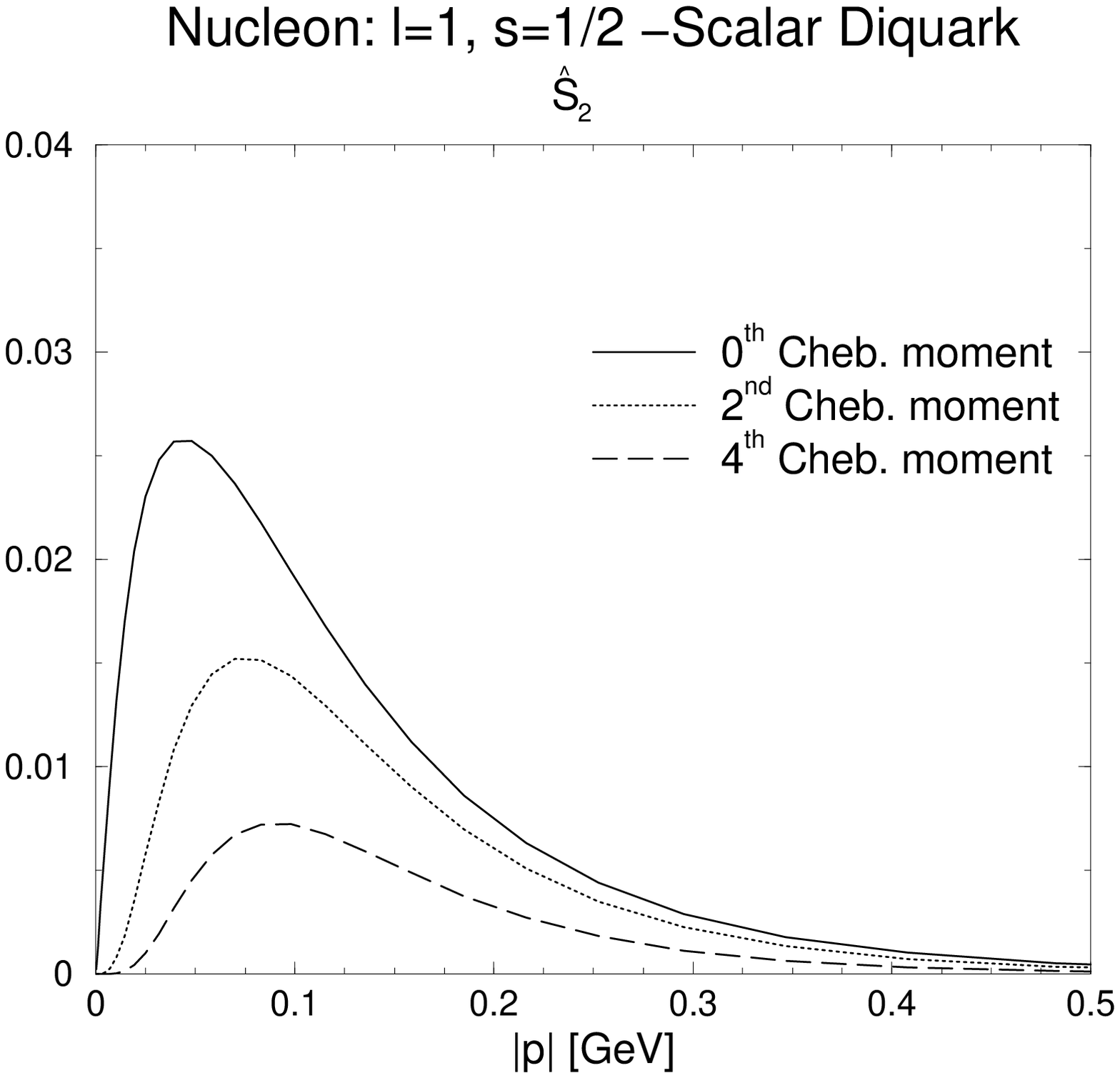,width=\figurewidtha} \hspace{5mm} 
   \epsfig{file=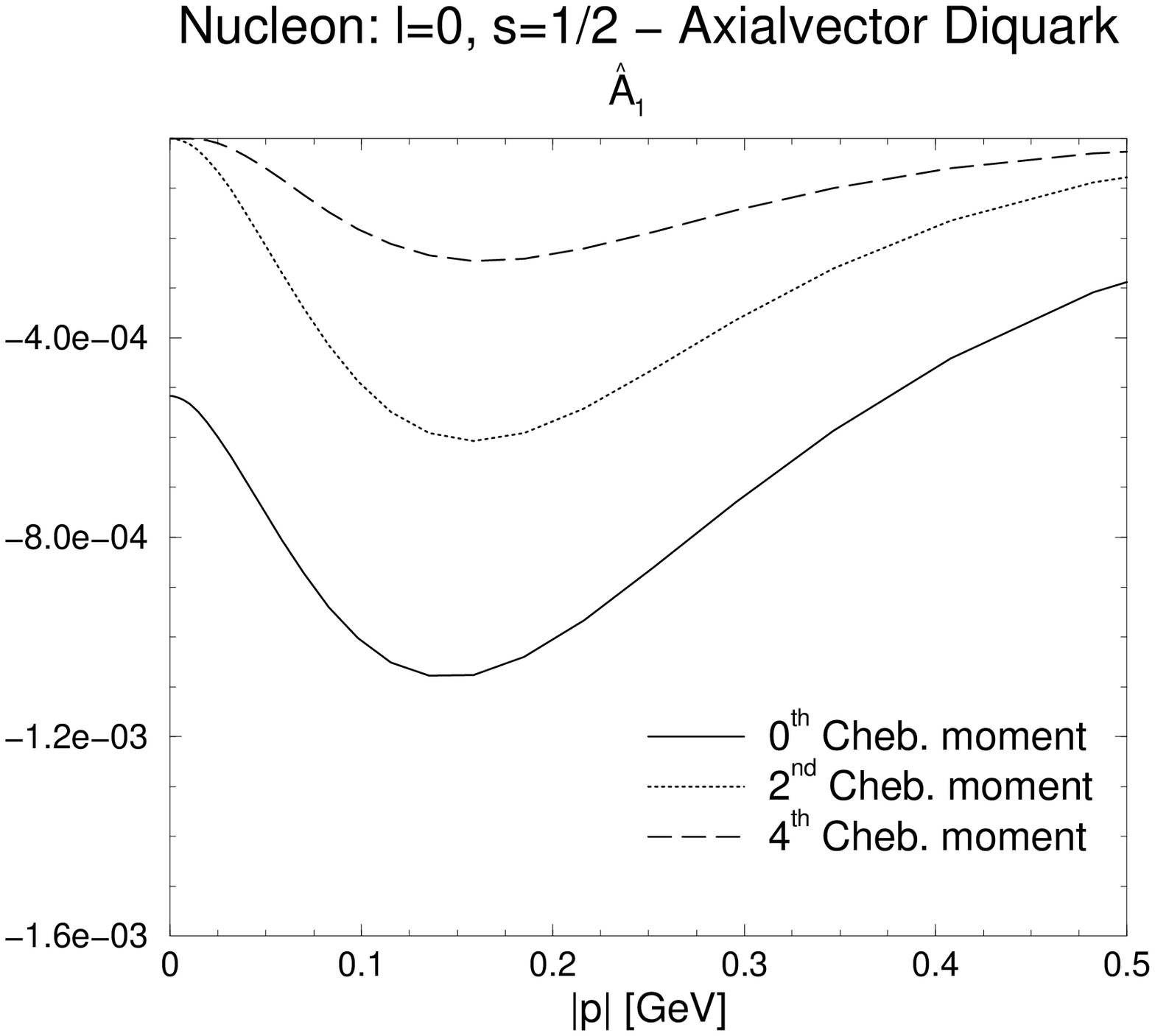,width=\figurewidtha} \hspace{5mm}
   \epsfig{file=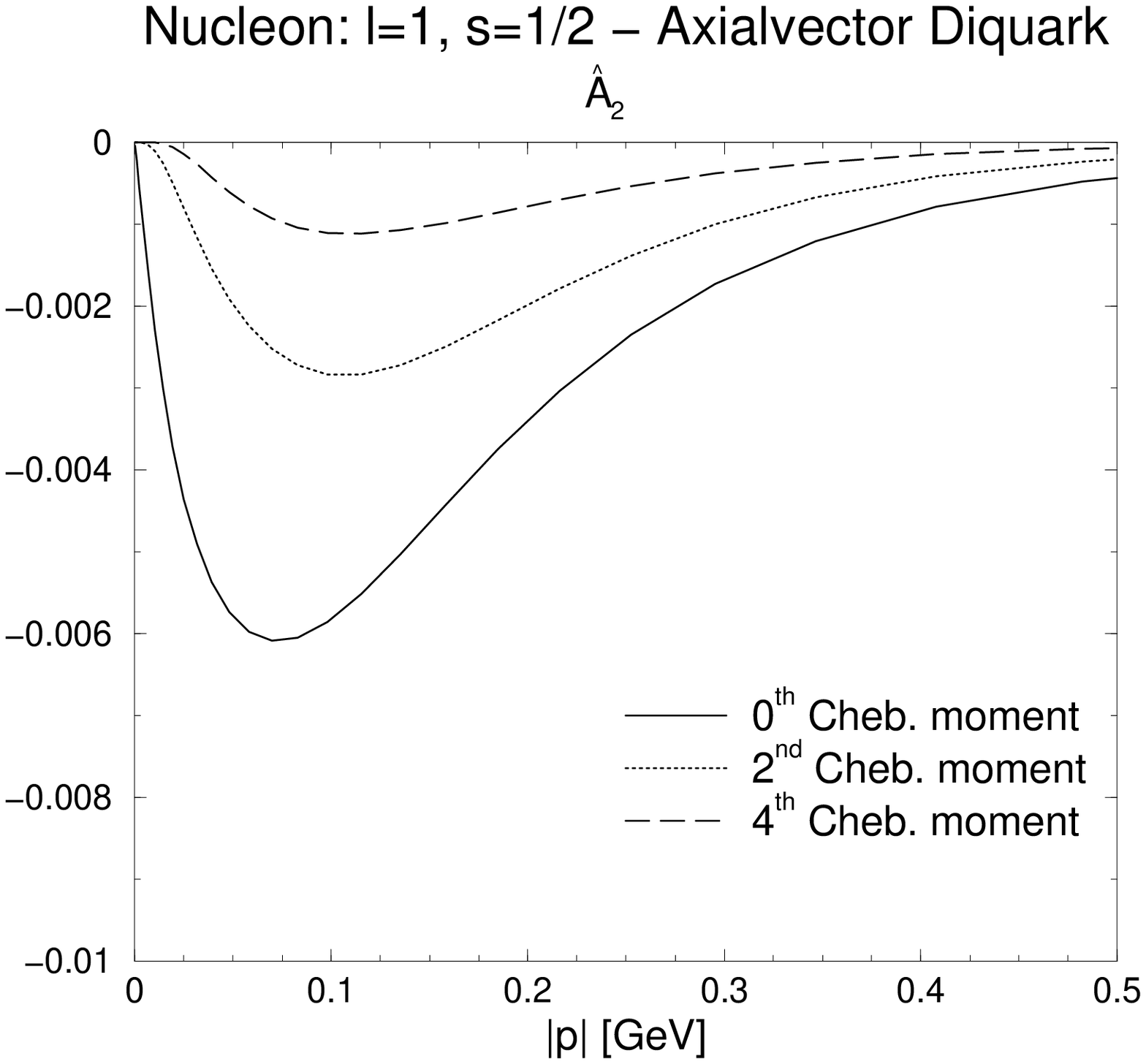,width=\figurewidtha} \hspace{5mm}
   \epsfig{file=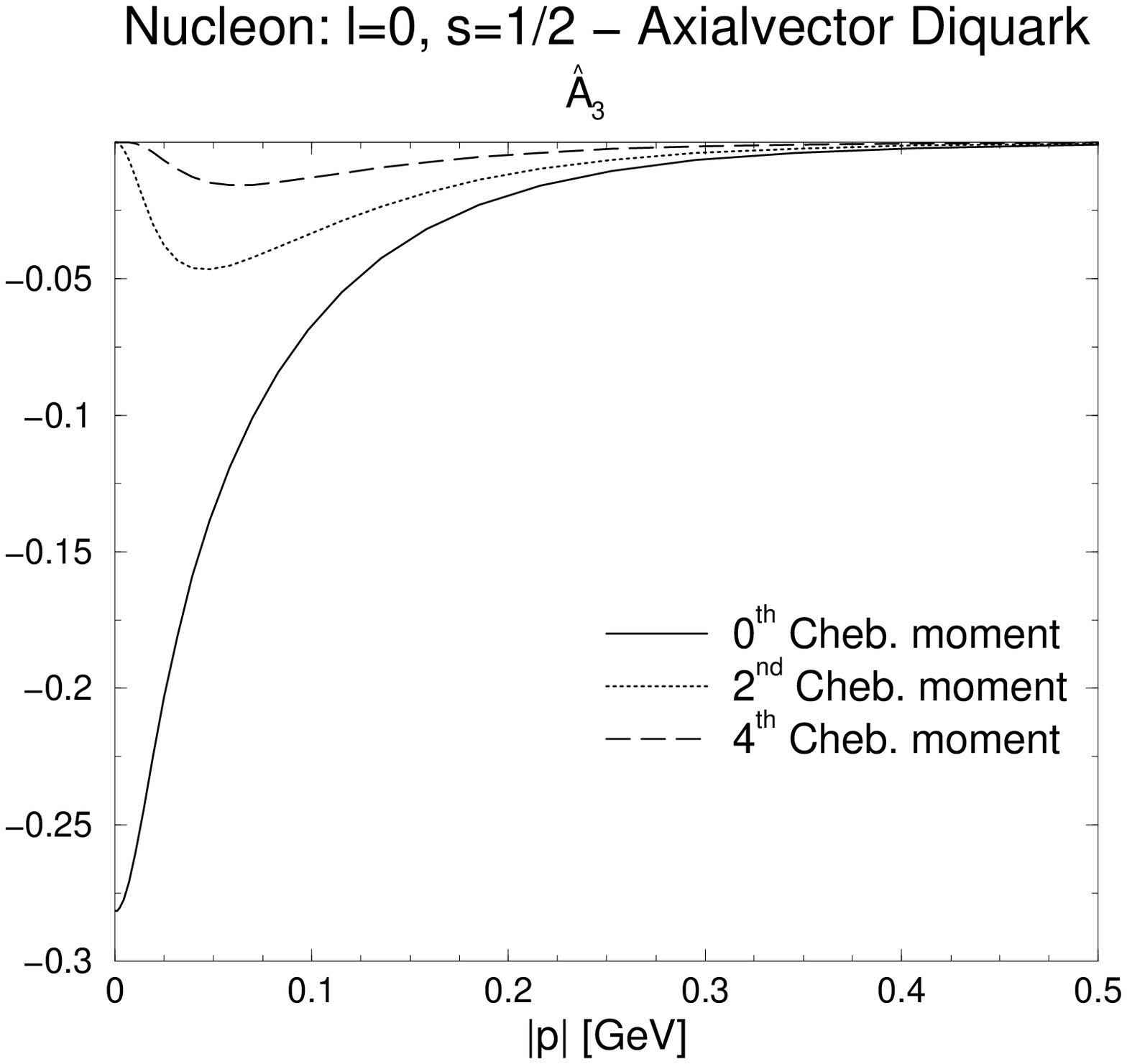,width=\figurewidtha} \hspace{5mm}
   \epsfig{file=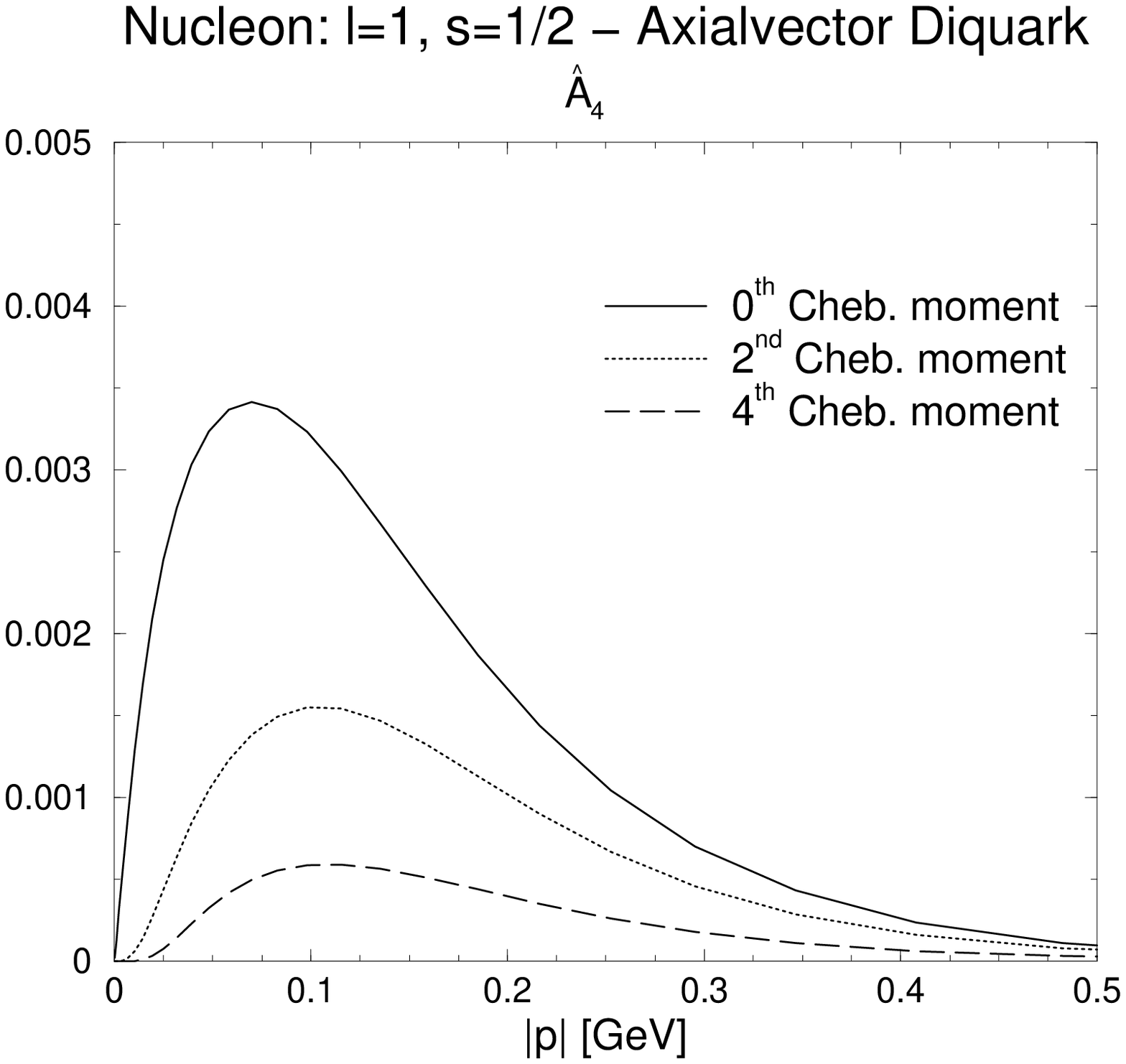,width=\figurewidtha} \hspace{5mm}
   \epsfig{file=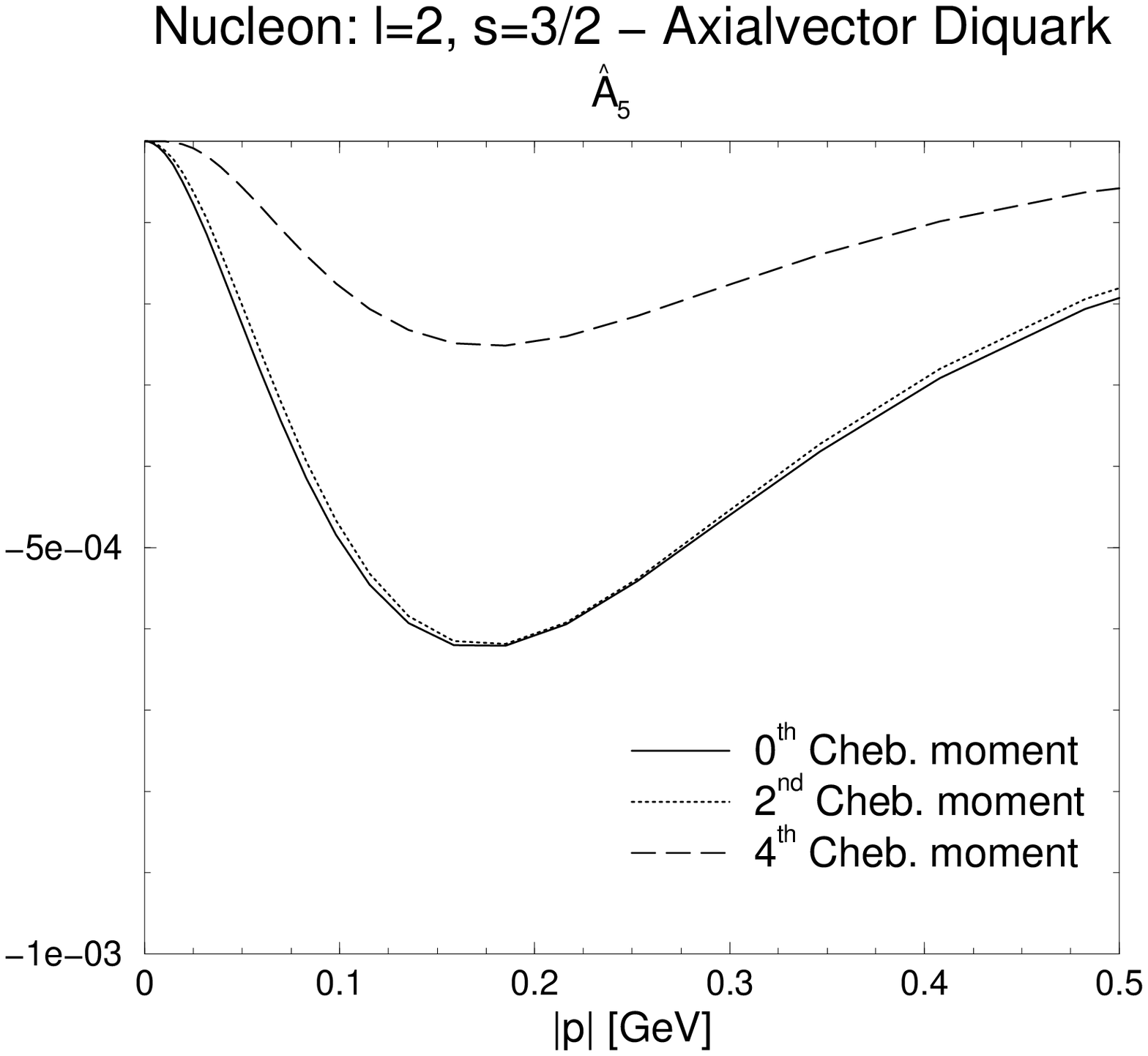,width=\figurewidtha} \hspace{5mm}
   \epsfig{file=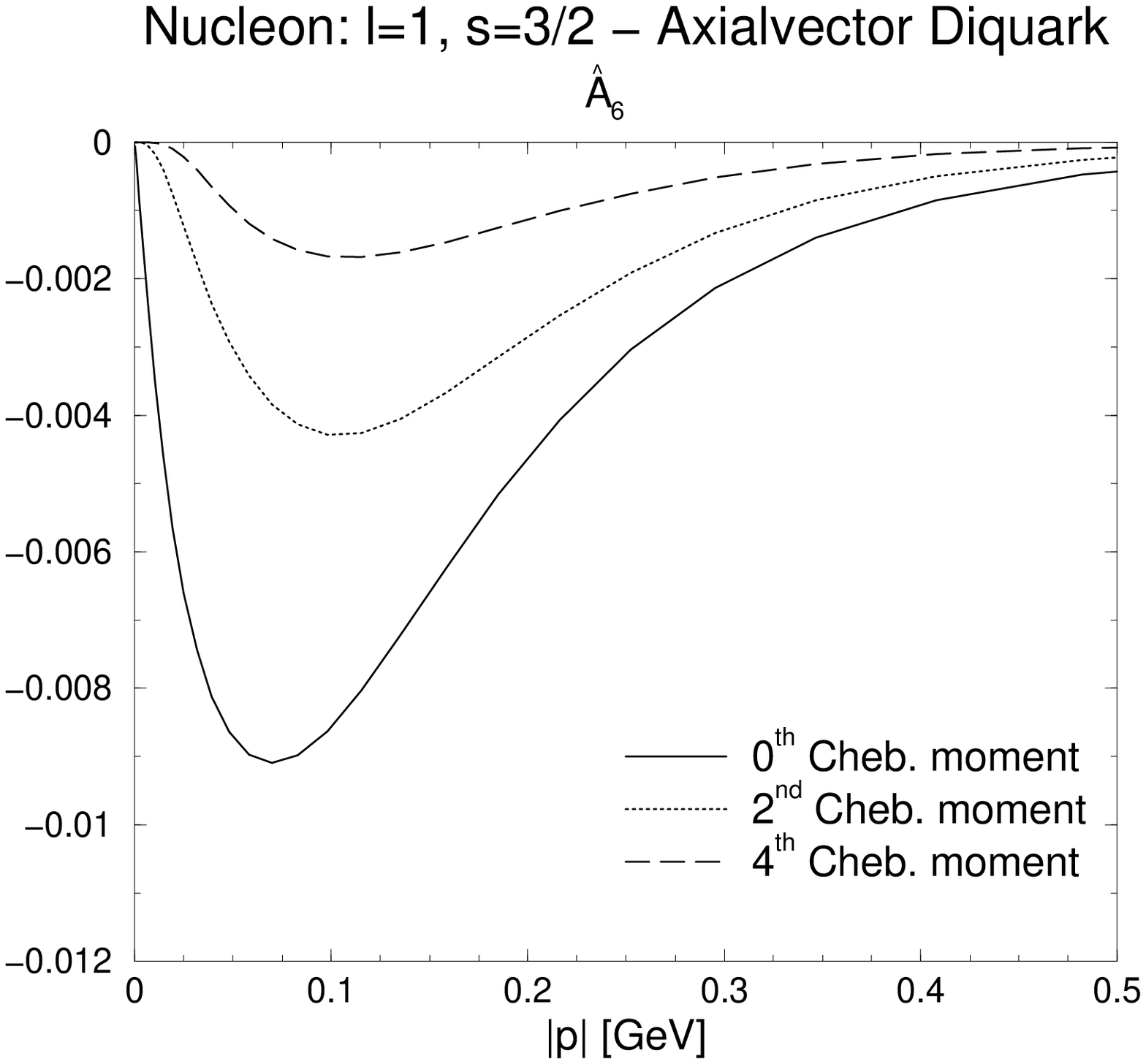,width=\figurewidtha} 
 \end{center}
 \caption{Chebyshev moments of the scalar functions $\hat S_i$ and
   $\hat A_i$ describing the nucleon wave function for parameter
   Set I.}
 \label{i}
\end{figure}

\begin{figure}[t]
 \begin{center}
   \epsfig{file=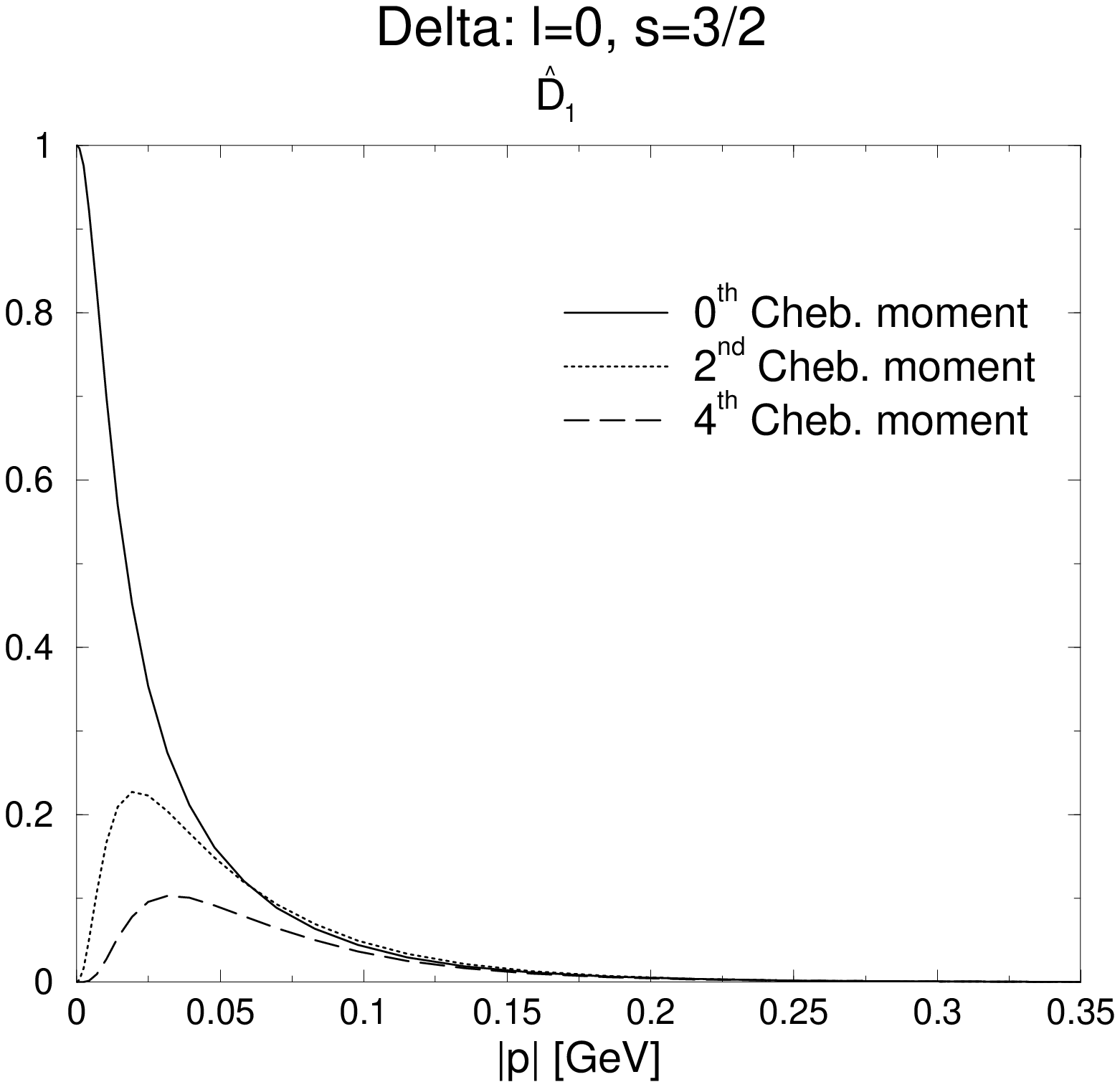,width=\figurewidtha} \hspace{5mm}
   \epsfig{file=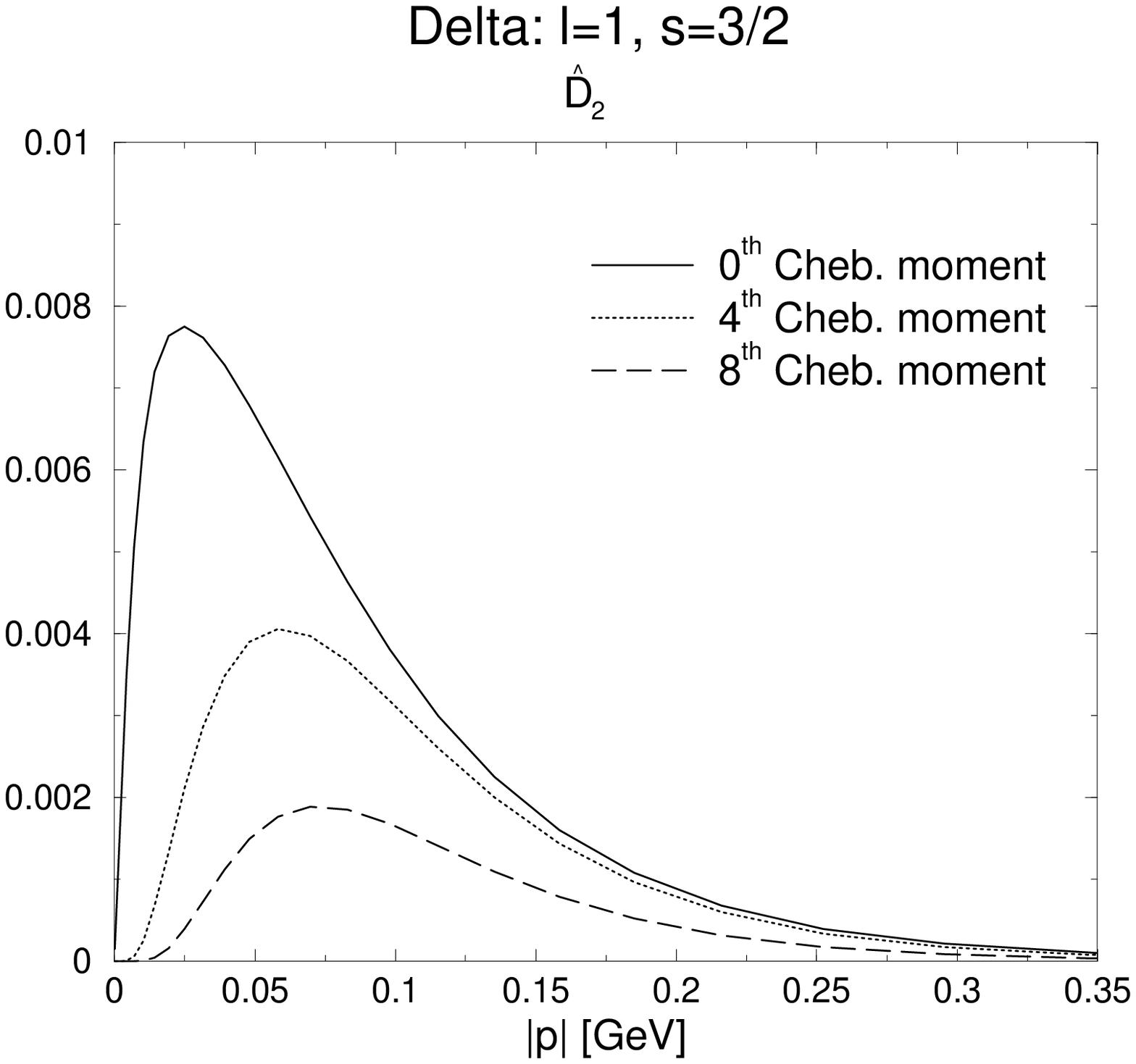,width=\figurewidtha} \hspace{5mm}
   \epsfig{file=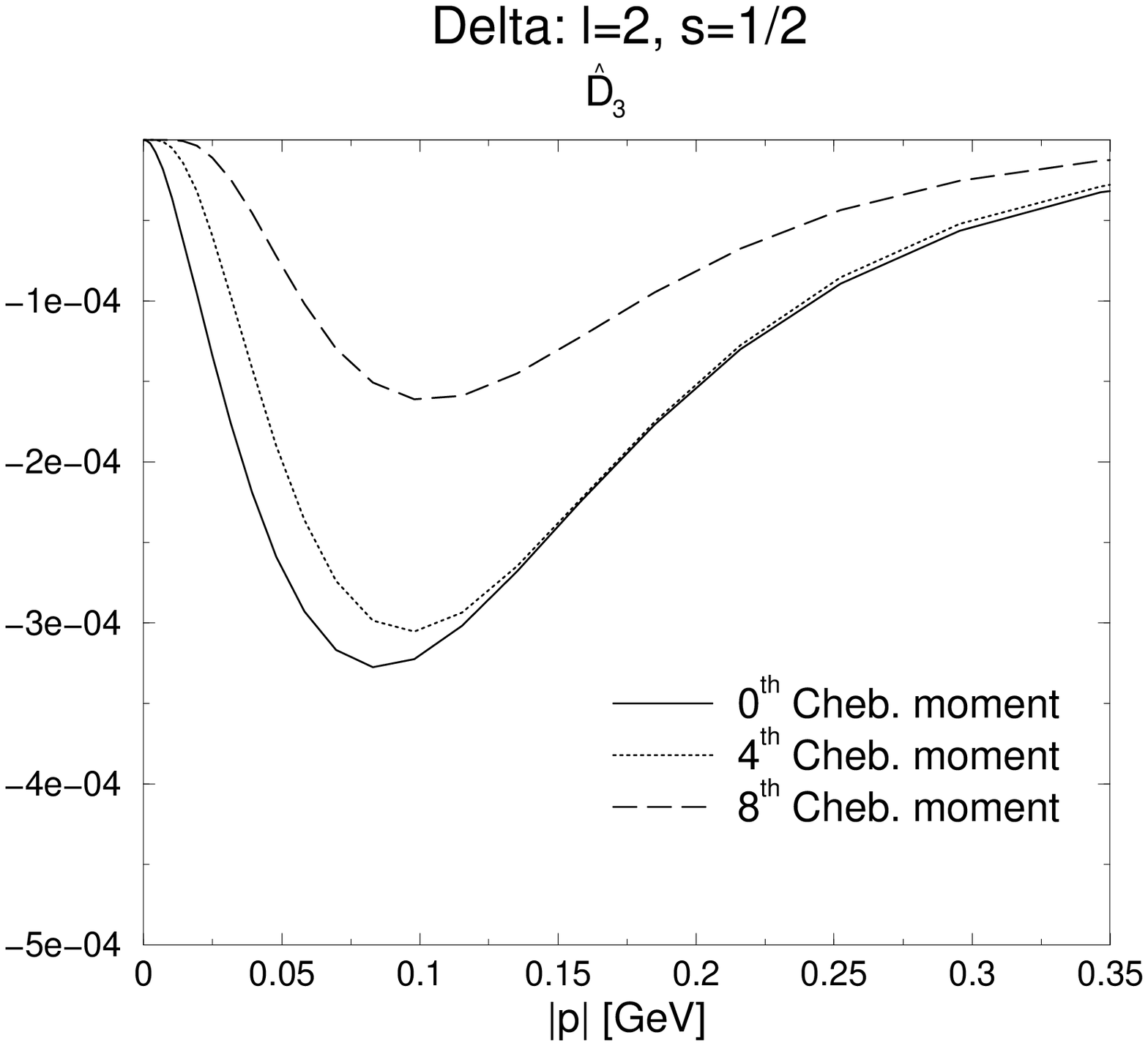,width=\figurewidtha} \hspace{5mm}
   \epsfig{file=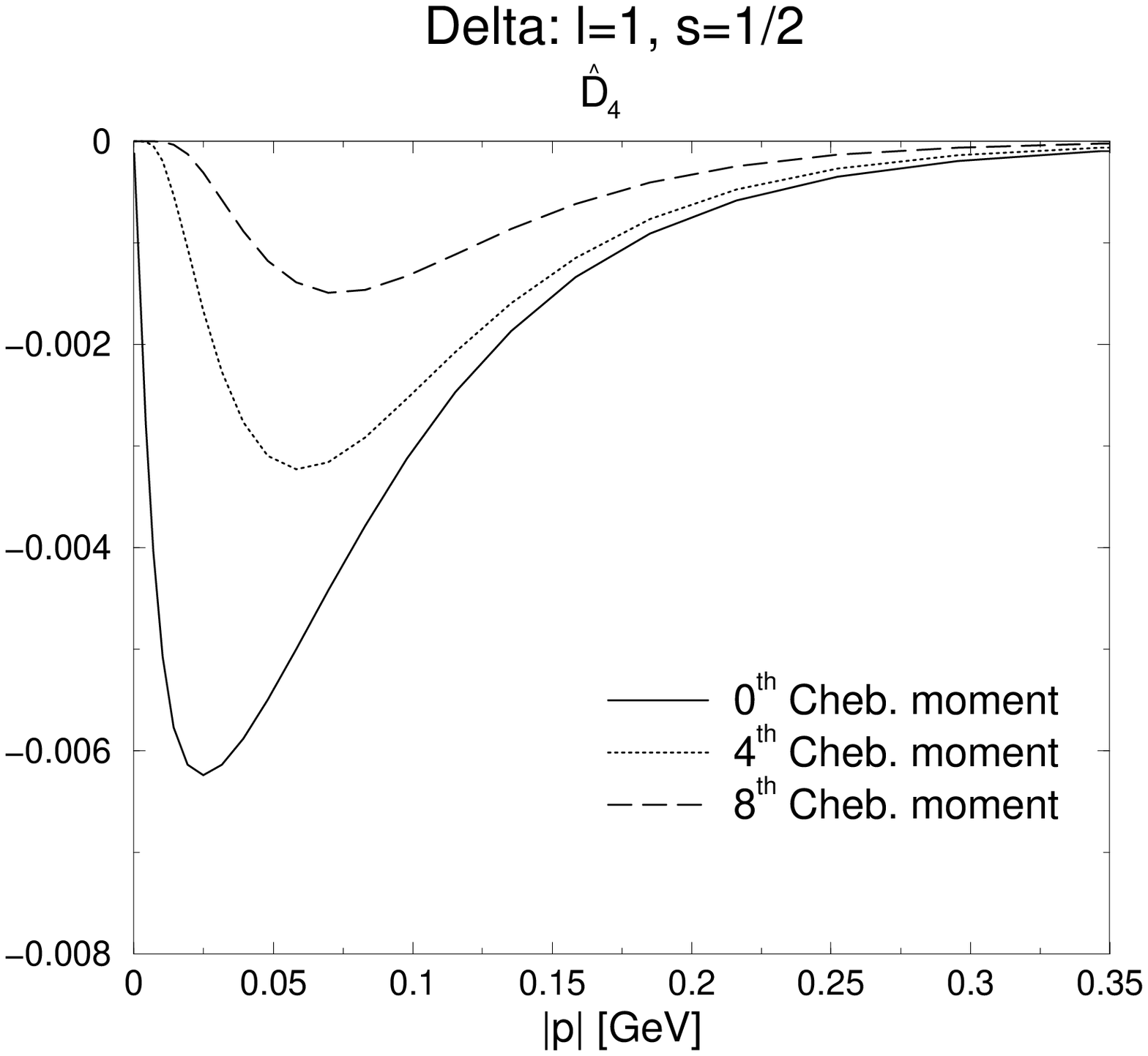,width=\figurewidtha} \hspace{5mm}
   \epsfig{file=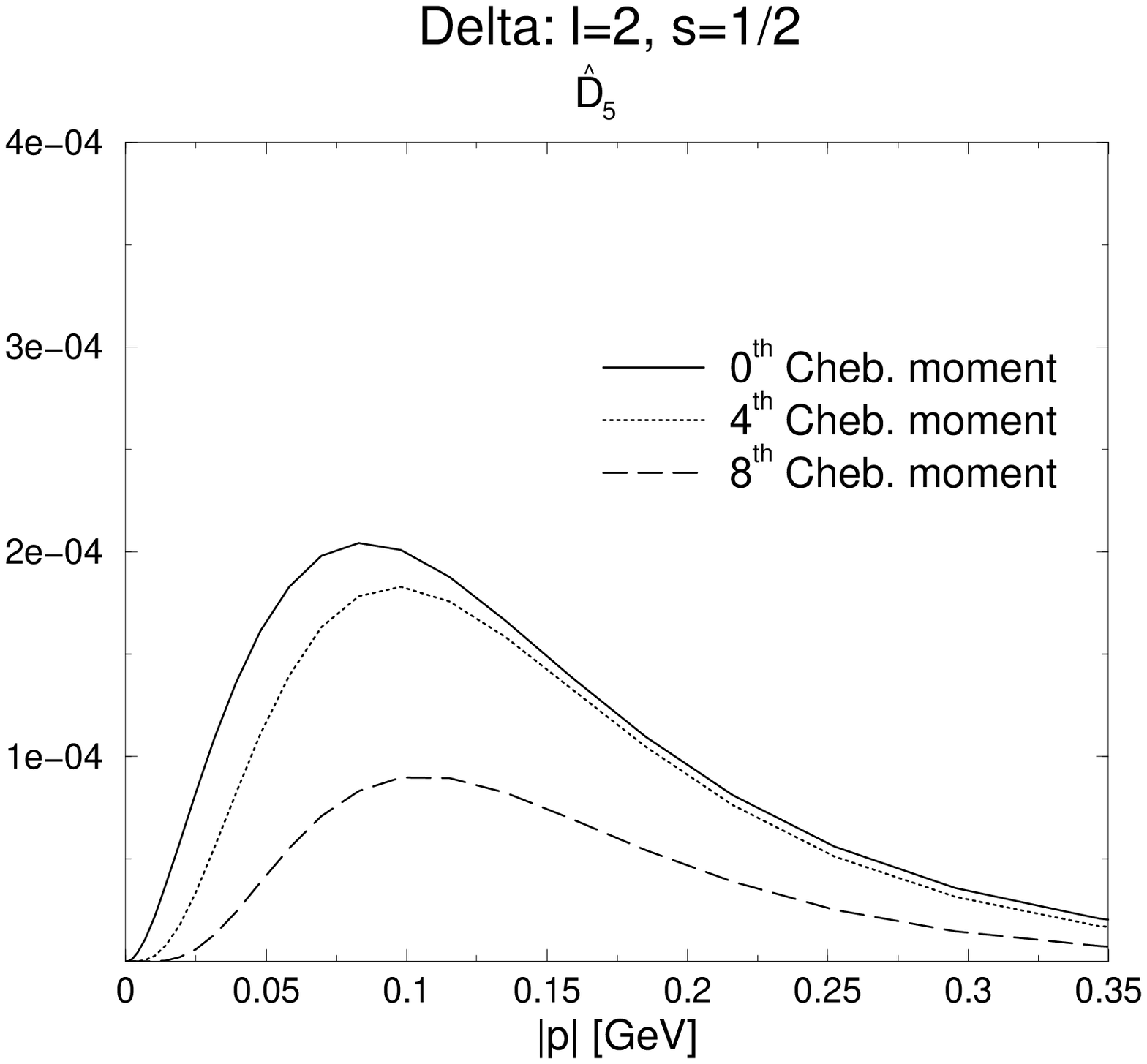,width=\figurewidtha} \hspace{5mm}
   \epsfig{file=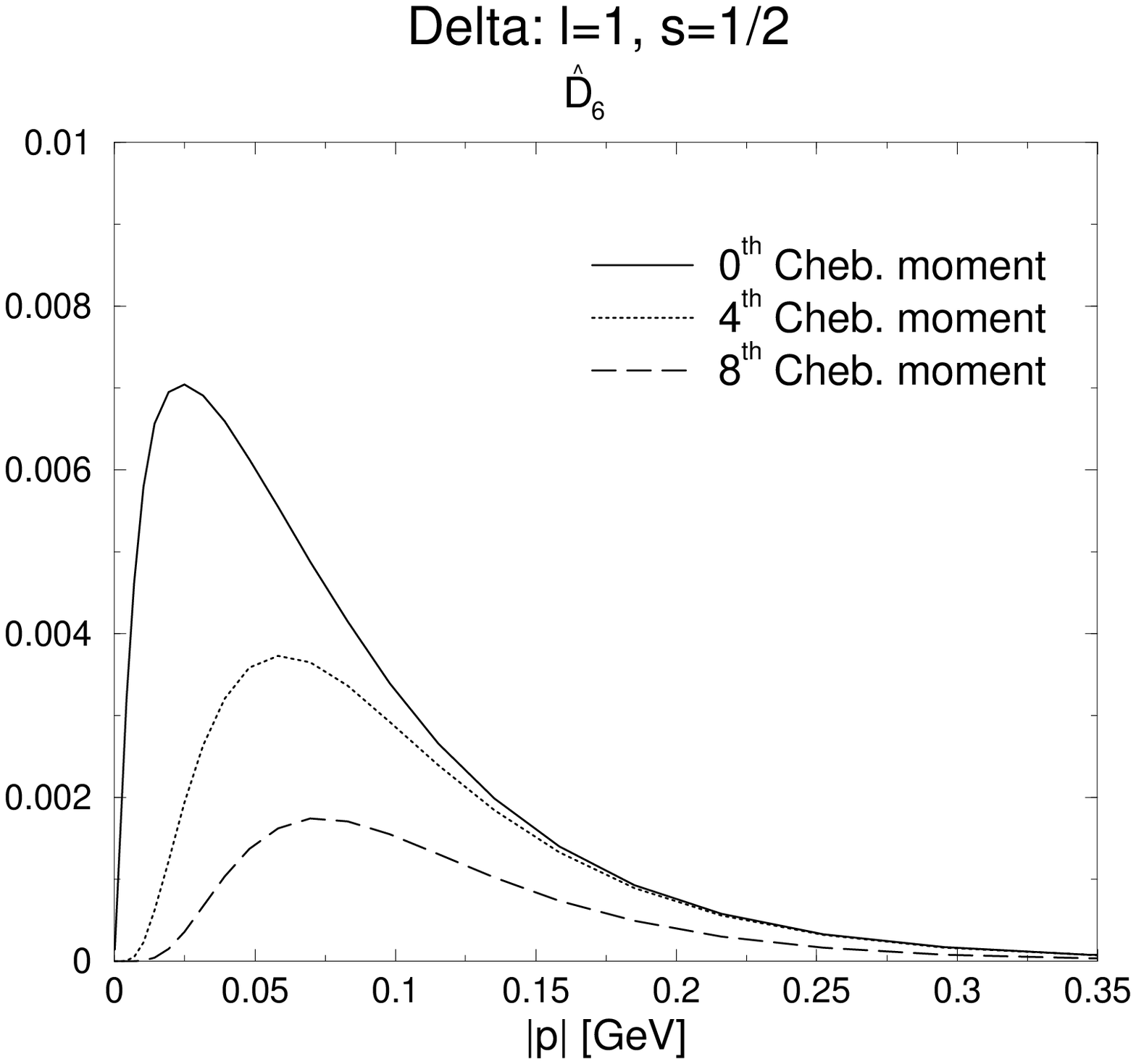,width=\figurewidtha} \hspace{5mm}
   \epsfig{file=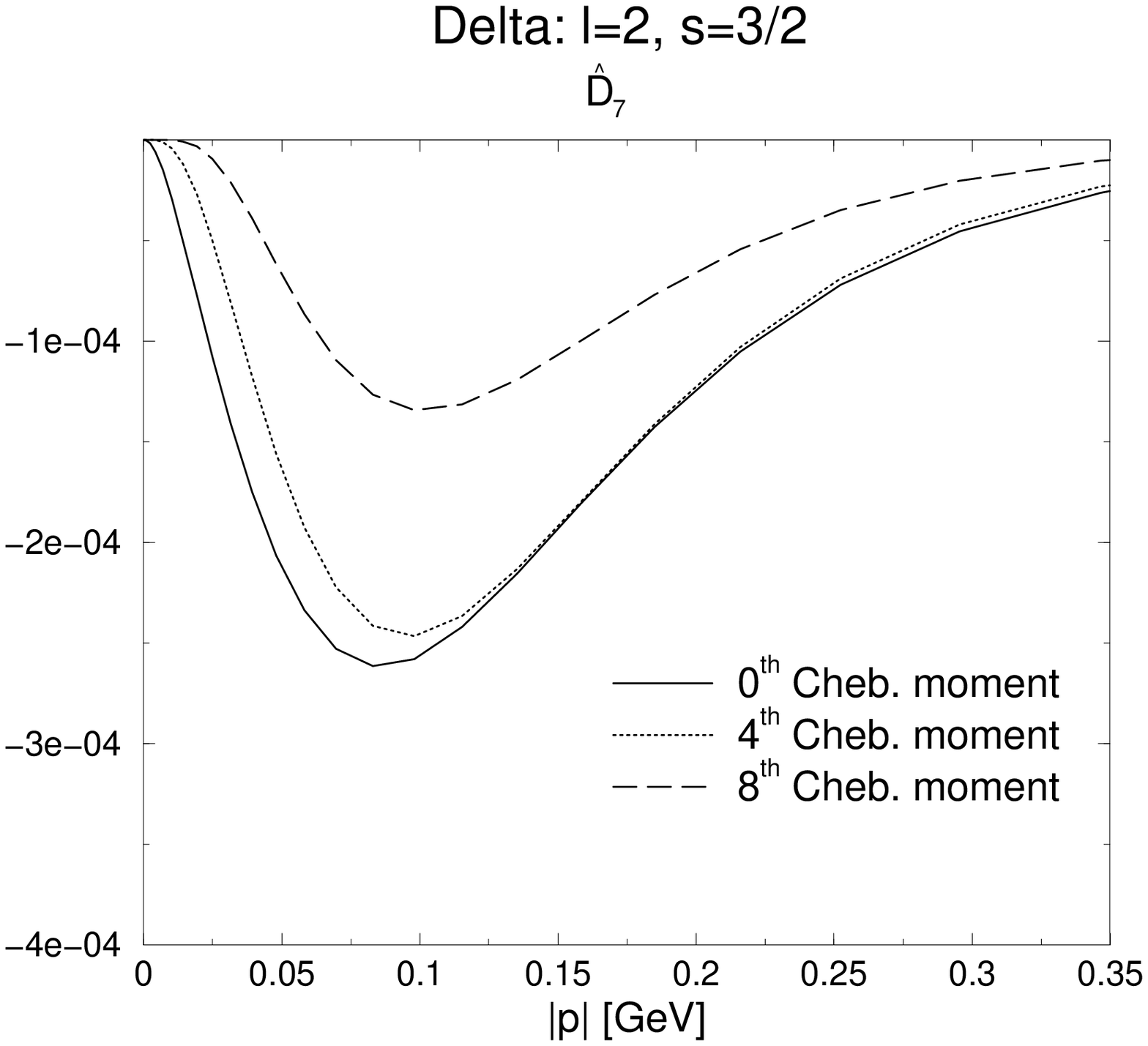,width=\figurewidtha} \hspace{5mm}
   \epsfig{file=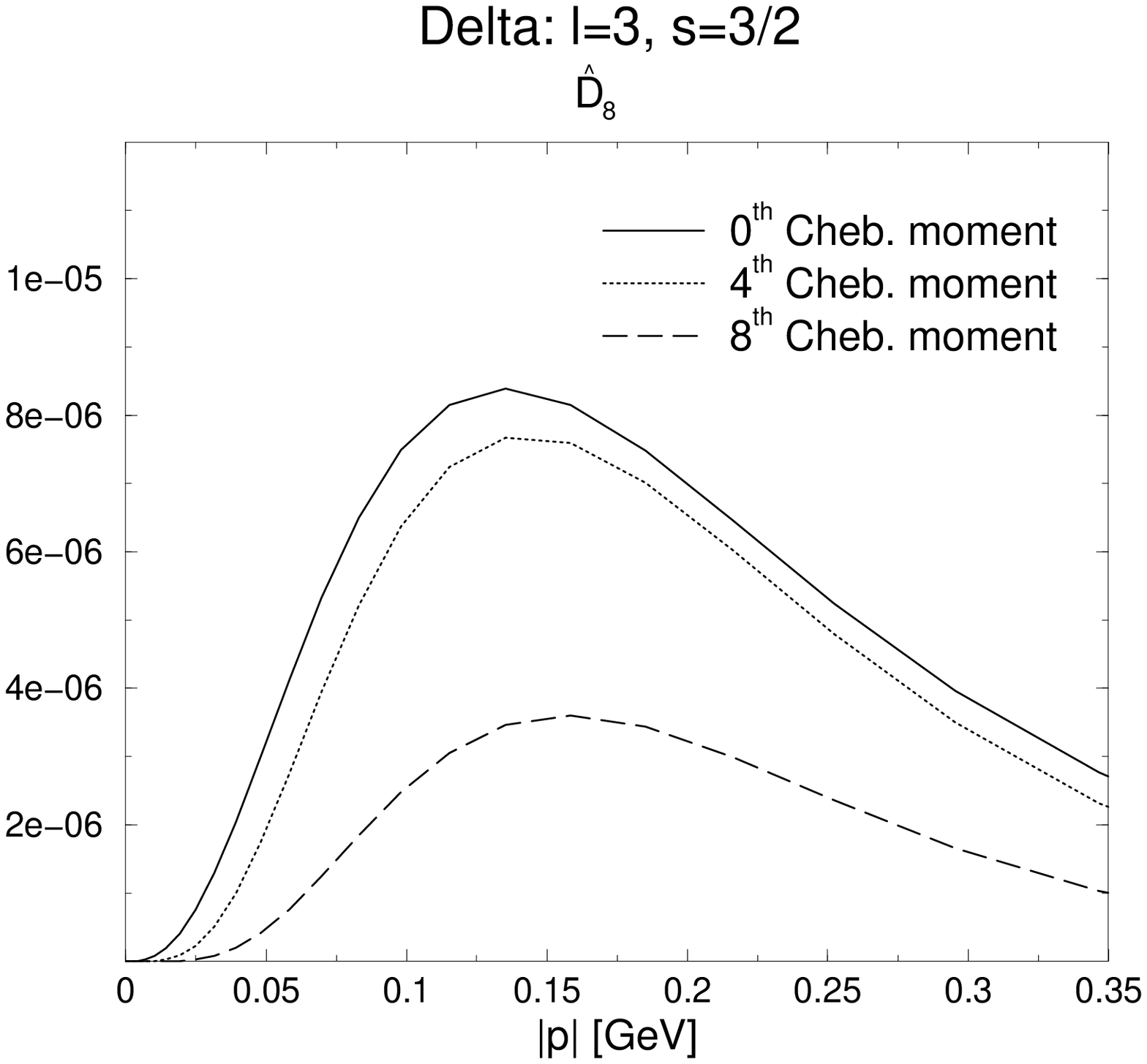,width=\figurewidtha} \hspace{5mm}
 \end{center}
 \caption{Chebyshev moments of the scalar functions $\hat D_i$
   describing the $\Delta$ wave function for parameter
   Set II.}
 \label{ii}
\end{figure}

\section{Bethe-Salpeter equations for octet and decuplet}
\label{od-app}

Here we list the Bethe-Salpeter equations for octet and decuplet
baryons needed in section \ref{od-sec}. First we introduce the notation:

\begin{center}
\begin{tabular}{p{2.0cm}p{11cm}}
$\Psi^5_{c(ab)},\Psi^\mu_{c[ab]}$ & Spin-1/2 wave function for
  flavor components with spectator quark $c$ and scalar $(ab)$ or 
  axialvector diquark $[ab]$, decomposed as in eq.~(\ref{wex_N})  \\[2mm]
$\Psi^{\mu\nu}_{c[ab]}$ & Spin-3/2 wave function for
  flavor components with spectator quark $c$ and 
  axialvector diquark $[ab]$, decomposed as in eq.~(\ref{wex_D}) \\[2mm]
$S_a$ & Quark propagator as in eq.~(\ref{qprop}) with $m_q=m_a$ \\[2mm]
$D_{(ab)}, D^{\mu\nu}_{[ab]}$ & Diquark propagators as in 
  eqs.~(\ref{Ds},\ref{Da}) with $m_{sc}=m_{(ab)}$ and $m_{ax}=m_{[ab]}$\\[2mm]
$K^{\sf ab}_a$ & Quark exchange kernel with momentum definitions as
  below eq.~(\ref{kbsdef}): 
  $K^{\sf ab}_a=-\half\fourint{k}\chi^{\sf a}\,S^T_a\,\bar\chi^{\sf b}$
  $\;({\sf a,b}=1, \dots, 5)$
\end{tabular}
\end{center}
\vskip 1mm
As we consider the isospin symmetric case, the flavor labels
$a,b,c$ will be chosen among $up$ and $strange$ ($u,s$).
We recapitulate the nucleon equation (\ref{BS4}),\\[1mm]
\begin{equation}
    \bp S^{-1}_u\, D^{-1}_{(uu)} & 0 \\[1mm] 
     0 & S^{-1}_u\, (D^{\mu\nu}_{[uu]})^{-1} \ep
    \bp \Psi^5_{u(uu)} \\[1mm] \Psi^\nu_{u[uu]} \ep
    = \bp K^{55}_u & -\sqrt{3}\, K^{\rho 5}_u \\[1mm] 
          -\sqrt{3}\, K^{5\mu}_u & -K^{\rho\mu}_u \ep
      \bp \Psi^5_{u(uu)} \\[1mm] \Psi^\rho_{u[uu]}  \ep \; .
\end{equation}
The $\Sigma$ hyperon is described by\\[1mm]
\begin{eqnarray}
  \bp S^{-1}_u\, D^{-1}_{(us)} & 0 & 0 \\[1mm]
     0 & S^{-1}_u\, (D^{\mu\nu}_{[us]})^{-1} & 0  \\[1mm]
     0 & 0 & S^{-1}_s\, (D^{\mu\nu}_{[uu]})^{-1} \ep
    \bp \Psi^5_{u(us)} \\[1mm] \Psi^\nu_{u[us]} \\[2mm] \Psi^\nu_{s[uu]} \ep 
    = \qquad \qquad \nonumber \\[1mm]
   \bp K^{55}_s & - K^{\rho 5}_s & \sqrt{2}\,K^{\rho 5}_u \\[1mm]
     -K^{5\mu}_s & K^{\rho\mu}_s &  \sqrt{2}\,K^{\rho\mu}_u \\[1mm]
     \sqrt{2}\, K^{5\mu}_u & \sqrt{2}\,K^{\rho\mu}_u & 0 \ep
     \bp \Psi^5_{u(us)} \\[1mm] \Psi^\rho_{u[us]} \\[2mm] \Psi^\rho_{s[uu]} \ep\; .
\end{eqnarray}\\[1mm]
The equation for the $\Xi$ hyperon is obtained by interchanging
$u \leftrightarrow  s$.
 $SU(3)$ symmetry breaking couples the eightfold way state of the $\Lambda$
to the flavor singlet, $s(ud)+u(ds)+d(su)$. We introduce
the flavor states $f_1=[d(us)-u(ds)]/\sqrt{2}$, 
$f_2=s(ud)$ and $l=(d[us]-u[ds])/\sqrt{2}$ and the equation
for the physical $\Lambda$ hyperon reads\\[1mm]
\begin{eqnarray}
    \bp S^{-1}_u\, D^{-1}_{(us)} & 0 & 0 \\[1mm]
     0 & S^{-1}_s\, D^{-1}_{(uu)} & 0  \\[1mm]
     0 & 0 & S^{-1}_u\, (D^{\mu\nu}_{[us]})^{-1} \ep
    \bp \Psi^5_{f_1} \\[1mm] \Psi^5_{f_2} \\[2mm] \Psi^\nu_{l} \ep 
    = \qquad \qquad \nonumber \\[1mm]
     \bp -K^{55}_s & \sqrt{2}\,K^{55}_u & K^{\rho 5}_s \\[1mm]
     \sqrt{2}\,K^{55}_u & 0 &  \sqrt{2}\,K^{\rho 5}_u \\[1mm]
      K^{5\mu}_s & \sqrt{2}\,K^{5\mu}_u & - K^{\rho\mu}_s\ep
 \bp \Psi^5_{f_1} \\[1mm] \Psi^5_{f_2} \\[2mm] \Psi^\rho_{l} \ep   \; .
\end{eqnarray}
The equation for the $\Delta$ reads, {\em cf.} eq.~(\ref{BS5}),
\begin{equation}
  S^{-1}_u \, (D^{\mu\nu}_{[uu]})^{-1}\;
  \Psi^{\nu\sigma}_{u[uu]} = 2\,K^{\rho \mu}_u\;
  \Psi^{\rho\sigma}_{u[uu]} \; .
\end{equation}
By interchanging $u \leftrightarrow  s$ one finds the equation
for the $\Omega$. Finally the $\Sigma^\ast$ obeys the equation\\[1mm]
\begin{equation}
 \bp  S^{-1}_u\, (D^{\mu\nu}_{[us]})^{-1} & 0  \\[1mm]
      0 & S^{-1}_s\, (D^{\mu\nu}_{[uu]})^{-1} \ep
 \bp \Psi^{\nu\sigma}_{u[us]} \\[1mm] \Psi^{\nu\sigma}_{s[uu]} \ep
  =
 \bp K^{\rho \mu}_s &  \sqrt{2}\,K^{\rho \mu}_u  \\[1mm]
     \sqrt{2}\,K^{\rho \mu}_u & 0       \ep
 \bp \Psi^{\rho\sigma}_{u[us]} \\[1mm] \Psi^{\rho\sigma}_{s[uu]} \ep\; .
 \label{sigma32}
\end{equation}
In eq.~(\ref{sigma32}) the interchange $u \leftrightarrow  s$ leads
to the $\Xi^\ast$ equation.

   \chapter{Form Factor Calculations}

\section{Resolving diquarks}
\subsection{Electromagnetic Vertices}
\label{dqres1}

Here we adopt an impulse approximation to couple the photon directly 
to the quarks inside the diquarks obtaining the scalar, axialvector 
and the photon-induced scalar-axialvector diquark transition 
couplings as represented
by the 3 diagrams in figure~\ref{emresolve}. For on-shell diquarks these
yield diquark form factors and
at the soft point ($Q=0$) the electric 
form factors of scalar and axialvector 
diquark are equivalent to the normalization conditions
(\ref{normsc},\ref{normax}). Hereby it is assumed that the diquark-quark
vertices $\chi^{5[\mu]}$ could be obtained from a quark-quark scattering
kernel which is independent of the total diquark momentum,
{\em cf.} also section \ref{normsec} where we discussed the 
equivalence of canonical and charge normalization 
in the nucleon case.

Due to the quark-exchange antisymmetry of the diquark amplitudes 
it suffices to calculate one diagram for each of the three contributions,
{\it i.e.}, those of figure~\ref{emresolve} in which the photon couples to the
``upper'' quark line. The color trace yields unity
as in the normalization integrals, eqs. (\ref{normsc}) and (\ref{normax}).
To perform the traces over the diquark flavor matrices 
with the charge matrix acting on the quark line 
it is advantageous to choose the diquark charge eigenstates,
{\em cf.} appendix \ref{cfn}. 
With the definition
\begin{equation}
  Q_c={\T \frac{1}{6}} \boldsymbol{1} + \half \tau^3
\end{equation}
we then obtain
the following flavor factors
\begin{eqnarray}
 (\tilde \Gamma^\mu_{sc})_{\rm flavor} &=& \half 
       {\rm Tr}\;\tau^2 Q_c \tau^2
      = {\T \frac{1}{6}} \; ,  \\
 (\tilde \Gamma^{\mu,\alpha\beta}_{ax})_{\rm flavor} &=&
   \half {\rm Tr}\; (\tau^2 \tau_m^\dagger)Q_c (\tau_{m} \tau^2)
   = \bp \quad{\T \frac{2}{3}} \\[1mm] \quad\third \\[1mm] -\third \ep \; , \\
 (\tilde \Gamma^{\mu,\beta}_{sa})_{\rm flavor} &=&
   -\half {\rm Tr}\; \tau^2 Q_c (\tau_{m} \tau^2) =-\delta_{m0}\; \half \; .
\end{eqnarray} 
For the flavor factors of the
scalar and axialvector diquarks we obtain  half their total charges.
The factor 1/2 is due to the antisymmetry of the vertices 
and also appears in the normalization conditions 
(\ref{normsc},\ref{normax}).  

We are left with the quark loop that involves just a trace 
over the Dirac indices. Including the minus sign for fermion loops,
we find for the Lorentz structure of the resolved vertices
\begin{eqnarray}
 \tilde \Gamma^\mu_{sc} &=&  -\text{Tr} \fourint{q}\,  
    \bar\chi^5 \left( \frac{\Sc p_2-p_3}{\Sc 2} \right) 
   \,S(p_2)\,\Gamma^\mu_q\,S(p_1)\, \chi^5 
  \left( \frac{\Sc p_1-p_3}{\Sc 2} \right)\,
    S^T(p_3)\, ,   \label{ressdqv} \\ 
 \tilde \Gamma^{\mu,\alpha\beta}_{ax} &= &
             -\text{Tr} \fourint{q}\,  
      \bar\chi^\alpha \left( \frac{\Sc p_2-p_3}{\Sc 2} \right)
      \,S(p_2)\,\Gamma^\mu_q\,S(p_1)\, 
            \chi^\beta \left( \frac{\Sc p_1-p_3}{\Sc 2}
        \right)\, S^T(p_3) \, ,\quad  \label{resadqv} \\
 \tilde \Gamma^{\mu,\beta}_{sa}  &=& -\text{Tr} \fourint{q}\, 
      \bar\chi^5 \left( \frac{\Sc p_2-p_3}{\Sc 2} \right)
      \,S(p_2)\,\Gamma^\mu_q\,S(p_1) \, 
           \chi^\beta \left( \frac{\Sc p_1-p_3}{\Sc 2} 
        \right)\, S^T(p_3)  \label{restv} \\
                      &=&
        4im_q \,\epsilon^{\mu\beta\rho\lambda}  (p_d+k_d)^\rho 
                Q^\lambda 
                      \fourint{q} \frac{g_s g_a \;V(q-Q/4)V(q+Q/4)}
                        {(p_1^2+m_q^2)(p_2^2+m_q^2)(p_3^2+m_q^2)} \;
    . \nonumber  
\end{eqnarray}
The quark momenta herein are,
\begin{eqnarray}
 p_1&=&\frac{p_d+k_d}{4}-\frac{Q}{2}+q\; ,\quad
 p_2=\frac{p_d+k_d}{4}+\frac{Q}{2}+q\;,\quad \nonumber \\
 p_3&=&\frac{p_d+k_d}{4}-q \;. 
\end{eqnarray}
Even though current conservation can be maintained with these vertices
on-shell, off-shell $\tilde \Gamma^\mu_{sc}$ and $\tilde
\Gamma^{\mu,\alpha\beta}_{ax}$ do not satisfy the Ward-Takahashi identities
for the free propagators in  
eqs.\ (\ref{Ds},\ref{Da}). Thus they cannot be directly 
employed to couple the photon to the diquarks inside the nucleon without
violating gauge invariance. 
For $Q=0$, however, they can be used to estimate the anomalous magnetic
moment $\kappa$ of the axialvector diquark and the strength of the
scalar-axialvector transition, denoted by $\kappa_{sa}$ in (\ref{sa_vert}),
as follows.  

First we calculate the contributions of the scalar and axialvector 
diquark to the proton charge, {\it i.e.} the contribution of the first  diagram in 
figure \ref{impulse}
to  $G_E (0)$, upon replacing the vertices $\Gamma^\mu_{sc}$ and
$\Gamma^{\mu,\alpha\beta}_{ax}$ given in 
eqs.~(\ref{vertsc},\ref{vertax}) 
by the resolved ones, $\tilde \Gamma^\mu_{sc}$ and $\tilde
\Gamma^{\mu,\alpha\beta}_{ax}$ in eqs.~(\ref{ressdqv}) and (\ref{resadqv}).
Since the {\em bare} vertices  satisfy the Ward-Takahashi
identities, and since current conservation is maintained in the calculation
of the electromagnetic form factors, the correct charges of both nucleons are
guaranteed to result from the contributions to $G_E (0)$ obtained with these
bare vertices, $\Gamma^\mu_{sc}$ and $\Gamma^{\mu,\alpha\beta}_{ax}$ of
eqs.~(\ref{vertsc},\ref{vertax}). In order to reproduce 
these correct contributions, we then adjust the values for the
diquark couplings, $g_s$ and $g_a$, to be used in connection with the
resolved vertices of eqs.~(\ref{ressdqv}) and (\ref{resadqv}). This yields
couplings $g_s^{\rm resc}$ and $g^{\rm resc}_a$, slightly rescaled (by a factor of
the order of one, {\em cf.} table~\ref{cc_1}).
We see that gauge invariance in form of the differential Ward identity
provides  an off-shell constraint on the diquark 
normalization where we have averaged over diquark off-shell
states by performing the four-dimensional integration for the
current matrix element.
Once the ``off-shell couplings'' $g_s^{\rm resc}$ and $g^{\rm resc}_a$ 
are fixed we can continue and calculate the contributions 
to the magnetic moment of the proton that arise from the resolved 
axialvector and transition couplings, 
$\tilde \Gamma^{\mu,\alpha\beta}_{ax}$ and
$\tilde \Gamma^{\mu,\beta}_{sa}$, respectively. 
These contributions determine
the values of the constants $\kappa$ and $\kappa_{sa}$ for the 
couplings in eqs.~(\ref{vertax}) and (\ref{sa_vert},\ref{as_vert}). 
The results are given in table \ref{cc_1}.
As can be seen, the values obtained for 
$\kappa$ and $\kappa_{sa}$ by this procedure are insensitive to the 
parameter sets for the nucleon amplitudes. 
In the calculations of observables we use $\kappa=1.0$ and $\kappa_{sa}=2.1$.

\begin{table}[t]
 \begin{center}
  \begin{tabular}{lllll} \hline \hline \\
    & $g_s^{\rm resc}/g_s$ & $g_a^{\rm resc}/g_a$ & $\kappa$ & $\kappa_{sa}$ \\ \hline
    & & & & \\
    Set I & 0.943 & 1.421  & 1.01 & 2.09\\
    Set II& 0.907  & 3.342  & 1.04 & 2.14 \\ \\ \hline \hline
  \end{tabular}
 \caption{
   Rescaled diquark normalizations 
   and constants of photon-diquark couplings.}
  \label{cc_1}
 \end{center}
\end{table}

\subsection{Pseudoscalar and pseudovector vertices}
\label{dqres2}

The pion and the pseudovector current do not couple to the scalar
diquark. Therefore, in both cases only those two contributions
have to be computed which are obtained from the middle and lower diagrams in
figure\ \ref{emresolve} with replacing the photon-quark vertex by
the pion-quark vertex of eq.~(\ref{vertpion}), and by the 
pseudovector-quark
vertex of eq.~(\ref{vertpv}), respectively. 

First we evaluate the flavor trace implicitly given in the diagrams
of figure \ref{emresolve}. As we employ only the third component
(in isospace)
of the pseudoscalar and the pseudovector currents to obtain
$g_{\pi NN}$ and $g_A$, {\em cf.} eqs. (\ref{gpNNtr},\ref{gatrace}),
we just need the following flavor factors,
\begin{eqnarray}
 (\tilde \Gamma^{\alpha\beta}_{5,ax})_{\rm flavor} &=&
   \half {\rm Tr}\; (\tau^2 \tau_m^\dagger) \tau^3 (\tau_{m} \tau^2)
   = \bp 1 \\ 0  \\ -1 \ep \; , \\
 (\tilde \Gamma^{\beta}_{5,sa})_{\rm flavor} &=&
   -\half {\rm Tr}\; \tau^2 \tau^3 (\tau_{m} \tau^2) =-\delta_{m0} \; .
\end{eqnarray} 
The flavor factors for the pseudovector vertices are 
half of the above factors.

In the quark loop calculations we use the rescaled couplings
from appendix \ref{dqres1} which proved to yield consistent results
with respect to the Ward identity for the nucleon.
For the Dirac part of the vertex
describing the pion coupling to the axialvector diquark we obtain,
\begin{eqnarray}
  \tilde \Gamma^{\alpha\beta}_{5,ax} &= -&2\frac{m_q^2}{f_\pi}
      \epsilon^{\alpha\beta\mu\nu} (p_d+k_d)^\mu Q^\nu \; \times
             \label{pionaxres} \\
                  & &       \fourint{q} \frac{(g_a^{\rm resc})^2\;V(q-Q/4)V(q+Q/4)}
             {(p_1^2+m_q^2)(p_2^2+m_q^2)(p_3^2+m_q^2)}  \; , \nonumber
\end{eqnarray}
and fixes its strength (at $Q^2=0$) to $\kappa_{ax}^5 \approx 4.5$, see
table~\ref{cc_2}.

For the effective pseudovector-axialvector diquark vertex in eq.\
(\ref{vert5muax}) it is sufficient to consider the regular part, since its
pion pole contribution is fully determined by eq.~(\ref{pionaxres}) already.
The regular part reads,
\begin{eqnarray}
  \tilde \Gamma^{\mu\alpha\beta}_{5,ax}& =&
            \fourint{q} \frac{(g_a^{\rm resc})^2\;V(q-Q/4)V(q+Q/4)}
                        {(p_1^2+m_q^2)(p_2^2+m_q^2)(p_3^2+m_q^2)} \;\times \\
        & & \left[ -4 m_q^2 \epsilon^{\mu\alpha\beta\nu}\;(p_1+p_2+p_3)^\nu
               - \text{Tr}\,\gamma_5 \gamma^\alpha \Slash{p}_2 \gamma^\mu
                          \Slash{p}_1 \gamma^\beta \Slash{p}_3 \right]. \nonumber
\end{eqnarray}
Although after the $q$-integration the terms in brackets
yield the four independent Lorentz structures discussed in the paragraph
above  eq.\ (\ref{vert5muax}), only the first term contributes to
$g_A(0)$ (with $p_1+p_2+p_3 = (3/4) (p_d+k_d) + q$).

The scalar-axialvector transition induced by the pion is described
by the vertex
\begin{eqnarray}
  \tilde \Gamma^{\beta}_{5,sa} & = & 4i\frac{m_q}{f_\pi}
   \fourint{q} g_s^{\rm resc} g_a^{\rm resc}\;V(q-Q/4)V(q+Q/4)\;\times \\
     & & \qquad \frac{(p_2\cdt p_3) p_1^\beta - (p_3\cdt p_1) p_2^\beta +
               (p_1\cdt p_2) p_3^\beta}
            {(p_1^2+m_q^2)(p_2^2+m_q^2)(p_3^2+m_q^2)},
         \nonumber
      \label{pionsares}
\end{eqnarray}
and the reverse (axialvector-scalar) transition is obtained by
substituting $Q \rightarrow -Q$ (or $p_1 \leftrightarrow p_2$) in
(\ref{pionsares}). The corresponding vertex for the pseudovector
current reads
\begin{eqnarray}
  \tilde \Gamma^{\mu\beta}_{5,sa} &= -4im_q&
     \fourint{q} \frac{g_s^{\rm resc} g_a^{\rm resc}\;V(q-Q/4)V(q+Q/4)}
     {(p_1^2+m_q^2)(p_2^2+m_q^2)(p_3^2+m_q^2)} \times \nonumber \\
     & &
    \left[ \delta^{\mu\beta}( m_q^2 - p_1\cdt p_2 -p_2\cdt p_3-p_3\cdt p_1)\;
     + \right. \label{pvsares} \\
    & & \left. \; \{p_1p_2\}_+^{\mu\beta} + \{p_1p_3\}_+^{\mu\beta}
     - \{p_2 p_3\}_-^{\mu\beta} \right]. \nonumber
\end{eqnarray}
The short-hand notation for a(n) (anti)symmetric product
used herein is defined as $ \{p_1  p_2\}_\pm^{\mu\nu}=p_1^\mu p_2^\nu \pm
p_1^\nu p_2^\mu$. The reverse transition is obtained
from $Q \rightarrow -Q$ together with an overall sign change
in (\ref{pvsares}). As already mentioned in the main text,
the term proportional to $\delta^{\mu\beta}$  provides  99 \%
of the value for $g_A$ as obtained with the full vertex. It
therefore clearly represents the dominant tensor structure.

\begin{table}
 \begin{center}
  \begin{tabular}{lllll}  \hline \hline \\
    &  $\kappa^5_{ax}$ & $\kappa^5_{\mu,ax}$ & $\kappa^5_{sa}$
    & $\kappa^5_{\mu,sa}$  \\ \hline
    & & & &  \\
 Set  I  & 4.53 & 4.41 & 3.97 & 1.97  \\
 Set  II & 4.55 & 4.47 & 3.84 & 2.13 \\ \\ \hline \hline
  \end{tabular}\\
  \caption{
          Strengths for
          pion- and pseudovector-diquark couplings.}
  \label{cc_2}
 \end{center}
\end{table}

As explained for the electromagnetic couplings of diquarks, we use these
resolved vertices in connection with the rescaled couplings $g_s^{\rm resc}$ and
$g_a^{\rm resc}$ to compute $g_{\pi NN}$ and $g_A$ in the limit $Q\rightarrow 0$.
In this way the otherwise unknown constants that occur in the
(pointlike) vertices of eqs.\ (\ref{vert5ax}--\ref{vert5musa}) are determined.

As seen from the results in table\ \ref{cc_2},
the values obtained for these effective coupling constants are only slightly
dependent on the parameter set (the only exception being
$\kappa^5_{\mu,sa}$ where the two values differ by 8 \%).
For the numerical calculations presented in section
\ref{num2} we employ
$\kappa^5_{ax}=4.5$, $\kappa^5_{\mu,ax}=4.4$, $\kappa^5_{sa}=3.9$ and
$\kappa^5_{\mu,sa}=2.1$.

\section{Calculation of the impulse approximation diagrams} 
\label{residue}

\begin{figure}
 \begin{center}
  \epsfig{file=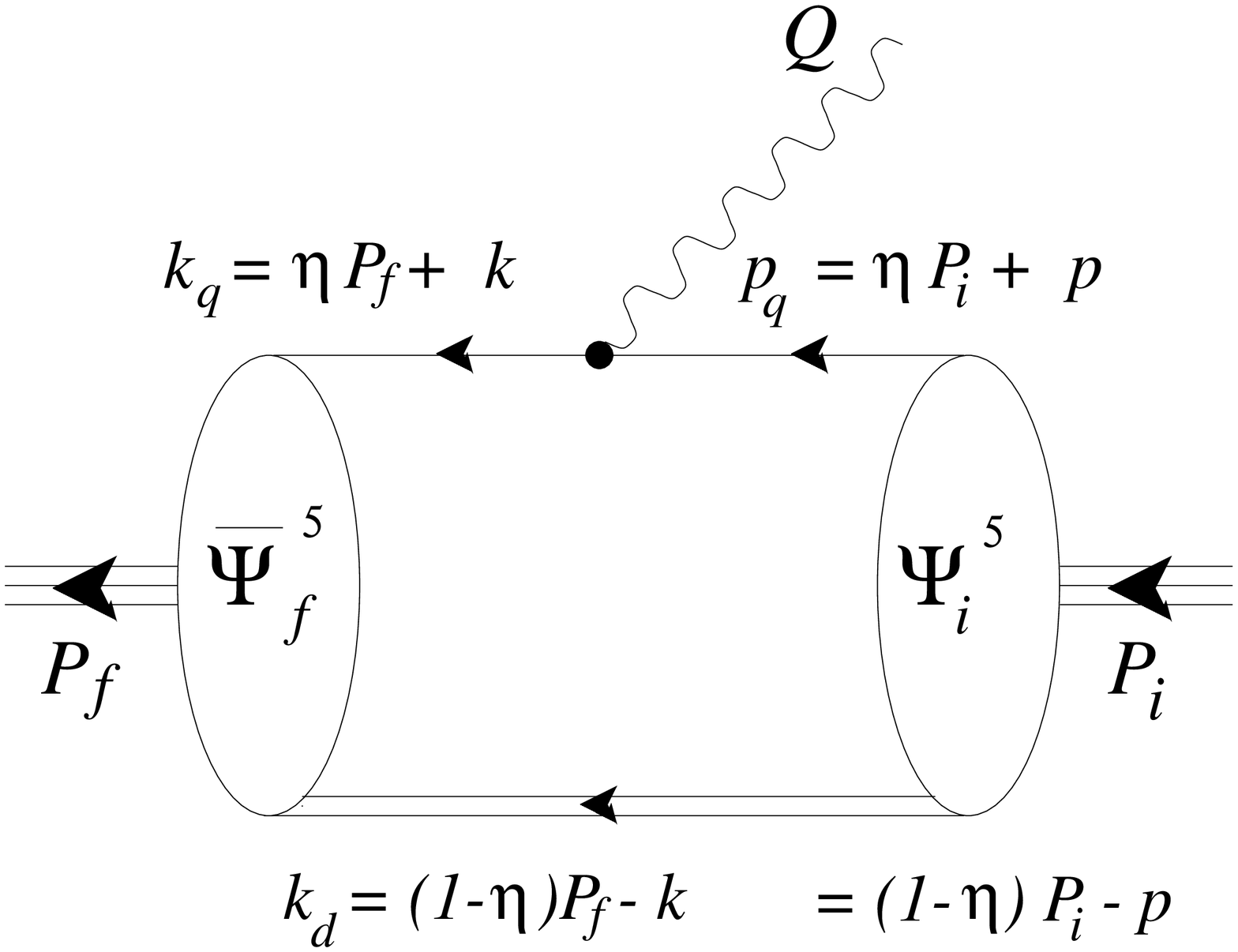,width=4.5cm}
 \end{center}
 \caption{The quark coupling diagram of the impulse approximation.}
 \label{imp-app}
\end{figure}

In this appendix we discuss the difficulties in the formal
transition from the Min\-kowski to the Euclidean metric. These are 
encountered
in the connection between Bethe-Salpeter wave function $\Psi$ and
vertex functions $\Phi$ in a general (boosted) frame of
reference of the nucleon bound state.

As the generic example for this discussion, we have chosen 
the second diagram 
in figure \ref{impulse} which describes the impulse-approximated 
contribution arising
from the coupling of the photon to the quark within the nucleon. 
Furthermore it is sufficient to restrict ourselves to
scalar diquark correlations, retaining only $\Psi^5$ and
$\Phi^5$. The principle of the calculation will nevertheless become clear.
We have singled out this diagram and, having added 
the appropriate momentum definitions, it is shown in figure \ref{imp-app}.
We evaluate the diagram in the Breit frame, {\em cf.} the
momentum definitions in eqs.~(\ref{BF_def}).

Let us consider  Mandelstam's formalism, eq.~(\ref{nuccurdef}),
in Minkowski space where it has been derived originally \cite{Mandelstam:1955}.
Here,
the matrix elements between bound states are 
related to the corresponding Bethe-Salpeter wave functions in Minkowski
space which for the moment are our 
nucleon Bethe-Salpeter wave functions $\Psi_M$.\footnote{In the
following the subscript $_M$ stands for definitions in Minkowski space and
$_E$ for the corresponding ones in Euclidean space.}  
Upon the transition to the Euclidean metric, the corresponding
contribution to the observable, here to the nucleon form factors, 
is determined by the ``Euclidean'' Bethe-Salpeter wave function $\Psi_E$. In the rest
frame of the nucleon bound state this transfer from $M \to E$ of the Bethe-Salpeter wave
functions commutes with the replacement of the wave by the 
vertex functions; that is, unique results are obtained from 
the Euclidean contributions based on either employing the Minkowski space wave
functions or the vertex functions which are related by the 
truncation of the 
propagators of the constituent legs, here
$\Phi_M=\left. (G_0^{\rm q-dq})^{-1}\right|_M \Psi_M$,  or, 
{\em vice versa},
$\Psi_M=\left. G_0^{\rm q-dq}\right|_M \Phi_M$. 

At finite momentum transfer $Q^2$ one needs to employ Bethe-Salpeter wave 
functions in a
more general frame of reference, here we use the Breit frame in which neither
the incoming nor the outgoing nucleon are at rest.
As described in section \ref{num_em}, the ``Euclidean wave function'' $\Psi_E$
in this frame is obtained from the solution to the Bethe-Salpeter equation in the rest frame by
analytic continuation, in particular, by inserting complex values for the
argument of the Chebyshev polynomials, see eqs.~(\ref{zi},\ref{zf}). 
This corresponds to the transition from left to right indicated by the arrow
of the upper line in figure \ref{box}.

\newcommand{\ampu}{\Phi_M=\left. (G_0^{\rm q-dq})^{-1}\right|_M \Psi_M}
\newcommand{\notampu}{\Phi_E \not =\left. (G_0^{\rm q-dq})^{-1}\right|_E
        \Psi_E}

\begin{figure}[t]
 \begin{center}
  \begin{equation*} \qquad
   \begin{CD}
     \int \bar\Psi_M \,\left. D^{-1}\Gamma^\mu_{q}\right|_M \,
          \Psi_M @>\text{Wick rotation}>>
     \int\bar\Psi_E \,\left. D^{-1}\Gamma^\mu_{q}\right|_E \, \Psi_E \\
     @V{ \begin{matrix} \; \\ \Sc \ampu \\ \; \end{matrix} }VV 
     @VV{ \begin{matrix} \; \\ \Sc \notampu \\ \; \end{matrix} }V \\
     \int \bar\Phi_M \,\left. S\Gamma^\mu_{q}G_0^{\rm q-dq} \right|_M \,
          \Phi_M @>\text{Wick rotation}>>
      \begin{array}[t]{c}
        \int \bar\Phi_E  \,\left. S\Gamma^\mu_{q}G_0^{\rm q-dq} \right|_E \,
          \Phi_E \\[2pt] 
       \hskip 1.5cm +\hskip .5cm \text{residue terms} 
      \end{array}
    \end{CD}
   \end{equation*}
  \end{center}
  \caption{
Interrelation of matrix elements in Minkowski and Euclidean space.
The integral sign is shorthand for the four-dimensional integration over
the relative momentum $p$, see equation (\ref{quark}).}
  \label{box}
\end{figure}

In the analogous transition on the other hand, when the truncated 
Bethe-Sal\-pe\-ter
amplitudes are employed, the possible presence of singularities in the
legs has to be taken into account explicitly. In the present
example, these are the single particle poles of the propagators of the 
constituent quark and diquark that might be encircled by the closed path in
the $p^0$-integration. The corresponding residues have to be included in the
transition to the Euclidean metric in this case, which is indicated in the
lower line of figure \ref{box}.  

The conclusion is therefore that the na{\"\i}ve relation between Bethe-Salpeter vertex and
wave functions cannot be maintained in the Chebyshev expansion of the 
Euclidean spherical momentum coordinates when singularities are encountered
in the truncation of the legs. Resorting to the Min\-kows\-ki space
definitions of vertex {\em vs.} wave functions, however, unique results are
obtained from either employing the domain of holomorphy of the Bethe-Salpeter wave
functions in the continuation to the Euclidean metric (with complex momenta)
or, alternatively and technically more involved, from keeping track of the
singularities that can occur in the Wick rotation when the truncated
amplitudes and explicit constituent propagators are employed. 

The rest of this section is concerned with the description of how to account for
these singularities which, for our present calculations are affected by 
constituent poles for quark and diquark, give rise to residue  
terms as indicated in the lower right corner of figure \ref{box}. 

To this end consider the quark contribution to the matrix elements of the
electromagnetic current which
is given by ({\em cf.} also eq.~(\ref{imp-example})) 
\begin{eqnarray} \label{quark}
 \langle J^\mu_{q} \rangle^{\rm sc-sc}  &=& 
  \int_M \!\frac{d^4 p}{(2\pi)^4}    
\bar\Phi^5(p+(1-\eta) Q,P_f)
 \; D(k_d)S(k_{q})\, \Gamma^\mu_{q}\, S(p_{q})\; \Phi^5(p,P_i) \; .
 \nonumber \\ &&
\end{eqnarray}

\begin{figure}
 \begin{center}
  \epsfig{file=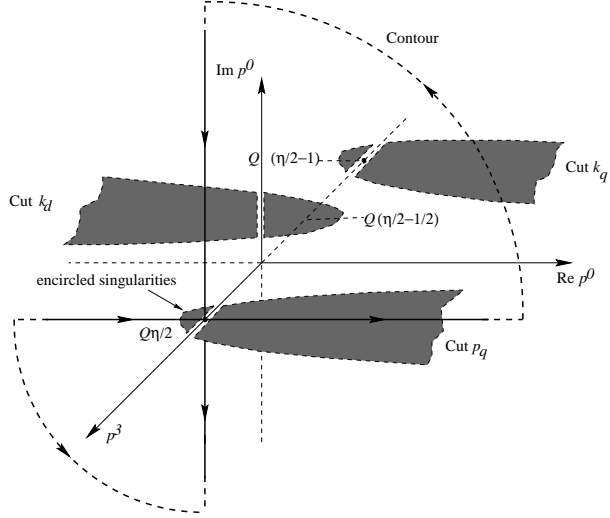,width=8cm}
 \end{center}
 \caption{Location of the relevant singularities in the impulse-approximate  
quark contribution to the form factors.
 The relative momentum $p$ is the integration variable in the loop diagram
corresponding to equation (\ref{quark}).}
\label{cuts}
\end{figure}

We are interested in the location of the propagator poles herein
(in Minkowski space). 
For these poles, solving the corresponding quadratic equation for the zeroth
component of the relative momentum $p^0$ yields
\begin{eqnarray}
 p_q^2-m_q^2-i\epsilon&=&0 \qquad  \Leftrightarrow   \\
 p^0_{{\rm pole},1} & =& 
               -\eta \, \omega_Q \, \pm \, W(m_{q},(\vect p -\eta/2 \,\vect
                                           Q)^2 ) 
                                         \nonumber\\
 k_q^2-m_q^2-i\epsilon&=&0 \qquad \Leftrightarrow    \label{k2} \\
 p^0_{{\rm pole},2} &=&  -\eta\,  \omega_Q
                      \, \pm\, W(m_{q}, (\vect p -(\eta/2-1) \vect Q)^2) 
   \nonumber\\
 k_d^2-m_{sc}^2-i\epsilon&=& 0 \qquad \Leftrightarrow   \label{k3}  \\
  p^0_{{\rm pole},3}  &  = &  
       (1-\eta)\, \omega_Q \, \pm\, W(m_{sc},(\vect p- 
   (\eta/2-1/2) \, \vect Q)^2) 
                   \nonumber 
\end{eqnarray}
with $W(m,\vect p^2) = \sqrt{\vect p^2+m^2-i\epsilon}$. We used
the momentum definitions given in 
figure \ref{imp-app} and in eqs.~(\ref{BF_def}).

For $Q=0$, {\it i.e.} in the rest frame of the nucleon in which the
Bethe-Salpeter equation was solved, the na{\"\i}ve Wick rotation is justified
for $1- \frac{m_{sc}}{M_n} < \eta < \frac{m_q}{M_n}$, since there is always a
finite gap between the cuts contained in the hypersurface Re$\,p^0=0$ of the
Re$\,p^0$ -- Im$\,p^0$ -- $\vect p$ space.
As $Q$ increases, these cuts are shifted along both,
the $p^3$ and the $p^0$-axis, as sketched in Figure \ref{cuts}. This 
eventually amounts to the effect that one of the two cuts arising from
each propagator crosses the Im$\,p^0$-axis. As indicated in the figure, 
the Wick rotation $p^0 \rightarrow ip^4$ is no longer possible for arbitrary 
values of $p^3$ without encircling singularities. The corresponding residues 
thus lead to 
\begin{eqnarray} 
  \langle J^\mu_{q} \rangle^{\rm sc-sc}  &\rightarrow&
  \int_E \!\frac{d^4 p}{(2\pi)^4}
 \bar\Phi^5(p+(1-\eta) Q,P_f)
 \; D(k_d)S(p_{q})\, \Gamma^\mu_{q}\, S(k_{q})\; \Phi^5(p,P_i) 
 \nonumber \\
 &&  +\; i \int\!\frac{d^3\vect p}{(2\pi)^3}
    \; \theta_{\vect p}
    \,\bar \Phi^5(p^4_{{\rm pole},1},\vect p +(1-\eta)\vect Q,P_f) \;
       D(k_d)S(k_{q})\; \times \nonumber \\
  & & \hskip 2.2cm \Gamma^\mu_{q}\, {\rm Res}(S(p_{q}))\; 
       \Phi^5(p^4_{{\rm pole},1},\vect p,P_i) \nonumber \\
 && + \quad\text{analogous terms for $S(k_{q})$ and $D(k_d)$} 
    \label{residueint}
\end{eqnarray}
upon transforming equation (\ref{quark}) to the Euclidean metric.
Here, the residue integral is evaluated
at the position of the pole in the incoming quark propagator $S(p_q)$
on the Euclidean $p^4$-axis
\begin{eqnarray}
 p_{{\rm pole},1}^4= -i\eta \sqrt{M_n^2+Q^2/4}+iW(m_q,(\vect p-\eta/2\, 
  \vect Q)^2) \, ,
\end{eqnarray}
where $\text{Res}(S(k_{q}))$ denotes the corresponding residue, and the
abbreviation 
\begin{eqnarray}
 \theta_{\vect p}  \equiv \theta\left(\eta \,\omega_Q -
 W(m_q,(\vect p- \eta/2 \, 
 \vect Q )^2)\right) \nonumber
\end{eqnarray}
was adopted to determine the integration domain for which the encircled
singularities of figure \ref{cuts} contribute. 

Analogous integrals over the spatial components of the relative momentum
$\vect p$ arise from the residues corresponding to the poles in the outgoing 
quark propagator $S(k_q)$ and the diquark propagator $D(k_d)$ as given in
eqs.~(\ref{k2},\ref{k3}).

One verifies that these cuts (as represented by the shaded areas
in Fig.~\ref{cuts}) never overlap. Pinching of the deformed contour does not
occur, since there are no anomalous thresholds for spacelike momentum
transfer $Q^2$ in these diagrams.

\begin{figure}
 \begin{center}
  \epsfig{file=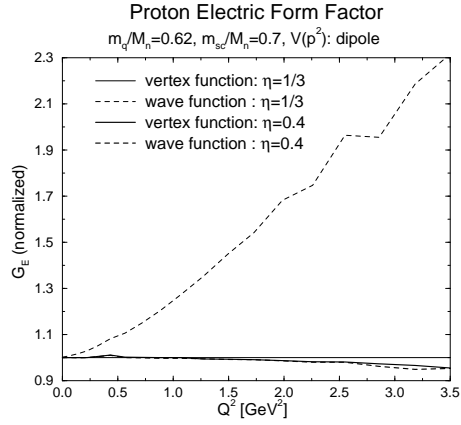,width=6cm}
 \end{center}
 \caption{The impulse-approximated contribution, corresponding to the
   two left diagrams
   of figure \ref{impulse}, to the electric form factor of the proton.
   We have employed the dipole-$V$ parameter set  
   from the scalar diquark sector,
   {\em cf.} table \ref{FWHM_table}.
   Results of the Bethe-Salpeter wave function 
   calculations for the $\eta$ values $1/3$ and $0.4$ are compared to the
   respective vertex function plus residue calculations. 
   The results are normalized to the latter that uses $\eta=1/3$.}
\label{vw}
\end{figure}

We have numerically checked the procedure described here using the parameter
set from the scalar diquark sector employing diquark-quark vertices 
of dipole shape, {\em cf.} table \ref{FWHM_table}.
We have calculated the contributions to the proton electric form factor
from the (quark and diquark) impulse approximation
diagrams for two different values of $\eta$, $\eta_1$ and $\eta_2$. 
Hereby $\eta_1=1/3$ is close to the left limit of the $\eta$ range
in which the (Euclidean) Bethe-Salpeter equation can be solved,
{\em cf.} eq.~(\ref{ebound1}). The second value, $\eta_2=0.4$,
lies safely in the middle of this range. For each $\eta$ value,
we calculated the form factor in Euclidean space
using ($i$) boosted wave functions $\Psi^5$ and ($ii$)
boosted vertex functions $\Phi^5$ along with the residue
prescription form eq.~(\ref{residueint}). The results are found
in figure \ref{vw}.
For $\eta_1$ the Chebyshev expansion of the Bethe-Salpeter wave function to 9 orders
still turns out insufficient to provide for stable numerical results. This is
due to being too close to the limit of the allowed range in $\eta$.
The considerably weaker suppression of higher orders in the Chebyshev
expansion of the wave function as compared to the expansion of the 
vertex function enhances the residual $\eta$-dependence of the observables
obtained from the former expansion at a given order, in particular, when it
has to reproduce close-by pole contributions in the constituent propagators.   
The impulse approximation contributions to $G_E$ deviate substantially from
those employing the vertex function and residue calculations in this case. 
On the other hand, 
for $\eta_2$ that has been employed to give  the other results of
figure \ref{vw}, unique results are obtained from both procedures. 
Both the Bethe-Salpeter wave function and vertex function calculation are in perfect
agreement for values of the momentum partitioning that are closer to
the middle of the range allowed to $\eta$. Furthermore the difference
between the vertex function calculations for $\eta_1$ and $\eta_2$ show
a good agreement and only deviate for larger momentum transfer ($Q^2 > 3$ GeV$^2$)
by a few per cent. Thus, the independence of the form factor on the
momentum distribution between quark and diquark holds
in actual numerical calculations.

 \end{appendix}

\setlength{\parskip}{-3mm}
\bibliographystyle{myunsrt}
\bibliography{cit1}

\end{document}